\documentclass[review]{elsarticle}

\pdfoutput=1
\usepackage{graphicx}
\usepackage{amsfonts, amsmath, amssymb}
\usepackage{booktabs,siunitx}
\usepackage[svgnames,table]{xcolor}
\usepackage[tableposition=above]{caption}
\usepackage{pifont}
\usepackage{bm}
\usepackage{cancel}
\usepackage{caption}
\usepackage{color}
\usepackage{enumerate}
\usepackage{float}
\usepackage{hyperref}
\usepackage[english]{babel}
\usepackage[margin=1in]{geometry}
\usepackage{mathtools}
\usepackage{multirow}
\usepackage{setspace}
\usepackage{subfigure}
\usepackage{lineno}
\renewcommand \d [2]{\frac{{\rm d} #1}{{\rm d} #2}}
\newcommand \D [2]{\frac{\partial #1}{\partial #2}}

\renewcommand{\vec}[1]{\bm{\mathrm{#1}}}
\newcommand{\V}[1]{\bm{\mathrm{#1}}}

\def \div{\nabla \cdot \mbox{}}
\def \grad{\nabla}

\def \x{\vec{x}}

\def \n{\vec{n}}

\def \u{\vec{u}}

\def \U{\vec{U}}
\def \L{\vec{L}}

\def \cM{\mathcal{M}}

\def \vcMG{\vec{\mathcal{M}}_{\text{G}}}

\def \Sb{S_\text{b}}

\def \Vb{V_\text{b}}

\def \C{\vec{C}}

\def \g{\vec{g}}

\def \Nx{N_x}
\def \Ny{N_y}
\def \Nz{N_z}

\def \Omegal{\Omega_{\text{l}}}
\def \Omegag{\Omega_{\text{g}}}

\def \U{\vec{U}}

\def \Ur{\U_{\text{r}}}
\def \W{\vec{W}}

\def \Wr{\W_{\text{r}}}

\def \X{\vec{X}}
\def \Xcom{\X_{\text{com}}}

\def \cH{\mathcal{H}}
\def \cT{\mathcal{T}}

\def \cO{\mathcal{O}}

\def \f{\vec{f}}

\def \fc{\f_{\text{c}}}

\def \half{\frac{1}{2}}
\def \3half{\frac{3}{2}}
\def \5half{\frac{5}{2}}

\def \n{\vec{n}}

\def \nref{n_{\text{ref}}}
\def \ncells{n_{\text{cells}}}
\def \ncycles{n_{\text{cycles}}}

\def \rhow{\rho_{\text{w}}}

\def \u{\vec{u}}
\def \ub{\u_{\text{b}}}

\def \x{\vec{x}}

\def \Re{\text{Re}}

\def \div{\nabla \cdot \mbox{}}
\def \grad{\nabla}

\def \dt{\Delta t}
\def \dx{\Delta x}
\def \dy{\Delta y}
\def \dz{\Delta z}

\def \dt{\Delta t}
\def \dx{\Delta x}

\newcommand{\upperRomannumeral}[1]{\uppercase\expandafter{\romannumeral#1}}

\begin{document}

\begin{frontmatter}
	
\title{The inertial sea wave energy converter (ISWEC) technology: device-physics, multiphase modeling and simulations}
\author[SDSU]{Kaustubh Khedkar}
\author[SDSU]{Nishant Nangia}
\author[SDSU]{Ramakrishnan Thirumalaisamy}
\author[SDSU]{Amneet Pal Singh Bhalla\corref{mycorrespondingauthor}}
\ead{asbhalla@sdsu.edu}

\address[SDSU]{Department of Mechanical Engineering, San Diego State University, San Diego, CA}
%\address[Northwestern]{Department of Engineering Sciences and Applied Mathematics, Northwestern University, Evanston, IL}
\cortext[mycorrespondingauthor]{Corresponding author}

\begin{abstract}
In this paper we investigate the dynamics of the inertial wave energy converter (ISWEC) device
using fully-resolved computational fluid dynamics (CFD) simulations. Originally prototyped by 
Polytechnic University of Turin, the device consists of a floating, boat-shaped hull
that is slack-moored to the sea bed. Internally, a gyroscopic power take off (PTO) unit converts
the wave-induced pitch motion of the hull into electrical energy. The CFD model is based on the incompressible
Navier-Stokes equations and utilizes the fictitious domain Brinkman penalization (FD/BP) technique to couple the device
physics and water wave dynamics.
A numerical wave tank is used to generate both regular waves based on fifth-order Stokes theory and irregular waves based on the JONSWAP spectrum to emulate realistic sea operating conditions.
A Froude scaling analysis is performed to enable two- and three-dimensional simulations for 
a scaled-down (1:20) ISWEC model. It is demonstrated that the scaled-down 2D model is sufficient to accurately simulate the
hull's pitching motion and to predict the power generation capability of the converter. 
A systematic parameter study of the ISWEC is conducted, and its optimal performance in terms of power 
generation is determined based on the hull and gyroscope control parameters. 
It is demonstrated that the device achieves peak performance when the gyroscope specifications are chosen based on 
the reactive control theory.
It is shown that a proportional control of the PTO control torque is required to generate continuous gyroscope precession 
effects, without which the device generates no power.
In an inertial reference frame, it is demonstrated that the yaw and pitch torques acting on the hull are of the same order of 
magnitude, informing future design investigations of the ISWEC technology.
Further, an energy transfer pathway from the water waves to the hull, the hull to the gyroscope, and
the gyroscope to the PTO unit is analytically described and numerically verified.
Additional parametric analysis demonstrates that a hull length to wavelength ratio between one-half and one-third yields
high conversion efficiency (ratio of power absorbed by the PTO unit to wave power per unit crest width).
Finally, device protection during inclement weather conditions is emulated by gradually reducing the gyroscope
flywheel speed to zero, and the resulting dynamics are investigated.
\end{abstract}

\begin{keyword}
\emph{renewable energy} \sep \emph{wave-structure interaction}  \sep \emph{Brinkman penalization method} \sep \emph{numerical wave tank}  \sep \emph{level set method}  \sep \emph{adaptive mesh refinement} 
\end{keyword}

\end{frontmatter}

%%%%%%%%%%%%%%%%%%%%%%%%%%%%%
\section{Introduction}
Ocean waves are a substantial source of renewable energy, with an estimated $2.11 \pm 0.05$ TW
available globally~\cite{Gunn12}. For perspective,  the United States generated $3.7$ TWy 
(terawatt years~\footnote{1 TWy = $8.76 \times 10^{12}$ kWh.}) worth of energy
in 2013, making up about 20\% of the world's total energy production. Of this amount, only about 9\% or 0.33 TWy was 
generated from renewable sources. It is estimated that the US will produce
approximately $8.65$ TWy by $2050$~\cite{Korde16}. There is an ever-increasing need to invest in renewable energy
harvesting techniques in order to accelerate economic growth while maintaining a safe and healthy planet Earth.
Wave energy conversion is one of the crucial strategies towards realizing future energy sustainability. 
It is estimated that about 230 TWh/year of wave energy can be extracted from the East Coast and about 590 TWh/year from the 
West Coast of the United States alone.  In spite of this abundantly available energy source, there is currently no
commercial-scale wave power operation that exists today. 

There are several unique challenges specific to wave energy extraction processes, including hostile ocean environments, 
saltwater corrosion, stochasticity of ocean and sea waves, and costly offshore wave farm setup. Nevertheless steady progress is 
being made both in the design and engineering analyses of wave energy extraction devices, which are known as wave energy converters (WECs). Consequently, several WEC designs have been proposed over the years after gaining popularity following the 1970s oil crisis. However unlike wind turbines, an ultimate WEC architecture has not yet been identified by researchers. 

\begin{figure}[h!]
 \centering
  \subfigure[]{
  \includegraphics[scale = 0.25]{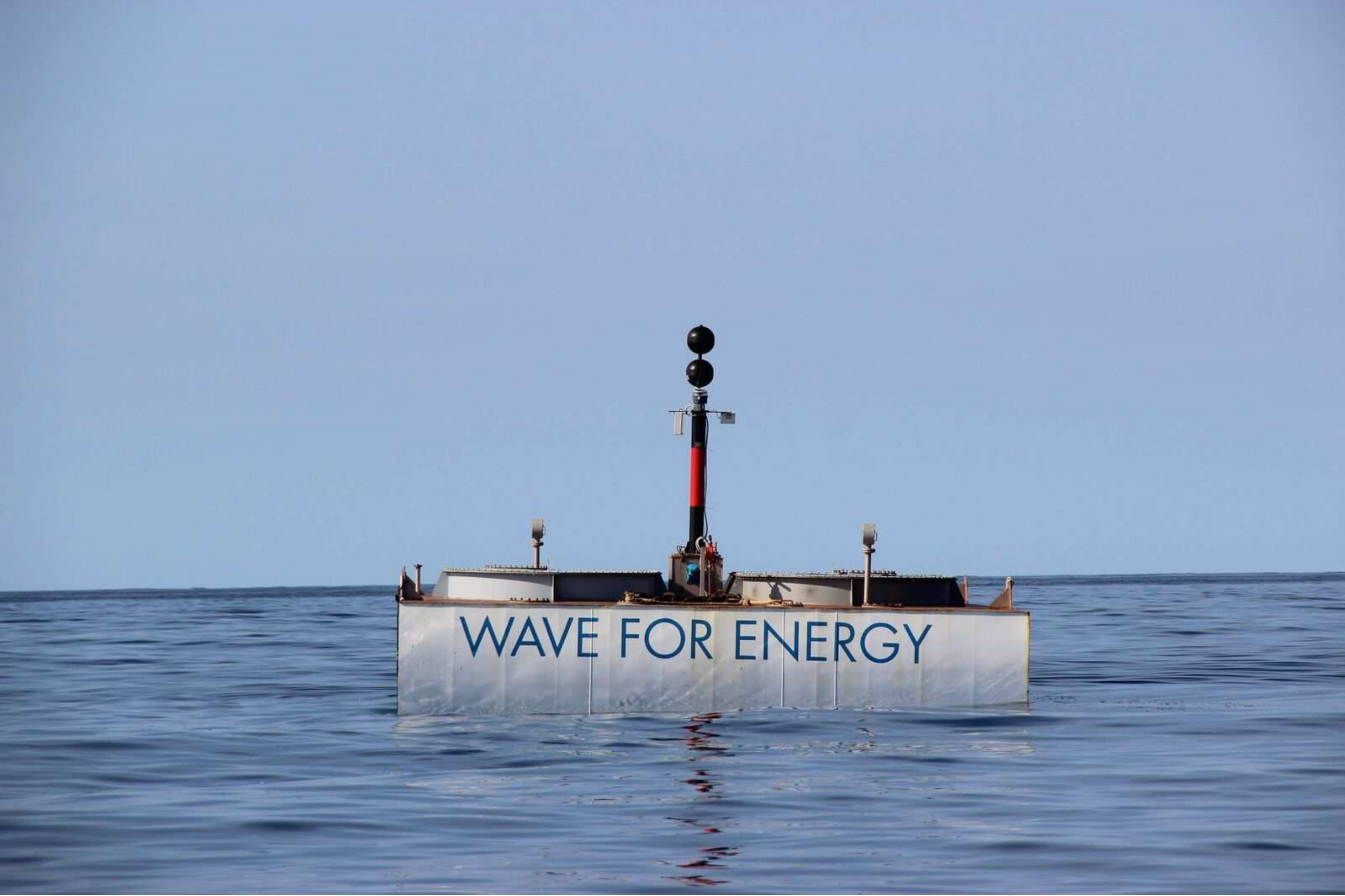}
  \label{fig_intro_iswec_front}
 }
    \subfigure[]{
  \includegraphics[scale = 0.29]{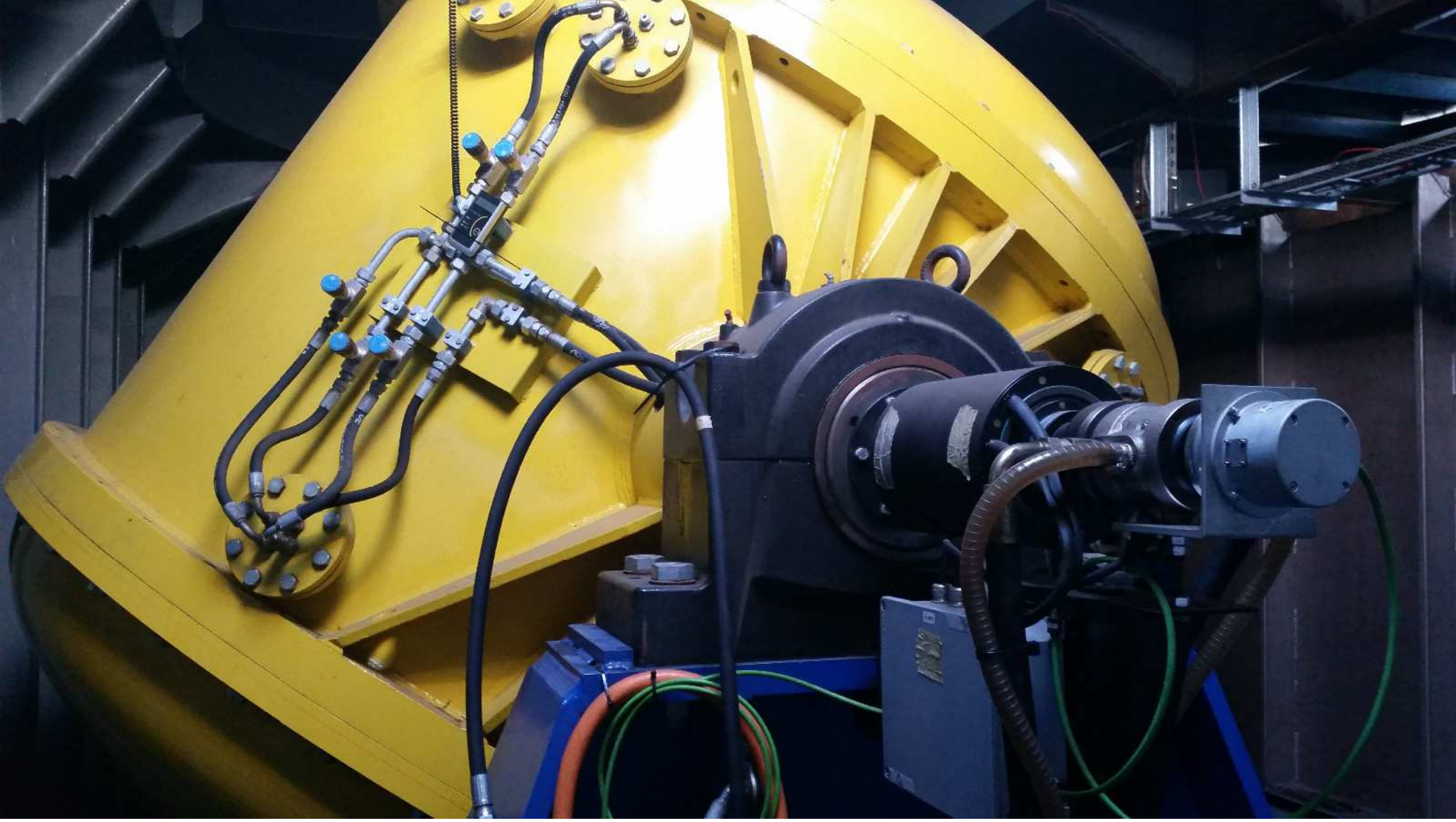}
  \label{fig_intro_yellow_casing}
 }
   \subfigure[]{
  \includegraphics[scale = 0.82]{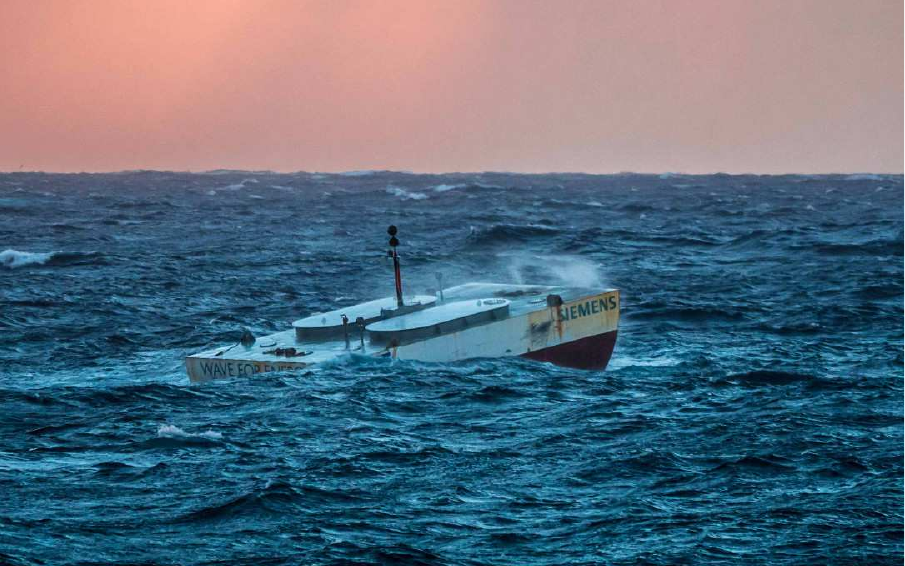}
  \label{fig_intro_iswec_pitching_front}
 }
  \subfigure[]{
  \includegraphics[scale = 0.55]{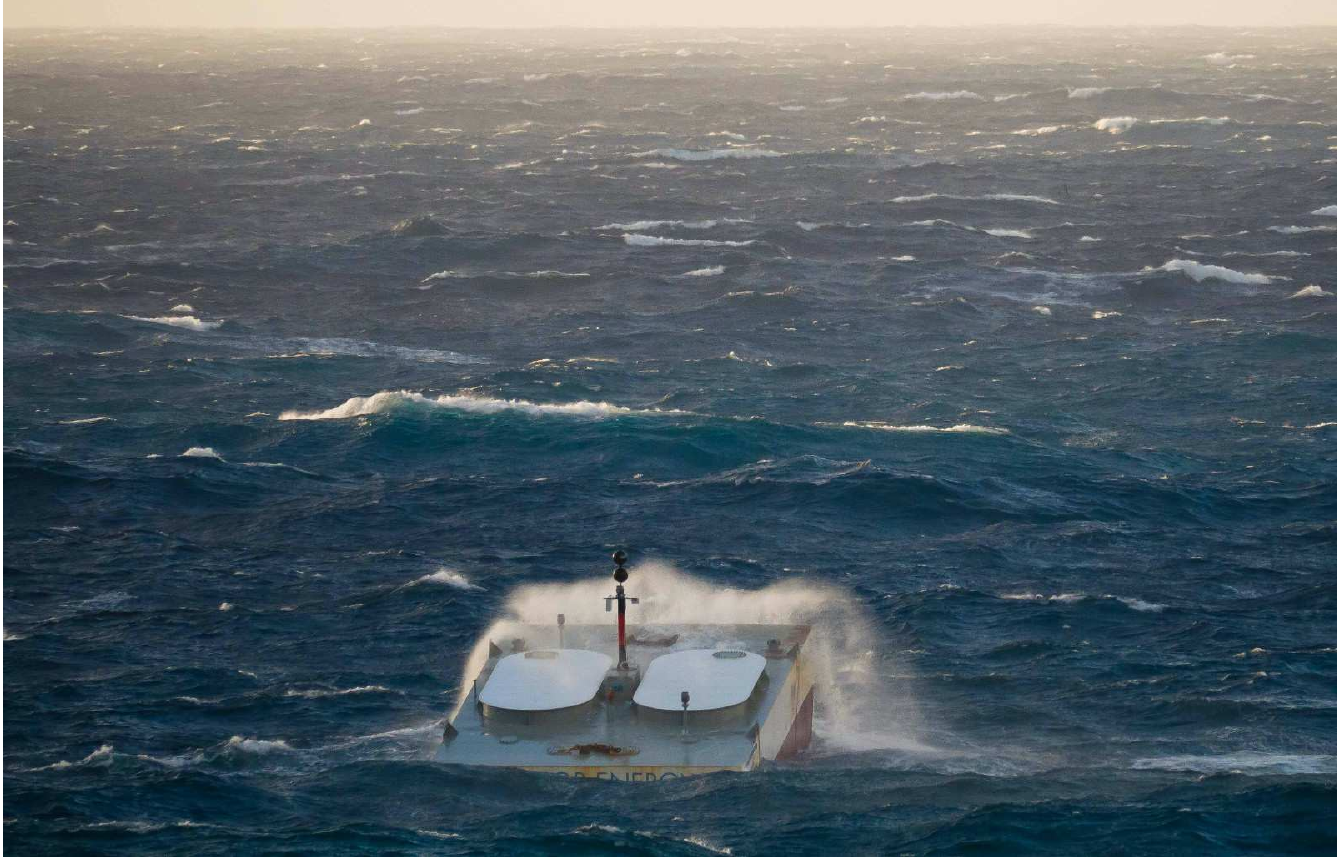}
  \label{fig_intro_iswec_pitching_side}
 }
 \caption{The inertial sea wave energy converter (ISWEC) device developed by the Mattiazzo Group at Polytechnic University of Turin. 
 \subref{fig_intro_iswec_front}  ISWEC device freely floating in relatively calm sea conditions.
 \subref{fig_intro_yellow_casing} Gyroscope casing mounted on the power take off (PTO) axis. The PTO system is housed inside the hull. 
 \subref{fig_intro_iswec_pitching_front} Front and \subref{fig_intro_iswec_pitching_side} side views of the ISWEC exhibiting 
 pitching motion during operation. Image courtesy of the Mattiazzo Group and Wave for Energy S.R.L., Turin.}
 \label{fig_intro_iswec}
\end{figure}

One WEC design that addresses some of the critical wave energy extraction challenges is the \emph{inertial sea wave energy converter} (ISWEC) device prototyped by Polytechnic University of Turin~\cite{Bracco2011, Cagninei2015,Vissio2017}. This device consists of a floating, boat-shaped hull that is slack-moored to the seabed, which internally houses a gyroscopic power take off unit (PTO); see Fig.~\ref{fig_intro_iswec}. The ISWEC can be classified as a pitching point-absorber whose dimensions are shorter than the length of the water waves. The device utilizes precession effects produced from the spinning gyroscope and pitching hull to drive a sealed electric generator/PTO. The rotational velocity of the spinning gyroscope and the PTO control torque act as sea-state tuning parameters that can be optimized/controlled (in real-time or via remote human-machine interfaces) to enhance the conversion efficiency of the device. Since all crucial electro-mechanical parts are sealed within the hull, the ISWEC is a robust and cost-effective wave energy conversion technology. Due to its simple design, devices can be produced by retrofitting abandoned ships, which can potentially reduce manufacturing costs and lead to easy adoption of the technology. Moreover, such devices could be lined up end-to-end just offshore, which would not only ensure maximal wave energy absorption but also protection of the coastline.

Although ISWEC devices have only recently been prototyped since their inception in 2011 by Bracco et al.~\cite{Bracco2011,Bracco2009,Bracco2010_ICOE,Bracco2010_ASME,Bracco2012}, their design and performance has been of much interest to the greater research community in the past few years. Medeiros and Brizzolara~\cite{Medeiros2018} used the boundary element method (BEM) based on linear potential flow equations to simulate the ISWEC and evaluate its power generation capabilities as a function of flywheel speed and derivative control of the PTO torque. They also demonstrated that the spinning gyroscopes can induce yaw torque on the hull. Faedo et al. used an alternative moment-matching-based approach to model the 
radiation force convolution integral, thereby overcoming the computational and representational drawbacks of simulating ISWEC devices using the BEM-based Cummins equation~\cite{Faedo2018}. Although these lower fidelity methods are able to simulate ISWEC dynamics at low computational costs, they are unable to resolve highly nonlinear phenomena often seen during practical operation such as wave-breaking and wave-overtopping. Unsurprisingly, the Turin group has extensively used carefully calibrated (with respect to wave tank experiments) BEM models to refine and optimize their preliminary designs~\cite{Bracco2015,Bracco2019,Raffero2015,Bracco2016}.  In contrast, simulations based on the incompressible Navier-Stokes (INS) equations are able to resolve the wave-structure interaction (WSI) quite accurately and without making small motion approximations employed by low-fidelity BEM models~\cite{Ruehl2014,Yu2013}. However, fully-resolved INS simulations are  computationally expensive and typically require high performance computing (HPC) frameworks. In a preliminary study, Bergmann et al. enabled fully-resolved simulation of the ISWEC's wave-structure interaction by making use of an INS-based flow solver coupled to an immersed boundary method~\cite{Bergmann2015}. The wave propagation in their channel followed the canonical ``dam-break" problem setup~\cite{Nangia2019WSI} --- a column of water is released from one end of the channel, which is then reflected from the opposite end, and so-forth. Although such simple wave propagation models are not suitable to study the device performance at a real site of operation, Bergmann et al. were nevertheless able to capture key device dynamics in their simulations.  
In addition to these research efforts, industry has become interested in piloting and manufacturing these devices. Recently, the 
multinational oil and gas corporation Eni installed an ISWEC device off the coast of Ravenna~\footnote{\url{https://www.eni.com/en-IT/operations/iswec-eni.html}} near their offshore assets. It is clear that there is a need to further investigate ISWEC dynamics and explore the design space to enable rapid adoption of this technology, possibly through an industry-academic partnership. 

In this work, we perform a comprehensive study of the ISWEC device using high fidelity simulations from
a previously developed fictitious domain Brinkman penalization (FD/BP) method based on the incompressible Navier-Stokes
equations~\cite{BhallaBP2019}. Although the methodology is similar to the work of Bergmann et al., we consider more 
realistic operating conditions by using a numerical wave tank (NWT) to generate both regular and irregular water waves.
We conduct a systematic variation of control parameters (i.e. PTO control torque, flywheel moment of inertia and speed, hull length)  
to determine the optimal performance of the device (in term of power generation) and study its dynamics as a function 
of these parameters. We also provide a theoretical basis to obtain the optimal control parameters for the device's design at a specific installation site. Moreover, we analytically describe an energy transfer pathway from water waves
to the hull, the hull to the gyroscope, and the gyroscope to the power take off (PTO) unit, and verify that it is numerically
satisfied by our simulations. A Froude scaling analysis is performed to reduce the computational cost of simulating a full-scale ISWEC device,
which is used to define the geometry and flow conditions for both two- and three-dimensional simulations of a scaled 
down 1:20 ISWEC device.  Additionally, we verify that the 2D ISWEC model produced similar dynamics to the 3D model, 
thereby allowing us to obtain accurate results at reduced simulation cycle times. We also simulate a possible device protection strategy during inclement weather conditions and study the resulting dynamics.

The rest of the paper is organized as follows. We first describe the dynamics, power generation, geometric properties, and
scaling analysis of the ISWEC device in Sec.~\ref{sec_iswec_eqs}. Next, we describe the numerical wave tank approach
used to generate both regular and irregular waves for our simulations in Sec.~\ref{sec_wave_eqs}. In Sec.~\ref{sec_wsi_eqs},
we describe the continuous and discrete equations for the multiphase wave-structure interaction system, and outline/validate 
the solution methodology for the FD/BP technique. In Sec.~\ref{sec_software}, we briefly describe the software 
implementation and computing hardware utilized in this study.
In Sec.~\ref{sec_spatiotemporal_tests}, we perform spatial and temporal
resolution tests to select a grid spacing and time step size that ensures adequate resolution of ISWEC dynamics.
Finally in Sec.~\ref{sec_results_and_discussion}, we conduct a systematic parameter study on the various hull and 
gyroscope parameters and evaluate the device performance in terms of generated power.

%%%%%%%%%%%%%%%%%%%%%%%%%%%%%%%%%
\section{ISWEC dynamics}\label{sec_iswec_eqs}
In this section, we mathematically describe the dynamics, power generation, and geometric properties of the ISWEC device.
%%%%%%%%%%%%%%%%%%%%%%%%%%%%%%%%%%%%%%%%%%
\subsection{ISWEC dynamics} \label{sec_iswec_dynamics}

\begin{figure}[]
  \centering
  \subfigure[] {
 \includegraphics[scale = 0.3]{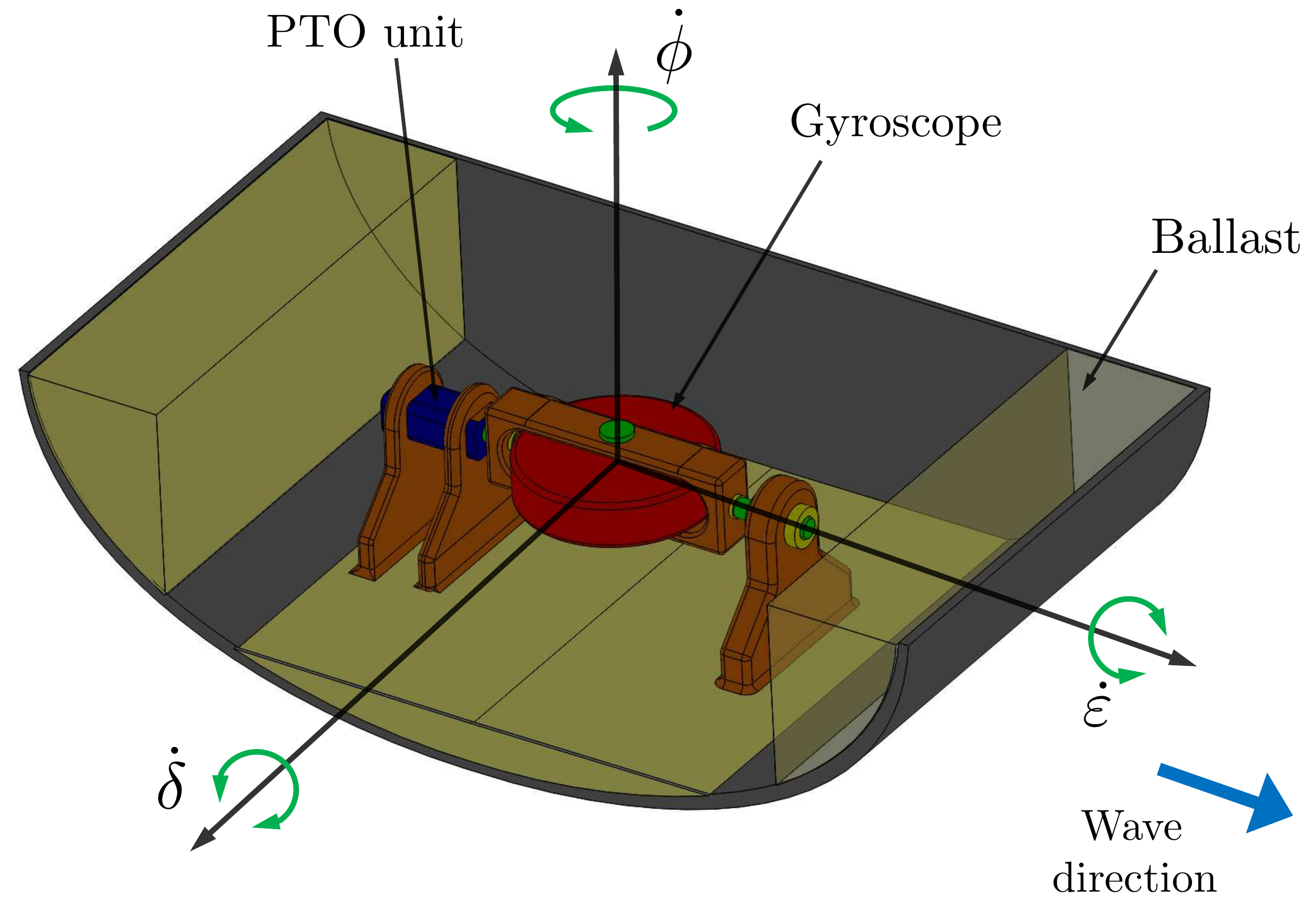}
 \label{fig_ISWEC3D_CAD_model}
 }
   \subfigure[] {
 \includegraphics[scale = 0.3]{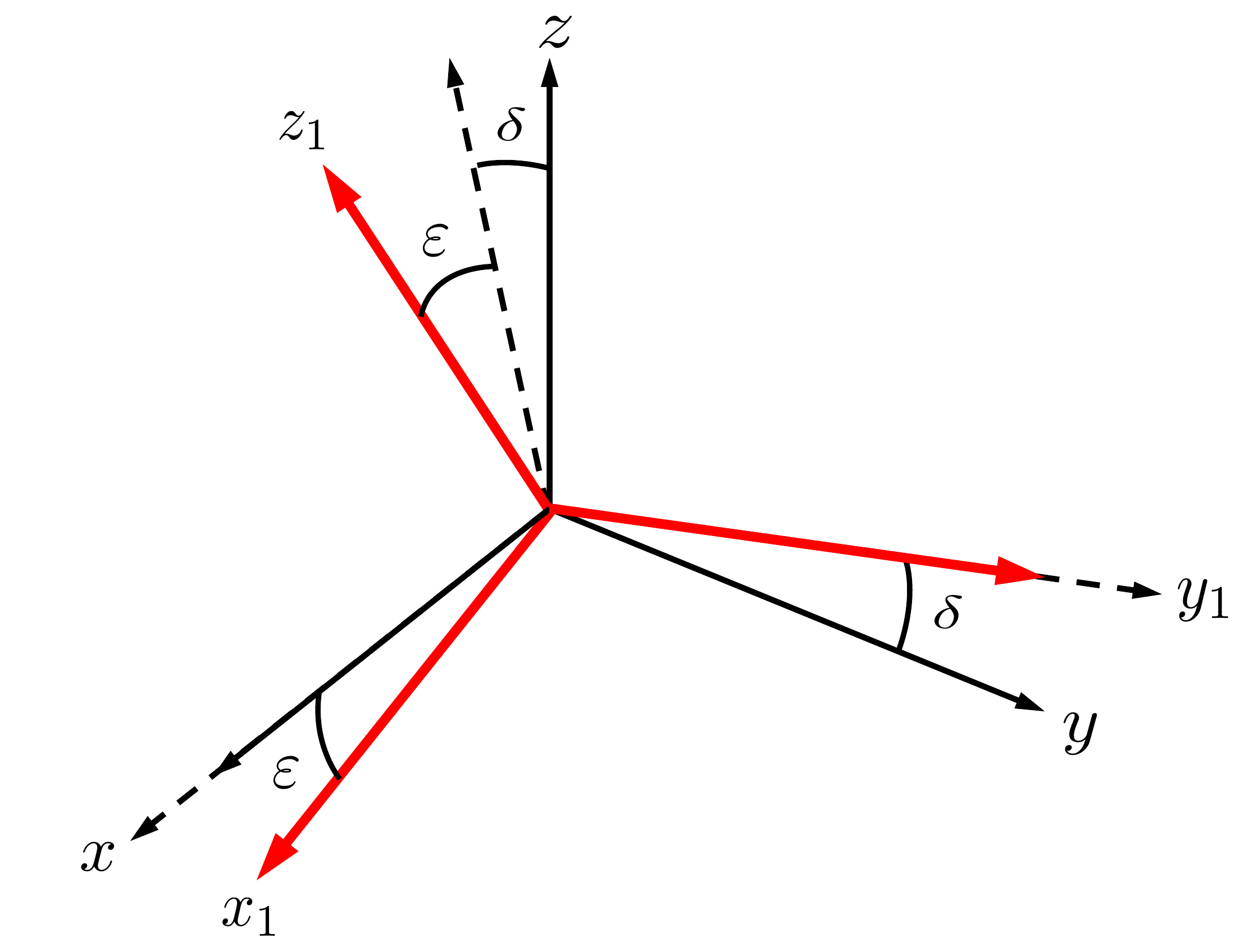}
 \label{fig_reference_frames}
 }
 \caption{
\subref{fig_ISWEC3D_CAD_model} ISWEC device schematic and the main rotational velocities of the system: hull's pitch velocity 
 $\dot{\delta}$, gyroscope's angular velocity $\dot{\phi}$, and the induced precession velocity of the PTO shaft  
 $\dot{\varepsilon}$.~\subref{fig_reference_frames} Hull and gyroscope reference frames.
 }
 \label{fig_schematic_frames}
\end{figure}

Externally, the ISWEC device appears as a monolithic hull that is slack-moored to the seabed. Internally, the device 
houses a spinning gyroscopic system that drives a sealed electric generator. The pitching velocity 
of the hull is mainly responsible for converting the wave motion into electrical output. To simplify the model and
discussion, the other remaining degrees of freedom of the hull are not considered in this study; 
see Appendix~\ref{sec_appendix} for a comparison of one and two degrees of freedom ISWEC models.
As the device operates, the combination of wave induced pitching velocity $\dot{\delta}$ 
and spinning gyroscope/flywheel velocity $\dot{\phi}$ induces a precession torque in the $\varepsilon$ coordinate direction. The wave energy conversion is made possible by damping the motion along the $\varepsilon$-direction by the electric generator, which is commonly referred to as the power take-off (PTO) unit.  Fig.~\ref{fig_ISWEC3D_CAD_model} shows the schematic of the ISWEC device, including the external hull, ballast, gyroscope, and PTO unit.

To derive the three-way coupling between the waves, hull, and gyroscopic system we consider an inertial reference frame 
$xyz$ attached to the hull and a rotating non-inertial reference frame $x_{1}y_{1}z_{1}$ attached to the gyroscope as shown in Fig.~\ref{fig_reference_frames}. The gyroscope reference frame is obtained from the hull reference frame by two subsequent finite rotations $\delta$ and $\varepsilon$. The origin of both reference frames is taken to be the center of gravity of the device.

In the absence of waves, $\delta = 0$ and $\varepsilon = 0$, and the flywheel rotates with a constant angular velocity $\dot{\phi}$ along the vertical $z_{1}$-axis. This configuration is taken to be the initial position of the device, in which the two reference frames also coincide. When the first wave reaches the hull location, it tilts the device by an angle $\delta$ and the hull attains a pitching velocity $\dot{\delta}$ along the $x$-axis. The gyroscope structure rotates by the same angle $\delta$
about the $x$- (or the $x_1$-) axis. The rotated configuration of the $x_{1}y_{1}z_{1}$ reference frame is shown by dashed lines in Fig.~\ref{fig_reference_frames}. As the hull begins to pitch, the gyroscope is subject to two angular velocities:
$\dot{\delta}$ along $x_1$-axis and $\dot{\phi}$ along $z_1$-axis. This velocity combination produces a precession torque 
in the third orthogonal direction $y_1$. This induced torque precesses the gyroscope by an angle $\varepsilon$ about the $y_1$-axis. As a result of the two subsequent rotations, the gyroscope frame attains an orientation shown by bold red lines in  Fig.~\ref{fig_reference_frames}. 

The evolution of the gyroscope's dynamics results in a gyroscopic torque $\vcMG = (\cM_{x_1}, \cM_{y_1}, \cM_{z_1})$, which 
can be related to the rotational kinematic variables using conservation of angular momentum. The angular velocity $\V{\Omega}_1$ of the gyroscope reference frame and the angular velocity $\V{\Omega}_{\text{G}}$ of the gyroscope are both written in the $x_{1}y_{1}z_{1}$ coordinate system and their evolution can be expressed in terms of $\delta$, $\varepsilon$, and $\dot{\phi}$ as
\begin{align}
\V{\Omega}_{1} &= \dot{\delta}\cos{\varepsilon} \,\hat{i}_{1} + \dot{\varepsilon} \, \hat{j}_{1} + \dot{\delta}\sin{\varepsilon} \, \hat{k}_{1}, \label{eq_omega1} \\
\V{\Omega}_{\text{G}} &= \dot{\delta}\cos{\varepsilon} \, \hat{i}_{1} + \dot{\varepsilon} \, \hat{j}_{1} + (\dot{\delta}\sin{\varepsilon}+\dot{\phi}) \, \hat{k}_{1}, \label{eq_omegaG}
\end{align}
in which $\hat{i}_{1}$, $\hat{j}_{1}$, and $\hat{k}_{1}$ are the unit vectors along $x_{1}$-, $y_{1}$-, and $z_{1}$-directions, respectively. The rate of change of the gyroscope's angular momentum with respect to time is related to the gyroscopic torque $\vcMG$ by
\begin{align}
\vcMG &= \d{\V{H}_{\text{G}}}{t},          \label{eqn_dHdt}  
\end{align}
in which $\V{H}_{\text{G}} = \mathbb{I}_{\text{G}} \V{\Omega}_{\text{G}}$ is the angular momentum of the gyroscope and $\mathbb{I}_{\text{G}}$ is the inertia matrix of the gyroscope. In the $x_{1}y_{1}z_{1}$ reference frame, $ \mathbb{I}_{\text{G}}$ reads as
\begin{equation}
 \mathbb{I}_{\text{G}}	= \begin{bmatrix}
   			    		I_{x_{1}x_{1}} & 0 & 0 \\ 
   			    		0 & I_{y_{1}y_{1}} & 0 \\ 
   			    		0 & 0 & I_{z_{1}z_{1}} 
    			   		\end{bmatrix}         \approx 
   									\begin{bmatrix}
   									I & 0 & 0 \\ 
   									0 & I & 0 \\ 
   									0 & 0 & J 
						        \end{bmatrix}.  \label{eq_flywheel_MI}
\end{equation}
The flywheel structure, including its support brackets etc., is typically designed such that $I_{x_{1}x_{1}} \approx I_{y_{1}y_{1}} = I$ and  $I_{z_{1}z_{1}} = J \gtrapprox I$. Using Eqs.~\eqref{eq_omegaG} and~\eqref{eq_flywheel_MI}, the angular momentum of the flywheel is given by
\begin{equation}
\V{H}_{\text{G}} = I \dot{\delta} \cos{\varepsilon} \, \hat{i}_{1} + I \dot{\varepsilon} \, \hat{j}_{1} + J (\dot{\delta} \sin{\varepsilon} + \dot{\phi}) \, \hat{k}_{1}. \label{eq_HG}
\end{equation}
Differentiating Eq.~\eqref{eq_HG} with respect to time in the inertial reference frame involves computing time derivatives of the unit vectors $\hat{i}_{1}$, $\hat{j}_{1}$, and $\hat{k}_{1}$:
\begin{align}
\label{eq_versor_time_derivative}
\d{\hat{i}_{1}}{t} &= \V{\Omega}_1 \, \times \, \hat{i}_{1} = -\dot{\varepsilon} \, \hat{k}_{1} + \dot{\delta}\sin{\varepsilon} \, \hat{j}_{1}, \\
\d{\hat{j}_{1}}{t} &= \V{\Omega}_1 \, \times \, \hat{j}_{1} = \dot{\delta}\cos{\varepsilon} \, \hat{k}_{1} - \dot{\delta}\sin{\varepsilon} \, \hat{i}_{1},  \\
\d{\hat{k}_{1}}{t} &= \V{\Omega}_1 \, \times \, \hat{k}_{1} = -\dot{\delta}\cos{\varepsilon} \, \hat{j}_{1} + \dot{\varepsilon} \, \hat{i}_{1}.
\end{align}
Finally after some algebraic simplification, a component-wise expression for the gyroscopic torque $\vcMG$ is obtained 
\begin{equation}
\vcMG  =  \begin{bmatrix}
		  \cM_{x_1} \\ 
		  \cM_{y_1} \\ 
		  \cM_{z_1}
		  \end{bmatrix}  =  \begin{bmatrix}
       I \ddot{\delta} \cos{\varepsilon} + \left(J- 2I \right) \dot{\delta} \dot{\varepsilon} \sin{\varepsilon} + J \dot{\phi} \dot{\varepsilon} \\
       I \ddot{\varepsilon} + (I - J) \dot{\delta}^{2} \sin{\varepsilon} \cos{\varepsilon} - J \dot{\phi} \dot{\delta} \cos{\varepsilon}  \\
      J \ddot{\delta} \sin{\varepsilon} + J \dot{\delta} \dot{\varepsilon} \cos{\varepsilon} + J\ddot{\phi}
      \end{bmatrix}.     
\end{equation}      

The  precession velocity $\dot{\varepsilon}$ of the generator shaft is damped using a proportional derivative (PD) control law implemented in the PTO unit. The PD control torque can be modeled as a spring-damper system with the following form
\begin{equation}
\cM_{\varepsilon} = \vcMG \cdot \hat{j}_1 = - k \varepsilon - c \dot{\varepsilon}.
 \label{eq_Me_balance}
\end{equation}
Here, $k$ is a spring-like stiffness parameter and $c$ is a damper-like dissipation parameter that can be adjusted in real-time (usually through feedback) to enhance the conversion efficiency of the device when the incoming waves change their characteristics. The wave power absorbed by the PTO unit (as a function of time) is 
\begin{equation}
P_{\text{PTO}} = c \dot{\varepsilon}^2.
\label{eq_PTO_power}
\end{equation}
Therefore, the precession component of the gyroscopic torque is balanced by the PD control torque, $\cM_{y_1} = \cM_{\varepsilon}$, which is also responsible for extracting the wave energy. The other components $\cM_{x_1}$ and $\cM_{z_1}$ of the gyroscopic torque are balanced/sustained by the hydrodynamic torques acting on the hull and the subsequent hull-gyroscope interactions. To understand this balance, we consider the hydrodynamic torque and motion of the hull about the pitch ($x$-direction) as observed from the inertial reference frame $xyz$
\begin{equation}
\cM_{\text{hydro}} = I_{\text{H}} \d{\dot{\delta}}{t} + \cM_{\delta}, \label{eq_MHydroGyro}
\end{equation} 
in which $\cM_{\text{hydro}}$ is the hydrodynamic torque acting on the hull, $I_{\text{H}}$ is the moment of inertia 
of the hull, and $\cM_{\delta}$ is the projection of the gyroscopic torque on the $x$-axis: 
\begin{align}
\cM_\delta &= \vcMG \cdot \hat{i} \nonumber \\
  	     &= \vcMG \cdot ( \hat{i}_1 \cos{\varepsilon} + \hat{k}_1 \sin{\varepsilon}) \nonumber \\
  	     &= (J \sin{^2\varepsilon} + I \cos{^2\varepsilon} ) \ddot{\delta} + J \dot{\phi} \dot{\varepsilon} \cos{\varepsilon} 
	     + 2 (J - I) \dot{\delta} \dot{\varepsilon} \sin{\varepsilon} \cos{\varepsilon} + J \ddot{\phi} \sin{\varepsilon}. 
  \label{eq_M_delta}
\end{align}
From Eq.~\eqref{eq_MHydroGyro} it can be seen that the gyroscopic reaction $\cM_\delta$ acting on the hull opposes the wave induced pitching motion. Similarly, a second reaction torque $\cM_\phi$ acts on the hull along 
the $z$-direction and opposes its wave induced yaw motion:
 \begin{align}
 \cM_\phi &= \vcMG \cdot \hat{k} \nonumber \\
                &= \vcMG \cdot \left[ (\hat{k}_1 \cos{\varepsilon} - \hat{i}_1 \sin{\varepsilon})\cos{\delta} + \hat{j}_1 \sin{\delta} \right] \nonumber \\
                 &= \left[(J - I) \ddot{\delta} \sin{\varepsilon} \cos{\varepsilon} + \dot{\delta} \dot{\varepsilon} [ J ( \cos^2{\varepsilon} - \sin^2{\varepsilon} ) + 2 I \sin^2{\varepsilon} ] - J \dot{\phi} \dot{\varepsilon} \sin{\varepsilon} + J \ddot{\phi} \cos{\varepsilon}\right]\cos{\delta} \nonumber \\
                 &+ \left[  I \ddot{\varepsilon} + (I - J) \dot{\delta}^{2} \sin{\varepsilon} \cos{\varepsilon} - J \dot{\phi} \dot{\delta} \cos{\varepsilon} \right]\sin{\delta}
  \label{eq_M_phi}
\end{align}
In Sec.~\ref{subsec_c_variation}, we show that this yaw torque is the same order of magnitude as the pitch torque $\cM_\delta$. In practice, however, its contribution is partially cancelled out by the mooring system of the device. 
Moreover, using an even number of gyroscopic units will cancel the yaw component of the gyroscopic torque acting on the hull if each flywheel pair spins with equal and opposite velocity, as described by Raffero~\cite{Raffero2014}. Therefore, we do not consider the effect of 
$\cM_\phi$ on the ISWEC dynamics in our (3D) model. 

%%%%%%%%%%%%%%%%%%%%%%%%%%%%%%%%%%%%%%%%%%%%%%

\subsection{Power transfer from waves to PTO} \label{sec_power_flow}

To understand the power transfer from waves to the hull and from the hull to the PTO unit, we derive the time-averaged kinetic energy equations of the ISWEC system. These equations highlight the coupled terms that are responsible for 
wave energy conversion through the ISWEC device.

First, we consider the rotation of the gyroscope around the PTO axis. The equation of motion in the $\varepsilon$ coordinate direction, as derived in the previous section is re-written below 
\begin{equation}
I \ddot{\varepsilon} + (I - J) \dot{\delta}^{2} \sin{\varepsilon} \cos{\varepsilon} - J \dot{\phi} \dot{\delta} \cos{\varepsilon} = \cM_{\varepsilon} = -k \varepsilon -c \dot{\varepsilon}. \label{eq_gyro_PTO}
\end{equation}    
Rearranging Eq.~\eqref{eq_gyro_PTO} with the approximation $I \approx J$ simplifies the equation to
 \begin{equation}
I \ddot{\varepsilon} + c \dot{\varepsilon} + k\varepsilon = J \dot{\phi} \dot{\delta} \cos{\varepsilon}. \label{eq_gyro_PTO_simple} 
\end{equation}
From the above equation, it can be seen that the product of the hull's pitch velocity $\dot{\delta}$ and the gyroscope's angular velocity $\dot{\phi}$ yields a forcing term that drives the PTO motion. Multiplying Eq.~\eqref{eq_gyro_PTO_simple} by the precession velocity $\dot{\varepsilon}$ and rearranging some terms, we obtain the kinetic energy equation for the PTO dynamics
\begin{equation}
I{\frac{\rm d}{ \rm dt }\left(\frac{\dot{\varepsilon}^{2}}{2}\right)}  + c{\dot{\varepsilon}^{2}} + k{\frac{ \rm d}{\rm dt }\left(\frac{{\varepsilon}^{2}}{2}\right)} = J \dot{\phi} \dot{\delta}  \dot{\varepsilon} \cos{\varepsilon}. \label{eq_PTO_KE}
\end{equation}
Taking the time-average of Eq.~\eqref{eq_PTO_KE} over one wave period, the first and third terms on the left hand side of the equation evaluate to zero. The remaining terms describe the transfer of power from the 
hull to the PTO unit:
 \begin{equation}
\langle{c \dot{\varepsilon}^{2} \rangle} = \langle{ J \dot{\phi} \dot{\delta} \dot{\varepsilon} \cos{\varepsilon} \rangle},\label{eq_hull_to_pto}
\end{equation}
in which $\langle \cdot \rangle = \frac{1}{\cT} \int_t^{t+\cT}(.) \, \rm dt$ represents the time-average of a quantity over one wave period $\cT$~\footnote{For irregular waves the time-average could be defined over one significant wave period.}. Here, $ \langle{c{\dot{\varepsilon}^{2}}\rangle} $ is the average power absorbed by the PTO, denoted $\bar{P}_{\text{PTO}}$, and $ \langle{J{\dot{\phi}}{\dot{\delta}}{\dot{\varepsilon}}\cos{\varepsilon}\rangle} $ is the average power generated due to the gyroscopic motion through its interaction with the hull, denoted by $\bar{P}_{\text{gyro}}$. Similarly, the kinetic energy equation of the hull can be derived by multiplying hull dynamics Eq.~\eqref{eq_MHydroGyro} by the pitch velocity $\dot{\delta}$
\begin{equation}
\cM_{\text{hydro}} \dot{\delta} = I_{\text{H}} \ddot{\delta} \dot{\delta} + \cM_{\delta} \dot{\delta}. \label{eq_MHydroGyro_deltadot}
\end{equation} 
Under the assumptions $I \approx J$ and a constant gyroscope spinning speed, $\cM_\delta$ in Eq.~\eqref{eq_M_delta} simplifies to
 \begin{equation}
\cM_\delta = J\ddot{\delta} + J \dot{\phi} \dot{\varepsilon} \cos{\varepsilon}. \label{eq_M_delta_simplified}
\end{equation}
Using Eqs.~\eqref{eq_MHydroGyro_deltadot} and~\eqref{eq_M_delta_simplified}, and rearranging some terms, we obtain
 \begin{equation}
\cM_{\text{hydro}} \dot{\delta} = I_{\text{H}} \frac{ \rm d}{\rm dt} \left(\frac{\dot{\delta}^2}{2}\right) + J \frac{\rm d}{\rm dt} \left(\frac{\dot{\delta}^2}{2}\right) + J \dot{\phi} \dot{\delta} \dot{\varepsilon} \cos{\varepsilon}. \label{eq_hull_KE}
\end{equation}
Taking the time-average of Eq.~\eqref{eq_hull_KE} over one wave period, the first and second terms on the right hand side evaluate to zero, and the expression reads
 \begin{equation}
\langle{\cM_{\text{hydro}}\dot{\delta}\rangle} = \langle{J \dot{\phi} \dot{\delta} \dot{\varepsilon} \cos{\varepsilon}\rangle}. \label{eq_waves_to_hull}
\end{equation}
Here, $ \langle{\cM_{\text{hydro}}\dot{\delta}\rangle}$ is the power transferred from the waves to the hull, denoted $\bar{P}_{\text{hull}}$, and $\langle{J\dot{\phi}\dot{\delta}\dot{\varepsilon}\cos{\varepsilon}\rangle}$ is the same expression on the right side of Eq.~\ref{eq_hull_to_pto}. Hence, combining Eqs.~\eqref{eq_hull_to_pto} and~\eqref{eq_waves_to_hull}, we obtain an equation describing the pathway of energy transfer from waves to the PTO:
 \begin{equation}
\underbrace{ \langle{\cM_{\text{hydro}} \dot{\delta} \rangle} }_{\text{waves} \rightarrow \text{hull}} = \underbrace{ \langle{J \dot{\phi} \dot{\delta} \dot{\varepsilon} \cos{\varepsilon} \rangle} }_{\text{hull} \rightarrow \text{gyroscope}}  = \underbrace{ \langle{c \dot{\varepsilon}^{2} \rangle} }_{\text{gyroscope} \rightarrow \text{PTO}}, \label{eq_pathway}
\end{equation}
which is written succinctly as $ \bar{P}_{\text{hull}} = \bar{P}_{\text{gyro}} = \bar{P}_{\text{PTO}}$. Eq.~\eqref{eq_pathway} is quantitatively verified for the ISWEC model under both regular and irregular wave environments in Sec.~\ref{sec_results_and_discussion}.

%%%%%%%%%%%%%%%%%%%%%%%%%%%%%%%%%%%%%%%%%%%%%%

\subsection{PTO and gyroscope parameters} \label{sec_PTO_params}

The energy transfer equation can be used to select the PTO and gyroscope parameters that achieve 
a rated power of the installed device $\bar{P}_{\text{R}}$. From Eq.~\eqref{eq_pathway} 
\begin{equation}
  c = \frac{ \bar{P}_{\text{R}}}{ \langle{\dot{\varepsilon}^2 \rangle} } =  \frac{2 \bar{P}_{\text{R}}}{\dot{\varepsilon}_0^2}, \label{eq_c}
\end{equation}
in which $\dot{\varepsilon}_0$ is the amplitude of the precession velocity, expressed in terms of the amplitude of the precession angle $\varepsilon_0$ as $\dot{\varepsilon}_0 =  \varepsilon_0 \omega$. Here, we assume that all of the ISWEC components are excited at the external wave frequency $\omega = 2\pi/\cT$ to achieve optimal performance. Based on experimental data of real ISWEC devices~\cite{Vissio2017,Cagninei2015}, we prescribe $\varepsilon_0$ in the range $ 40^{\circ} \le \varepsilon_0 \le 80^{\circ}$ to obtain the damping parameter $c$ from Eq.~\eqref{eq_c}. To prescribe the rest of the gyroscope parameters, we make use of Eq.~\ref{eq_hull_to_pto}. Since this expression is nonlinear, we linearize it about $\varepsilon=0^{\circ}$ (a reasonable approximation for relatively calm conditions) to estimate the gyroscope angular momentum as
\begin{equation}
	J \dot{\phi} = \frac{c\varepsilon_0}{\delta_0},  \label{eq_Jphi}
\end{equation}
in which the amplitude of the hull pitch velocity $\dot{\delta}_0 = \delta_0  \omega$, expressed in terms of the amplitude of the hull pitch angle 
$\delta_0$, is used. Again based on the experimental data, we prescribe $\delta_0$ in the range $ 2^{\circ} \le \delta_0 \le 20^{\circ}$, and $\dot{\phi}$ in the range $250 \le \dot{\phi} \le 1000$ RPM~\footnote{This range of $\dot{\phi}$ is for the full-scale ISWEC device, which can be scaled by an appropriate factor for the scaled-down model. See Sec.~\ref{sec_Fr_scaling} for scaling analyses.} to obtain the $J$ value of the gyroscope from Eq.~\eqref{eq_Jphi}. The $I$ value of the gyroscope is set as a scaled value of $J$, i.e. $I = \gamma J$ where $\gamma \le 1$. We study the effect of varying $\gamma$ in Sec.~\ref{subsec_J_variation}.

The only remaining free parameter is the PTO stiffness $k$ used in the control torque. We make use of reactive control theory~\cite{Korde16} and prescribe $k$ as 
\begin{equation}
	 k = \omega^2 I, \label{eq_k}
\end{equation}
ensuring that the gyroscopic system oscillates at the wave frequency around the PTO axis. Using the linearized version of Eq.~\eqref{eq_gyro_PTO_simple}, it can be shown that if the gyroscope oscillates with the external wave forcing frequency, a resonance condition is achieved along the PTO axis and the output power is maximized~\cite{Korde16}. In this state, both the hull and gyroscopic systems oscillate at the external wave frequency and their coupling is maximized.

%%%%%%%%%%%%%%%%%%%%%%%%%%%%%%%%%%%%%%%%%%

\subsection{Hull shape} \label{sec_hull_shape}

\begin{figure}[h!]
 \centering
  \subfigure[Idealized ISWEC hull geometry]{
  \includegraphics[scale = 0.31]{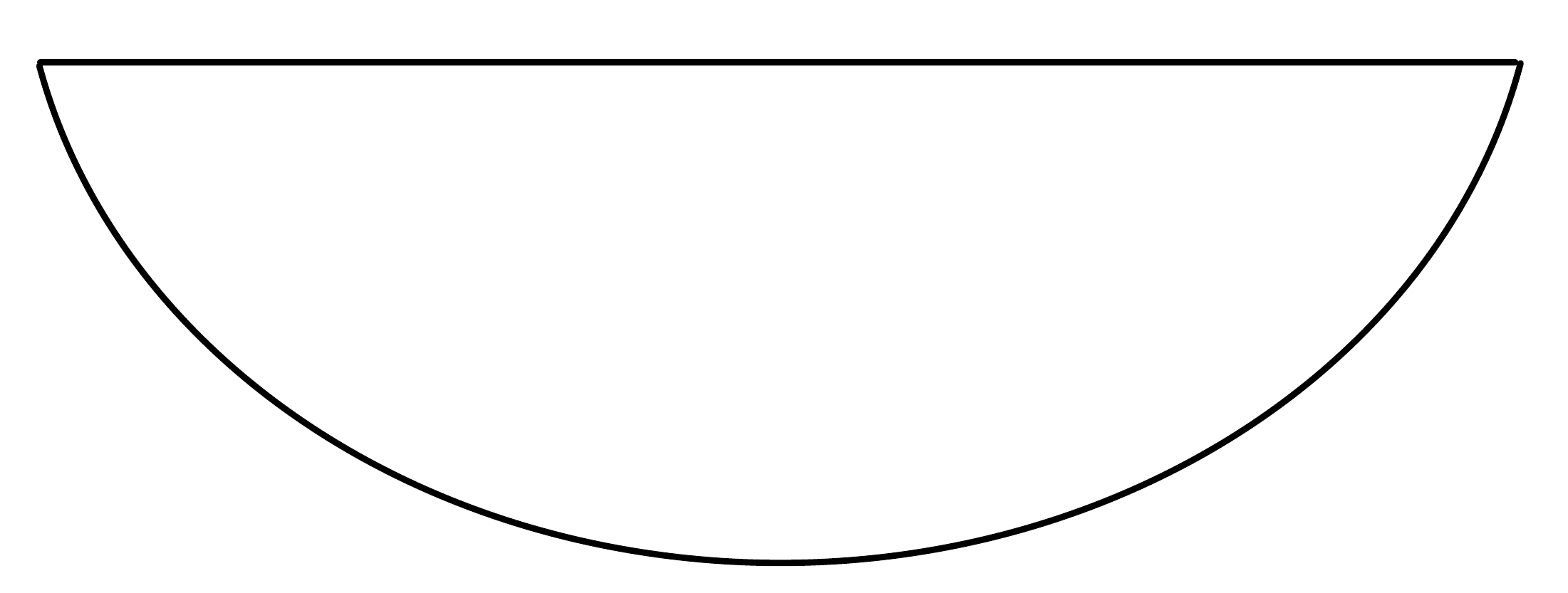}
  \label{fig_iswec_hull_geometry}
 }
   \subfigure[Experimental ISWEC hull geometry]{
  \includegraphics[scale = 0.315]{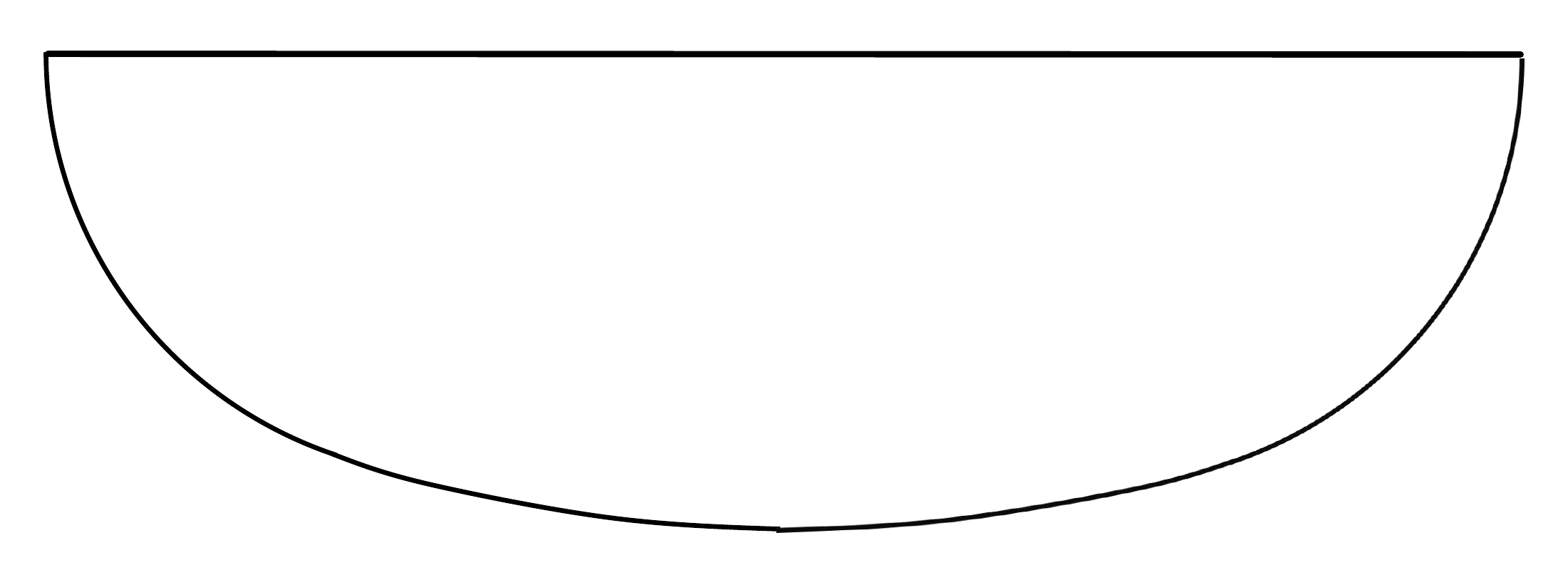}
  \label{fig_exp_hull_geometry}
 }
 \caption{ISWEC hull shapes. 
 }
 \label{fig_hull_geometry}
\end{figure}

The ISWEC's external hull is a boat-shaped vessel, which we idealize by a half-cylinder of length $L$, height $H$, and width $W$. For the actual device, a part of the outer periphery is flattened out to ease the installation of mechanical and electrical parts near the bottom-center location (see Fig.~\ref{fig_exp_hull_geometry}). We neglect these geometric details in our model shown in Fig.~\ref{fig_iswec_hull_geometry}. The inside of the device is mostly hollow and the buoyancy-countering ballast is placed around the outer periphery.   

The hull length $L$ is a function of $\lambda$, the wavelength of the ``design" wave at device installation site. As analyzed by Cagninei et al.~\cite{Cagninei2015}, the optimal hull length is between $ \lambda/ 3 \leq L \leq \lambda/2$ for an ISWEC device that is mainly excited in the pitch direction. The hull width W is decided based on the rated power of the installed device $\bar{P}_{\text{R}}$, relative capture width (RCW) of the device (or the device conversion efficiency) $\eta$, and wave power per unit crest width $\bar{P}_{\text{wave}}$. These quantities are related through the expressions
\begin{equation}
W = \frac{\bar{P}_{\text{R}}} { \eta \cdot \bar{P}_{\text{wave}} } \quad \quad \text{and} \quad \quad \eta = \bar{P}_{\text{PTO}}/\bar{P}_{\text{wave}},
  \label{eqn_W_eta}
\end{equation}
in which $\bar{P}$ denotes time-averaged power. Sec.~\ref{sec_wave_eqs} provides closed-form expressions of $\bar{P}_{\text{wave}}$ for both regular and irregular waves. For the 2D ISWEC model we use $W = 1$, which corresponds to a unit crest width of the wave.

%%%%%%%%%%%%%%%%%%%%%%%%%%%%%%%%%%%%%%%%%%%%%

\subsection{Scaled ISWEC model} \label{sec_Fr_scaling}

In order to reduce the computational cost of simulating a full-scale ISWEC device operating in high Reynolds number $(\Re)$ regimes, we use Froude scaling~\cite{book_offshore_hydromechanics} to scale the model problem down by a 1:20 ratio. 
The Froude number (Fr) measures the resistance of a partially submerged object moving through water and is defined as 
\begin{equation}
\text{Fr} = \frac{\text {characteristic velocity}}{\text{gravitational wave velocity}} = \frac{U_\text{c}}{\sqrt{gL_\text{c}}}, \label{eq_Fr}
\end{equation}
in which $U_\text{c}$ is the characteristic velocity, $L_\text{c}$ is the characteristic length, and $g$ is the gravitational acceleration constant.  In offshore 
marine hydromechanics, Froude scaling allows us to compare the dynamics of two vessels even if their sizes are different (since they produce a similar wake). Two vessels having the same Froude number may not be operating in the same Reynolds number regime. In the present study, the scaled-down model operates in lower $\Re$ conditions and it does not capture fine scale details such as extreme wave breaking and spray dynamics that occur at higher Reynolds numbers. These small scale features are mostly dictated by viscous and surface tension effects, and a very fine computational mesh is needed to adequately resolve them. However, the main quantities of interest such as power generated for the full-scale device can be inferred from scaled-down simulations by using appropriate scaling factors, some of which we derive next. 

\begin{itemize}

\item Length scaling: The geometric parameters such as length, width or height are simply scaled by a factor of $\alpha$.
In the present study, we use $\alpha = 20$. An exception to this length scaling is hull width in 2D, which is taken to be unity in the scaled model. Therefore in 2D, the scaling factor for hull width is $W$ rather than $\alpha$. 

\item Acceleration scaling: The full-scale and scaled-down models operate under the same gravitational force field. Therefore, the gravitational constant $g$ (or any other acceleration) remains unchanged. 

 \item Density scaling: Density is an intrinsic material property, and thus it remains the same for both the full-scale and scaled-down models.
 
 \item Volume scaling: Since volume is proportional to the length cubed, it is scaled by $\alpha^3$.
 
 \item Mass scaling: Mass can be expressed as a product of density $\rho$ and volume, and its scaling for 2D and 3D ISWEC models are obtained as
 \begin{equation}
\frac{M_{\textrm{model}}}{M_{\textrm{full-scale}}} = \frac{ \rho \left(L \times H \times W\right)\big|_{\textrm{model}} }{ \rho \left(L \times H \times W\right)\big|_{\textrm{full-scale}} } = \frac{W_{\textrm{model}}}{\alpha^{2} \cdot W_{\textrm{full-scale}}}.  \label{eq_Fr_scale_mass}
\end{equation}

\item Velocity scaling: Velocity scaling is obtained by equating the Froude numbers 
\begin{align}
\frac{U_\text{c}}{\sqrt{gL_\text{c}}} \Bigg|_{\text{model}} &=  \frac{U_\text{c}}{\sqrt{gL_\text{c}}} \Bigg|_{\textrm{full-scale}} \\
\Rightarrow \frac{ U_{\text{c},{\textrm{model}}} }{U_{\text{c},{\textrm{full-scale}}} } &= \sqrt{\frac{L_{\text{c},{\textrm{model}}} }{L_{\text{c},{\textrm{full-scale}}}} } = 1/\alpha^{\half}. \label{eq_Fr_scale_vel} 
\end{align}

\item Time scaling: Letting $t_\text{c}$ represent a characteristic time, time scaling can be obtained from the length and velocity scalings as
\begin{align}
\frac{U_{\text{c},\textrm{model}}}{U_{\text{c},\textrm{full-scale}}} &= \frac{L_\text{c}/t_\text{c}\Big|_{\textrm{model}}}{L_\text{c}/t_\text{c}\Big|_{\textrm{full-scale}}} \\
\Rightarrow \frac{t_{\text{c},\textrm{model}}}{t_{\text{c},\textrm{full-scale}}} &= 1/\alpha^{\half}.  \label{eq_Fr_scale_time}
\end{align}

\end{itemize}
Similarly, scaling factors of other quantities of interest such as force and power can be obtained in terms of $\alpha$, and are enumerted in Table~\ref{tab_Fr_scaling} for both two and three spatial dimensions. Full-scale (scaled-down) quantities should be divided (multiplied) by factors in the third and fourth columns to obtain the scaled-down (full-scale) quantities, in three and two spatial dimensions, respectively.

\begin{table}[]
 \centering
 \caption{Froude scaling of various quantities for the 3D and 2D ISWEC models. Dimensional units for the quantities used in this work are shown in column 2.}
 \rowcolors{2}{}{gray!10}
 \begin{tabular}{*6c}
 \toprule
 Quantity & Units & Scaled 3D model & Scaled 2D model \\
 \midrule
 Length  & m & $\alpha$  & $\alpha$  \\
 Area  & m$^2$ & $\alpha^2$ & $\alpha^2$  \\
 Volume  & m$^3$ & $\alpha^3$ & $-$    \\
 Time  & s & $\alpha^{\half}$ & $\alpha^{\half}$ \\
 Velocity  & m/s & $\alpha^{\half}$ & $\alpha^{\half}$ \\
 Acceleration  & m/s$^2$ & $1$ & $1$ \\
 Frequency & s$^{-1}$ & $\alpha^{-\half}$ & $\alpha^{-\half}$ \\
 Angular velocity& s$^{-1}$ & $\alpha^{-\half}$ & $\alpha^{-\half}$ \\
 Mass   & kg  & $\alpha^{3}$ & $\alpha^{2}\cdot W$ \\
 Density  & kg/m$^3$  & $1$ & $1$ \\
 Force  & kg $\cdot$ m/s$^2$ & $\alpha^{3}$ & $\alpha^{2}\cdot W$\\
 Moment of inertia & kg $\cdot$ m$^2$ & $\alpha^{5}$ & $\alpha^{4}\cdot W$ \\
 Torque  & kg $\cdot$ m$^2$/s$^{2}$ & $\alpha^{4}$ & $\alpha^{3}\cdot W$\\
 Power &  kg $\cdot$ m$^2$/s$^{3}$ & $\alpha^{\frac{7}{2}}$ & $\alpha^{\frac{5}{2}}\cdot W$ \\
 \bottomrule
 \end{tabular}
 \label{tab_Fr_scaling}
\end{table}

%%%%%%%%%%%%%%%%%%%%%%%%%%%%%%%%%%%%%%%%%%%%%

\subsection{Scaled hull parameters} \label{sec_hull_params}

In this section, we use the Froude scaling derived in the previous section to derive the scaled-down hull parameters required 
for our simulations. The scaled-down parameters of the gyroscope will be presented in Sec.~\ref{sec_results_and_discussion}, where they are systematically varied to study their effect on device performance. The hull properties of the full-scale ISWEC device are taken from an experimental campaign~\cite{Cagninei2015,Vissio2017} conducted at the Pantelleria test site in the Mediterranean Sea. 
%We also verify the scaled-down hull parameters through ``back-of-the-envelope" calculations by assuming a half-cylinder geometry of the hull. 

\begin{table}[]
 \centering
 \caption{ISWEC hull full-scale and scaled-down parameters. Freeboard (FB) is the distance between the hull top surface and the still waterline, which is found experimentally.}
 \rowcolors{2}{}{gray!10}
 \begin{tabular}{*6c}
 \toprule
 Hull property & Notation & Units & Full-scale & Scaled-down 3D model & Scaled-down 2D model \\
 \midrule
 Length & $L$ & m & $15.33$ & $0.7665$ & $0.7665$ \\
 Height & $H$ & m & $4.5$ & $0.225$ & $0.225$ \\
 Width & $W$ &  m & $8$  & $0.4$ & $1$  \\
 Freeboard & FB &   m & $1.52$ & $0.076$ & $0.076$ \\
 Center of gravity &  $Z_{\text{CG}}$ & m & $0.57$ & $0.0285$ & $0.0285$ \\
 Mass & $M_\text{H}$        &  kg & $288000$ & $36$ & $90$  \\
 Pitch moment of inertia & $I_\text{H}$ & kg $\cdot$ m$^{2}$ & $7.712\times 10^{6}$ & $2.41$ & $6.025$ \\
 \bottomrule
 \end{tabular}
  \label{tab_scaled_hull_values}
\end{table}

\begin{figure}[h!]
 \centering
 \includegraphics[scale = 0.4]{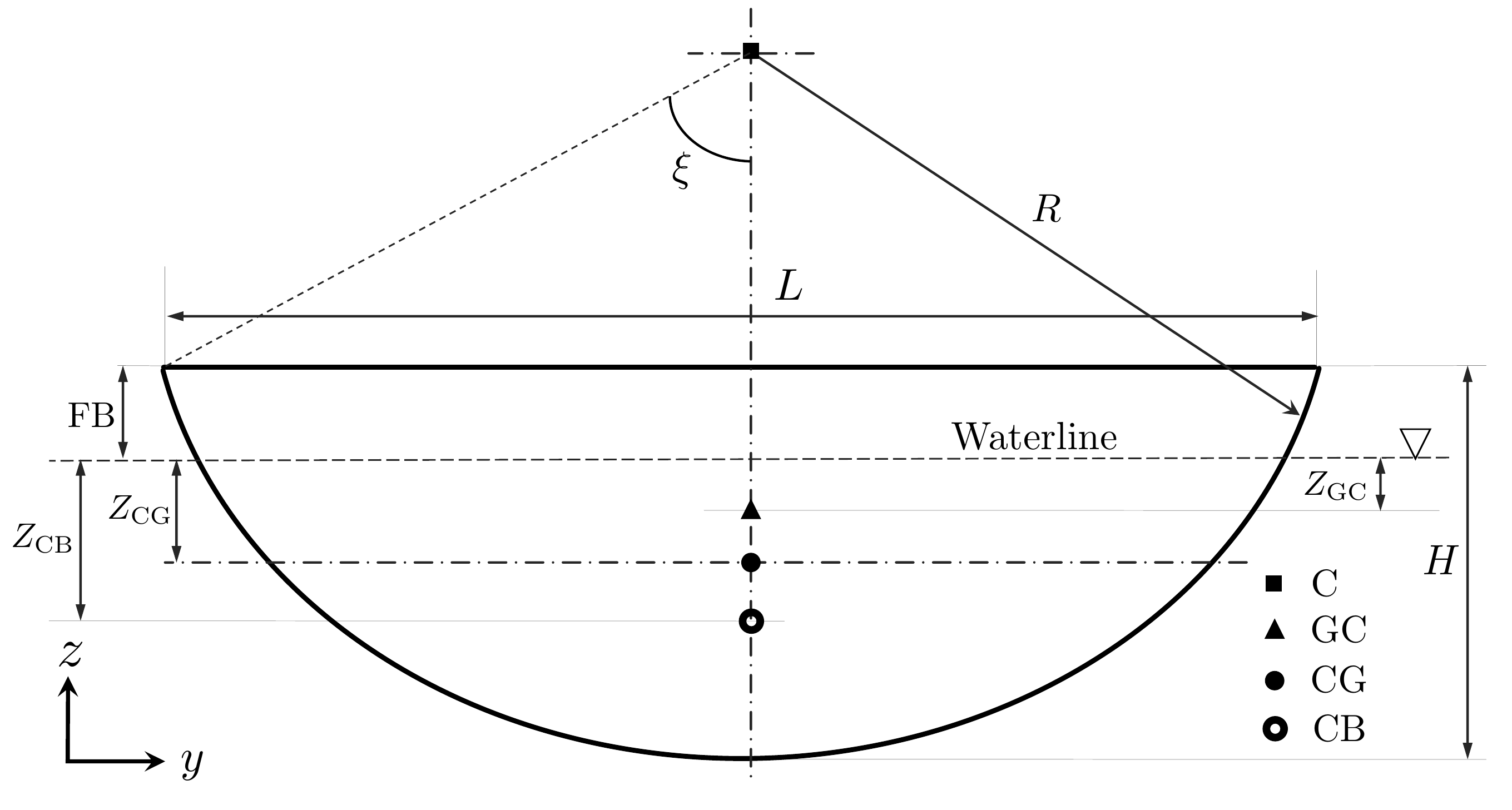}
 \caption{Hull geometry with properties: $L$ = 0.7665 m, $H$ = 0.225 m, $R$ = 0.4389 m and $\xi$ = 60.83$^\circ$.}
 \label{fig_hull_properties}
\end{figure}

The scaled-down (1:20) values of the hull properties are tabulated in Table~\ref{tab_scaled_hull_values}. To verify that the scaled-down values correlate well to the model geometry, we perform geometric estimation of the hull properties by assuming the hull to be a filled sector of a circle in two spatial dimensions. The geometric center (GC) and the center of buoyancy (CB) of the submerged sector can be calculated geometrically, and are found to be located at a distance $Z_{\text{GC}}$ = 0.0163 m and $Z_{\text{CB}}$ = 0.0605 m below the still waterline, respectively (see Fig.~\ref{fig_hull_properties}). From Table~\ref{tab_scaled_hull_values}, the scaled distance between the center of gravity of the device and waterline is $Z_{\text{CG}}$ = 0.0285 m. It can be seen that the CB lies below the CG and GC, satisfying the stability condition for floating bodies. Additionally CG lies below GC because in the real device, more mass is distributed towards the lower half portion. 

Similarly, the scaled-down moment of inertia of the hull $I_\text{H}$ can be argued geometrically. We first estimate the density of the hull from the scaled mass (90 kg) and the area of the sector (0.1225 m$^2$) to be $\rho_{\text{estimate}} = 734.69$ kg/m$^3$. Then we use $\rho_{\text{estimate}}$ to calculate the moment of inertia of the filled sector about its geometric center as $I_{\text{GC}}$ = 3.1768  kg $\cdot$ m$^{2}$. In the real device, most of the mass is concentrated along the outer periphery, resembling a ring rather than a filled sector. Since, the moment of inertia of a ring is twice as that of a filled circle,  $I_{\text{estimate}} \approx 2 I_{\text{GC}} = 6.3536$ kg $\cdot$ m$^{2}$, which is close to what we obtain from Table~\ref{tab_scaled_hull_values}.

 %%%%%%%%%%%%%%%%%%%%%%%%%%%%%%%%%%%%%%%%%%

%%%%%%%%%%%%%%%%%%%%%%%%%%%%%%%%%
\section{Wave dynamics}\label{sec_wave_eqs}
%%%%%%%%%%%%%%%%%%%%%%%%%%%%%%%%%%%%%%%%%%%%%%%
This section describes the types of waves, both regular and irregular, and the numerical tank approach used to simulate the ISWEC dynamics.

\subsection{Regular waves} \label{sec_regular_waves}

We use Fenton's fifth-order wave theory~\cite{Fenton1985} to generate regular waves of height $\cH$, time period $\cT$, and wavelength $\lambda$. According to fifth-order Stokes theory and assuming that the waves propagate in the positive $y$-direction, the wave elevation $\eta(y,t)$ from a still water surface at depth $d$ above the sea floor is
\begin{equation}
	\eta (y,t) = s \, \eta_1(y, t) + s^2 \, \eta_2 (y,t ) + s^3 \, \eta_3 (y,t) + s^4 \, \eta_4 (y,t) + s^5 \, \eta_5(y,t), \label{eq_Stokes_fifth}
\end{equation} 
in which, $s = \kappa \cH/2$ is the wave steepness, $\eta_1 =\kappa^{-1}\cos(\omega t - \kappa y)$ is the basic harmonic component, $\kappa= 2\pi/{\lambda}$ is the wavenumber, and $\omega = 2\pi/\cT$ is the wave frequency. The remaining terms in Eq.~\eqref{eq_Stokes_fifth} are higher-order corrections to linear wave theory, whose details are given in~\cite{Fenton1985}. The velocity and pressure solutions to the fifth-order Stokes wave can also be found in Fenton~\cite{Fenton1985}.

The (fifth-order) Stokes waves satisfy the dispersion relationship given by 
\begin{equation}
	\omega^2 =  g \kappa \tanh{(\kappa d)}, \label{eq_dispersion_relation}
\end{equation}
which relates the wavenumber $\kappa$ to the wave frequency $\omega$.  Eq.~\eqref{eq_dispersion_relation} is an implicit equation requiring an iterative process to compute $\kappa$ given $\omega$, or vice versa. Instead, an explicit equation can be used with sufficient accuracy in all water depth regimes~\cite{Fenton1988}:
\begin{equation}
	\label{eq_explicit_dispersion_relation}
   \kappa d \approx \frac{\Gamma+ \beta^{2}\left(\cosh\beta\right)^{-2}}{\tanh\beta + \beta\left(\cosh\beta\right)^{-2}},
\end{equation}
in which, $\beta = \Gamma \left(\tanh \Gamma \right)^{-\half}$, and $\Gamma = \omega^2 d / g$.

A converter's efficiency $\eta$ is measured relative to the available wave energy at the installation site. The traveling water waves transport (kinetic and potential) energy as they move along the sea or ocean surface, which is partially absorbed by the converter. The time-averaged wave power per unit crest width carried by regular waves in the propagation direction is given by~\cite{book_offshore_hydromechanics} 
\begin{equation}
	\label{eq_regular_wave_power}
	\bar{P}_{\rm wave} = \frac{1}{8} \rhow g \cH^2 c_\text{g},
\end{equation}
in which $\rhow$ is the density of water and $c_\text{g}$ is the group velocity of waves (the velocity with which wave energy is transported) given by
\begin{equation}
	\label{eq_group_velocity}
	c_\text{g} = \half \frac{\lambda}{\cT} \left(1 + \frac{2 \kappa d}{\sinh (2 \kappa d )} \right).
\end{equation}
In the deep water limit, where $d > \lambda/2$ and $\kappa d\rightarrow \infty$, Eqs.~\eqref{eq_dispersion_relation} and \eqref{eq_group_velocity} become
\begin{equation}
	\label{eq_deep_water_limit}
	\omega^2 = g \kappa  \quad \text{or} \quad \lambda = \frac{g\mathcal{T}^2}{2\pi} \qquad \text{and} \qquad c_\text{g} = \frac{\lambda}{2 \cT}.  \qquad \text{(deep water limit)}
\end{equation}
Using Eq.~\eqref{eq_deep_water_limit} in Eq.~\eqref{eq_regular_wave_power}, the wave power in the deep water limit can be expressed as
\begin{equation}
		\label{eq_regular_wave_power_deepwater}
		\bar{P}_{\rm wave} = \frac{\rhow g^2 \cH^2 \cT}{32 \pi} \approx \cH^2 \cT \;\; \text{kW/m}, \qquad \text{(deep water limit)}
\end{equation}
in which the constant numerical factor $\rhow g^2/32\pi \approx 10^3$ when evaluated with SI units.  

When simulating a scaled-down model, both $\cH$ and $\lambda$ are scaled-down by the length scale $\alpha$ to generate waves similar to the full-scale model. The scaled time period is obtained from the dispersion relationship between $\lambda$ and $\cT$.  

%%%%%%%%%%%%%%%%%%%%%%%%%%%%%%%%%%%%%%%%%%%%%%

\subsection{Irregular waves} \label{sec_irregular_waves}

Irregular waves depict a more realistic sea state and are modeled as a superposition of a large number of first-order regular wave components. Using the superposition principle, the sea surface elevation is expressed as 
\begin{equation}
     \label{eq_irregular_elevation}
     \eta(y,t) = \sum_{i=1}^{N} a_i \cos (\kappa_i y - \omega_i t + \theta_i ),
\end{equation}
in which $N$ is the number of wave components, each having its own amplitude $a_i$, angular frequency $\omega_i$, wavenumber $\kappa_i$, and a random phase $\theta_i$.  The wavenumber $\kappa_i$ is related to the angular frequency $\omega_i$ by the dispersion relationship given by Eq.~\eqref{eq_dispersion_relation}. The phases $\theta_i$ of each wave component are random variables following a uniform distribution in the range $\left[0, 2\pi\right]$.

The linear superposition of wave components also implies that the energy carried by an irregular wave is the sum of the energy transported by individual wave components. When the number of wave components $N$ tends to infinity, a continuous wave spectral density function $S(\omega)$ is used to describe the energy content of the wave components in an infinitesimal frequency bandwidth $\text{d}\omega$. The area under the curve gives the total energy of an irregular wave, modulo the factor $\rhow g$. Discretely, the component wave frequencies are typically chosen at an equal interval of $\Delta \omega$ between a narrow bandwidth of frequencies. The wave spectrum $S(\omega)$ approaches zero for frequencies outside the narrow bandwidth and peaks at a particular value of frequency $\omega_\text{p}$~\footnote{Here we consider only singly-peaked wave spectra.}. The amplitude of each wave component is related to the spectral density function by
\begin{equation}
		\label{eq_wave_spectrum_amplitude}
		a_{i} = \sqrt{ 2 \cdot S (\omega_i) \cdot \Delta{\omega} } \; .
\end{equation}

\begin{figure}[h!]
 \centering
 \includegraphics[scale = 0.45]{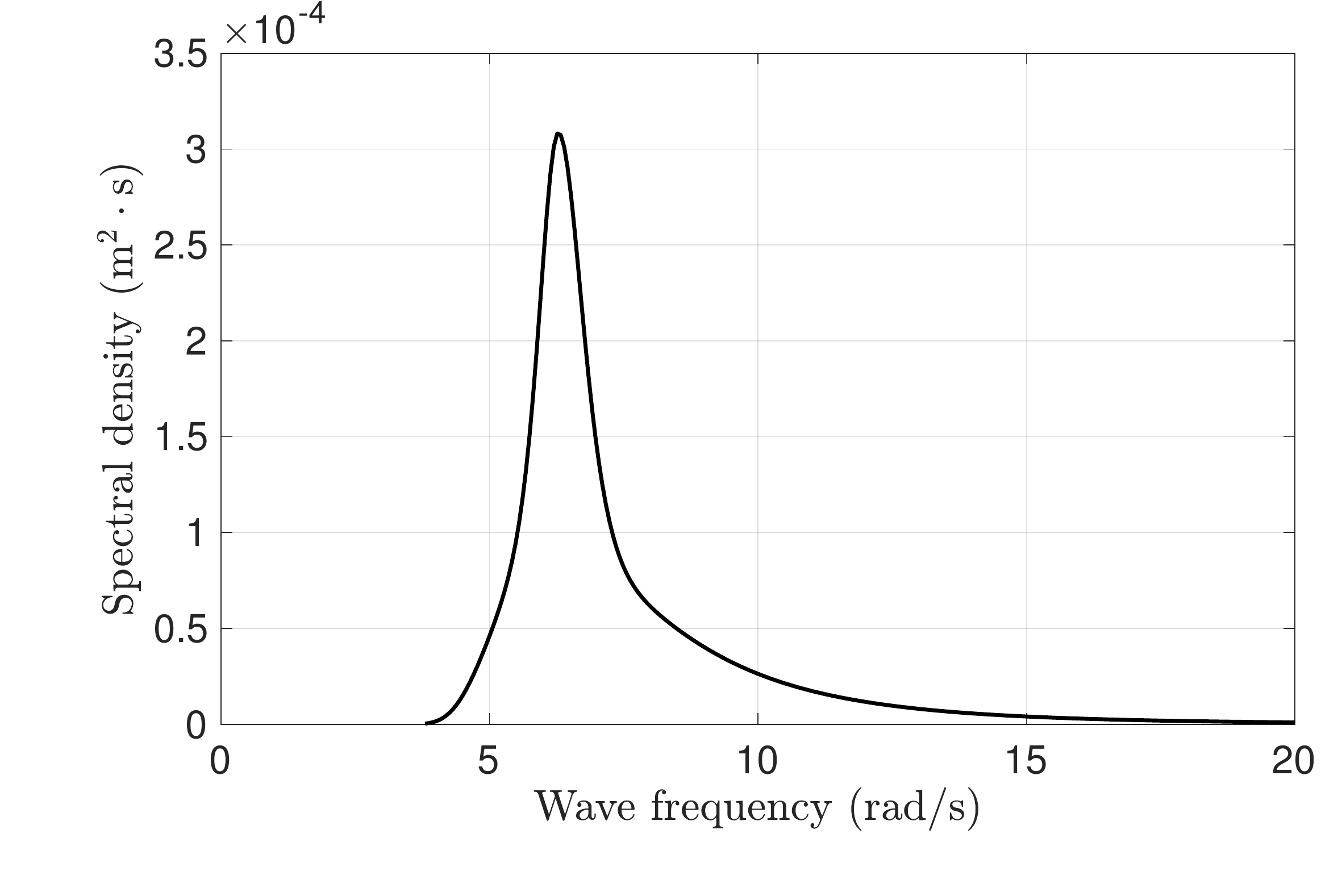}
 \caption{The JONSWAP wave spectrum obtained using $\cH_\text{s} = 0.1$ m, and $\cT_\text{p} = 1$ s ($\omega_\text{p} = 2 \pi$ rad/s).}
 \label{fig_JONSWAP_spectrum}
\end{figure}

We use the JONSWAP spectrum~\cite{book_offshore_hydromechanics} to generate irregular waves, which reads as 
\begin{equation}
	\label{eq_JONSWAP_wave_spectrum}
	S(\omega) = \frac{320 \cdot \cH^2_\text{s}}{\cT^4_\text{p}} \cdot \omega^{-5} \cdot \exp \left(\frac{-1950}{\cT^4_\text{p}} \cdot \omega^{-4}\right) \cdot \gamma^{A}, 
\end{equation}
in which $\cH_\text{s}$ is the significant wave height, and $\cT_\text{p}$ is the peak time period, i.e., the time period with the highest spectral peak (see Fig.~\ref{fig_JONSWAP_spectrum}). The remaining parameters in Eq.~\ref{eq_JONSWAP_wave_spectrum} are:

\begin{align}
	\label{eq_JONSWAP_parameters}
	\gamma &= 3.3 \qquad   (\text{peakedness factor}) \\
	A &= \exp \left[-\left(\frac{\frac{\omega}{\omega_\text{p}}-1}{\sigma \sqrt{2}}\right)^2\right]	 \nonumber \\
	\omega_\text{p} &= \frac{2\pi}{\cT_\text{p}}  \qquad \left(\text{angular frequency at spectral peak}\right)	\\
	\sigma &= \left\{\begin{matrix}
 			 0.07 \qquad \text{if} \quad \omega \leq \omega_\text{p}  \\
		         0.09  \qquad \text{if} \quad \omega > \omega_\text{p}	
         \end{matrix}\right. 
\end{align}

The mean wave power per unit crest width carried by an irregular wave is obtained from $S(\omega)$ as
\begin{equation}
	\label{eq_irreg_wave_power}
	\bar{P}_{\rm wave} = \rhow g \left(\int_{0}^{\infty} S(\omega) \; \text{d}\omega \right) c_\text{g} \approx   \rhow g \left(\sum_{i=1}^{N}\frac{1}{2}a^2_{i}\right) c_\text{g},
\end{equation}
in which the group velocity $c_\text{g}$ is calculated from Eq.~\eqref{eq_group_velocity} using the significant wavelength and peak time period of the spectrum. In the deep water limit, Eq.~\eqref{eq_irreg_wave_power} simplifies to
\begin{equation}
	\label{eq_irreg_wave_power_deep}
	\bar{P}_{\rm wave} \approx 0.49 \cH_\text{s}^2 \cT_\text{p} \;\; \text{kW/m}. \qquad \text{(deep water limit)}.
\end{equation}

%%%%%%%%%%%%%%%%%%%%%%%%%%%%%%%%%%%%%%%%%%%%%%

\subsection{Wave steepness} \label{sec_wave_steepness}

As discussed in Sec.~\ref{sec_PTO_params}, if the oscillation frequencies of the hull and gyroscope system are synchronized with that of the wave, the coupling between the hull and the gyroscope system (and therefore the output power) can be increased. Along with frequency synchronization, the oscillation amplitude of the hull can also be increased to enhance the device performance. This will result in more power transfer from the hull to the gyroscope system. The wave steepness ($s$) defined in Eq.~\ref{eq_Stokes_fifth}, which gives a relation between the wave height $\cH$ and wavelength $\lambda$, plays an important role in deciding the PTO and gyroscope system parameters such that the hull exhibits larger pitching motion. This is achieved by adjusting the gyroscope and PTO parameters such that the maximum hull pitch angle (amplitude) $\delta_0$ is expected to reach the maximum wave steepness angle $\delta_\text{s}$ of the incoming wave. An expression for $\delta_\text{s}$ can be obtained by approximating the incoming wave as a regular (simple harmonic) wave with elevation $\eta(y, t)$ given by
\begin{equation}
	\label{eq_simple_harmonic_wave}
	\eta(y, t) = a\cdot\cos(\kappa y - \omega t),
\end{equation}
where $a$ = $\cH$/2, is the wave amplitude. Differentiating the above equation with respect to $y$, we obtain
\begin{equation}
	\label{eq_slope}
	\frac{{\rm d}\eta(y, t)}{{\rm d} y} = a\cdot(-\kappa \sin(\kappa y - \omega t)).
\end{equation}
The maximum wave steepness (i.e. the slope) is obtained when $\sin(\kappa y - \omega t) = -1$, 
\begin{equation}
	\label{eq_max_slope}
	\left(\frac{{\rm d} \eta(y, t)}{{\rm d} y}\right)_{\rm max} = \kappa \cdot \frac{\cH}{2} = s.
\end{equation}
Finally, the maximum wave steepness angle is then given by
 \begin{equation}
	\label{eq_max_wave_steepness}
	\delta_\text{s} = \tan^{-1}\left(\frac{\kappa \cH}{2}\right) = \tan^{-1}\left(\frac{\pi \cH}{\lambda}\right)
\end{equation}
When the condition $\delta_0 = \delta_\text{s}$ is used to calculate the gyroscope and PTO parameters, the ISWEC device is observed to have maximum efficiency. A study on the variation of $\delta_0$ relative to $\delta_\text{s}$ for different wave heights is conducted in  Sec.~\ref{subsec_presc_angle_selec}.

%%%%%%%%%%%%%%%%%%%%%%%%%%%%%%%%%%%%%%%%%%%%%%

\subsection{Numerical wave tank} \label{sec_NWT}

\begin{figure}[]
 \centering
 \includegraphics[scale = 0.5]{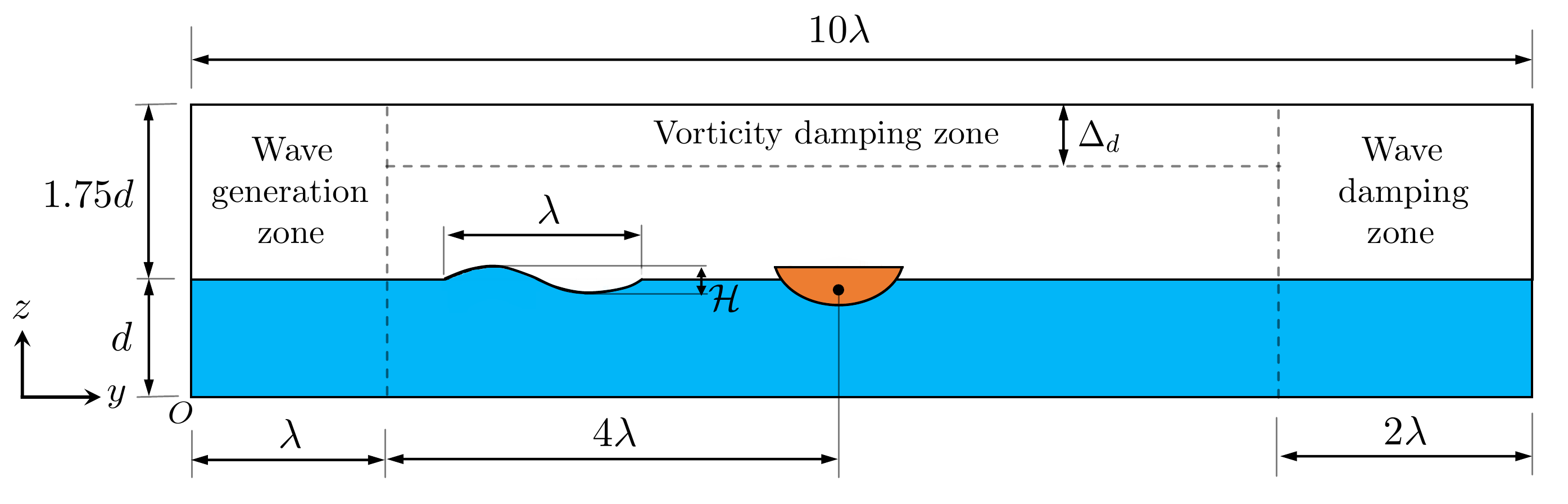}
 \caption{
 Numerical wave tank (NWT) schematic showing wave generation, wave damping, and vorticity damping zones. The ISWEC device is placed in the working zone of length $7\lambda$. 
 }
 \label{fig_NWT_schematic}
\end{figure}

The wave-structure interaction of the scaled-down ISWEC device is simulated in a numerical wave tank (NWT) as shown in Fig.~\ref{fig_NWT_schematic}. Water waves are generated at the left boundary of the domain using Dirichlet boundary conditions for the velocity components. The waves traveling in the positive $y$-direction are reflected back towards the inlet side from the device surface and also from the right boundary of the domain.  This results in wave distortion and wave interference phenomena, which reduces the ``quality" of waves reaching the device to study its performance. Several techniques have been proposed in the literature~\cite{Miquel2018,Windt2018,Windt2019} to mitigate these effects, including the relaxation zone method~\cite{Jacobsen2012}, the active wave absorption method~\cite{Higuera2013,Frigaard1995,Schaffer2000}, the momentum damping method~\cite{Choi2009,Ha2013}, the viscous beach method~\cite{Ghasemi2014}, the porous media method~\cite{Dong2009,Jacobsen2015}, and the mass-balance PDE method~\cite{Hu2016}. In this work, we use the relaxation zone method at inlet and outlet boundaries. The purpose of the relaxation zone near the channel inlet (the wave generation zone) is to smoothly extend the Dirichlet velocity boundary conditions into the wave tank up to a distance of one wavelength, so that the reflected waves coming from the ISWEC device do not interfere with the left boundary. In contrast, the relaxation zone near the right boundary (the wave damping zone) smoothly damps out the waves reaching the domain outlet near the right end. The wave damping zone is taken to be two wavelengths wide in our simulations. More details on the implementation of the relaxation zone method and level set based NWT can be found in our prior work~\cite{Nangia2019WSI}.    

We impose zero-pressure boundary condition at the channel top boundary, $z_{\rm max} = 2.75d$. To mitigate the interaction between shed vortices (due to the device motion) and the top boundary of the channel, we use a vorticity damping zone to dissipate the vortex structures reaching the boundary; see Fig.~\ref{fig_NWT_schematic}. The vorticity damping zone is implemented in terms of a damping force $\f_\text{d}$ in the momentum equation 
\begin{equation}
	\label{eq_spongelayer}
	\f_\text{d} =  -g(\tilde{z}) \u,
\end{equation}
in which, $g(\tilde{z}) = \rho_{\rm air} \, (\cos (\pi \tilde{z} )+1)/ (4 \Delta t)$ is the smoothed damping coefficient, $\rho_{\rm air}$ is the density of the air phase, $\Delta t$ is the time step size, $\tilde{z} = (z -z_{\rm max} )/ \Delta_\text{d}$ is the normalized $z$ coordinate, and $\Delta_\text{d}$ is the vorticity damping zone width, which is taken to be six cells wide in our simulations.

%%%%%%%%%%%%%%%%%%%%%%%%%%%%%%%%%
\section{Numerical model based on the incompressible Navier-Stokes equations}\label{sec_wsi_eqs}

We use a fully-Eulerian \emph{fictitious domain} Brinkman penalization (FD/BP) method~\cite{BhallaBP2019} to perform 
fully-resolved wave-structure interaction simulations. In FD/BP methods, the governing equations for the fluid 
are extended \emph{into} the region occupied by the solid structure,
yielding a single set of PDEs for the entire domain. Additional constraints are imposed in the structure domain to 
ensure that the velocity field within acts like a rigid body.
This is in contrast to body-conforming grid methods, in which the fluid
equations are solved only on a domain surrounding the immersed body.
For applications involving moving body fluid-structure interaction (FSI), fictitious domain methods are less computationally expensive
than body-conforming grid techniques due to the absence of expensive re-meshing operations.

In this section, we first describe the continuous governing equations for the FD/BP formulation and the interface
tracking approach for the multiphase FSI system. Next, we briefly describe the spatiotemporal discretization, overall
solution methodology, and time-stepping scheme. Finally, we describe the coupling used to simulate the dynamics of
an inertial sea wave energy converter device, which involves modeling the effect of a rigid body pitch torque.
A validation case for vortex induced vibration of a rectangular plate exhibiting galloping motion is presented at the end
of this section.
We refer readers to the references by Nangia et al.~\cite{Nangia2019WSI,Nangia2019MF}, Bhalla et al.~\cite{BhallaBP2019} and  
Dafnakis et al.~\cite{Dafnakis2019} for more details on the Cartesian grid fluid solver, FD/BP formulation, and simulating
wave energy converters within this framework, respectively.

\subsection{Continuous equations of motion} \label{sec_cont_eqs}

Let $\Omega \subset \mathbb{R}^d$ with $d = 3$ denote a fixed three-dimensional region in space. 
The dynamics of a coupled multiphase fluid-structure system occupying this domain are governed by the incompressible
Navier-Stokes (INS) equations

\begin{align}
  \D{\rho \u(\x,t)}{t} + \div \rho\u(\x,t)\u(\x,t) &= -\grad p(\x,t) + \div \left[\mu \left(\grad \u(\x,t) + \grad \u(\x,t)^T\right) \right]+ \rho\g + \fc(\x,t), \label{eqn_momentum}\\
  \div \u(\x,t) &= 0, \label{eqn_continuity} 
\end{align}
which describe the momentum and incompressibility of a fluid with velocity $\u(\x,t)$ and pressure $p(\x,t)$ using an Eulerian 
coordinate system $\x = (x,y,z) \in 
\Omega$. Note that Eqs.~\eqref{eqn_momentum} and~\eqref{eqn_continuity} are written for the entire fixed region $
\Omega$, which can be further decomposed into two regions occupied by the fluid $\Omega_\text{f}(t) \subset \Omega$ and the 
immersed body $\Omega_\text{b}(t) \subset \Omega$. These regions are non-overlapping, i.e. $\Omega = \Omega_\text{f}(t) \cup 
\Omega_\text{b}(t)$, and $\fc(\x,t)$ represents a rigidity-enforcing constraint force density that vanishes outside $\Omega_\text{b}(t)$;
this ensures a rigid body velocity $\ub(\x,t)$ is attained within the solid region.
The spatiotemporally varying density and viscosity fields are denoted by $\rho(\x,t)$ and $\mu(\x,t)$, 
respectively. An indicator function $\chi(\x,t)$ is further used to track the location of the solid body, which is nonzero only
within $\Omega_\text{b}(t)$. The acceleration due to gravity is directed towards the negative $z$-direction: $\g = (0, 0, -g)$.

The immersed structure is treated as a porous region with vanishing permeability $\kappa_\text{p} \ll 1$, yielding the following
formula for the Brinkman penalized constraint force

\begin{align}  
\fc(\x,t)  &= \frac{\chi(\x,t)}{\kappa_\text{p}}\left(\ub(\x,t) - \u(\x,t)\right). \label{eqn_brinkman_force}
\end{align}
Sec.~\ref{sec_fsi_coupling} describes the fluid-structure coupling algorithm, and Sec.~\ref{sec_iswec_coupling} describes
the external ISWEC torque specification, which together are used to determine the rigid body velocity $\ub(\x,t)$ 
applied to $\Omega_\text{b}(t)$.

\subsection{Interface tracking}

All of the cases described in the present work involve a single rigid structure interacting with an air-water interface.
We briefly describe the interface tracking methodology here, and refer readers to Nangia et al.~\cite{Nangia2019MF,Nangia2019WSI}
for further details on its implementation.
A scalar level set function $\sigma(\x,t)$ is used to demarcate liquid (water) and gas (air) regions, $\Omegal \subset \Omega$ and $\Omegag \subset \Omega$, respectively, in the computational domain.
The air-water interface $\Gamma(t) = \Omegal \cap \Omegag$ is implicitly defined
by the zero-contour of $\sigma$.
The same methodology is employed to track the surface of the immersed body $\Sb(t) = \partial \Vb(t)$ using the
zero-contour of a level set function $\psi(\x,t)$; the aforementioned indicator function for the solid domain is computed
based on $\sigma$. The evolution of these level set fields is governed by linear advection via the local fluid velocity field
\begin{align}
\D{\sigma}{t} + \div \sigma \u &= 0, \label{eq_ls_fluid_advection} \\
\D{\psi}{t} + \div \psi \u &= 0. \label{eq_ls_solid_advection}
\end{align}
Making use of the signed distance property, the density and viscosity across the entire computational domain
can be conveniently expressed as a functions of $\sigma(\x,t)$ and $\psi(\x,t)$    
\begin{align}
\rho (\x,t) &= \rho(\sigma(\x,t), \psi(\x,t)), \label{eq_rho_ls}\\
\mu (\x,t) &= \mu(\sigma(\x,t), \psi(\x,t)) \label{eq_mu_ls}.
\end{align}
After every time step, both level set functions are reinitialized to maintain signed distance functions to their 
respective interfaces. Standard approaches for computing a steady-state solution to the Hamilton-Jacobi equation
is used to reinitialize $\sigma$, whereas an analytical distance computation to the immersed body is used to reinitialize
$\psi$.

\subsection{Spatial discretization} 

The continuous equations of motion Eqs.~\eqref{eqn_momentum}-\eqref{eqn_continuity}  are discretized on a staggered
Cartesian grid made up of $\Nx \times \Ny \times \Nz$ rectangular cells covering the domain $\Omega$.
%The continuous equations of motion Eqs.~\eqref{eqn_momentum}-\eqref{eqn_continuity} are discretized on a staggered 
%Cartesian grid as shown in Fig.~\ref{fig_cfd_domains}. The discretized domain is made up of $\Nx \times \Ny$ rectangular cells that cover the physical 
%domain $\Omega$. 
The mesh spacings in the three spatial directions are denoted by $\dx$, $\dy$, and $\dz$ respectively.
Without loss of generality, let the lower left corner of the rectangular 
domain be the origin $(0,0,0)$ of the coordinate system. Using this reference point, each cell center of the grid has
position $\x_{i,j,k} = \left((i + \half)\dx,(j + \half)\dy,(k + \half)\dz\right)$ for $i = 0, \ldots, \Nx - 1$, $j = 0, \ldots, \Ny - 1$, 
and $k = 0, \ldots, \Nz - 1$.
The physical location of the cell face that is half a grid cell away from $\x_{i,j,k}$ in the $x$-direction 
is given by $\x_{i-\half,j,k} = \left(i\dx,(j + \half)\dy,(k + \half)\dz\right)$. Similarly, $\x_{i,j-\half,k} =\left((i + \half)\dx,j\dy,(k + \half)\dz\right)$ and $\x_{i,j,k-\half} =\left((i + \half)\dx,(j + \half)\dy,k\dz\right)$
are the physical locations of the cell faces that are half a grid cell away from $\x_{i,j,k}$ in the $y$- and
$z$-directions, respectively. The level set fields,
pressure degrees of freedom, and the material properties are all approximated at cell centers and are denoted by 
$\sigma_{i,j,k}^{n} \approx \sigma \left(\x_{i,j,k}, t^n\right)$, $\psi_{i,j,k}^{n} \approx \psi\left(\x_{i,j,k}, t^n\right)$,  
$p_{i,j,k}^{n} \approx p\left(\x_{i,j,k},t^{n}\right)$, $\rho_{i,j,k}^{n} \approx \rho\left(\x_{i,j,k},t^{n}\right)$
and  $\mu_{i,j,k}^{n} \approx \mu\left(\x_{i,j,k},t^{n}\right)$, respectively; some of these quantities are interpolated onto the required 
degrees of freedom as needed (see~\cite{Nangia2019MF} for further details).
Here, the time at time step $n$ is denotes by $t^n$.
The velocity degrees of freedom are approximated on cell faces, with notation $u_{i-\half,j,k}^{n} \approx u\left(\x_{i-\half,j,k}, 
t^{n}\right)$, $v_{i,j-\half,k}^{n} \approx v\left(\x_{i,j-\half,k}, t^{n}\right)$,
and $w_{i,j,k-\half}^{n} \approx w\left(\x_{i,j,k-\half}, t^{n}\right)$. The gravitational body force and constraint force density on the
right-hand side of the momentum equation \eqref{eqn_momentum} are also approximated on the cell faces.  

Second-order finite differences are used to discretize all spatial derivatives. 
Subsequently, the discretized version of these differential operators are denoted with $h$ subscripts; i.e., 
$\grad \approx \grad_h$. For more details on the spatial discretization, we refer readers to 
prior studies~\cite{Nangia2019MF,Cai2014,Griffith2009,Bhalla13}.

\begin{figure}[]
  \centering
  \subfigure[Continuous domain]{
    \includegraphics[scale = 0.45]{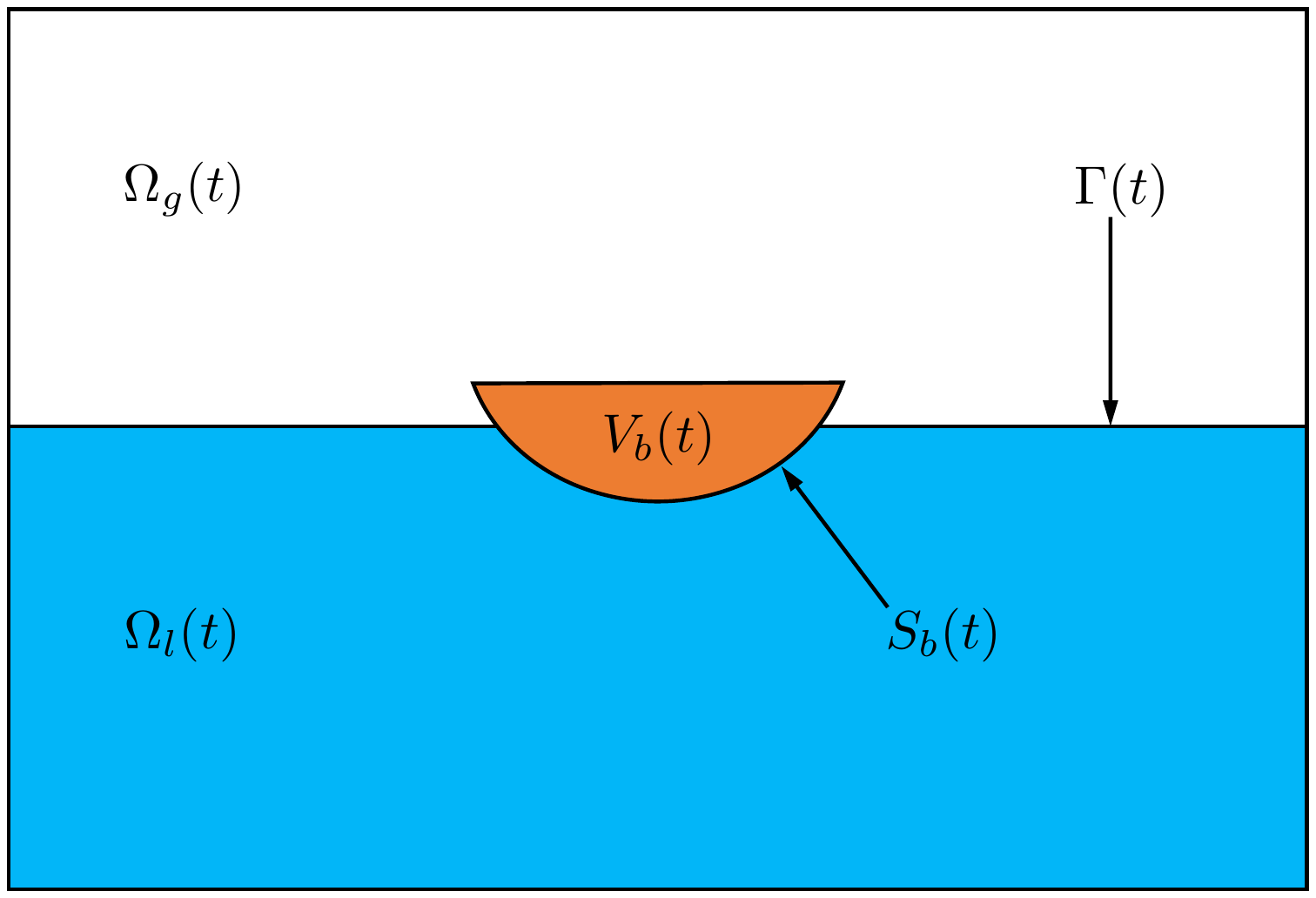} 
    \label{fig_cont_domain}
  }
   \subfigure[FD/BP discretized domain]{
    \includegraphics[scale = 0.45]{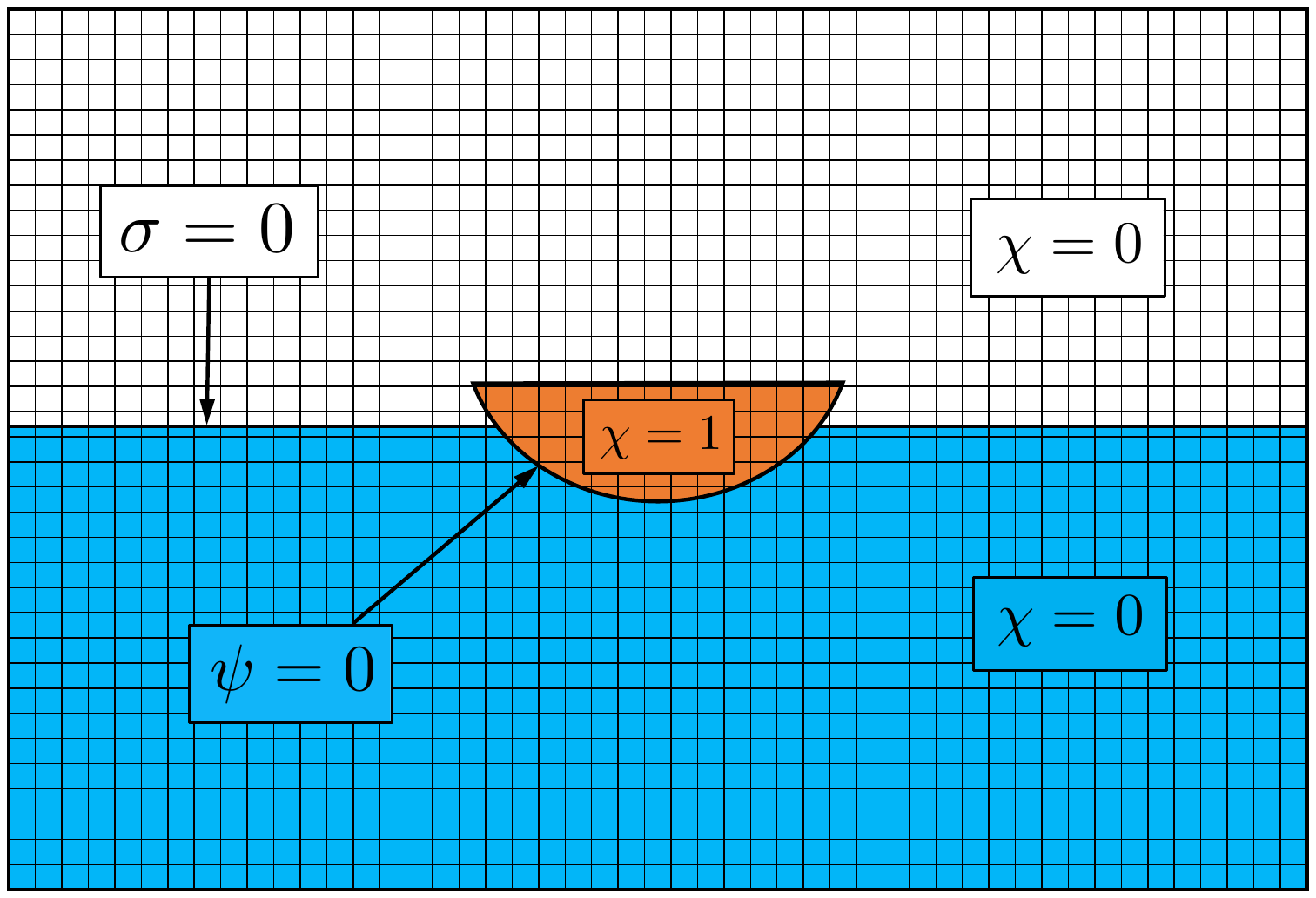}
    \label{fig_discrete_domain}
  }
  \caption{\subref{fig_cont_domain} Schematic of a two-dimensional slice through the computational domain $\Omega$, in which an immersed body interacts with an air-water interface.~\subref{fig_discrete_domain} Cartesian mesh discretization of the domain $\Omega$ and the value of the indicator 
  function $\chi(\x,t)$ used to differentiate between the fluid and solid regions in the FD/BP method; $\chi(\x,t)= 1$ inside the solid domain 
  and $\chi(\x,t) = 0$ in air and water domains. The zero contour of $\sigma(\x,t)$ tracks the air-water interface $\Gamma(t)$, while the zero contour of $\psi(\x,t)$ tracks the solid-fluid interface $\Sb(t)$.
  }
  \label{fig_cfd_domains}
\end{figure}

\subsection{Solution methodology} \label{sec_sol_method}
Next, we describe the methodology used to solve the discretized equations of motion.
At a high level, this involves three major steps:
\begin{enumerate}
\item Specifying the material properties $\rho(\x, t)$ and $\mu(\x, t)$ throughout the computational domain.
\item Calculating the Brinkman penalization rigidity constraint force density $\fc$ based on the ISWEC dynamics
\item Computing the updated solutions to $\sigma$, $\psi$, $\u$, and $p$.
\end{enumerate}
In the present work, we briefly review the computations above in the context of ISWEC devices with a single
unlocked rotational degree of freedom. For a more general treatment of the FD/BP method, we refer readers
to previous work by Bhalla et al.~\cite{BhallaBP2019} and references therein.

\subsubsection{Density and viscosity specification}

At the air-water $\Gamma$ and fluid-solid $\Sb$ interfaces, a smoothed Heaviside
function is used to transition between the three phases. In this transition region, $\ncells$ grid cells
are used on either side of the interfaces to provide a smooth variation of material properties.
A given material property $\Im$ (such as density $\rho$ or viscosity $\mu$) is prescribed throughout the computational
domain first by computing the \emph{flowing} phase (i.e. air and water)
\begin{equation}
\label{eqn_ls_flow}
\Im^{\text{flow}}_{i,j,k} = \Im_\text{l} + (\Im_\text{g} - \Im_\text{l}) \widetilde{H}^{\text{flow}}_{i,j,k},
\end{equation}
and then correcting $\Im^{\text{flow}}$ to account for the solid region
\begin{equation}
\label{eqn_ls_solid}
\Im_{i,j,k}^{\text{full}} = \Im_\text{s} + (\Im^{\text{flow}}_{i,j,k} - \Im_\text{s}) \widetilde{H}^{\text{body}}_{i,j,k}. 
\end{equation}
Here, $\Im^{\text{full}}$ is the final scalar material property field throughout $\Omega$. Standard numerical
Heaviside functions are used to facilitate the transition specified by Eqs.~\eqref{eqn_ls_flow} and~\eqref{eqn_ls_solid}:
\begin{align}
\widetilde{H}^{\text{flow}}_{i,j,k} &= 
\begin{cases} 
       0,  & \sigma_{i,j,k} < -\ncells \dx,\\
        \frac{1}{2}\left(1 + \frac{1}{\ncells \dx} \sigma_{i,j,k} + \frac{1}{\pi} \sin\left(\frac{\pi}{ \ncells \dx} \sigma_{i,j,k}\right)\right) ,  & |\sigma_{i,j,k}| \le \ncells \dx,\\
        1,  & \textrm{otherwise},
\end{cases}       \label{eqn_Hflow} \\
\widetilde{H}^{\text{body}}_{i,j,k} &= 
\begin{cases} 
       0,  & \psi_{i,j,k} < -\ncells \dx,\\
        \frac{1}{2}\left(1 + \frac{1}{\ncells \dx} \psi_{i,j,k} + \frac{1}{\pi} \sin\left(\frac{\pi}{ \ncells \dx} \psi_{i,j,k}\right)\right) ,  & |\psi_{i,j,k}| \le \ncells \dx,\\
        1,  & \textrm{otherwise},  \label{eqn_Hbody}
\end{cases}
\end{align}
We use the same number of transition cells $\ncells = 2$  for both air-water and fluid-solid interfaces in our simulations, although this is not an
inherent limitation of our method. We refer readers to Nangia et al.~\cite{Nangia2019WSI} for more discussion.
%Consistent with Eqs.~\eqref{eqn_Hflow}-\eqref{eqn_ls_solid} and without loss of generality,  $\phi$ signed distance values
%are taken to be negative in the liquid phase and positive in the air phase. Similarly, $\psi$ signed distance values are taken 
%to be negative inside the solid body, whereas they are taken to be positive outside the solid region.    

\subsubsection{Time stepping scheme}

A fixed-point iteration time stepping scheme using $\ncycles = 2$ cycles per time step is used to evolve quantities 
from time level $t^n$ to time level $t^{n+1} = t^n + \Delta t$. A $k$ superscript is used to denote the
cycle number of the fixed-point iteration. At the beginning of each time step, the solutions
from the previous time step are used to initialize cycle $k = 0$:
$\u^{n+1,0} = \u^{n}$, $p^{n+\half,0} = p^{n-\half}$, $\sigma^{n+1,0} = \sigma^{n}$, and
$\psi^{n+1,0} = \psi^{n}$. At the initial time $n = 0$, the physical quantities are prescribed via initial condition. 
 
\subsubsection{Level set advection}
An standard explicit advection scheme is used to evolve the two level set functions
\begin{align}
\frac{\sigma^{n+1,k+1} - \sigma^{n}}{\dt} + Q\left(\u^{n+\half,k}, \sigma^{n+\half,k}\right) &= 0, \\
\frac{\psi^{n+1,k+1} - \psi^{n}}{\dt} + Q\left(\u^{n+\half,k}, \psi^{n+\half,k}\right) &= 0,
\end{align}
in which $Q(\cdot,\cdot)$ represents an explicit piecewise parabolic method (xsPPM7-limited) approximation to the 
linear advection terms on cell centers~\cite{Griffith2009,Rider2007}.

\subsubsection{Incompressible Navier-Stokes solution}

The following spatiotemporal discretization of the incompressible Navier-Stokes Eqs.~\eqref{eqn_momentum}-\eqref{eqn_continuity} (in \emph{conservative} form) is employed
\begin{align}
	\frac{\breve{\V \rho}^{n+1,k+1} \u^{n+1,k+1} - { \V \rho}^{n} \u^n}{\dt} + \C^{n+1,k} &= -\grad_h p^{n+\half, k+1}
	+ \left(\L_{\mu} \u\right)^{n+\half, k+1}
	+  \V \wp^{n+1,k+1}\g +  \fc^{n+1,k+1}, \label{eqn_c_discrete_momentum}\\
	 \grad_h \cdot \u^{n+1,k+1} &= 0, \label{eqn_c_discrete_continuity}
\end{align}
in which the discretized convective derivative $\C^{n+1,k}$ and the density approximation $\breve{\V \rho}^{n+1,k+1}$
are computed using a consistent mass/momentum transport scheme; this is vital to ensure numerical stability
for cases involving air-water density ratios. This scheme is described in detail in previous studies by Nangia et al.
and Bhalla et al.~\cite{Nangia2019MF,BhallaBP2019}.
A standard semi-implicit approximation to the viscous strain rate
 $\left(\L_{\mu} \u\right)^{n+\half, k+1} =  \half\left[\left(\L_{\mu} \u\right)^{n+1,k+1} + \left(\L_{\mu} \u\right)^n\right]$
is employed, in which
$\left(\L_{\mu}\right)^{n+1} = \grad_h \cdot \left[\mu^{n+1} \left(\grad_h \u + \grad_h \u^T\right)^{n+1}\right]$.
The two-stage process described by Eqs.~\eqref{eqn_ls_flow} and~\eqref{eqn_ls_solid} is used to
obtain the newest approximation to viscosity $\mu^{n+1,k+1}$. The flow density field is used to construct the gravitational 
body force term $\V \wp \g = \V{\rho}^{\text{flow}} \g$, which avoids spurious currents due to large density variation near the 
fluid-solid interface~\cite{Nangia2019WSI}. 

\subsubsection{Fluid-structure coupling} \label{sec_fsi_coupling}

Next, we describe the Brinkman penalization term that imposes the rigidity constraint in the solid region,
and the overall fluid-structure coupling scheme. In this work, we simplify the treatment of the FSI coupling
by only considering immersed bodies with a single unlocked rotational degree of freedom (DOF); a more
general approach is described in Bhalla et al.~\cite{BhallaBP2019}.

The Brinkman penalization term is treated implicitly and computed as
\begin{align}
\fc^{n+1,k+1} = \frac{\widetilde{\chi}}{\kappa_\text{p}}\left(\ub^{n+1,k+1} - \u^{n+1,k+1}\right),  \label{eqn_bp_discrete}
\end{align}
in which the discretized indicator function $\widetilde{\chi} = 1 - \widetilde{H}^{\text{body}}$ is $1$ only inside the body 
domain and defined using the structure Heaviside function $ \widetilde{H}^{\text{body}}$ from Eq.~\eqref{eqn_Hbody}.
A permeability value of $\kappa_\text{p} \sim \cO(10^{-8})$ has been shown to be sufficiently small enough to effectively enforce
the rigidity constraint~\cite{Gazzola2011,BhallaBP2019}.
In general, the solid body velocity $\ub = \Ur + \Wr \times \left(\x-\Xcom\right)$ can be expressed as a sum of translational $\Ur$ and rotational $\Wr$ velocities. In this work, $\Ur = \mathbf{0}$ and we simply have
\begin{equation}
\ub^{n+1,k+1} = \Wr^{n+1,k+1} \times \left(\x - \Xcom^{n+1,k+1}\right).
\end{equation}
Moreover, two of the rotational DOFs are locked in the present study, i.e. they are constrained to zero. Hence, the
expression for $\Wr$ can be simplified even further,
\begin{equation}
\Wr^{n+1,k+1}  = \left(\dot{\delta}^{n+1,k+1}, 0, 0\right), \label{eq_rot_vel}
\end{equation}
in which $\dot{\delta}$ is the rotational velocity of the structure about its pitch axis.

The rigid body velocity can be computed by integrating Newton's second law of motion for the pitch axis rotational velocity:
\begin{align}
%		\Mb \frac{\Ur^{n+1,k+1} - \Ur^n}{\dt} &=  \cF^{n+1,k} + \Mb \g + \F_m + \F_{\text{PTO}},  \label{eqn_newton_u} \\
		I_\text{H} \frac{\dot{\delta}^{n+1,k+1} - \dot{\delta}^n}{\dt} &=  \cM_{\text{hydro}} ^{n+1,k} - \cM_\delta^{n+1,k},  \label{eqn_newton_w}
\end{align}
in which $I_\text{H}$ is the moment of inertia of the hull, $\cM_{\text{hydro}}$ is the net hydrodynamic torque, and
$\cM_\delta^{n+1,k}$ is the projection of the gyroscopic torque about the $x$-axis.  The net 
hydrodynamic torque is computed by summing the contributions from pressure 
and viscous forces acting on the hull and taking the pitch component

\begin{align}
%\cF^{n+1,k}  &= \sum_f    \left(-p^{n+1,k} \n_f + \mu_f \left(\grad_h \u^{n+1,k} +  \left(\grad_h \u^{n+1,k}\right)^T \right)\cdot \n_f \right) \Delta A_f,   \label{eqn_int_fh} \\
\cM_{\text{hydro}}  &= \hat{i} \cdot \left[\sum_\text{f}  \left(\x - \Xcom^{n+1,k}\right) \times  \left(-p^{n+1,k} \n_\text{f} + \mu_\text{f} \left(\grad_h \u^{n+1,k} +  \left(\grad_h \u^{n+1,k}\right)^T \right)\cdot \n_\text{f} \right) \Delta A_\text{f}\right].  \label{eqn_int_mh}
\end{align}
The hydrodynamic traction in the above equation is evaluated on Cartesian grid faces near the hull that define a stair-step 
representation of the body on the Eulerian mesh~\cite{BhallaBP2019}, with $\n_\text{f}$ and $\Delta A_\text{f}$ representing the unit normal vector and the area of a 
given cell face, respectively.
The computation of the gyroscopic action $\cM_\delta$ is described in the following section.

%Since the net gravitational force on the body is already included in Eq.~\eqref{eqn_newton_u}, we exclude it from the body region 
%in the INS momentum Eq.~\eqref{eqn_c_discrete_momentum}. 

\subsubsection{Coupling ISWEC dynamics} \label{sec_iswec_coupling}
The ISWEC is allowed to freely rotate about its pitch axis and its motion depends on the hydrodynamic and external torques
acting on it. The external torque $\cM_\delta$ generated by the gyroscopic action is unloaded on the hull and \emph{opposes} 
the wave induced pitching motion. Therefore, $\cM_\delta$ appears with negative sign on the right side of 
Eq.~\ref{eqn_newton_w}. The analytical expression for this pitch torque is given by Eq.~\ref{eq_M_delta},
while its discretization is written as
\begin{align}
\cM_\delta^{n+1,k} = \left(J\sin{^2\varepsilon^{n+1,k}}	+I\cos{^2\varepsilon^{n+1,k}}\right){\ddot{\delta}}^{n+1,k}+J{\dot{\phi}}{\dot{\varepsilon}		^{n+1,k}}\cos{\varepsilon^{n+1,k}} \nonumber \label{eq_T_delta_discretized} \\
 	+ 2\left(J-I\right){\dot{\delta}^{n+1,k}}{\dot{\varepsilon}^{n+1,k}}		\sin{\varepsilon^{n+1,k}}\cos{\varepsilon^{n+1,k}}+J{\ddot{\phi}}				\sin{\varepsilon^{n+1,k}},
\end{align}
in which the pitch acceleration term $\ddot{\delta}^{n+1,k}$ is calculated using a standard finite difference (explicit forward Euler) of
the hull's pitch velocity:
\begin{align}
  \ddot{\delta}^{n+1,k} &= 
  \begin{cases} 
       \frac{\dot{\delta}^{n+1,k} - \dot{\delta}^{n}}{\Delta t},  & k>0,\\ \\
       \frac{\dot{\delta}^{n} - \dot{\delta}^{n-1}}{\Delta t},  & k=0.\\ 
  \end{cases} \label{eq_pitch_velocity_computation}
\end{align}
%\begin{align}
%  \ddot{\delta}^{n+1,k} &= 
%  \begin{cases} 
%       \frac{\dot{\delta}^{n+1,k}-\dot{\delta}^{n}}{\Delta t},  & k>0,\\ \\
%       \frac{\dot{\delta}^{n+1,0}-\dot{\delta}^{n,\ncycles-1}}{\Delta t},  & k=0,\\ 
%  \end{cases} \label{eq_pitch_velocity_computation}
%\end{align}
We set $\delta^{n+1,0} = \delta^n$, $\varepsilon^{n+1,0} = \varepsilon^{n}$, $\dot{\delta}^{n+1,0} = \dot{\delta}^{n}$, and $\dot{\varepsilon}^{n+1,0} = \dot{\varepsilon}^{n}$ for cycle $k = 0$. 

The precession acceleration $\ddot{\varepsilon}$ is given analytically by Eq.~\ref{eq_gyro_PTO}, which in discretized form
reads
\begin{equation}
	\ddot{\varepsilon}^{n+1,k} = \frac{1}{I}\left[-k\varepsilon^{n+1,k-1}-c{\dot{\varepsilon}}^{n+1,k-1}-\left(I-J\right)\left({\dot{\delta}^{n+1,k}}\right)^2\sin\varepsilon^{n+1,k-1}\cos\varepsilon^{n+1,k-1}+J\dot{\phi}\dot{\delta}^{n+1,k}\cos\varepsilon^{n+1,k-1}\right].
\end{equation}
This newest approximation to the precession acceleration $\ddot{\varepsilon}^{n+1,k}$ is explicitly calculated using only
the \emph{prior} cycle's values of precession velocity $\dot{\varepsilon}^{n+1,k-1}$ and angle $\varepsilon^{n+1,k-1}$.
New approximations to $\dot{\varepsilon}$ and $\varepsilon$ at cycle $k$ are computed using the Newmark-$\beta$
method~\cite{Newmark1959} as follows:
\begin{align}
	\dot{\varepsilon}^{n+1,k} &= \dot{\varepsilon}^{n}+\frac{\Delta t}{2}\left(\ddot{\varepsilon}^{n}+\ddot{\varepsilon}^{n+1,k}\right) \label{eq_newmark_vel}\\
	\varepsilon^{n+1,k}&=\varepsilon^{n}+{\Delta t}\dot{\varepsilon}^{n}+\frac{\Delta t^2}{4}\left(\ddot{\varepsilon}^{n}+\ddot{\varepsilon}^{n+1,k}\right) \label{eq_newmark_ang}
\end{align}
As described in Sec.~\ref{sec_iswec_eqs}, the PTO stiffness $k$ and damping $c$ parameters in the control torque and the 
gyroscope's angular velocity $\dot{\phi}$, acceleration $\ddot{\phi} = 0$, and moments of inertia $I$ and $J$ are known \emph{a priori} 
and do not represent additional unknowns in the calculation of $\cM_\delta^{n+1,k}$.
Hence the procedure outlined by Eqs.~\eqref{eq_T_delta_discretized} to~\eqref{eq_newmark_ang} enables the
calculation of the external pitch torque shown on the right-hand side of Eq.~\ref{eqn_newton_w}, thus coupling the ISWEC 
dynamics to the FD/BP methodology.

\subsection{FSI validation}

To validate our implementation of the method described in this section, we simulate the vortex induced vibration
of a rectangular plate undergoing galloping motion.
This single rotational degree of freedom case has been widely used as a benchmark problem for FSI algorithms in prior literature. It also mimics the ISWEC model well, which primarily oscillates in the pitch direction. The governing equation for the spring-mass-damper plate model reads as
\begin{equation}
	I_{\vartheta}\ddot{\vartheta} + C_{\vartheta}\dot{\vartheta} + K_{\vartheta}\vartheta = \cM_{\rm hydro},
\end{equation}
in which $\vartheta$ is the pitch angle of the plate measured from the horizontal axis, $I_\vartheta$ is the pitch moment of inertia about the center of mass, $C_\vartheta$ is the torsional damping constant, $K_\vartheta$ is the torsional spring constant, and  $\cM_{\rm hydro}$ is the hydrodynamic moment acting on the plate due to the external fluid flow.  

\begin{figure}[]
  \centering
  \subfigure[]{
    \includegraphics[scale = 0.22]{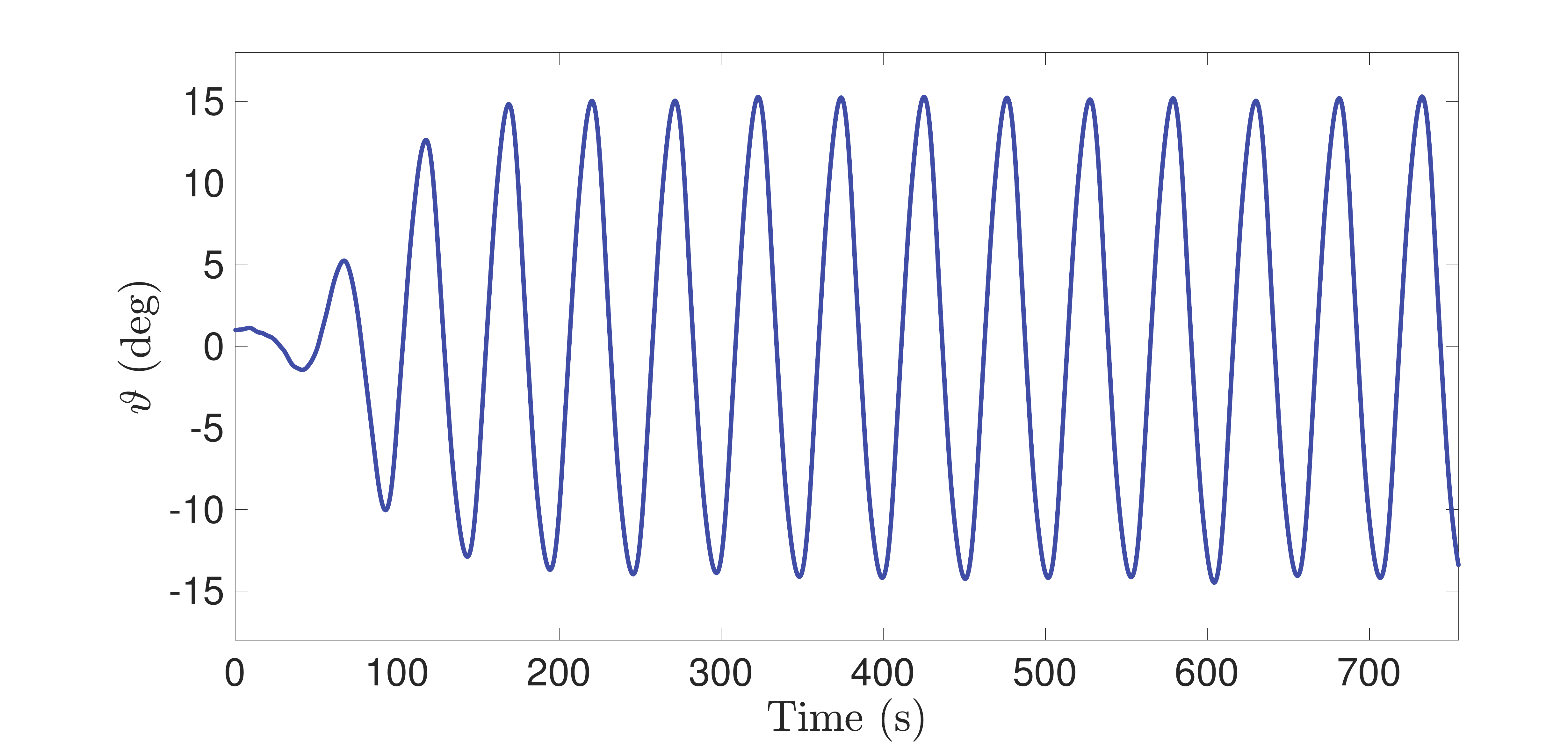}
    \label{fig_gallop_vs_time}
  } \\
  \subfigure[]{
    \includegraphics[scale = 0.2]{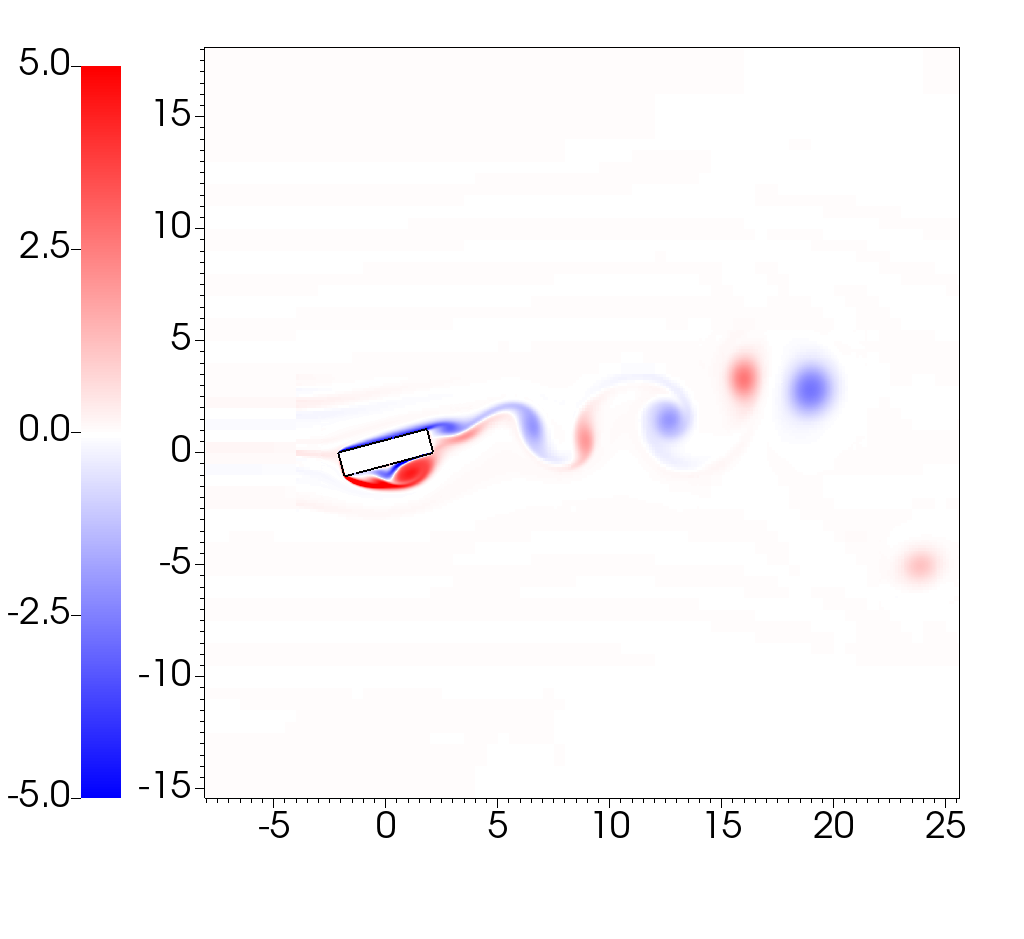}
    \label{fig_gallop_omega_221_25}
  }
    \subfigure[]{
    \includegraphics[scale = 0.2]{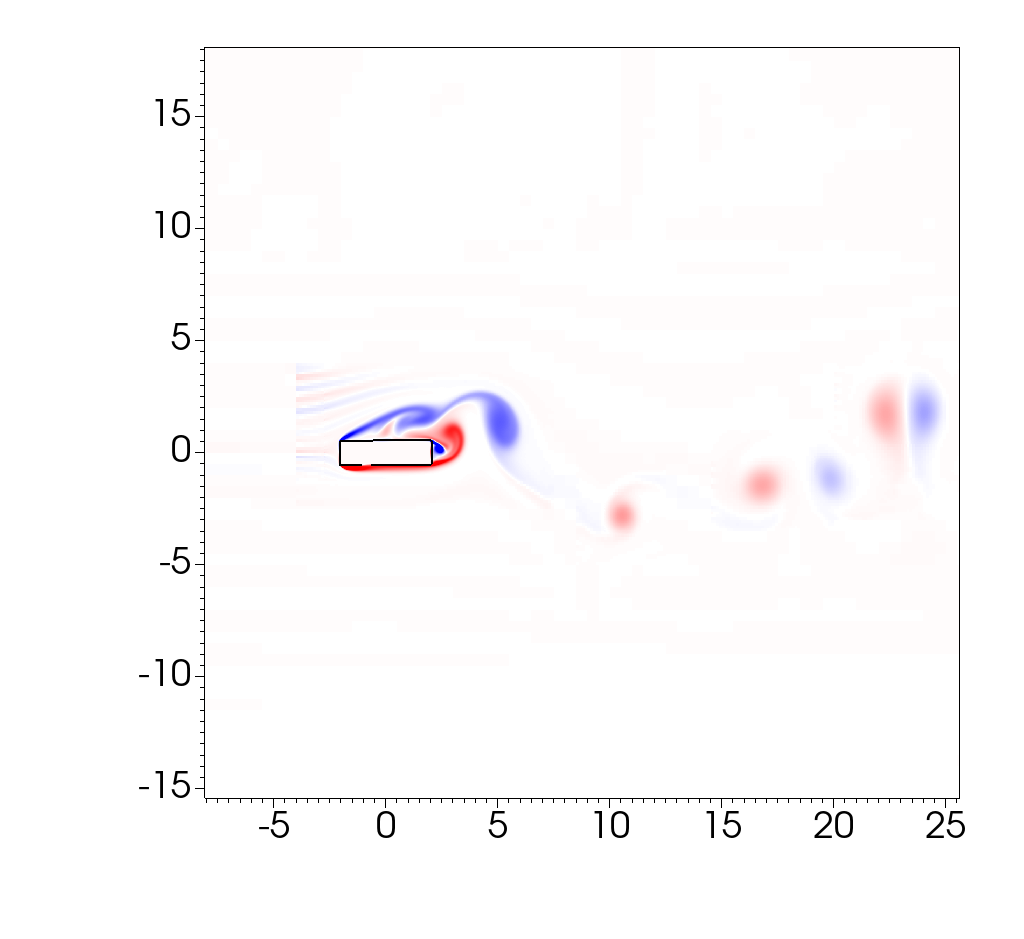}
    \label{fig_gallop_omega_309}
  }
     \subfigure[]{
    \includegraphics[scale = 0.2]{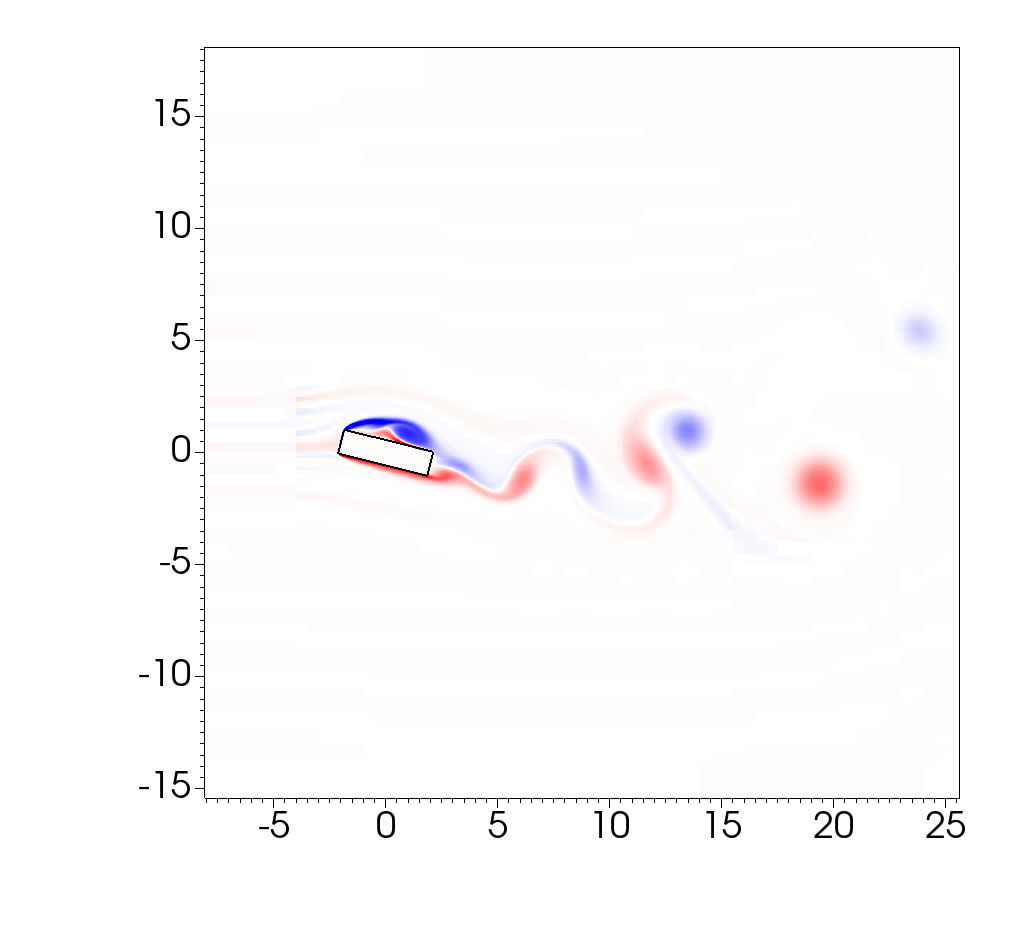}
    \label{fig_gallop_omega_349_5}
  }  
     \subfigure[]{
    \includegraphics[scale = 0.2]{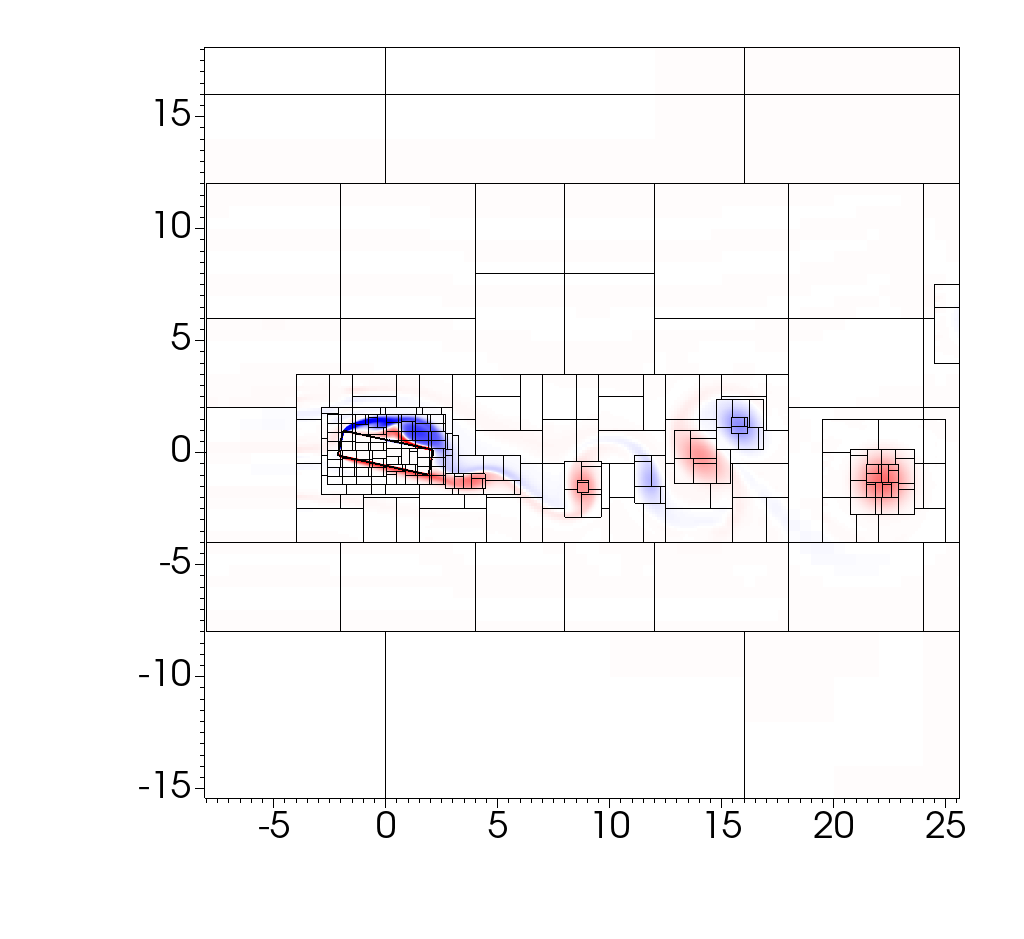}
    \label{fig_gallop_amr_352_5}
  }    
\caption{Galloping motion of a rectangular plate with $\Lambda^* = 4$, $I^* = 400$, $\zeta_\vartheta^* = 0.25$ and $U^* = 40$.\subref{fig_gallop_vs_time} Temporal evolution of pitch angle $\vartheta$; \subref{fig_gallop_omega_221_25}---\subref{fig_gallop_omega_349_5} Vorticity (1/s) plots at three representative time instants t = 221.25 s, t = 309 s, and t = 349.5 s, respectively. \subref{fig_gallop_amr_352_5} Dynamic mesh/patch distribution in the domain at a representative time instant t = 352.5 s.}
  \label{fig_galloping}
\end{figure}

\begin{table}[]
 \centering
 \caption{Comparison of maximum pitch angle $\vartheta_{\rm max}$ and galloping frequency $f_{\vartheta}$ with prior numerical studies.} 
 \rowcolors{2}{}{gray!10}
 \begin{tabular}{*6c}
 \toprule
              & $\vartheta_{\rm max}$ & $f_{\vartheta}$ \\
 \midrule
 Robertson et al.~\cite{Robertson2003}  & $15^{\circ}$ & 0.0191   \\
 Yang \& Stern~\cite{Yang2012}  & $15.7^{\circ}$  & 0.0198   \\
 Yang \& Stern~\cite{Yang2015}  & $16.1^{\circ}$ & 0.0197     \\
 Kolahdouz et al.~\cite{Kolahdouz2020}  & $15^{\circ}$ & 0.0198  \\
 Present  & $15.2^{\circ}$ & 0.0197  \\
 \bottomrule
 \end{tabular}
 \label{tab_gallop}
\end{table}

To compare of our results with prior simulations, we consider a plate with a width-to-thickness ratio of $\Lambda^* = L_{\rm p}/H_{\rm p} = 4$, a non-dimensional moment of inertia of $I^{*}_\vartheta = I_\vartheta/(\rho_{\rm s} H_{\rm p}^4)  = 400$, a non-dimensional damping ratio of $\zeta_\vartheta^* = C_\vartheta/\left(2 \sqrt{K_\vartheta I_\vartheta} \right) = 0.25$, and a reduced velocity of $U^* = U_\infty/(f_\vartheta H_{\rm p}) = 40$. Here, $U_\infty$ is the free stream velocity and $f_\vartheta = \sqrt{K_\vartheta/ I_\vartheta}/2\pi$ is the natural frequency of the spring-mass-damper system. The rectangular plate is centered at the origin with an initial non-zero pitch of $\vartheta = 1^{\circ}$. The computational domain is taken to be $\Omega = [-32 \, \text{cm}, 96\, \text{cm}] \times [-32 \, \text{cm}, 32\, \text{cm}]$, a rectangular domain of size $L_x \times L_y$ = 128 cm $\times$ 64 cm. Five grid levels are used to discretize the domain, with the structure embedded on the finest grid level. A coarse grid spacing of $h_{\rm coarsest} = L_y/32$ is used on the coarsest level. The finest level is refined with a refinement ratio of $\nref = 2$, whereas the rest of the finer levels are refined using a refinement ratio of $\nref = 4$ from their next coarser level. A uniform inflow velocity $\U = \left( U_\infty  = 1\, \text{cm/s}, 0\, \text{cm/s} \right)$ is imposed on the left boundary (x = -32 cm), whereas zero normal traction and zero tangential velocity boundary conditions are imposed on the right boundary (x = 96 cm). The bottom (y = -32 cm) and top (y = 32 cm) boundaries satisfy zero normal velocity and zero tangential traction boundary conditions. The Reynolds number of the flow based on the inlet velocity is set to Re = $\rho_{\rm f} U_\infty H_{\rm p}/\mu_{\rm f} = 250$. A constant time step size of $\Delta t = 0.048\, h_{\rm finest}$ is used for the simulation. After the initial transients, a vortex shedding state is established, which results in a periodic galloping of the rectangular plate. Fig.~\ref{fig_gallop_vs_time} shows the pitch angle of the plate as a function of time. Figs.~\ref{fig_gallop_omega_221_25}-\ref{fig_gallop_omega_349_5} show three representative snapshots of the FSI dynamics and the vortex shedding pattern. Fig.~\ref{fig_gallop_amr_352_5} shows a typical AMR patch distribution in the domain due to the evolving structural and vortical dynamics. Table~\ref{tab_gallop} compares the maximum pitch angle $\vartheta_{\max}$ and galloping frequency of the plate $f_{\vartheta}$ with values obtained from previous numerical studies; an excellent agreement with prior studies is obtained for both these rotational quantities.

%%%%%%%%%%%%%%%%%%%%%%%%%%%%%%%%%
\section{Software implementation}
\label{sec_software}
The FD/BP algorithm and the numerical wave tank method described here is implemented within the IBAMR library~\cite{IBAMR-web-page}, which is an open-source 
C++ simulation software focused on immersed boundary methods with adaptive mesh refinement;
the code is publicly hosted at \url{https://github.com/IBAMR/IBAMR}.
IBAMR relies on SAMRAI~\cite{HornungKohn02, samrai-web-page} for Cartesian grid management and the AMR framework. Linear and
nonlinear solver support in IBAMR is provided by the PETSc library~\cite{petsc-efficient, petsc-user-ref, petsc-web-page}. All of the example cases
in the present work made use of distributed-memory parallelism using the Message Passing Interface (MPI)
library. 
Simulations were carried out on both the XSEDE Comet cluster at the San Diego
Supercomputer Center (SDSC) and the Fermi cluster at San Diego State University (SDSU).
Comet houses 1,944 Intel Haswell standard compute nodes consisting of Intel Xeon 
E5-2680v3 processors with a clock speed of 2.5 GHz, and 24 CPU cores per node.
Fermi houses 45 compute nodes with different generations of Intel Xeon processors.

Between $64$ and $128$ cores were used for the 2D computations presented here, while $128$ cores were used for the 3D
computations. The 2D ISWEC model using the medium grid resolution described in Sec.~\ref{subsec_grid_convergence}
required approximately 6,129 seconds to execute 15,000 time-steps on Comet using $80$ cores. The 3D ISWEC model
using the coarse grid resolution described in Sec.~\ref{subsec_3D_2D_models} required approximately 81,998 seconds
to execute 10,000 time-steps on Fermi using $128$ cores of Intel Xeon E5-2697Av4 Broadwell processors with a clock speed of 2.6 GHz.

%%%%%%%%%%%%%%%%%%%%%%%%%%%%%%%%%
\section{Spatial and temporal resolution tests}
\label{sec_spatiotemporal_tests}
%%%%%%%%%%%%%%%%%%%%%%%%%%%%%%%%%%%%%%%%%%%
In this section, we perform a grid convergence study on the 2D ISWEC model in a NWT with regular waves using three different spatial resolutions. We also conduct a temporal resolution study to determine a time step size $\Delta t$ that is able to adequately resolve the high-frequency wave components of irregular waves. Although our implementation is capable of adaptive mesh refinement, we use \emph{static} grids for all cases presented in this section. As mentioned in Sec.~\ref{sec_wsi_eqs}, we lock all the translational degrees of freedom of the hull and only consider its pitching motion. Appendix~\ref{sec_appendix} compares the rotational dynamics in the presence of heaving motion of the device, and justifies the accuracy of the 1-DOF model to calculate the main quantities of interest such as power output and conversion efficiency of the device.   

The size of the computational domain is $\Omega$ = [0,$10\lambda$] $\times$ [0, $2.75d$] with the origin located at the bottom left corner (see Fig.~\ref{fig_NWT_schematic}). The hull parameters for the 2D model are given in Table~\ref{tab_scaled_hull_values}, and the CG of the hull is located at ($5\lambda, d-Z_{\text{CG}}$). The quiescent water depth is $d = 0.65$ m, acceleration due to gravity is $g = 9.81$ m/s (directed in negative $z$-direction), density of water is $\rhow  = 1025$ kg/m$^{3}$, density of air is $\rho_\text{air} = 1.2$ kg/m$^{3}$, viscosity of water $\mu_\text{w} =10^{-3}$ Pa$\cdot$s and viscosity of air is $\mu_{\text{air}} = 1.8\times 10^{-5}$ Pa$\cdot$s. Surface tension effects are neglected for all cases as they do not affect the wave and converter dynamics at the scale of these problems. 

%%%%%%%%%%%%%%%%%%%%%%%%%%%%%%%%%%%%%%%%%%%%

\subsection{Grid convergence study} \label{subsec_grid_convergence}

To ensure the wave-structure interaction dynamics are accurately resolved, we conduct a grid convergence study to determine an adequate mesh spacing.
The dynamics of the ISWEC hull interacting with regular water waves are simulated on three meshes: coarse, medium, and fine.
Each mesh consists of a hierarchy of $\ell$ grids; the computational domain is discretized by a coarsest grid of size $N_{y} \times N_{z}$ and then locally refined $\ell - 1$ times by an integer refinement ratio $n_{\text{ref}}$ ensuring that the ISWEC device and air-water interface are covered by the finest grid level.
The grid spacing on the finest level are calculated using the following expressions: $\Delta y_\text{min} = \Delta y_{0}/n^{\ell-1}_\text{ref}$ and $\Delta z_\text{min} = \Delta z_{0}/n^{\ell-1}_\text{ref}$, where $\Delta y_{0}$ and $\Delta z_{0}$ are the grid spacings on the coarsest level.
The time step size $\dt$ is chosen to ensure the maximum Courant-Friedrichs-Levy (CFL) number $= 0.12$ for each grid resolution.
The mesh parameters and time step sizes considered here are shown in Table~\ref{tab_grid_convergence_study}.

\begin{figure}[h!]
 \centering
 \subfigure[Hull pitch angle]{
  \includegraphics[scale = 0.3]{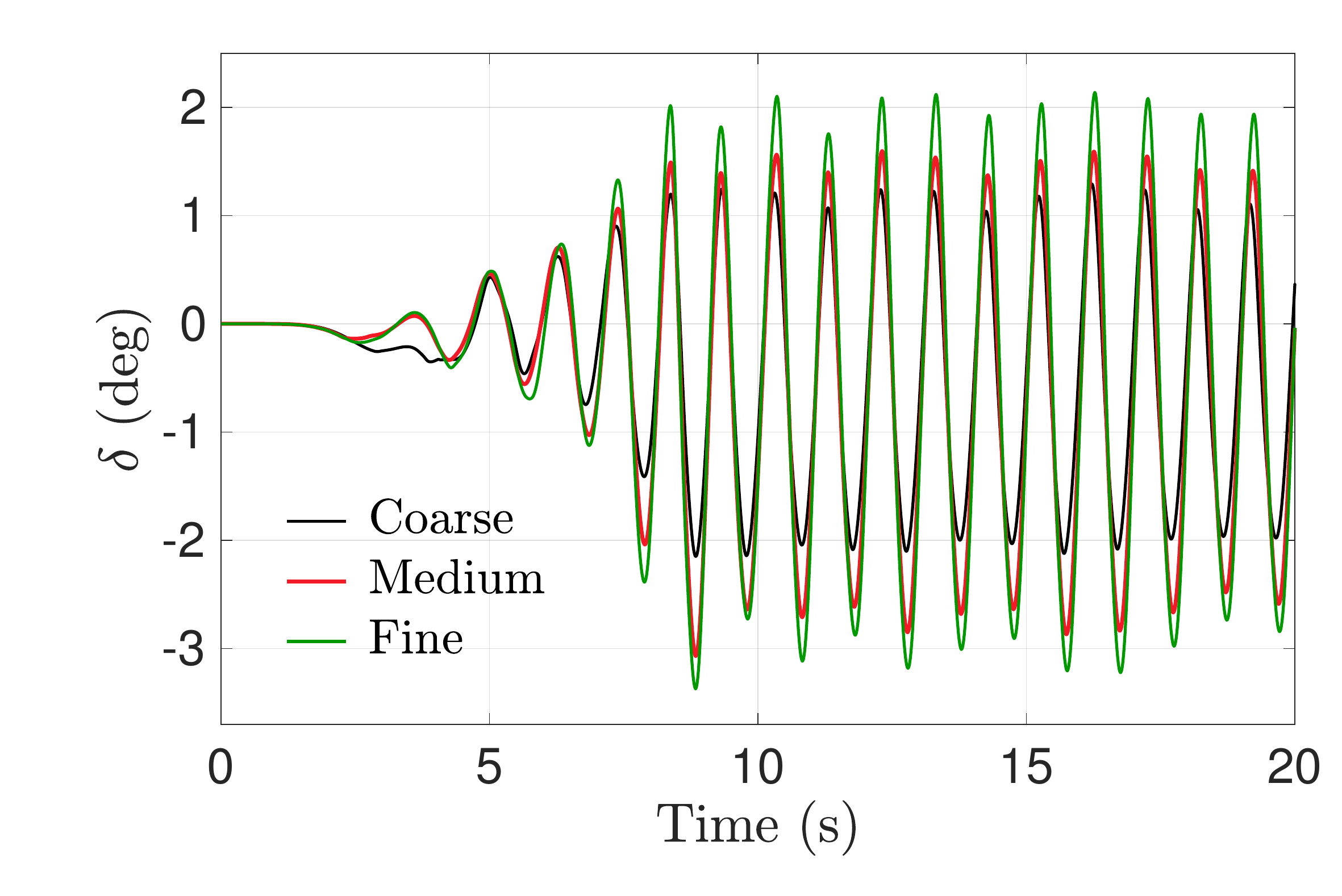}
  \label{pitch_angle_vs_time}
 }
   \subfigure[Gyroscope precession angle]{
  \includegraphics[scale = 0.3]{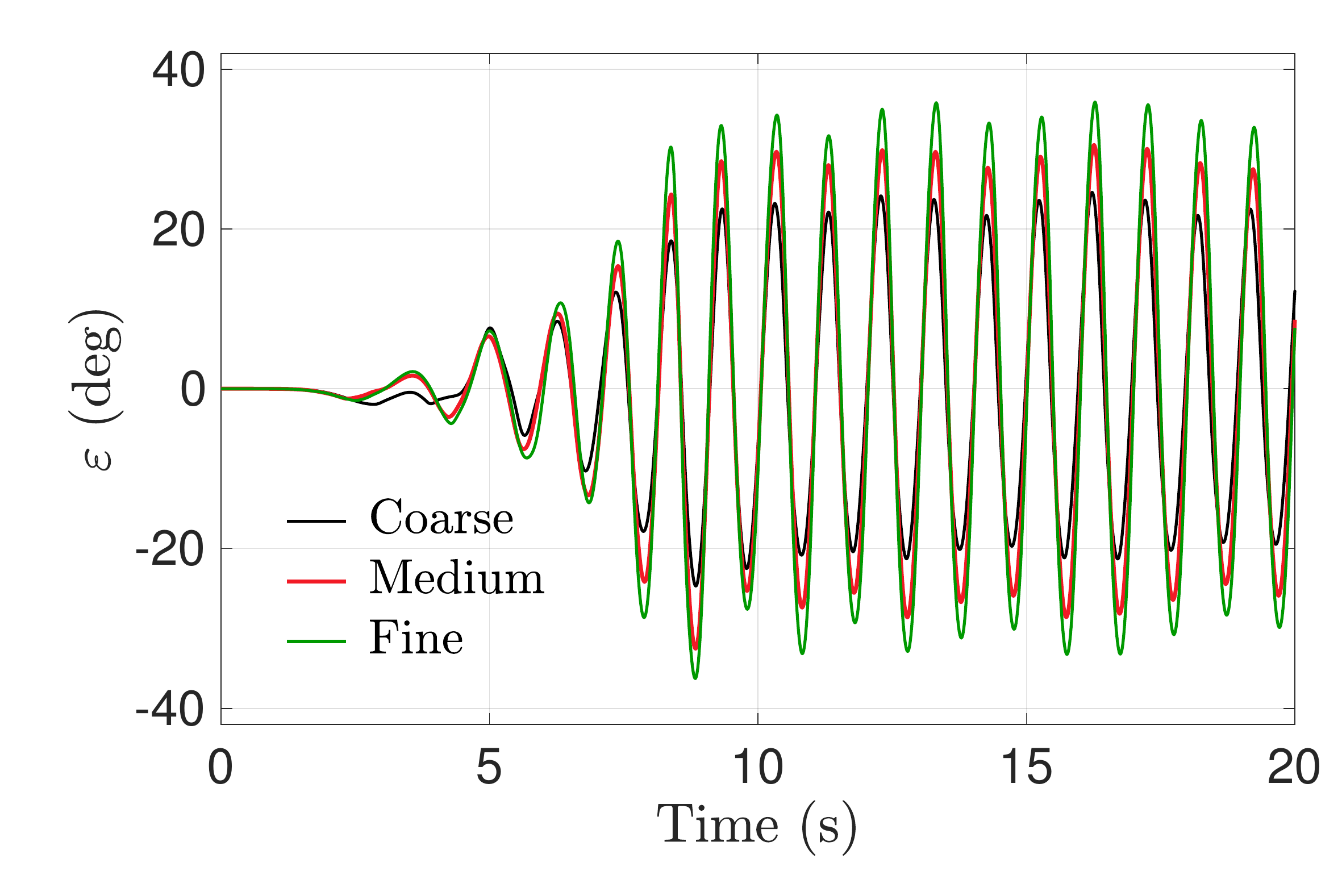}
  \label{percession_angle_vs_time}
 }
\caption{Temporal evolution of~\subref{pitch_angle_vs_time} hull pitch angle $\delta$ 
and~\subref{percession_angle_vs_time} gyroscope precession angle $\varepsilon$, for coarse (\textcolor{black}{\textbf{-----}}, black), medium (\textcolor{red}{\textbf{-----}}, red) and fine (\textcolor{ForestGreen}{\textbf{-----}}, green) grid resolutions.
Fifth-order regular water waves are generated with $\cH$ = 0.1 m, $\cT$ = 1 s and $\lambda$ = 1.5456 m,
satisfying the dispersion relation given by Eq.~\ref{eq_dispersion_relation}.
A maximum ISWEC pitch angle $\delta_0$ = 5$^\circ$ and a maximum gyroscope precession angle of $\varepsilon_0$ = 70$^\circ$ are used to calculate the rest of the parameters following the procedure described in
 Sec.~\ref{sec_PTO_params}: PTO damping coefficient $c$ = 0.3473 N$\cdot$m$\cdot$s/rad, gyroscope moment of inertia 
 $J = 0.0116$ kg$\cdot$m$^{2}$ and PTO stiffness coefficient $k$ = 0.4303 N$\cdot$m/rad.
The speed of the flywheel is $\dot{\phi} = 4000$ RPM, and $I = 0.94 \times J = 0.0109$ kg
 $\cdot$m$^{2}$.
}
 \label{fig_grid_convergence_plots}
\end{figure}

\begin{figure}[h!]
   \centering
   \subfigure[Locally refined Cartesian grid (medium grid resolution)]{
   	\includegraphics[scale = 0.48]{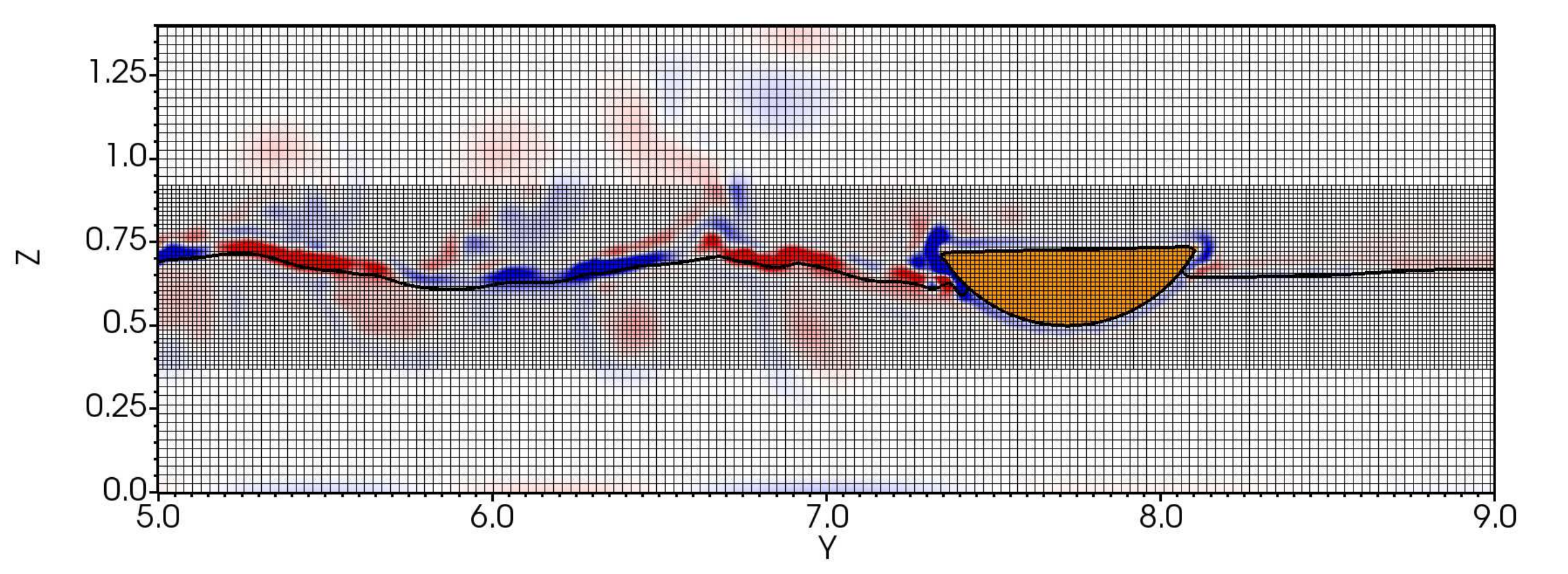}
	\label{fig_iswec2d_mesh}
   }
   \subfigure[Vorticity (medium grid resolution)]{
   	\includegraphics[scale= 0.5]{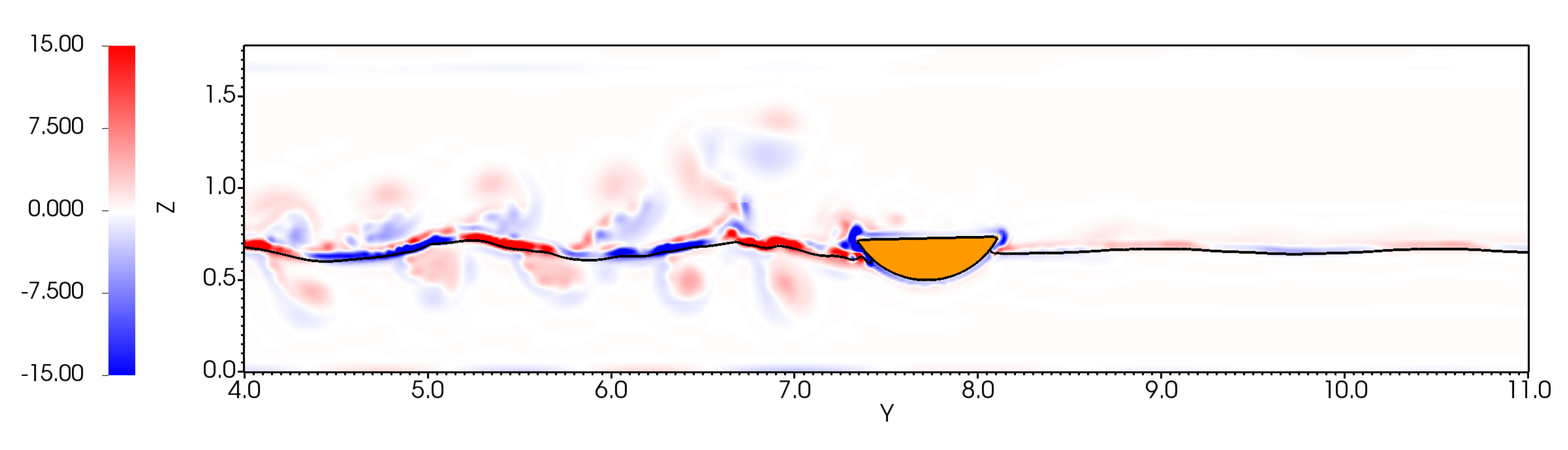}
	\label{fig_iswec2d_vorticity}
   }
   \caption{Wave-structure interaction of the 2D ISWEC model at $t = 27$ s using the medium grid resolution:~\subref{fig_iswec2d_mesh} locally refined mesh with two levels ($\ell = 2$) and~\subref{fig_iswec2d_vorticity} representative vortical and air-water interfacial dynamics resulting from the WSI.
}
   \label{fig_2d_regular_medium}
\end{figure}

Regular water waves, generated based on the fifth-order wave theory presented in Sec.~\ref{sec_regular_waves}, enter the left side of
the domain and interact with the ISWEC hull. Temporal evolution of the hull's pitch angle $\delta$ and the gyroscope's precession angle 
$\varepsilon$ are the primary outputs used to evaluate mesh convergence. The results and the specification of the wave, ISWEC, and gyroscope parameters are shown in Fig.~\ref{fig_grid_convergence_plots}. 
%Between the medium and fine meshes, the maximumpercent changes in $\delta$ and $\varepsilon$ are $xxx$ and $yyy$, respectively \NN{TODO: KK, please calculate these from the data}.
Fig.~\ref{fig_iswec2d_mesh} shows a close-up of the medium resolution grid and the 2D ISWEC model.
A minimum of $8$ grid cells vertically span the height of the wave, indicating that the wave elevation is adequately resolved; for the coarse (fine) grid resolution, approximately $4$ ($15$) grid cells span the wave height (results not shown).
Additionally, the number of cells covering the ISWEC hull length is approximately $30$, $60$, and $119$ for the coarse, medium, and fine grid resolutions, respectively. 
Fig.~\ref{fig_iswec2d_vorticity} shows well-resolved vortical structures produced by the interaction of the ISWEC device and air-water interface on the medium resolution grid. From the quantitative and qualitative results shown in Fig.~\ref{fig_grid_convergence_plots} and Fig.~\ref{fig_2d_regular_medium}, we conclude that the medium grid resolution can capture the WSI dynamics with reasonable accuracy. Therefore for the remaining cases studied here, we make use of the medium grid resolution.

\begin{table}[]
 \centering
 \caption{Refinement parameters used for the 2D ISWEC dynamics grid convergence study.}
 \rowcolors{2}{}{gray!10}
 \begin{tabular}{*6c}
 \toprule
 Parameters & Coarse & Medium & Fine \\
 \midrule
  $n_{\text{ref}}$ & 2   & 2   & 4    \\
  $\ell$         & 2   & 2   & 2    \\
 $N_{y}$       & 300 & 600 & 600  \\
 $N_{z}$       & 34  & 68  & 68   \\
 $\Delta t$ (s)   & $2\times 10^{-3}$ & $1\times 10^{-3}$ & $5\times 10^{-4}$ \\
 \bottomrule
 \end{tabular}
 \label{tab_grid_convergence_study}
\end{table}

%%%%%%%%%%%%%%%%%%%%%%%%%%%%%%%%%%%%%%%%%%%%

\subsection{Temporal resolution study} \label{temporal_resol_study}
Next, we conduct a temporal resolution study to ensure that WSI dynamics of irregular waves and the ISWEC device are adequately resolved. 
As described in Sec.~\ref{sec_irregular_waves}, irregular water waves are modeled as a superposition of $N$ harmonic wave components. The energy carried by each wave component is related to its frequency $\omega_i$ (see Eq.~\ref{eq_JONSWAP_wave_spectrum} and Fig.~\ref{fig_JONSWAP_spectrum}).
Moreover as shown in Fig.~\ref{fig_JONSWAP_spectrum}, high frequency waves with $\omega_i$ in the range of 10 rad/s to 20 rad/s carry considerable amounts of energy.
Hence, the time step $\dt$ should be chosen such that these high frequency (small wave period $\cT_i$) components are well-resolved since they contribute significantly to the power absorbed by the device.

\begin{figure}[h!]
   \centering
   \subfigure[Hull pitch angle]{
   	\includegraphics[scale = 0.3]{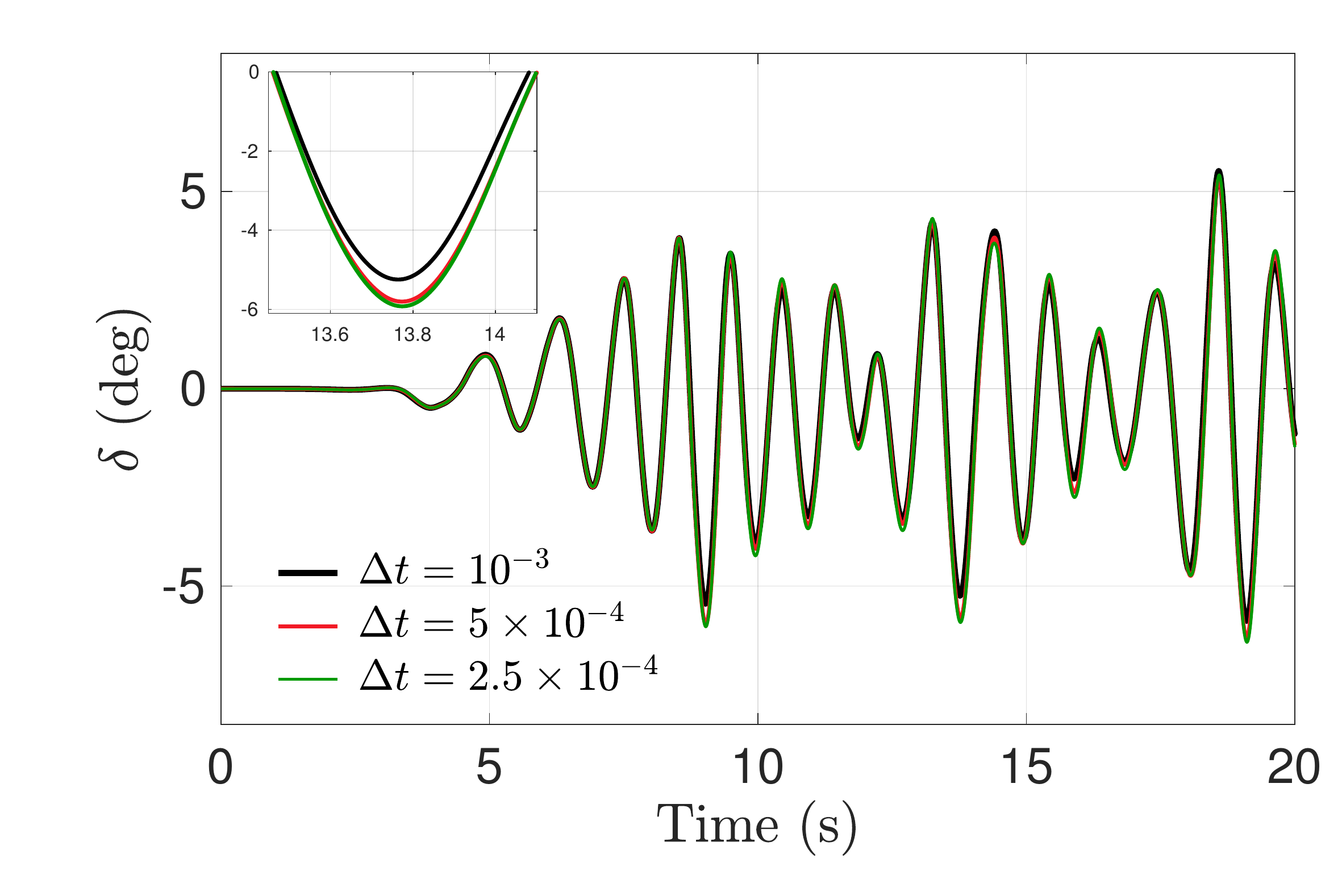}
	\label{fig_temporal_resol_pitch}
   }
   \subfigure[Power absorbed by PTO]{
   	\includegraphics[scale= 0.3]{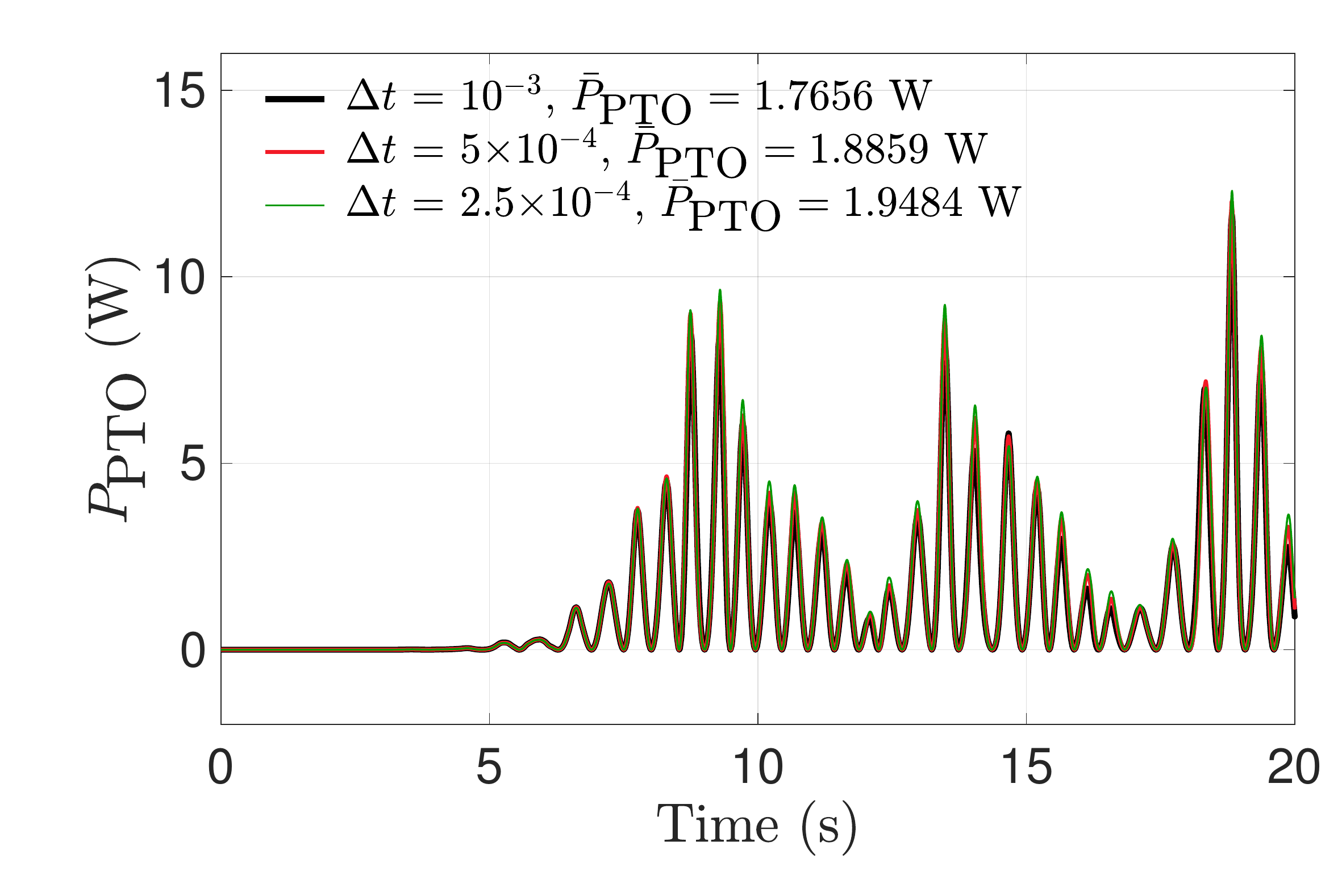}
	\label{fig_temporal_resol_power}
   }
   \caption{Temporal evolution of~\subref{fig_temporal_resol_pitch} hull pitching dynamics and~\subref{fig_temporal_resol_power} power absorbed by PTO, for three different time step sizes: $\dt = 10^{-3}$ s (\textcolor{black}{\textbf{-----}}, black), $\dt = 5 \times 10^{-4}$ s (\textcolor{red}{\textbf{-----}}, red) and $\dt = 2.5 \times 10^{-4}$ s (\textcolor{ForestGreen}{\textbf{-----}}, green).
Irregular water waves (satisfying the dispersion relation Eq.~\ref{eq_dispersion_relation}) are generated with $\cH_\text{s}$ = 0.1 m, $\cT_\text{p}$ = 1 s and 
 $N$ = 50 wave components, with frequencies in the range $3.8$ rad/s to $20$ rad/s distributed at equal $\Delta \omega$ intervals.
A maximum ISWEC pitch angle $\delta_0 = 5^\circ$ and a maximum gyroscope precession angle of $\varepsilon_0 = 70^{\circ}$ are used to calculate the rest of the parameters following the procedure described in
 Sec.~\ref{sec_PTO_params}: PTO damping coefficient $c = 0.1724$ N$\cdot$m$\cdot$s/rad, gyroscope moment of inertia 
 $J = 0.0057$ kg$\cdot$m$^{2}$ and PTO stiffness coefficient $k$ = 0.2138 N$\cdot$m/rad.
The speed of the flywheel is $\dot{\phi} = 4000$ RPM, and $I = 0.94 \times J = 0.0054$ kg$\cdot$m$^{2}$.
 The mean wave power per unit crest width carried by the irregular waves calculated by Eq.~\ref{eq_irreg_wave_power} is
 $\bar{P}_{\rm wave} = 5.0798$ W.}
   \label{fig_irregwave_temporal_resol}
\end{figure}

The dynamics of the ISWEC hull interacting with irregular water waves are simulated using three different time step sizes: $\dt =$ $10^{-3}$ s, $5 \times 10^{-4}$ s and $2.5 \times 10^{-4}$ s. For all three cases, we use a medium resolution grid with refinement parameters given by Table~\ref{tab_grid_convergence_study}. 
Temporal evolution of the hull's pitch angle $\delta$ and the power absorbed by the PTO unit $P_\text{PTO}$ are the primary outputs
used to evaluate temporal convergence. The results and the specification of the irregular wave, ISWEC, and gyroscope parameters are shown in Fig.~\ref{fig_irregwave_temporal_resol}. It is observed that the hull's pitching motion is relatively insensitive to the chosen time step size $\dt$. Since its dynamics are governed mainly by those waves carrying the largest energy, we can conclude that the higher frequency wave components are adequately resolved.
The difference between the three temporal resolutions is more apparent in Fig.~\ref{fig_temporal_resol_power},
in which we calculate the average power absorbed by the PTO unit $\bar{P}_\text{PTO}$ over the interval $t = 10$ s and $t = 20$ s.
For $\dt = 10^{-3}$ s, $5 \times 10^{-4}$ s and $2.5 \times 10^{-4}$ s, the power absorbed is $\bar{P}_\text{PTO} = 1.7656$ W, $1.8859$ W and $1.9484$ W, respectively.
It is seen that smaller time step sizes allow for the resolution of higher-frequency wave peaks, which directly increases the absorbed power.

Based on these results, we hereafter use the medium grid spatial resolution, and time step sizes of $\dt = 1 \times 10^{-3}$ s and $\dt = 5 \times 10^{-4}$ s for regular and irregular wave WSI cases, respectively.

%%%%%%%%%%%%%%%%%%%%%%%%%%%%%%%%%
\section{Results and discussion}
\label{sec_results_and_discussion}
%%%%%%%%%%%%%%%%%%%%%%%%%%%%%%%%%%%%%%%%%%%
In this section, we investigate several aspects of the dynamics of the inertial sea wave energy converter device:
\begin{itemize}

\item First, we compare the PTO power predictions by the 3D and 2D ISWEC models under identical wave conditions. Utilizing the scaling factors presented in Table~\ref{tab_Fr_scaling}, we show that the power predicted by the 3D model can be inferred from the power predicted by the 2D model reasonably well. 

\item Next, we study the effect of the maximum hull pitch angle parameter $\delta_0$ and make recommendations on how to select it based on the maximum wave steepness $\delta_\text{s}$. We consider different ``sea states" characterized by regular waves of different heights $\cH$, and consequently of different steepnesses. 
%The wave conditions and the corresponding PTO and gyroscope parameters are summarized in Table~\ref{tab_gyro_parameters_regularwave}.

\item Thereafter, a parametric analysis for the 2D ISWEC model is performed using both regular and irregular water waves to study its dynamics. We vary the following parameters to recommend ``design" conditions for the device: PTO damping coefficient $c$, flywheel speed $\dot{\phi}$, moment of inertia $J$ and $I$, and PTO stiffness coefficient $k$. 

\item Afterwards, the effect of varying hull length to wavelength ratios is studied. 
%Additionally, throughout this work we have assumed that the hull is constrained to pitching motion and its remaining degrees of freedom are locked; in this section, we further explore this assumption. 

\item Finally, we simulate a possible device protection strategy during inclement weather conditions and study the resulting dynamics. 
\end{itemize}

All the 2D simulations are conducted in a NWT with computational domain size $\Omega$ = [0,$10\lambda$] $\times$ [0, $2.75d$] as shown in Fig.~\ref{fig_NWT_schematic}. For 3D cases the computational domain size is same as in 2D, with the additional dimension having length $5W$; $W$ is the width of 3D model of the hull. The domain sizes are large enough to ensure that the ISWEC dynamics are undisturbed by boundary effects. The origin of the NWT is taken to be the bottom left corner of the domain and shown by the point $O$ in Fig.~\ref{fig_NWT_schematic}. The CG of the ISWEC hull is located at ($2.5\,W, 5\lambda, d-Z_{\text{CG}}$) for 3D cases and ($5\lambda, d-Z_{\text{CG}}$) for the 2D cases. The rest of the hull parameters are presented in Table \ref{tab_scaled_hull_values}.
The water and air material properties are the same as those described in Sec.~\ref{sec_spatiotemporal_tests}.

\begin{table}[]
 \centering
 \caption{The PTO and gyroscope parameters for various regular wave heights $\cH$ and $\delta_0$ values, as calculated by the procedure described in Sec.~\ref{sec_PTO_params}. The rated power of the device $\bar{P}_\text{R}$ is taken to be the available wave power $\bar{P}_\text{wave}$ for these calculations. The prescribed gyroscope parameters are $\varepsilon_0$ = 70$^\circ$, $\dot{\phi}$ = 4000 RPM, and $I = 0.94 \times J$. The parameter units for $c$ are N$\cdot$m$\cdot$s/rad, $J$ and $I$ are kg$\cdot$m$^2$, and $k$ are N$\cdot$m/rad.}
\begin{tabular}{c c c c c c c c}
\toprule
\multirow{2}{*}{Regular wave properties}   & \multirow{2}{*}{Parameters} & \multicolumn{6}{c}{Prescribed hull pitch angle $\delta_0$}     \\
\cline{3-8}
                             &       &  2$^\circ$  &  5$^\circ$  &  10$^\circ$  &  15$^\circ$  &  20$^\circ$  &  $\delta_s$    \\
\midrule
\multirow{3}{*}{$\cH$ = 0.025 m and $\cT$ = 1 s} 
& \cellcolor{gray!10} $c$    & \cellcolor{gray!10} 0.0217   & \cellcolor{gray!10} 0.0217   & \cellcolor{gray!10} 0.0217   & \cellcolor{gray!10}  0.0217   & \cellcolor{gray!10} 0.0217    & \cellcolor{gray!10} 0.0217     \\
&                             $J$      & 0.0018   & 0.00072   & 0.00036   & 0.00024   & 0.00018   & 0.0012     \\ 
& \cellcolor{gray!10}$k$     & \cellcolor{gray!10} 0.0673   & \cellcolor{gray!10} 0.0269   & \cellcolor{gray!10} 0.0134   & \cellcolor{gray!10} 0.0089   & \cellcolor{gray!10} 0.0067    & \cellcolor{gray!10} 0.0464      \\
\midrule
\multirow{3}{*}{$\cH$ = 0.05 m and $\cT$ = 1 s}  
&                               $c$   & 0.0868   & 0.0868    & 0.0868    & 0.0868    & 0.0868    & 0.0868       \\
& \cellcolor{gray!10} $J$   & \cellcolor{gray!10} 0.0072   & \cellcolor{gray!10} 0.0029   & \cellcolor{gray!10} 0.0014   & \cellcolor{gray!10} 0.00090   &\cellcolor{gray!10} 0.00072   & \cellcolor{gray!10} 0.0025       \\
&                               $k$   & 0.2692    & 0.1076    & 0.0538    & 0.0358    & 0.0269    & 0.0928     \\
\midrule
\multirow{3}{*}{$\cH$ = 0.1 m and $\cT$ = 1 s}   
& \cellcolor{gray!10} $c$    & \cellcolor{gray!10} 0.3473    & \cellcolor{gray!10} 0.3473    & \cellcolor{gray!10} 0.3473    & \cellcolor{gray!10} 0.3473    & \cellcolor{gray!10} 0.3473    &\cellcolor{gray!10}  0.3473     \\
&                              $J$     & 0.0290   & 0.0116    & 0.0058    & 0.0039    & 0.0029    & 0.0050     \\
& \cellcolor{gray!10} $k$    & \cellcolor{gray!10} 1.0777   & \cellcolor{gray!10} 0.4303    & \cellcolor{gray!10} 0.2171    & \cellcolor{gray!10} 0.1421    & \cellcolor{gray!10} 0.1065    & \cellcolor{gray!10} 0.1876     \\
\midrule
\multirow{3}{*}{$\cH$ = 0.125 m and $\cT$ = 1 s} 
&                               $c$     & 0.5427 & 0.5427  & 0.5427  & 0.5427  & 0.5427  & 0.5427    \\
& \cellcolor{gray!10} $J$     & \cellcolor{gray!10} 0.0453 & \cellcolor{gray!10} 0.0181  & \cellcolor{gray!10} 0.0090  & \cellcolor{gray!10} 0.0060  & \cellcolor{gray!10} 0.0045  & \cellcolor{gray!10} 0.0063    \\
&                              $k$     & 1.6827  & 0.6731   & 0.3365   & 0.2243   & 0.1682   & 0.2361   \\
\bottomrule                             
 \end{tabular}
 \label{tab_gyro_parameters_regularwave}
\end{table}

%%%%%%%%%%%%%%%%%%%%%%%%%%%%%%%%%%%%
\subsection{3D and 2D ISWEC models} \label{subsec_3D_2D_models}

\begin{figure}[]
   \centering
   \subfigure[Locally refined Cartesian mesh]{
   	\includegraphics[scale= 0.35]{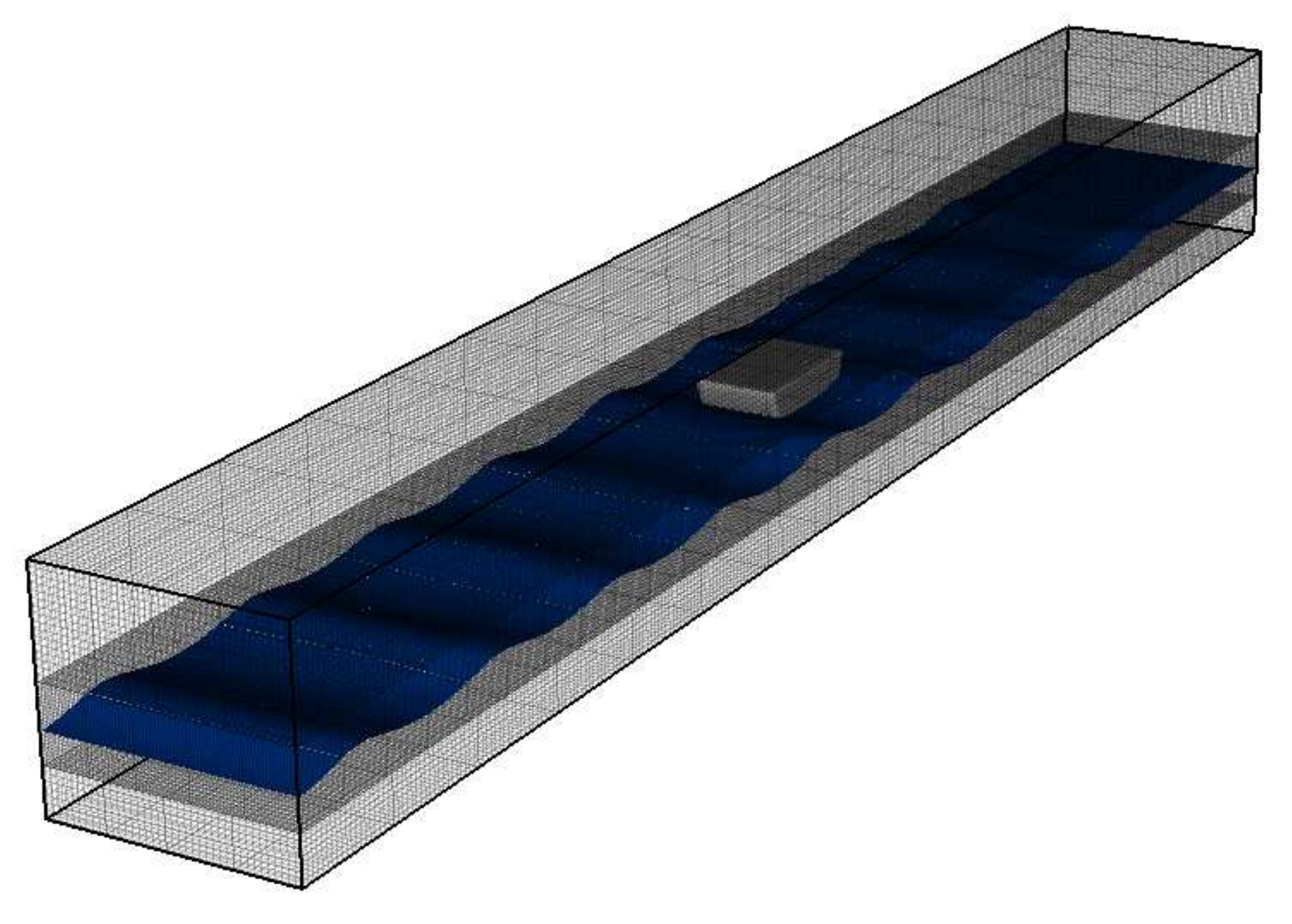}
	\label{fig_iswec3d_mesh}
   }
   \subfigure[Regular waves]{
   	\includegraphics[scale = 0.35]{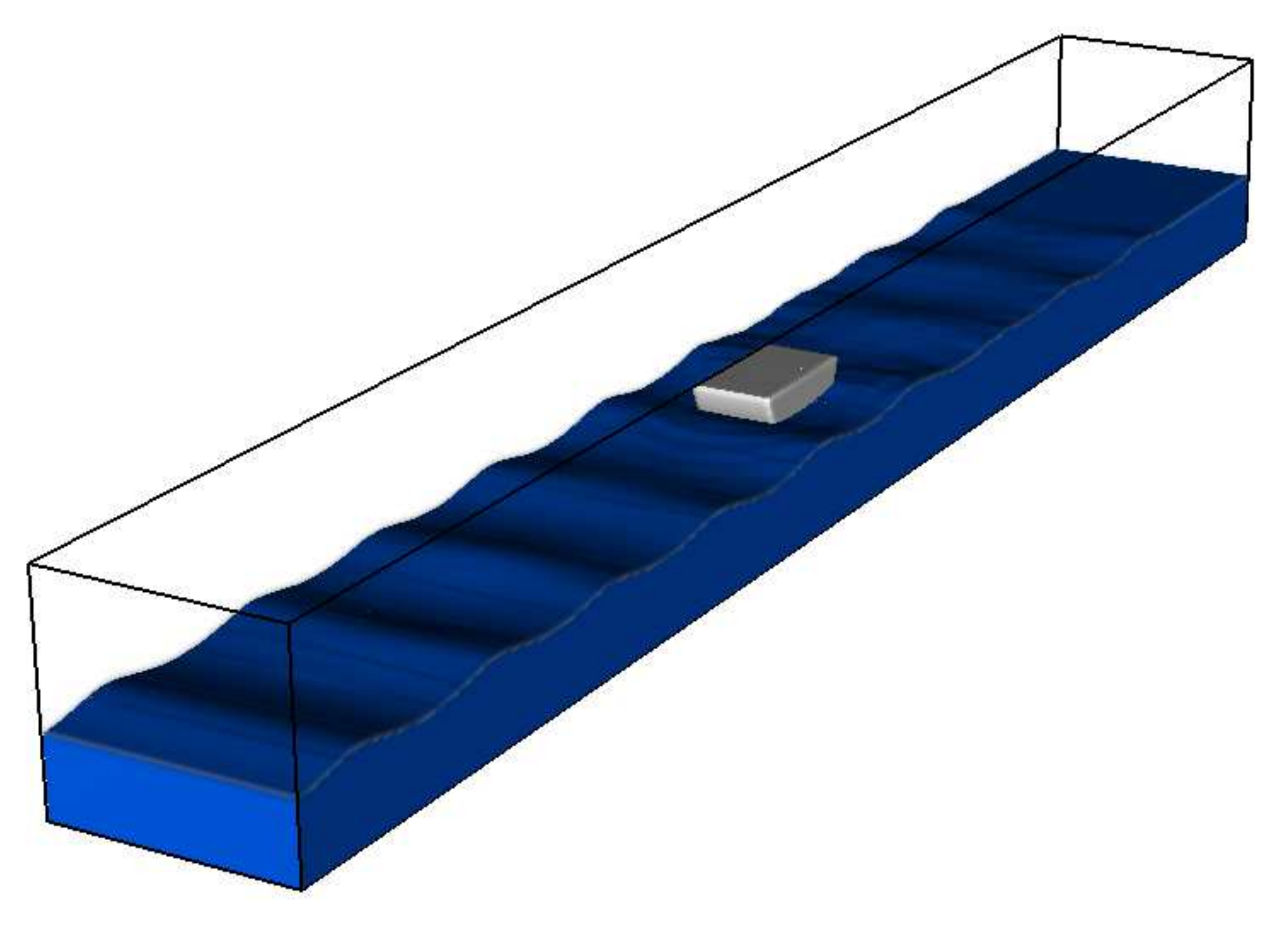}
	\label{fig_iswec3d_regular}
   }
   \subfigure[Irregular waves]{
   	\includegraphics[scale= 0.35]{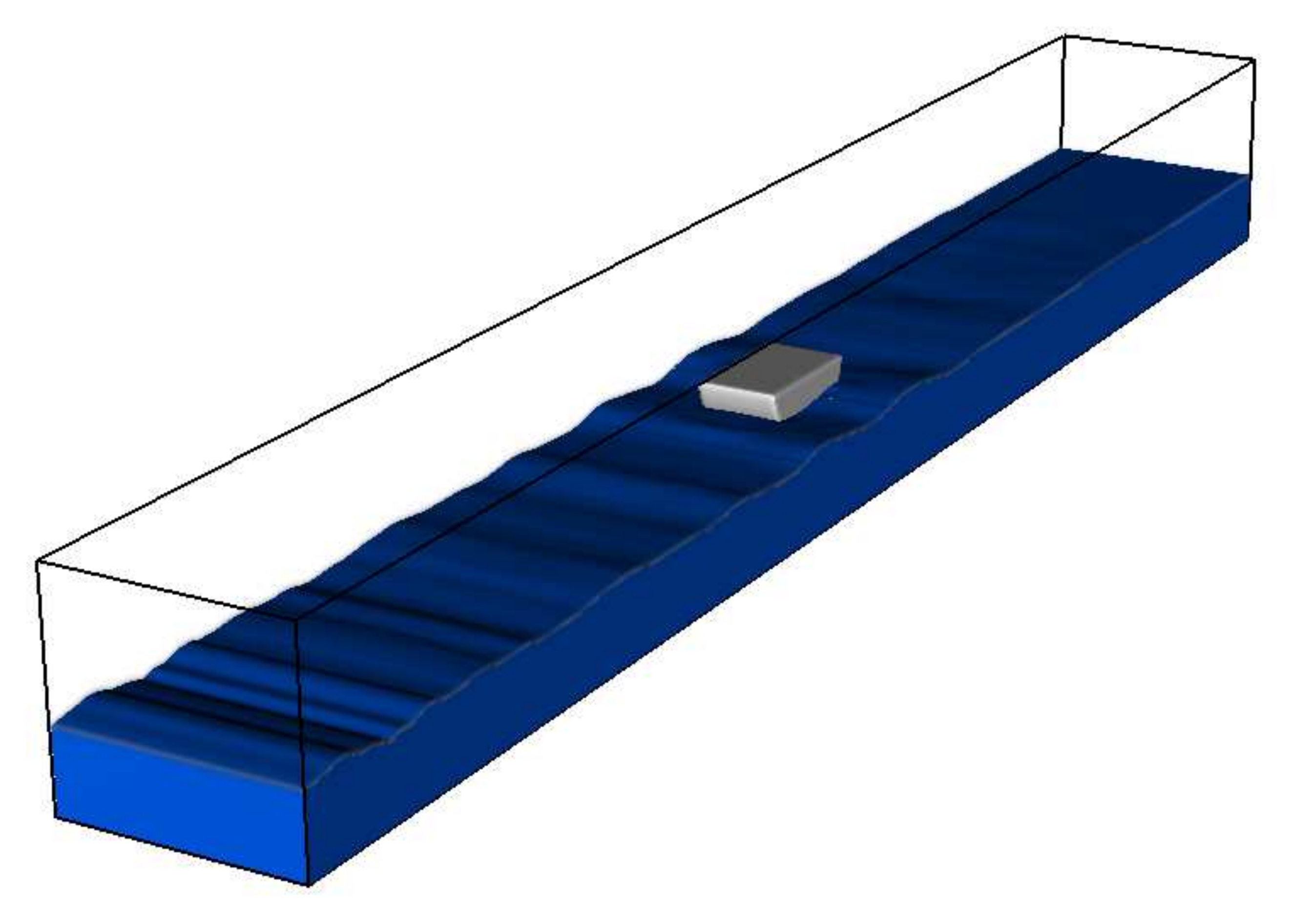}
	\label{fig_iswec3d_irregular}
   }
   \caption{\subref{fig_iswec3d_mesh} Locally refined Cartesian mesh with two levels of mesh refinement used for the 3D NWT. Representative WSI of the 3D ISWEC model at $t$ = 28.8 s: ~\subref{fig_iswec3d_regular} regular waves, and~\subref{fig_iswec3d_irregular} irregular waves.}
   \label{fig_3d_regular_irregular_coarse}
\end{figure}

In this section, we investigate the dynamics of the 2D and 3D ISWEC models interacting with regular and irregular water waves. We compare the motion of the hull and the power absorption capabilities of each model.
The 2D model is simulated on a medium grid resolution and the 3D model on a coarse grid resolution using the refinement parameters specified in Table~\ref{tab_grid_convergence_study}. The third dimension is discretized with $N_x$ = 38 grid cells for 3D cases. Fig.~\ref{fig_iswec3d_mesh} shows the configuration of the locally refined mesh ($\ell = 2$), along with visualizations of regular and irregular waves for the three-dimensional NWT.

\begin{figure}[]
  \centering
   \subfigure[Hull motion for $\delta_0$ = 5$^\circ$]{
    \includegraphics[scale = 0.3]{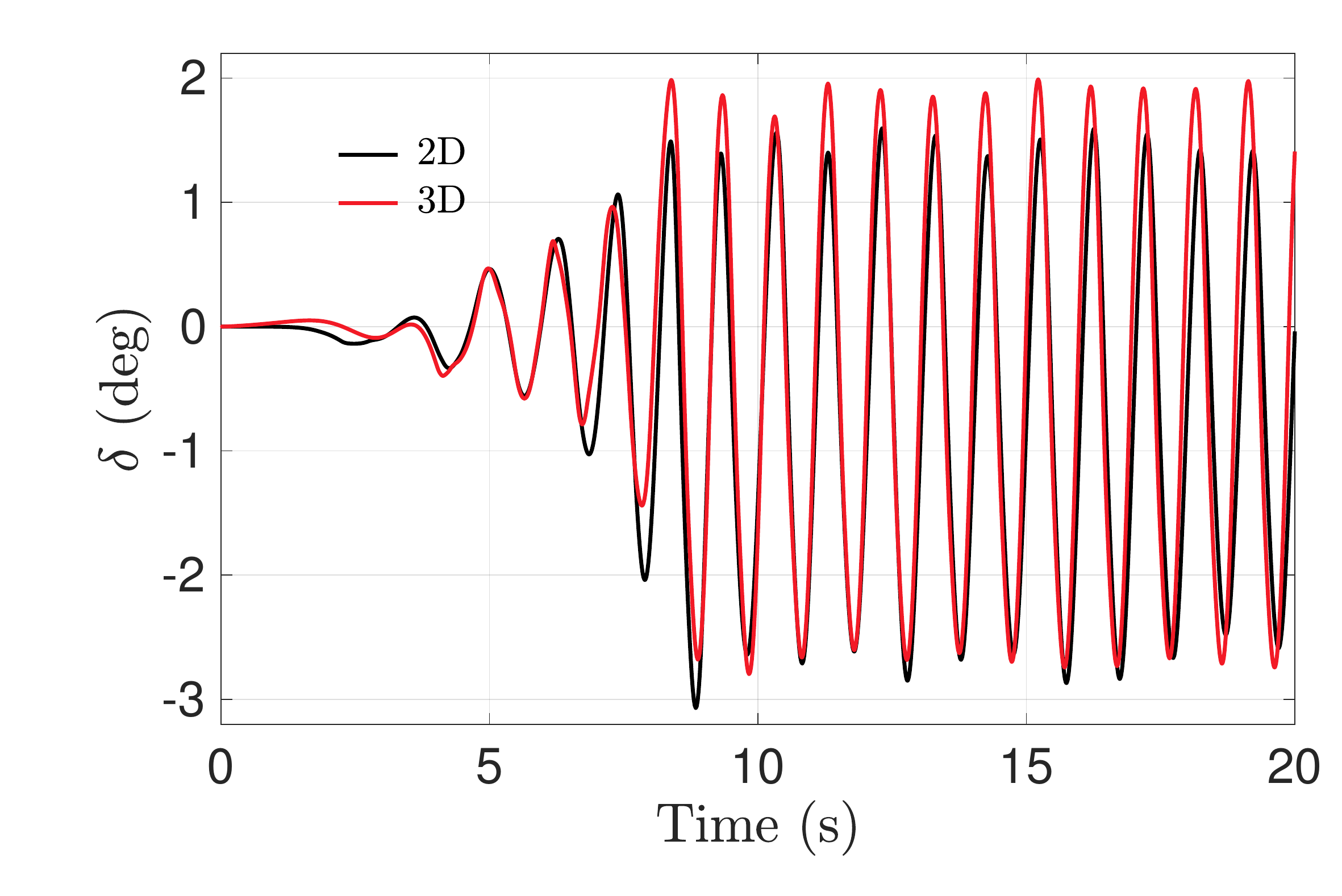}
    \label{fig_regwave_delta5deg_2d3d}
  }
     \subfigure[Hull motion for $\delta_0$ = 20$^\circ$]{
    \includegraphics[scale = 0.3]{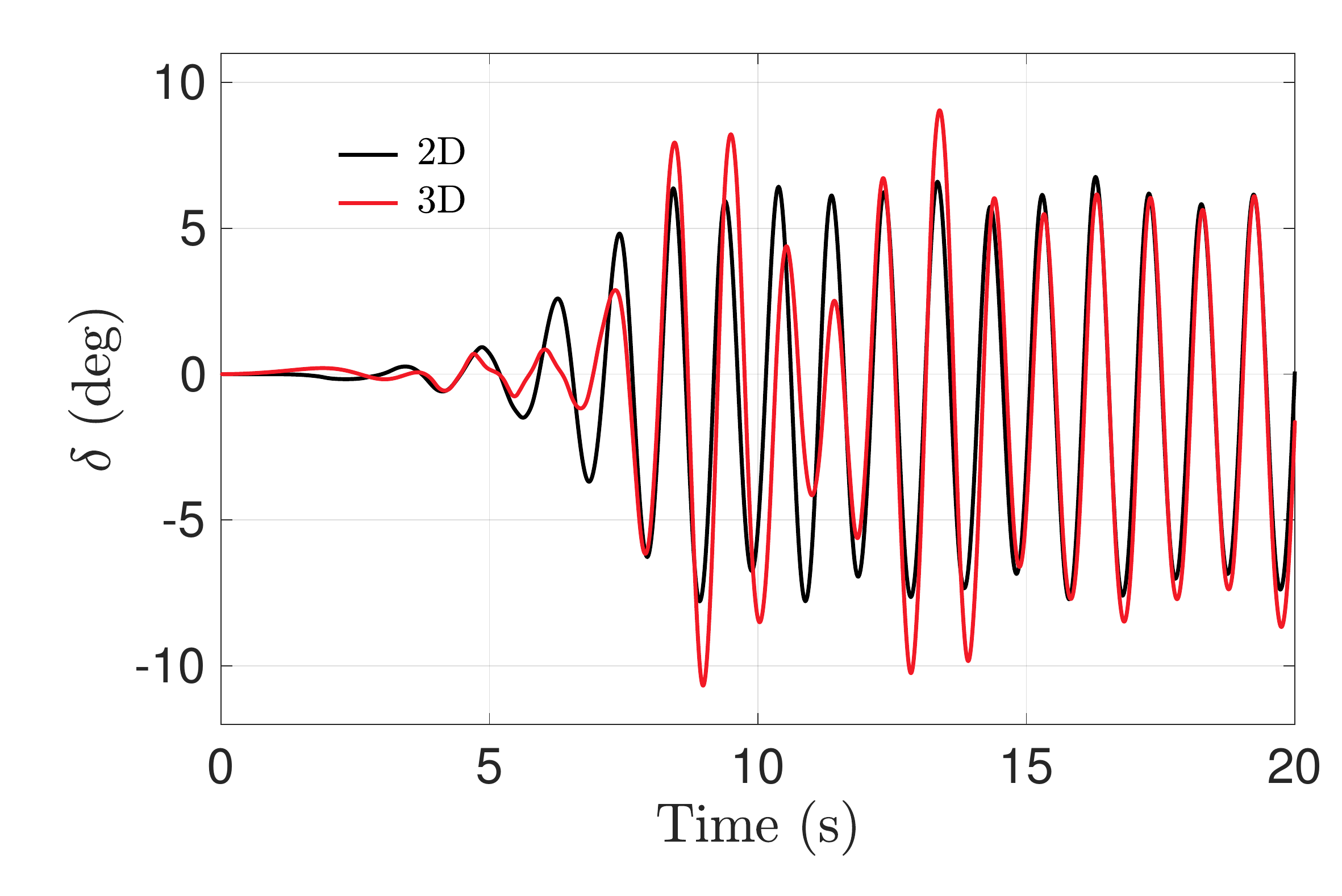}
    \label{fig_regwave_delta20deg_2d3d}
  }
   \subfigure[Powers for $\delta_0$ = 5$^\circ$ (2D model)]{
    \includegraphics[scale = 0.3]{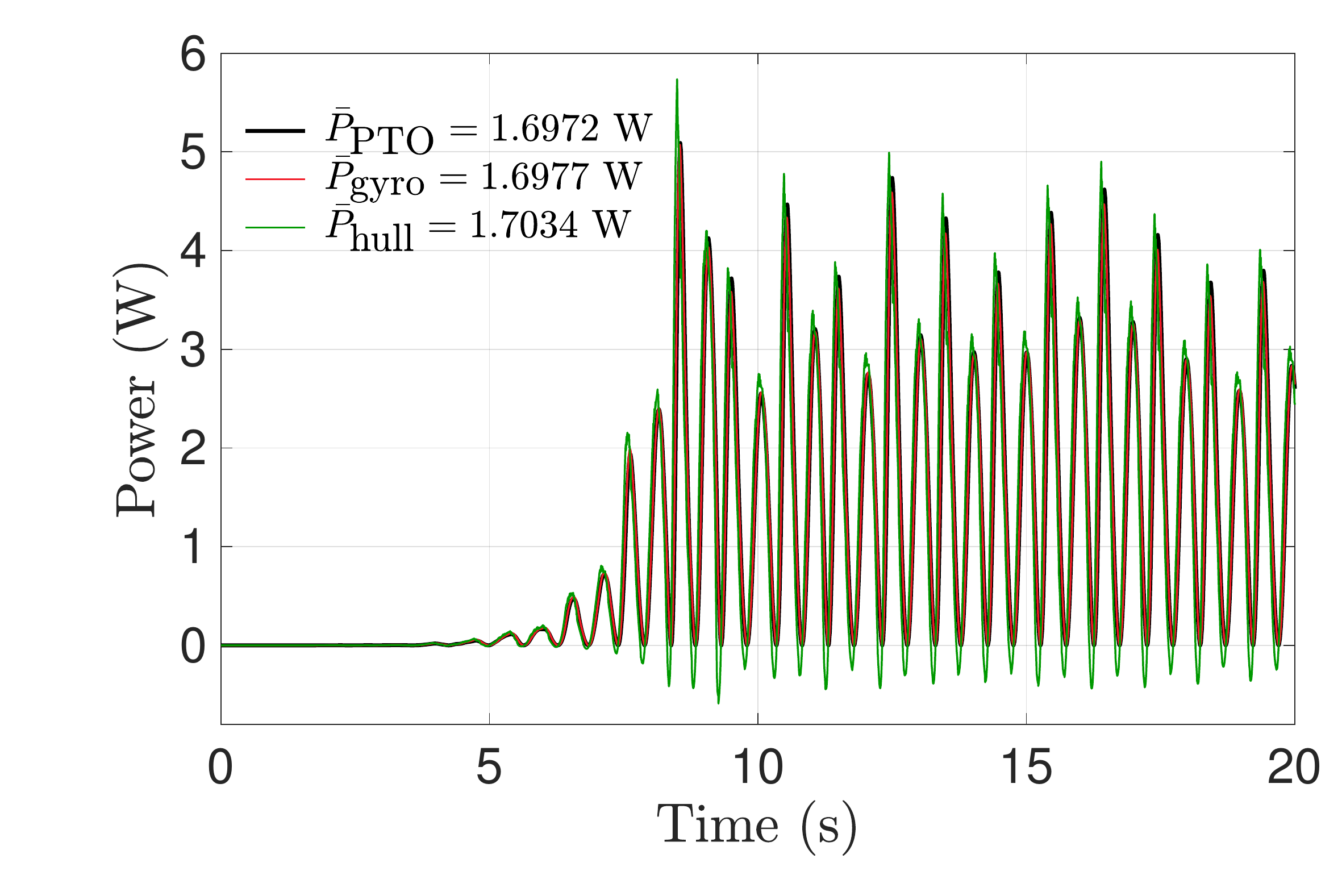}
    \label{fig_regwave_delta5deg_2dpower}
  }
    \subfigure[Powers for $\delta_0$ = 5$^\circ$ (3D model)]{
    \includegraphics[scale = 0.3]{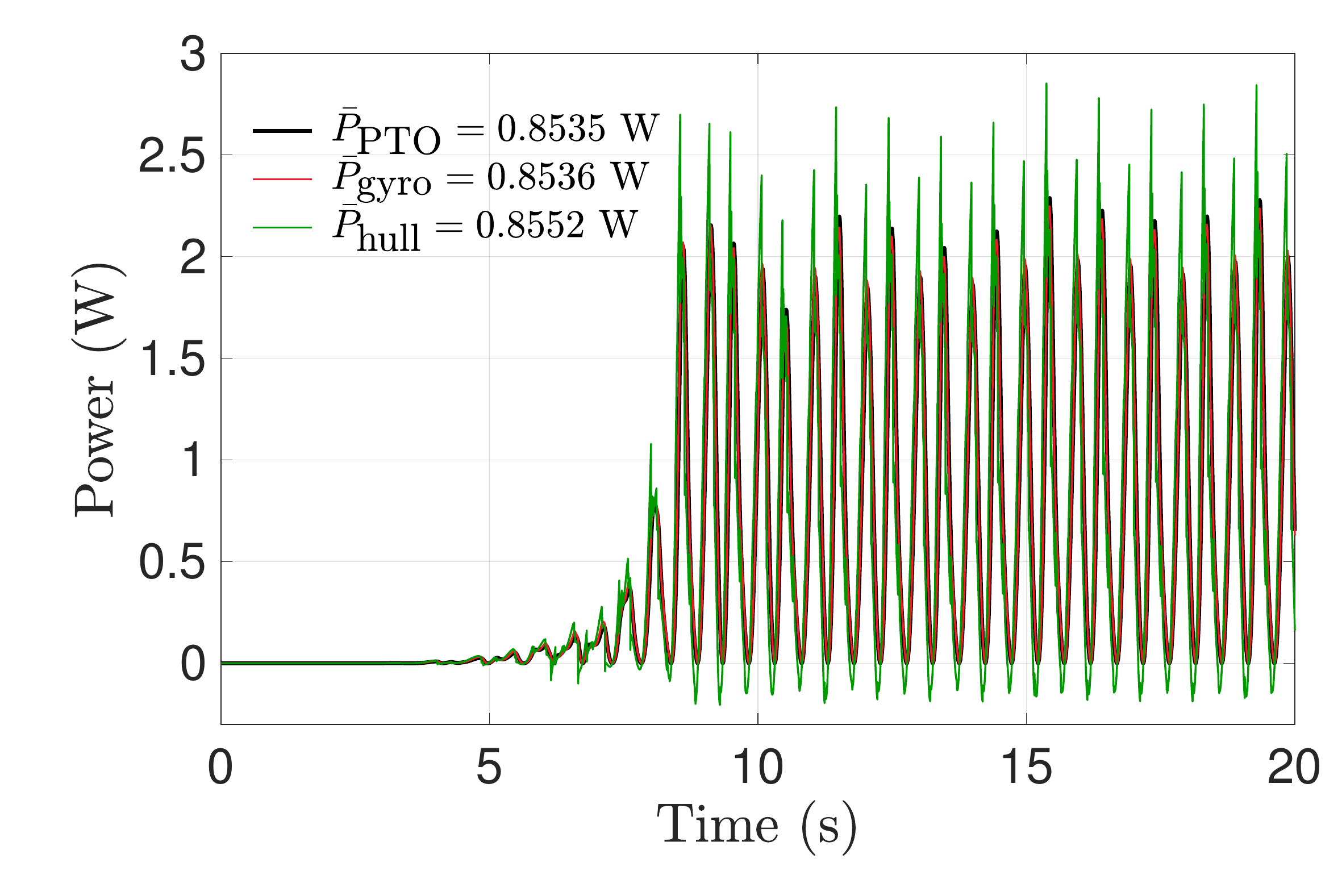}
    \label{fig_regwave_delta5deg_3dpower}
  }
    \subfigure[Powers for $\delta_0$ = 20$^\circ$ (2D model)]{
    \includegraphics[scale = 0.3]{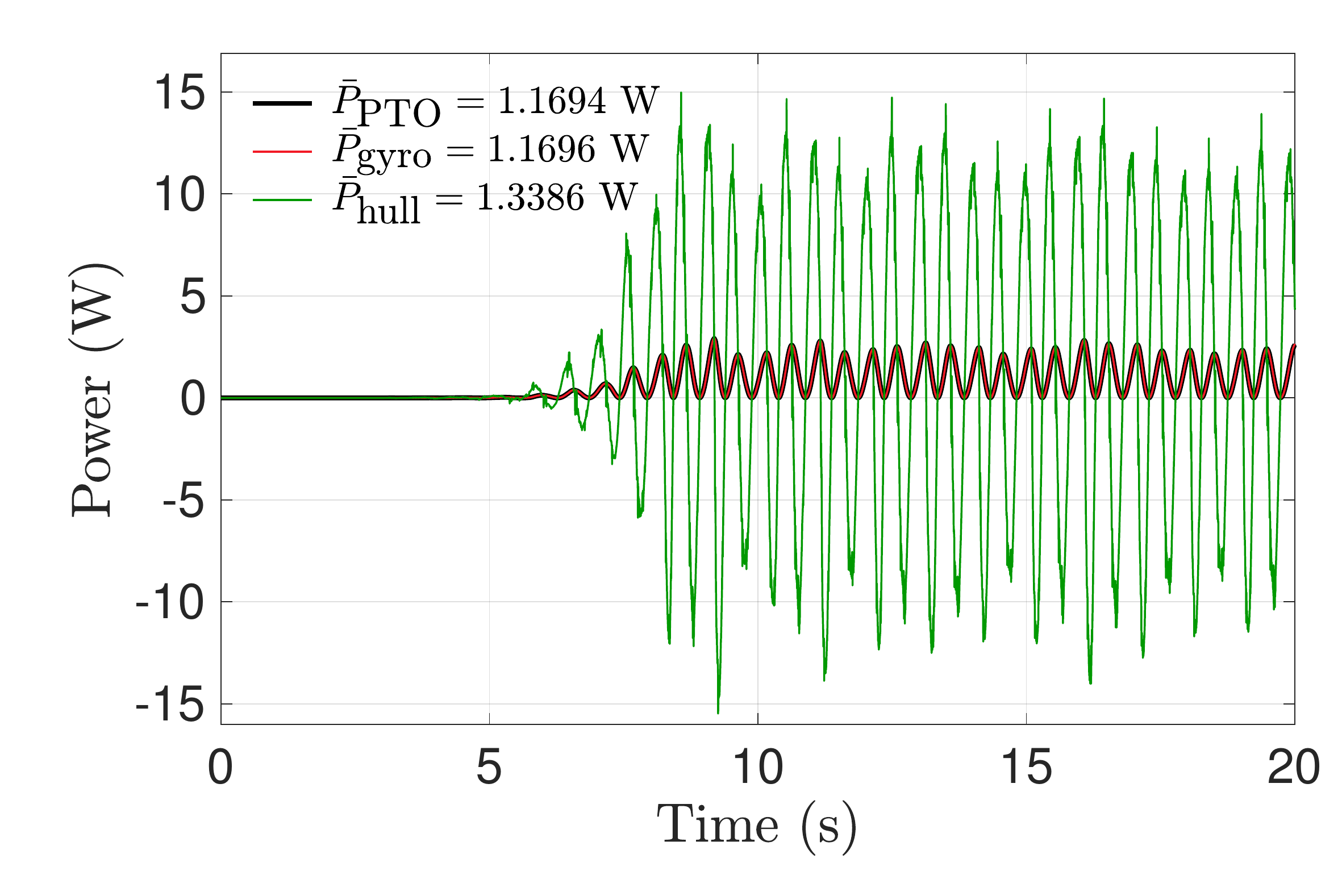}
    \label{fig_regwave_delta20deg_2dpower}
  }
    \subfigure[Powers for $\delta_0$ = 20$^\circ$ (3D model)]{
    \includegraphics[scale = 0.3]{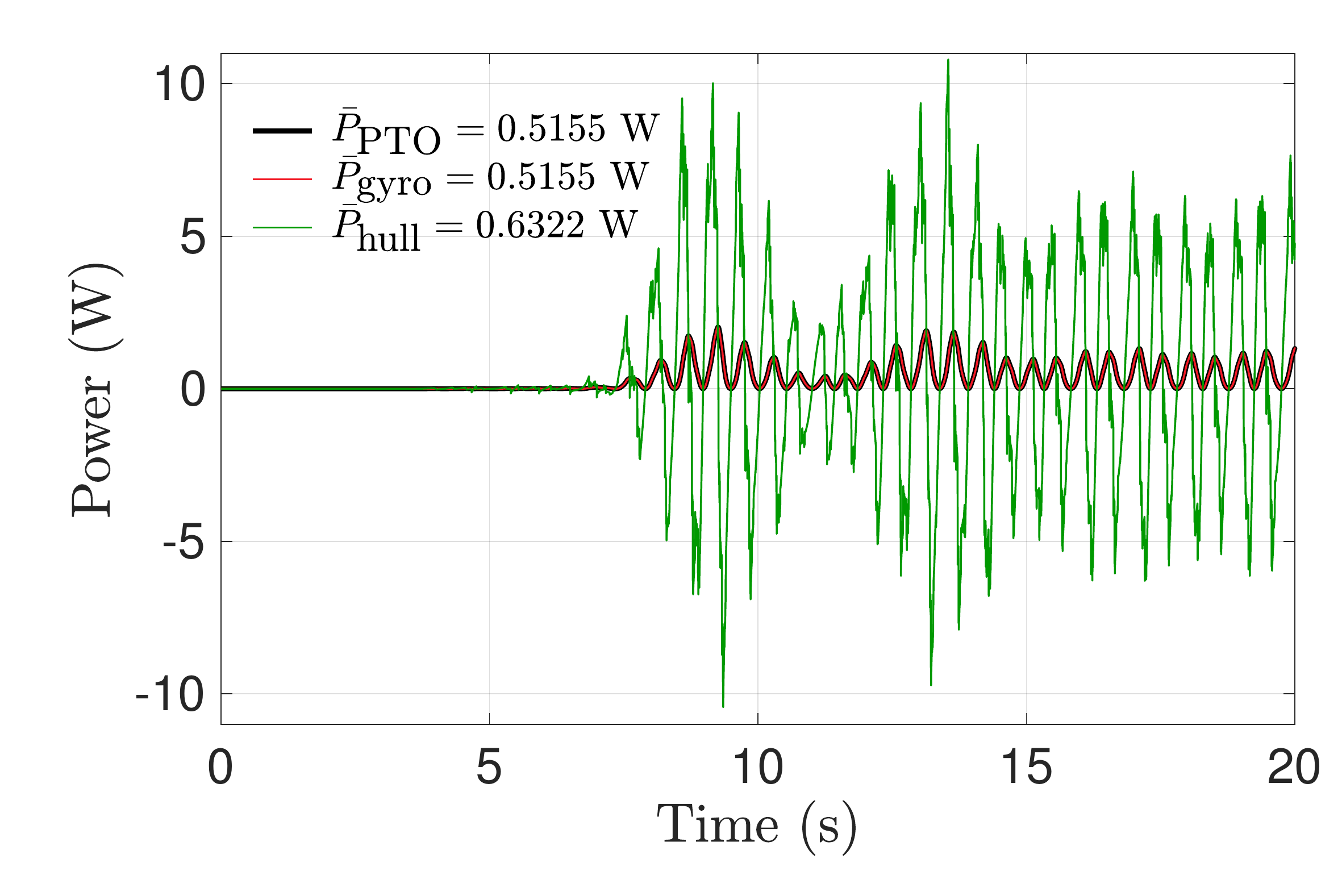}
    \label{fig_regwave_delta20deg_3dpower}
  }
  \caption{WSI of the 2D and 3D ISWEC models in regular water wave conditions ($\cH$ = 0.1 m and $\cT$ = 1 s). Temporal evolution of the hull pitch angle ($\delta$) for the 2D and 3D ISWEC models for \subref{fig_regwave_delta5deg_2d3d} $\delta_0$ = 5$^\circ$ and \subref{fig_regwave_delta20deg_2d3d} $\delta_0$ = 20$^\circ$. Power absorbed by the PTO unit $P_\text{PTO}$ (\textcolor{black}{\textbf{-----}}, black), power generated through the hull-gyroscope interaction $P_\text{gyro}$ (\textcolor{red}{\textbf{-----}}, red) and power transferred to the hull from the regular water waves $P_\text{hull}$ (\textcolor{ForestGreen}{\textbf{-----}}, green) for the \subref{fig_regwave_delta5deg_2dpower} 2D model with $\delta_0$ = 5$^\circ$, \subref{fig_regwave_delta5deg_3dpower} 3D model with $\delta_0$ = 5$^\circ$, \subref{fig_regwave_delta20deg_2dpower} 2D model with $\delta_0$ = 20$^\circ$ and \subref{fig_regwave_delta20deg_3dpower} 3D model with $\delta_0$ = 20$^\circ$.
  Averaged power over the time period $t = 10$ s and $t = 20$ s are shown in the legends.
}
  \label{fig_regwave_2d3d_comparison}
\end{figure}

First, we consider two different prescribed maximum pitch angles $\delta_0$ = 5$^\circ$ and 20$^\circ$ for each model.  Regular waves are generated with properties $\cH$ = 0.1 m and $\cT$ = 1 s. The values for the gyroscope and PTO parameters for this choice of $\delta_0$ are given in Table~\ref{tab_gyro_parameters_regularwave}. The rated power of the device $\bar{P}_\text{R}$ is taken to be the available wave power $\bar{P}_\text{wave}$ for calculating the parameters reported in Table~\ref{tab_gyro_parameters_regularwave}. The hull undergoes pitching motion as the regular waves impact the device, as shown in Fig.~\ref{fig_iswec3d_regular}. The temporal evolution of the hull pitch angle $\delta$ for the 2D and the 3D ISWEC models are shown in Figs.~\ref{fig_regwave_delta5deg_2d3d} and \ref{fig_regwave_delta20deg_2d3d}. From these results, it is observed that the dynamics for the 2D case match well with the 3D case after the initial transients. The power transferred to the hull from the waves $P_\text{hull}$, the power generated through the hull-gyroscope interaction $P_\text{gyro}$, and the power absorbed by the PTO unit $P_\text{PTO}$ at $\delta_0$ = 5$^\circ$ ($\delta_0$ = 20$^\circ$) for the 2D and 3D models are shown in Figs.~\ref{fig_regwave_delta5deg_2dpower} and~\ref{fig_regwave_delta5deg_3dpower} (Figs.~\ref{fig_regwave_delta20deg_2dpower}and~\ref{fig_regwave_delta20deg_3dpower}), respectively. The time-averaged powers $\bar{P}_\text{PTO}$, $\bar{P}_\text{gyro}$ and $\bar{P}_\text{hull}$ over the time interval $t = 10$ s and $t = 20$ s (after the hull's motion achieved a periodic steady state) are also shown in Figs.~\ref{fig_regwave_delta5deg_2dpower}-\ref{fig_regwave_delta20deg_3dpower}. From these results, it is seen that the energy transfer pathway Eq.~\eqref{eq_pathway} is numerically verified. Furthermore, the power absorbed by the PTO unit for the full-scale device can be calculated by multiplying the power absorbed by the 2D model by the Froude scaling given in Table~\ref{tab_Fr_scaling}
\begin{equation}
	\label{eq_2dpower_to_fullscale}
	\bar{P}_\text{full-scale} = \alpha^\frac{5}{2} \, \cdot \, W \, \cdot \, \bar{P}_\text{2D}, \qquad \text{(PTO unit)}.
\end{equation}
Similarly for the 3D model,
\begin{equation}
	\label{eq_3dpower_to_fullscale}
	\bar{P}_\text{full-scale} = \alpha^\frac{7}{2} \, \cdot \, \bar{P}_\text{3D},  \qquad \text{(PTO unit)}.
\end{equation}
Finally, combining the two expressions above yields
\begin{equation}
	\label{eq_2dpower_scaling_to_3d}
	\bar{P}_\text{3D} = \frac{W_\text{full-scale}}{\alpha} \cdot \bar{P}_\text{2D} =  0.4 \times \bar{P}_\text{2D},  \qquad \text{(PTO unit)},
\end{equation}
in which $W_\text{full-scale}$ = 8 m is the width of the full-scale model and $\alpha = 20$ is the length scaling factor. For the 2D cases, the average power absorbed by the PTO unit is $\bar{P}_\text{2D} = 1.6972$ W for $\delta_0 = 5^\circ$, and
$\bar{P}_\text{2D} = 1.1694$ W for $\delta_0 = 20^\circ$. For the 3D cases, the average powers absorbed by the PTO unit are $0.8535$ W and $0.5155$ W for $\delta_0 = 5^\circ$ and $\delta_0 = 20^\circ$, respectively, which are close to the expected values of
$0.6788$ W and $0.4677$ W predicted by Eq.~\ref{eq_2dpower_scaling_to_3d}.
Note that better agreement between the simulated and expected average powers in 3D can be obtained by increasing the spatial and temporal resolutions. Nevertheless, we are confident that the dynamics are reasonably resolved for the chosen grid spacing and time step size.
%to arrive at this conclusion.

\begin{figure}[]
  \centering
   \subfigure[Hull motion for $\delta_0$ = 5$^\circ$]{
    \includegraphics[scale = 0.3]{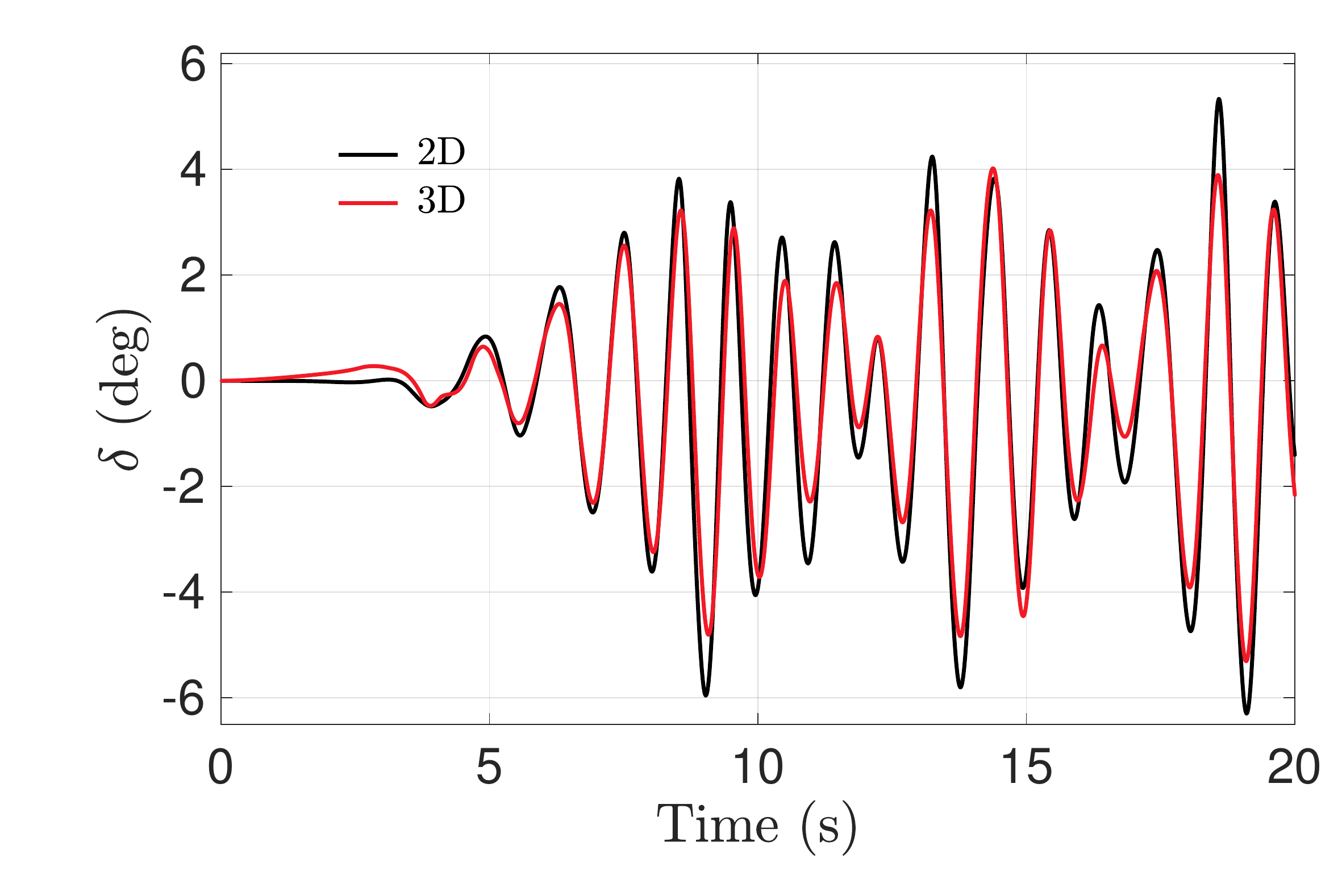}
    \label{fig_irregwave_delta5deg_2d3d}
  } \\
     \subfigure[Powers for $\delta_0$ = 5$^\circ$ (2D model)]{
    \includegraphics[scale = 0.3]{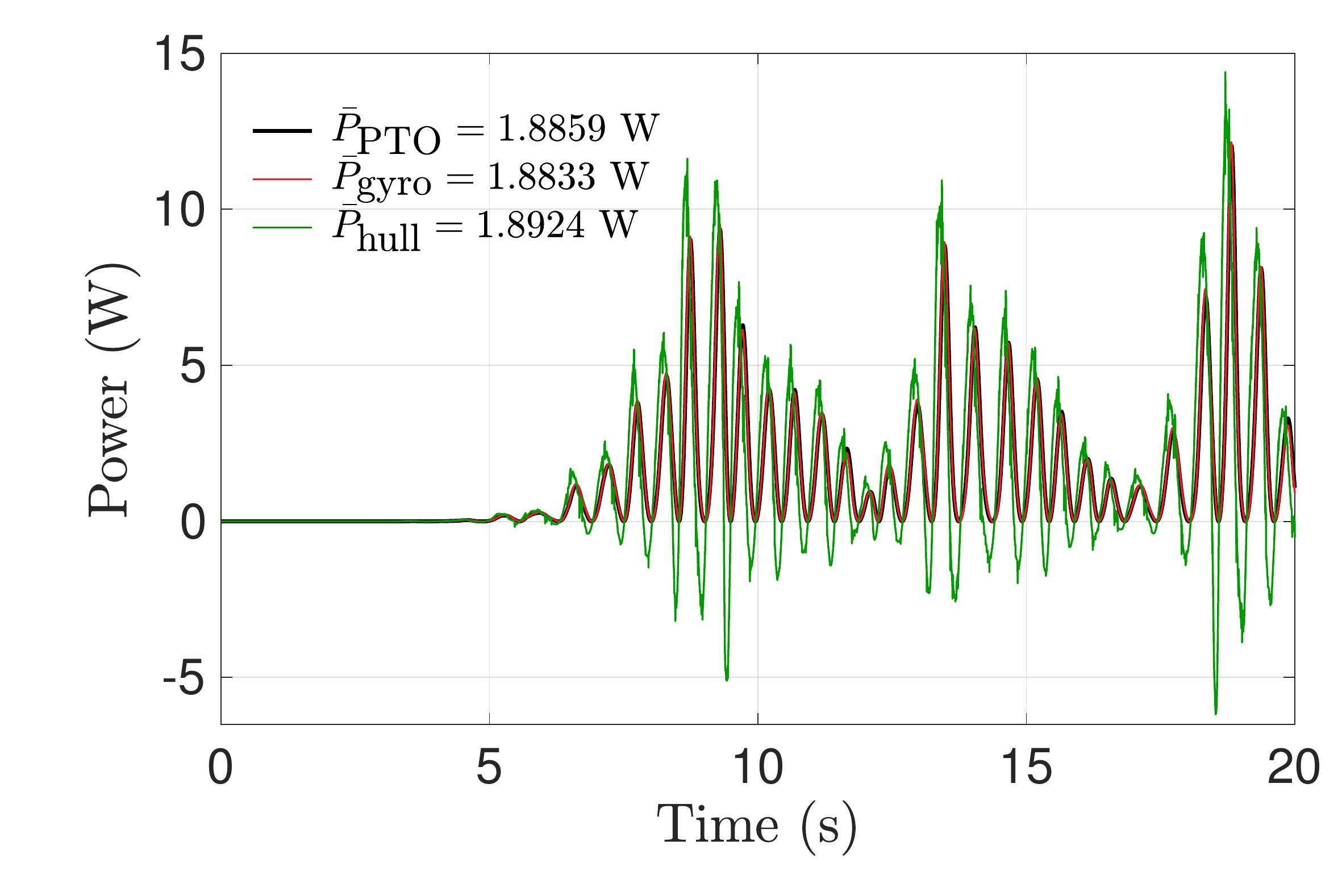}
    \label{fig_irregwave_delta5deg_2dpower}
  }
   \subfigure[Powers for $\delta_0$ = 5$^\circ$ (3D model)]{
    \includegraphics[scale = 0.3]{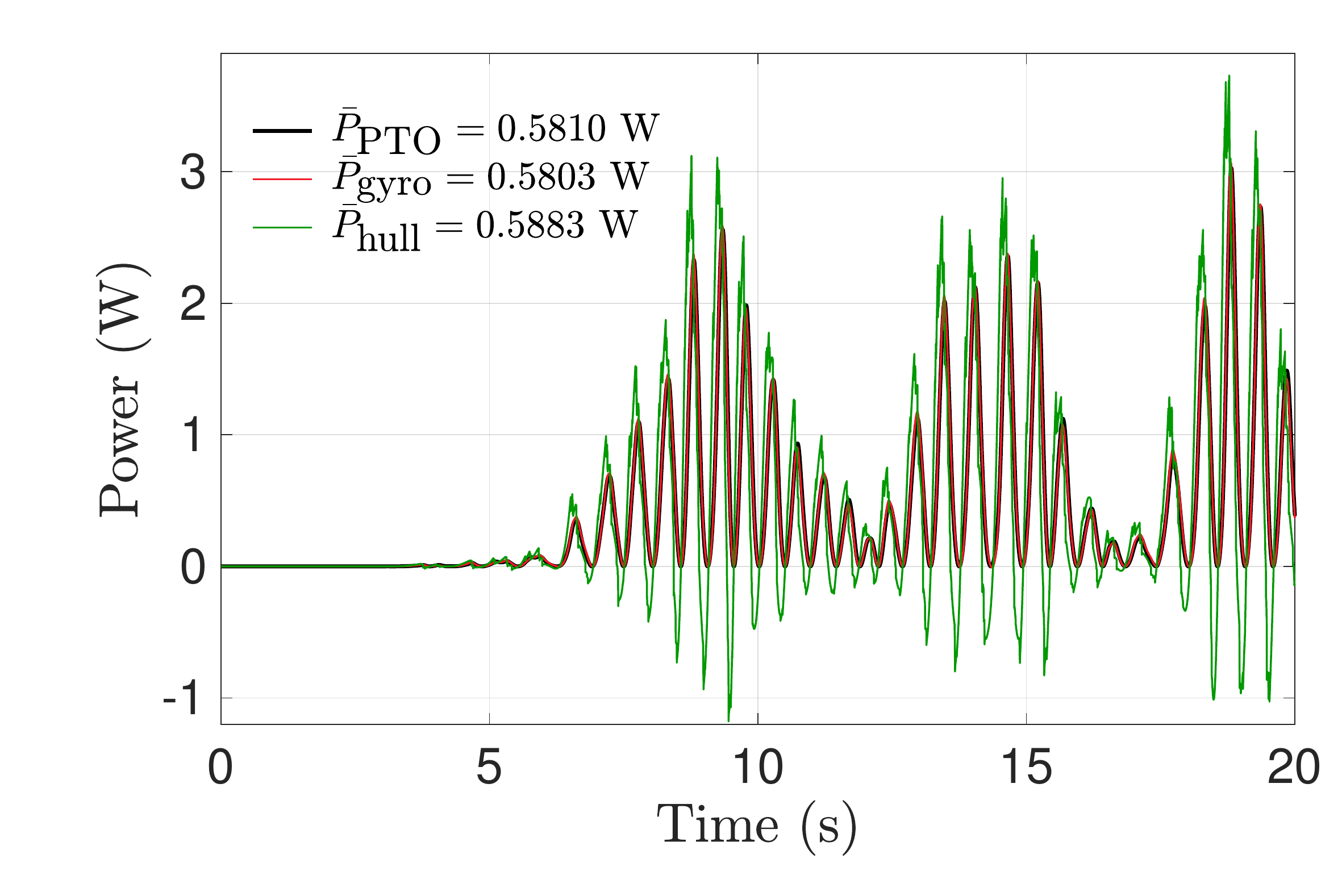}
    \label{fig_irregwave_delta5deg_3dpower}
  }
  \caption{WSI of the 2D and 3D ISWEC models with $\delta_0$ = 5$^\circ$ in irregular water wave conditions ($\cH_\text{s}$ = 0.1 m and $\cT_\text{p}$ = 1 s, $N$ = 50, and $\omega_i$ in the range 3.8 rad/s to 20 rad/s). \subref{fig_irregwave_delta5deg_2d3d} Temporal evolution of the hull pitch angle ($\delta$) for the 2D and 3D ISWEC models. Power absorbed by the PTO unit $P_\text{PTO}$ (\textcolor{black}{\textbf{-----}}, black), power generated through the hull-gyroscope interaction $P_\text{gyro}$ (\textcolor{red}{\textbf{-----}}, red) and power transferred to the hull from the irregular water waves $P_\text{hull}$ (\textcolor{ForestGreen}{\textbf{-----}}, green) for the \subref{fig_irregwave_delta5deg_2dpower} 2D and \subref{fig_irregwave_delta5deg_3dpower} 3D models.
}
  \label{fig_irregwave_2d3d_comparison}
\end{figure}

Next, we perform a similar scaling analysis for 2D and 3D ISWEC models in irregular wave conditions. Irregular water waves are generated with properties $\cH_\text{s}$ = 0.1 m, $\cT_\text{p}$ = 1 s and $50$ wave components with frequencies $\omega_i$ in the range $3.8$ rad/s to $20$ rad/s (see Fig.~\ref{fig_iswec3d_irregular}). Through empirical testing, fifty wave components were found to be sufficient to represent the energy of the JONSWAP spectrum. We consider a maximum pitch angle of  $\delta_0$ = 5$^\circ$ for the device. The evolution of $\delta$ for the two models are compared in Fig.~\ref{fig_irregwave_delta5deg_2d3d}. Similar to the regular wave case presented above, the dynamics of the 2D and 3D models numerically agree and the energy transfer pathway Eq.~\eqref{eq_pathway} is nearly satisfied. Moreover, the average power absorbed by the PTO unit for the 2D model is $\bar{P}_\text{2D} = 1.8859$ W, yielding an expected 3D power of $0.7543$ W according to Eq.~\eqref{eq_2dpower_scaling_to_3d}; this is close to the power value of $0.5810$ W 
obtained by the 3D simulation. 

Based on these results, we ultimately conclude that the 2D model is sufficient to accurately simulate ISWEC dynamics and to predict the power generation/absorption capability of the converter. Hereafter, we focus on further investigating dynamics and parameter choices for the 2D model.

%%%%%%%%%%%%%%%%%%%%%%%%%%%%%%%%%%%%%%%%%%
\subsection{Selection of prescribed hull pitch angle $\delta_0$} \label{subsec_presc_angle_selec}

In this section, we investigate the relationship between the prescribed hull pitch angle parameter $\delta_0$, the maximum pitch angle actually attained by the hull $\delta_\text{max}$ through WSI, and the maximum wave steepness of the incoming waves $\delta_\text{s}$.
Recall that the maximum wave steepness was calculated in Sec.~\ref{sec_wave_steepness} by approximating the fifth-order wave as a linear harmonic wave.
We consider the ISWEC dynamics on four regular water waves with same time period $\cT$ = 1 s (i.e. $\lambda$ = 1.5456 m) but varying wave heights: $\cH = 0.025$ m, $0.05$ m, $0.1$ m and $0.125$ m, each having maximum wave steepness $\delta_\text{s} = 2.9^\circ$, $5.8^\circ$, $11.48^\circ$ and $14.25^\circ$, respectively (see Eq~\ref{eq_max_wave_steepness}).  The prescribed PTO and gyroscope system parameters for each sea state and six maximum pitch angle values $\delta_0 = 2^\circ$, $5^\circ$, $10^\circ$, $15^\circ$, $20^\circ$ and $\delta_\text{s}$ are shown in Table~\ref{tab_gyro_parameters_regularwave}. Additionally, $\delta_0 = 1^\circ$ and $30^\circ$ cases 
are also simulated, but the parameter values are not tabulated for brevity. 

\begin{figure}[]
 \centering
  \subfigure[Maximum attained hull pitch angle]{
  \includegraphics[scale = 0.3]{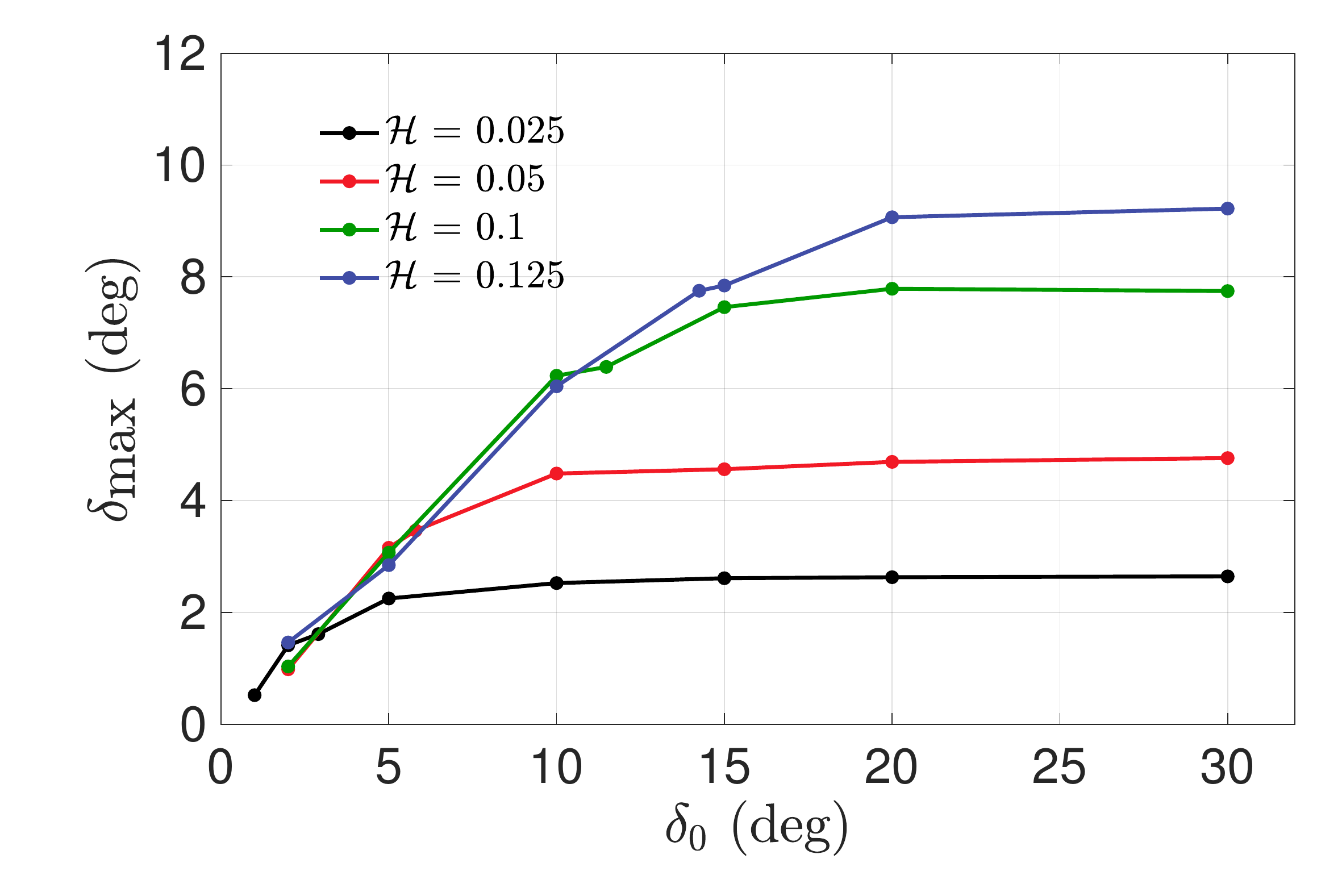}
  \label{fig_delta_actual_H_variation}
 }
  \subfigure[Maximum attained gyroscope precession angle]{
  \includegraphics[scale = 0.3]{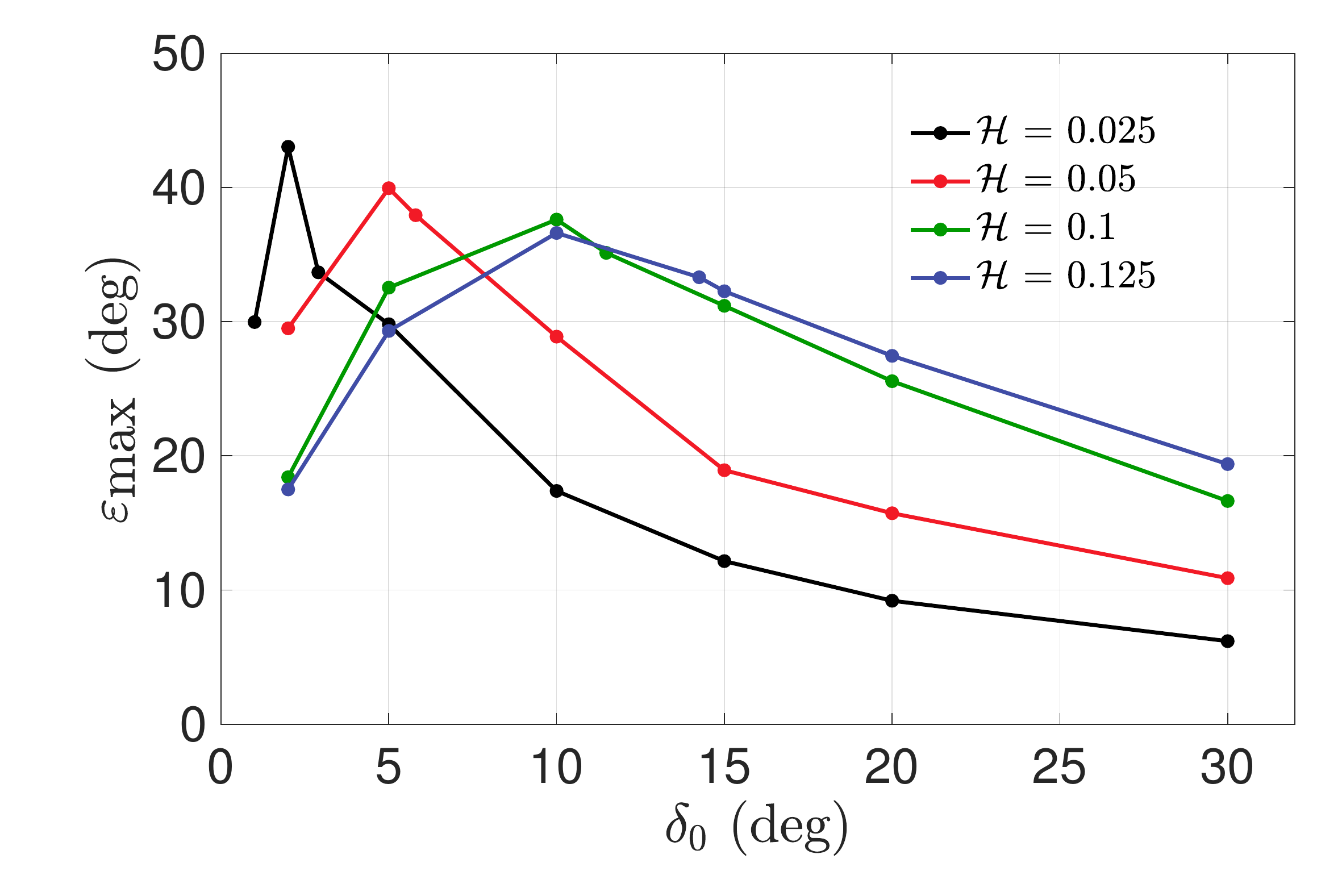}
  \label{fig_e_actual_H_variation}
 }
   \subfigure[Relative capture width (RCW)]{
  \includegraphics[scale = 0.3]{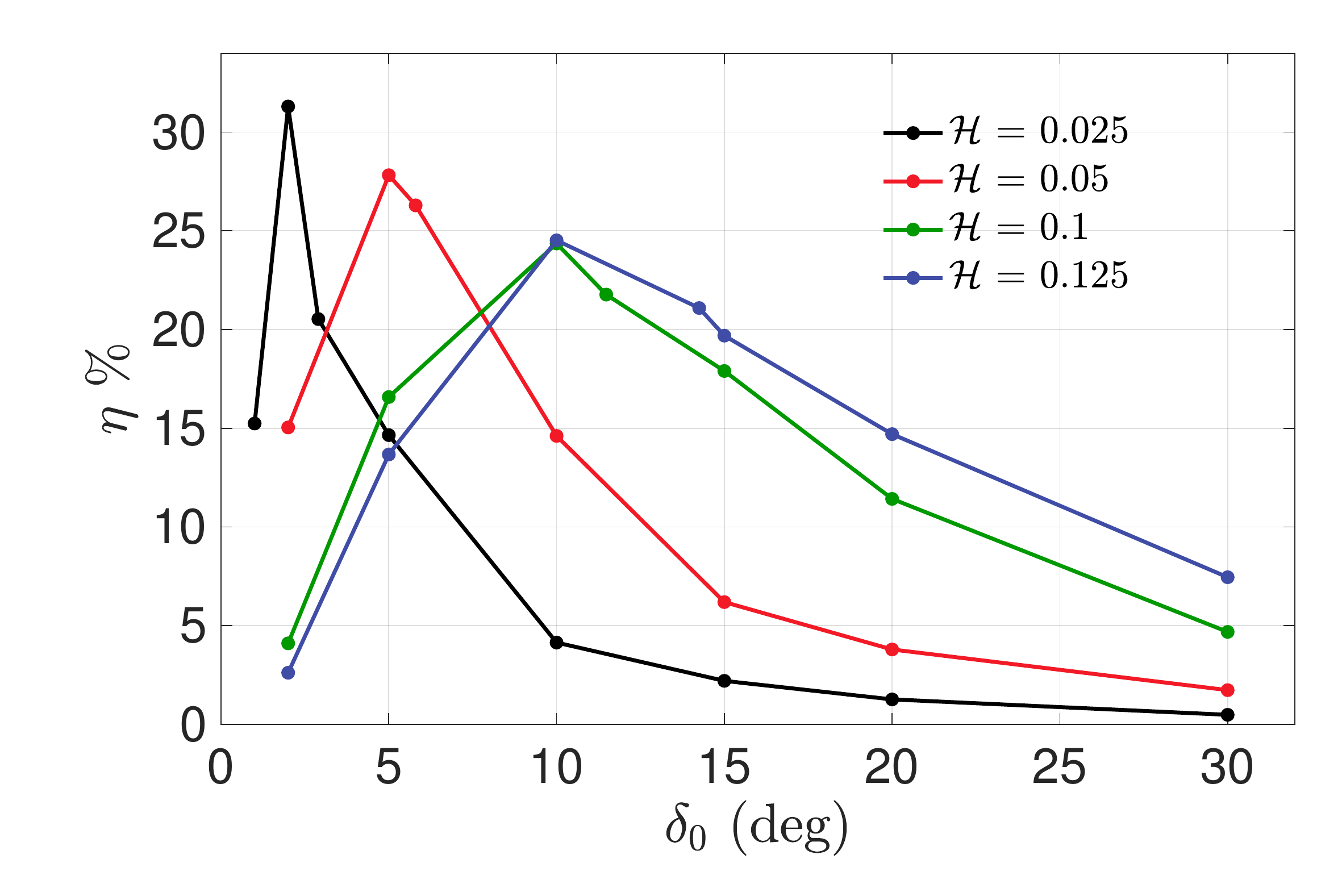}
  \label{fig_RCW_H_variation}
 }
 \caption{\subref{fig_delta_actual_H_variation} Maximum hull pitch angle $\delta_\text{max}$, \subref{fig_e_actual_H_variation} maximum gyroscope precession angle $\varepsilon_\text{max}$, and \subref{fig_RCW_H_variation} relative capture width $\eta$ of the ISWEC device for various regular wave sea states and prescribed pitch angles $\delta_0$: $\cH$ = 0.025 m (\textcolor{black}{\textbf{-----}}, black), $\cH$ = 0.05 m (\textcolor{red}{\textbf{-----}}, red), $\cH$ = 0.1 m (\textcolor{ForestGreen}{\textbf{-----}}, green), and $\cH$ = 0.125 m (\textcolor{blue}{\textbf{-----}}, blue).
 RCW is calculated from time-averaged powers over the interval $t = 10$ s to $t = 20$ s
}
 \label{fig_prescr_pitch_Hs_variation}
\end{figure}

The results of this parameter study are shown in Fig.~\ref{fig_prescr_pitch_Hs_variation}.
It is observed that when $\delta_0 < \delta_\text{s}$, $\delta_\text{max}$ increases linearly with $\delta_0$  (Fig.~\ref{fig_delta_actual_H_variation}), illustrating that the hull's maximum oscillation amplitude correlates well with $\delta_0$. When the prescribed $\delta_0$ is greater than $\delta_\text{s}$, it is seen that $\delta_\text{max}$ no longer
increases; rather it maintains a constant value with respect to $\delta_0$. This indicates that further increasing $\delta_0$
will not lead to larger pitch oscillations, i.e. the $\delta_\text{max}$ attained by the hull is the largest value permitted by the slopes of the
wave. In Figs.~\ref{fig_e_actual_H_variation} and~\ref{fig_RCW_H_variation}, we show trends in the 
maximum precession angle attained by the gyroscope $\varepsilon_\text{max}$ and the relative capture width (RCW) $\eta$,
which measures the device efficiency as a ratio of the average power absorbed by the PTO unit to the average wave power per unit crest width (see Eq.\eqref{eqn_W_eta}). Maximization of both these quantities is achieved when $\delta_0$ is set close to $\delta_\text{s}$.
As the hull achieves the maximum pitch angle physically permitted by the slopes of the wave, further increasing $\delta_0$ amounts to reducing $J\dot{\phi}$ (Eq.~\eqref{eq_Jphi}) or the hull-gyroscope coupling, which explains the reduction in both maximum precession and device efficiency.
Hereafter, we prescribe $\delta_0$ based on the value maximizing $\eta$ as we conduct further parametric analyses of the 2D ISWEC model.

%%%%%%%%%%%%%%%%%%%%%%%%%%%%%%%%%%%%%%%%%%%%
\subsection{Parametric analyses of gyroscope parameters} \label{subsec_PA}

In this section, we conduct a parameter sweep around the energy-maximizing PTO and gyroscope parameters estimated by the theory presented in Sec.~\ref{sec_PTO_params}. We test the theory's predictive capability and describe the effect of these parameters on the converter's performance and dynamics. In each of the following subsections, only a single parameter is varied at a time. 

Simulations are conducted using both regular water waves with $\cH$ = 0.1 m and $\cT$ = 1 s, and irregular waves with $\cH_\text{s}$ = 0.1 m, $\cT_\text{p}$ = 1 s and $50$ wave components with frequencies $\omega_i$ in the range $3.8$ rad/s to $20$ rad/s. These wave conditions serve as device ``design" conditions at its installation site. For regular waves, the prescribed pitch angle is taken to be $\delta_0 = 10^\circ$, and the PTO and gyroscope parameters are given in Table~\ref{tab_gyro_parameters_regularwave}.  For irregular waves, the prescribed pitch angle $\delta_0 = 5^\circ$ is used. The PTO and gyroscope parameters remain the same as those used in the temporal resolution study (see Sec.\ref{temporal_resol_study}). These particular values of $\delta_0$ were found to maximize the RCW of the converter at design conditions; for an example, see Fig.~\ref{fig_prescr_pitch_Hs_variation} for regular waves with $\cH$ = 0.1 m and $\cT$ = 1 s.

%%%%%%%%%%%%%%%%%%%%%%%%%%%%%
\subsubsection{PTO damping coefficient $c$} \label{subsec_c_variation}

\begin{figure}[]
 \centering
  \subfigure[Hull pitch angle]{
  \includegraphics[scale = 0.3]{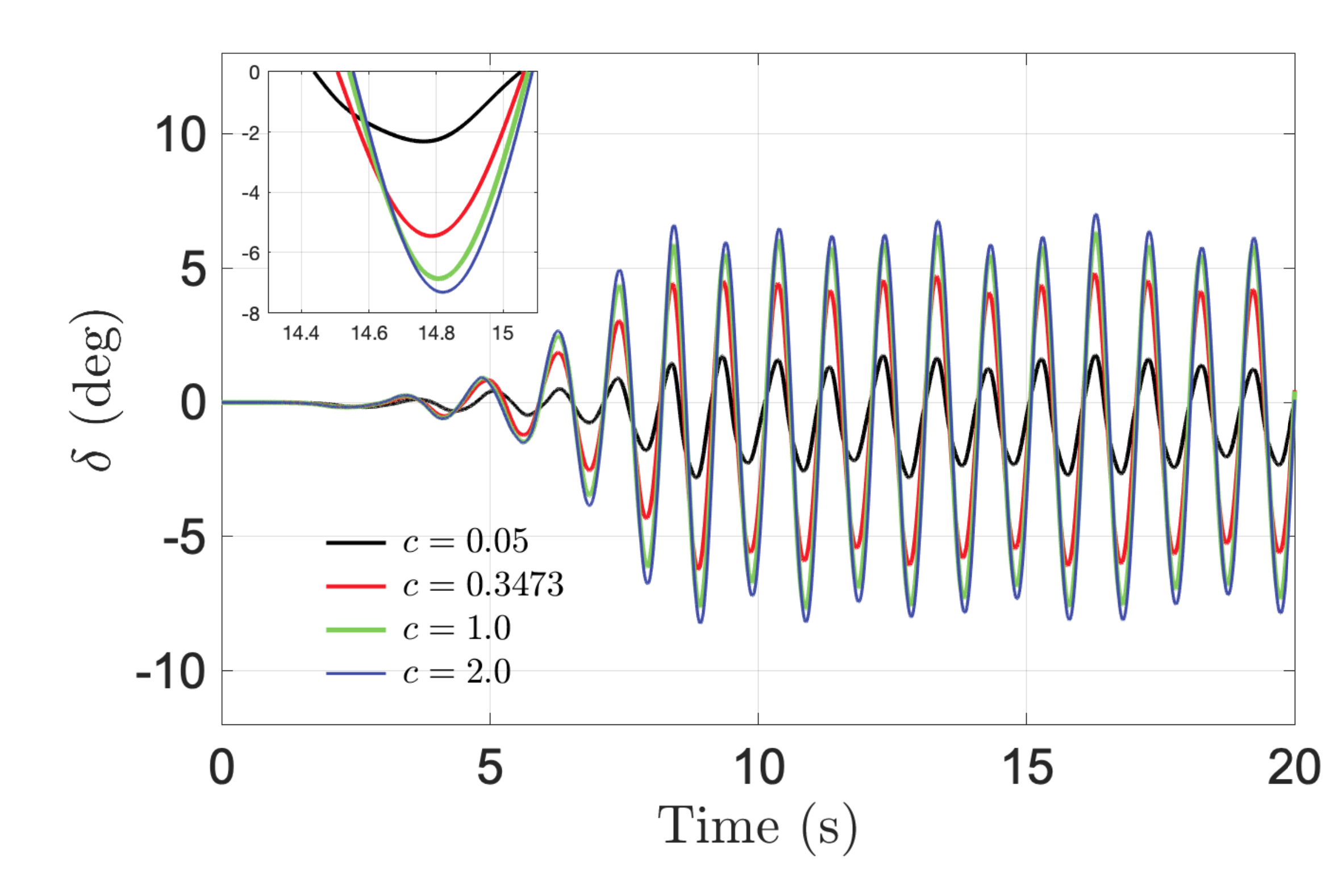}
  \label{fig_delta_c_compare_regwave}
 }
   \subfigure[Gyroscope precession angle]{
  \includegraphics[scale = 0.3]{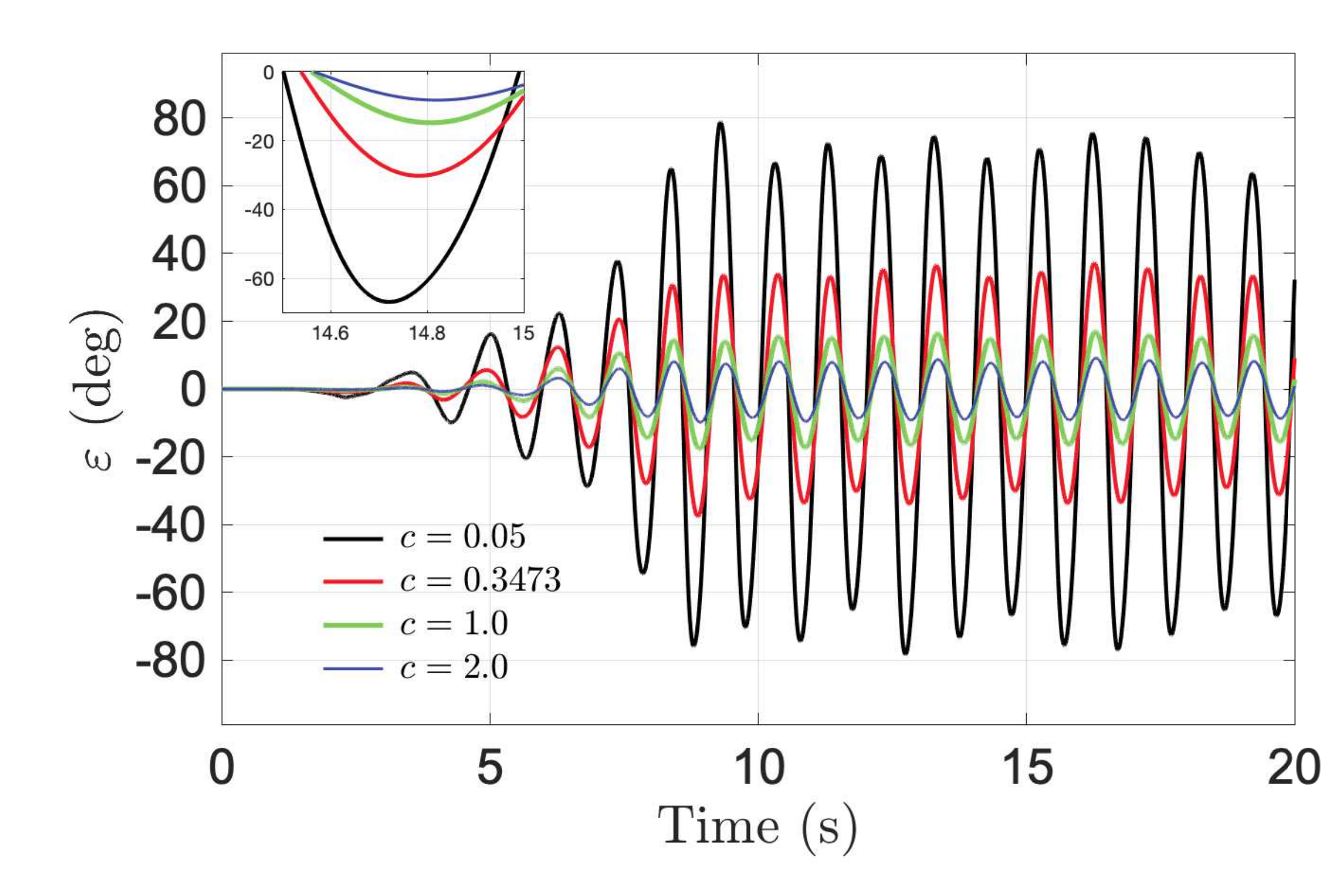}
  \label{fig_e_c_compare_regwave}
 }
  \subfigure[Pitch torque unloaded on the hull]{
  \includegraphics[scale = 0.3]{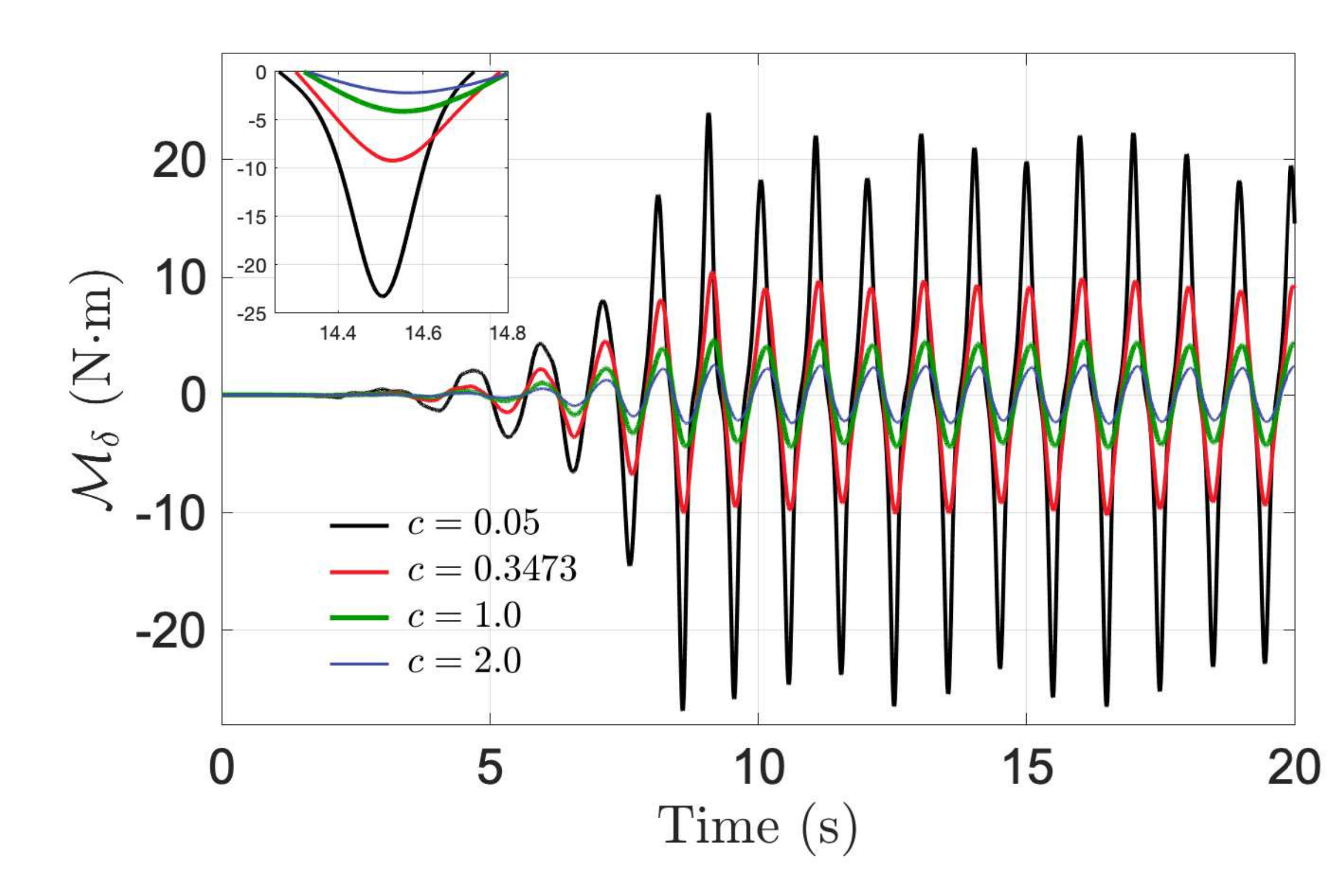}
  \label{fig_Md_c_compare_regwave}
 }
  \subfigure[Precession torque on the PTO axis]{
  \includegraphics[scale = 0.3]{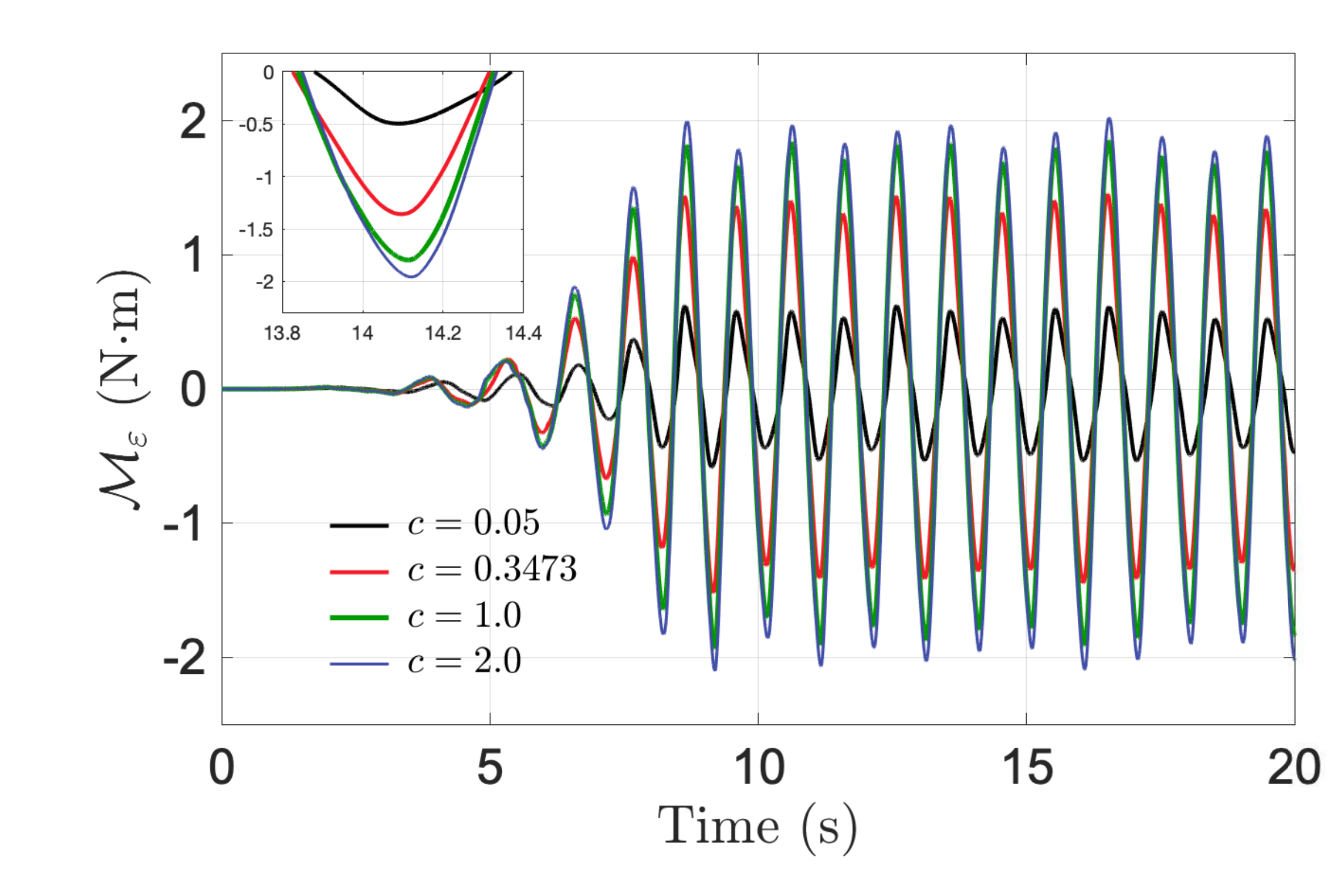}
  \label{fig_Me_c_compare_regwave}
 }
   \subfigure[Time-averaged powers in the system]{
  \includegraphics[scale = 0.3]{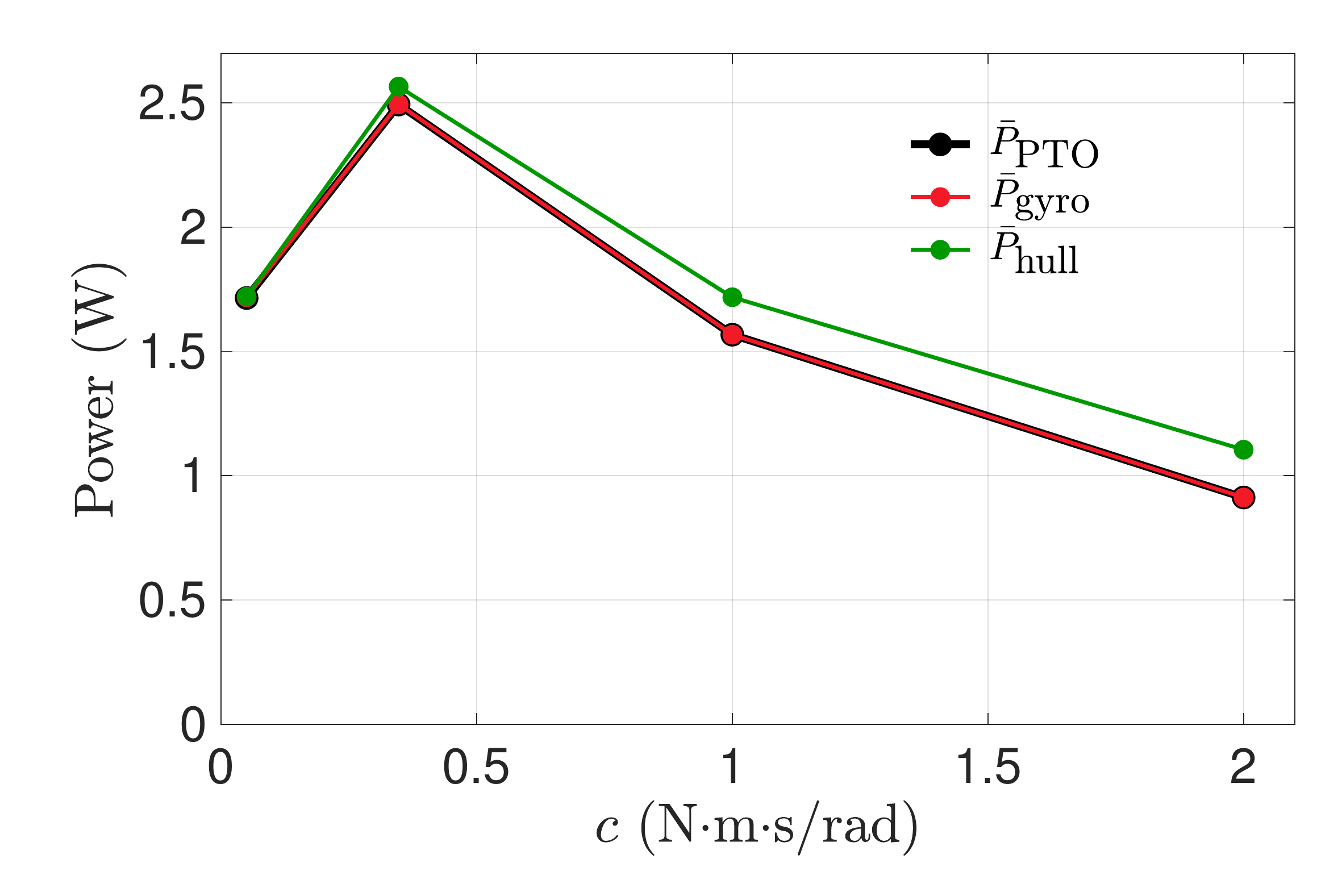}
  \label{fig_Powers_c_compare_regwave}
  }
  \subfigure[Yaw torque]{
  \includegraphics[scale = 0.3]{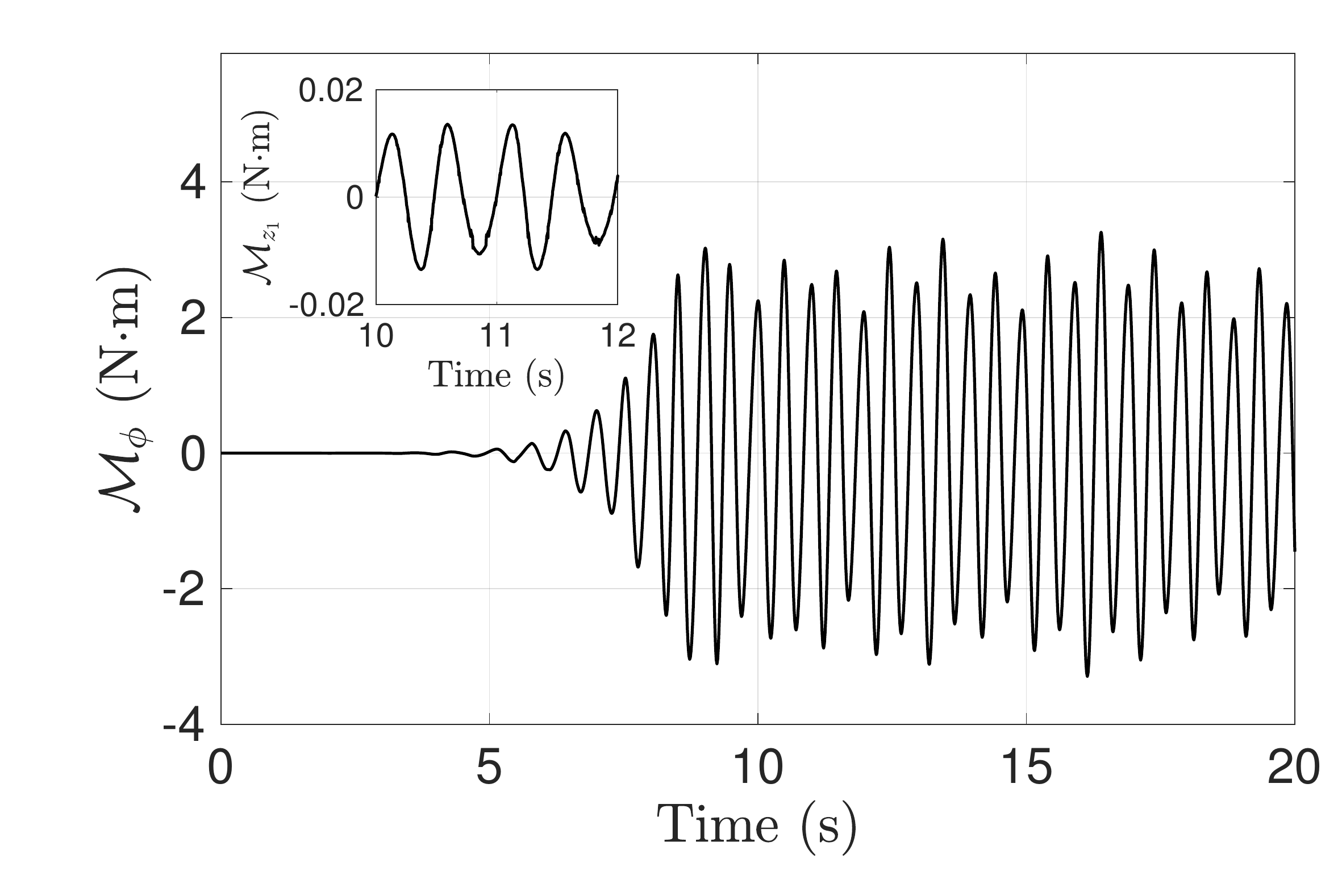}
  \label{fig_yaw_torque}
 }
 \caption{Dynamics of the 2D ISWEC model for four different values of PTO damping coefficient $c$, with regular wave properties $\cH$ = 0.1 m and $\cT$ = 1 s. Temporal evolution of \subref{fig_delta_c_compare_regwave} hull pitch angle $\delta$, \subref{fig_e_c_compare_regwave} gyroscope precession angle $\varepsilon$, \subref{fig_Md_c_compare_regwave} pitch torque $\cM_\delta$, and \subref{fig_Me_c_compare_regwave} precession torque $\cM_\varepsilon$ for $c$ = 0.05 N$\cdot$m$\cdot$s/rad (\textcolor{black}{\textbf{-----}}, black), $c$ = 0.3473 N$\cdot$m$\cdot$s/rad (\textcolor{red}{\textbf{-----}}, red), $c$ = 1.0 N$\cdot$m$\cdot$s/rad (\textcolor{ForestGreen}{\textbf{-----}}, green), and $c$ = 2.0 N$\cdot$m$\cdot$s/rad (\textcolor{blue}{\textbf{-----}}, blue);
 \subref{fig_Powers_c_compare_regwave} comparison of time-averaged powers from the interval $t = 10$ s to $t = 20$ s for each value of $c$; \subref{fig_yaw_torque} yaw torques $\cM_\phi$ and $\cM_{z_1}$ produced in the inertial reference frame and gyroscope reference frame (inset), respectively.
}
 \label{fig_c_compare}
\end{figure}

We first consider the PTO unit damping coefficient $c$, which directly impacts the power absorption capability of the device.
We prescribe four different values, $c = 0.05$, $0.3473$, $1.0$ and $2.0$ N$\cdot$m$\cdot$s/rad, to evaluate its impact on ISWEC dynamics.
The optimal damping coefficient value of $c = 0.3473$ is predicted by the theory. Results for the hull interacting with regular waves are shown in Fig.~\ref{fig_c_compare}.
As expected for smaller damping coefficients, the gyroscope is able to attain larger precession angles $\varepsilon$ and velocities 
$\dot{\varepsilon}$, as seen in Fig.~\ref{fig_e_c_compare_regwave}. Higher precession velocities yield larger pitch torque $\cM_\delta$ values
(see Eq.~\eqref{eq_M_delta_simplified}), which opposes the motion of the hull and restrict its maximum pitch oscillation; this is consistent
with the dynamics shown in Figs.~\ref{fig_delta_c_compare_regwave} and~\ref{fig_Md_c_compare_regwave}. Moreover the hull's pitch
velocity $\dot{\delta}$ is reduced with decreasing $c$, leading to a smaller (in magnitude) precession torque $\cM_\varepsilon$ acting on 
the PTO shaft (see Eq.~\eqref{eq_gyro_PTO}); our simulations show this behavior as observed in Fig.~\ref{fig_Me_c_compare_regwave}.

In Fig.~\ref{fig_Powers_c_compare_regwave}, we compare the time-averaged powers $\bar{P}_\text{hull}$, 
$\bar{P}_\text{gyro}$, and $\bar{P}_\text{PTO}$ as a function of varying PTO damping coefficient. It can be seen that these three powers
are in reasonable agreement with each other, indicating that the energy transfer pathway Eq.~\eqref{eq_pathway} is approximately 
satisfied. In terms of power generation, it is observed that the device achieves peak performance when a PTO damping coefficient
$c = 0.3473$ is prescribed, which validates the theoretical procedure. The reason for an optimum value of $c$ is as follows: as the damping coefficient increases, the precession 
velocity decreases. The power absorbed by the PTO unit is the product of $c$ and $\dot{\varepsilon}^2$ (Eq.~\eqref{eq_PTO_power}), and
therefore these competing factors must be balanced in order to achieve maximum power generation.

Finally in Fig.~\ref{fig_yaw_torque}, we show the evolution of the the yaw torque $\cM_\phi$ acting on the hull for $c = 0.3473$,
noting that its magnitude is approximately one-fifth of the pitch torque $\cM_\delta$. Although this is not insignificant,
we do not consider the effect of $\cM_\phi$ for the 3D ISWEC model (see Sec.~\ref{sec_iswec_dynamics}) since its contribution will be cancelled out 1)~by using an even number of gyroscopic units (if each flywheel pair spins with equal and opposite velocities)~\cite{Raffero2014}, and 2)~partially by the mooring system. Discounting $\cM_\phi$ during the ISWEC design  phase would misalign the converter with respect to the main wave direction, which will reduce its performance. It is also interesting to note that the yaw torque in the gyroscopic frame of reference  $\cM_{z1}$ is at least two orders of magnitude lower than the yaw torque in the inertial reference frame, as evidenced by the inset of Fig.~\ref{fig_yaw_torque}.

\begin{figure}[]
 \centering
  \subfigure[Hull pitch angle]{
  \includegraphics[scale = 0.3]{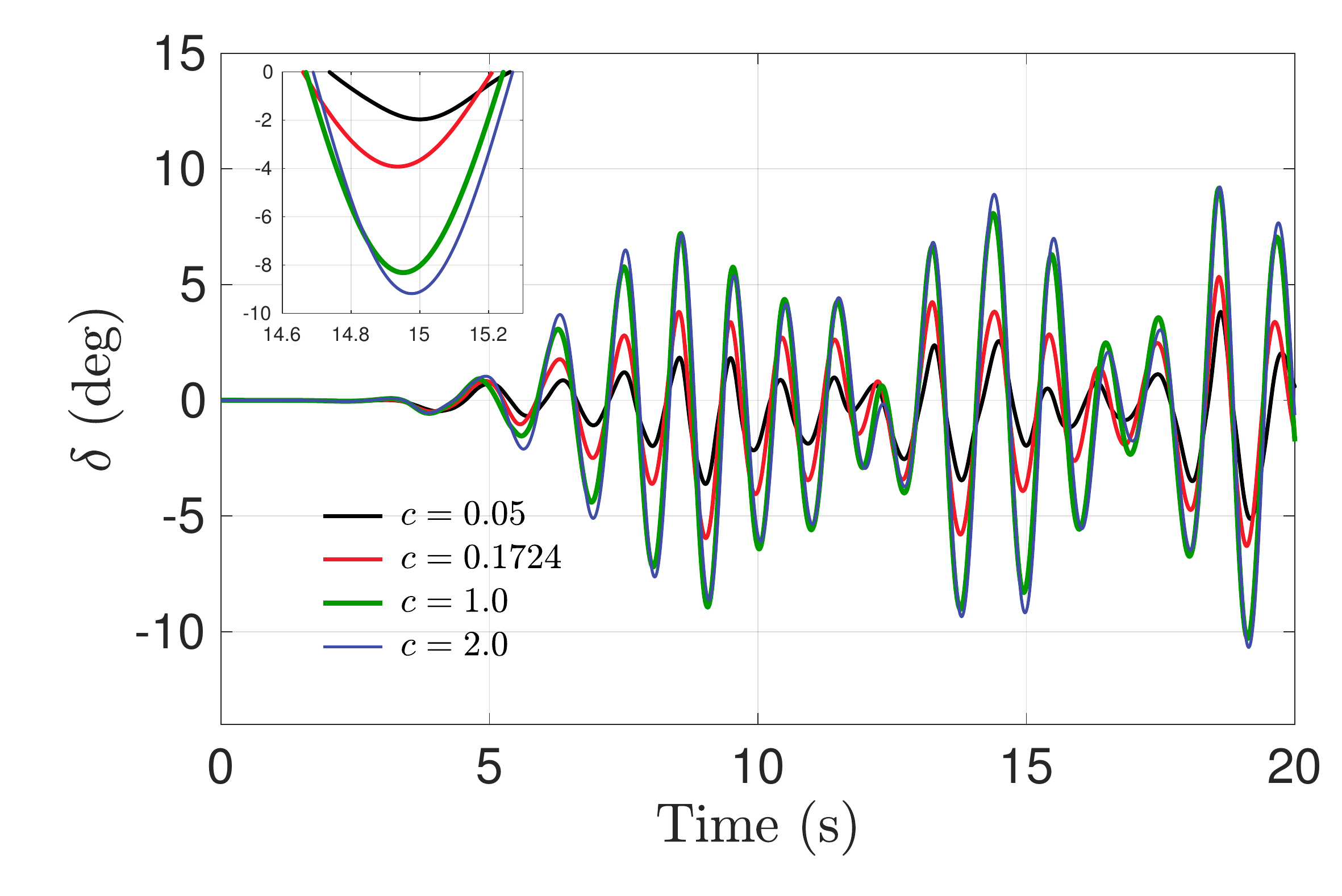}
  \label{fig_delta_c_compare_irregwave}
 }
   \subfigure[Gyroscope precession angle]{
  \includegraphics[scale = 0.3]{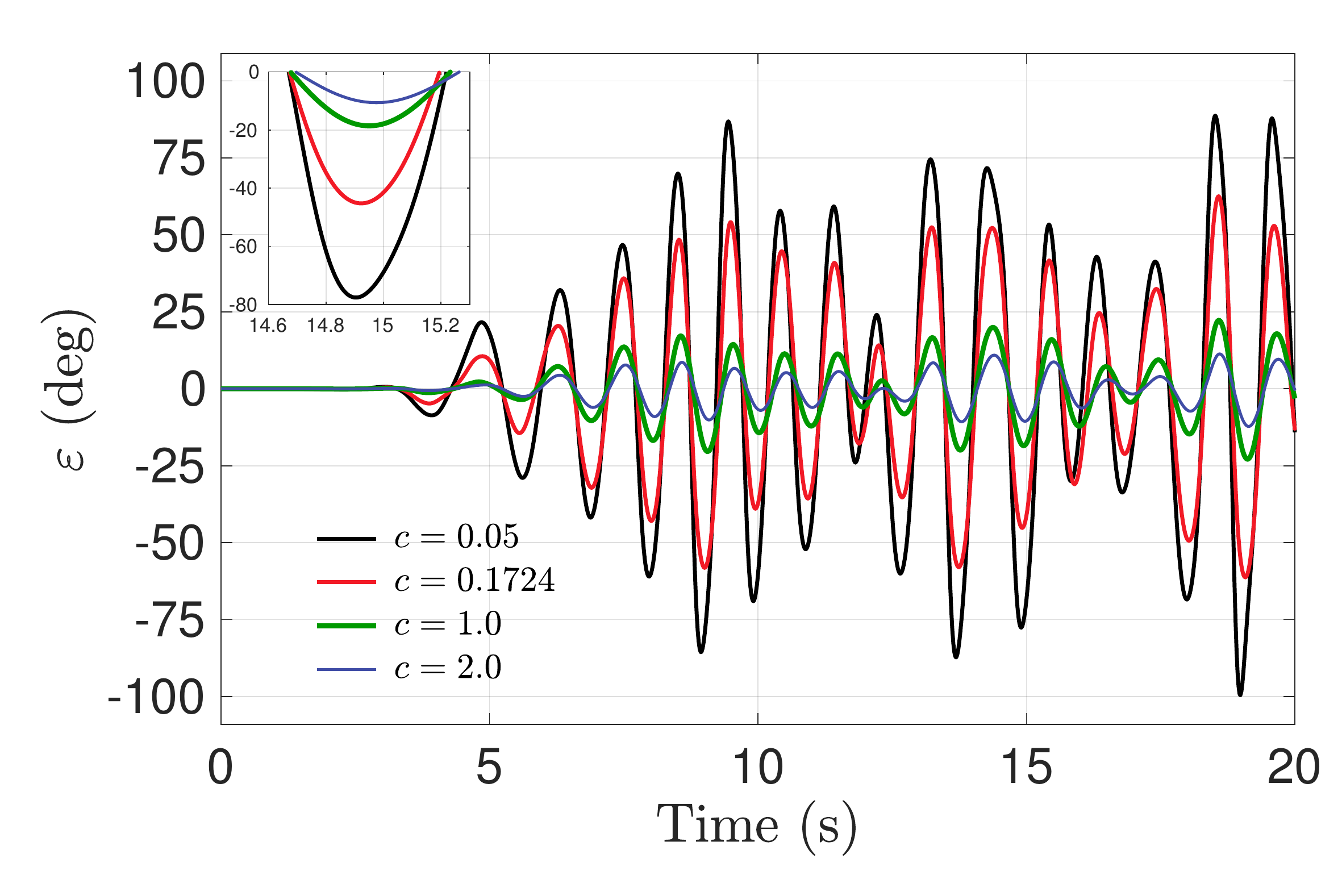}
  \label{fig_e_c_compare_irregwave}
 }
  \subfigure[Pitch torque unloaded on the hull]{
  \includegraphics[scale = 0.3]{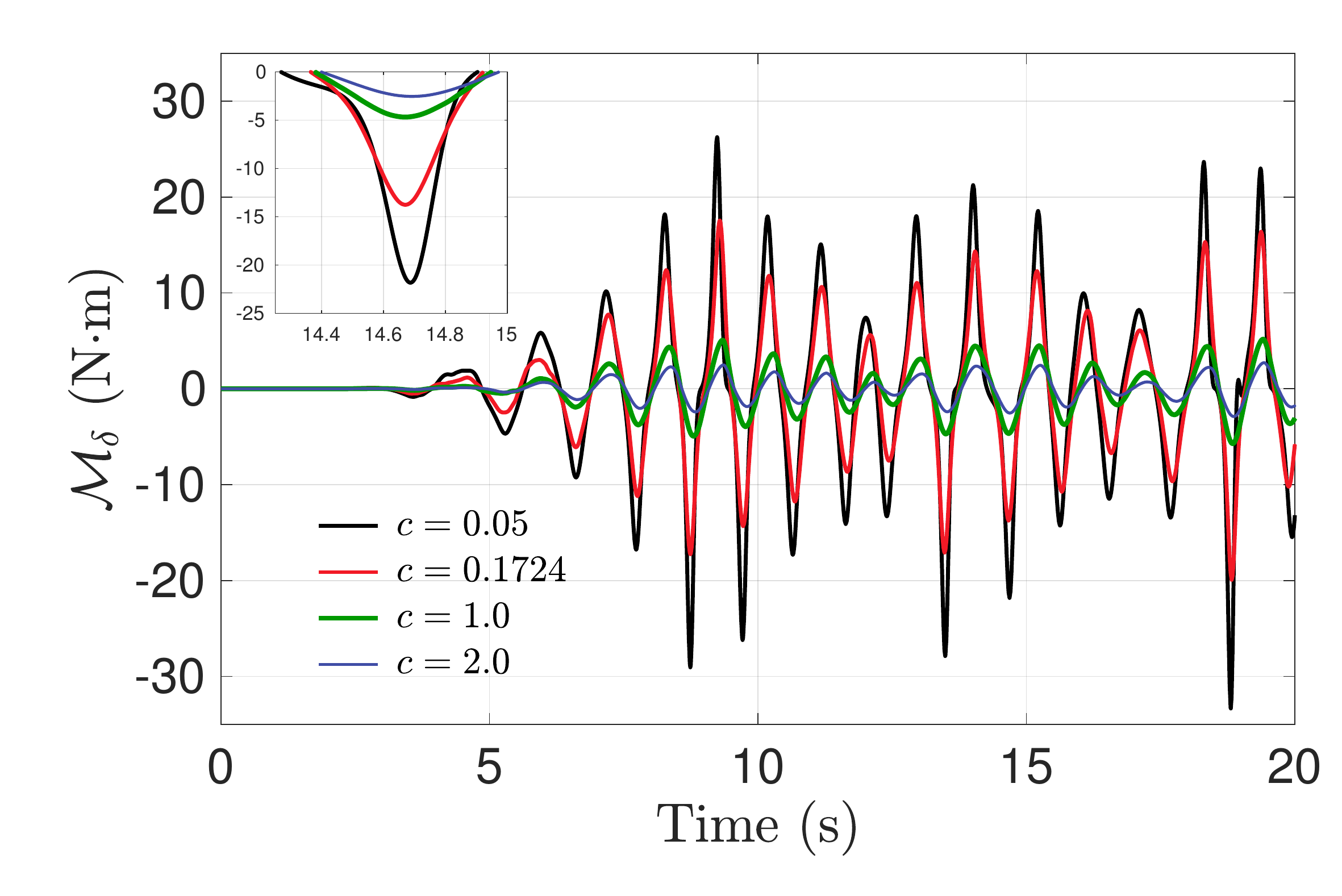}
  \label{fig_Md_c_compare_irregwave}
 }
  \subfigure[Precession torque on the PTO axis]{
  \includegraphics[scale = 0.3]{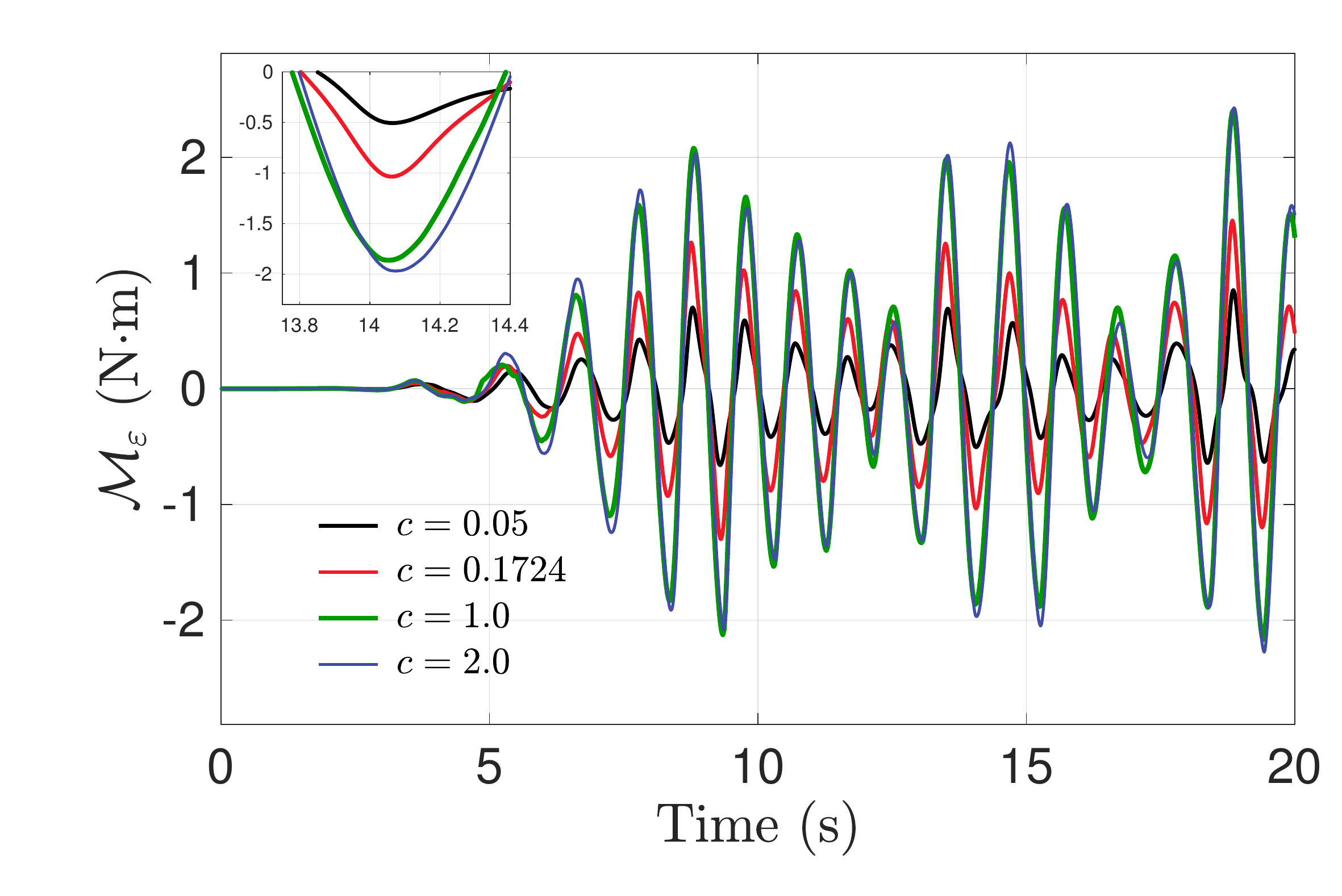}
  \label{fig_Me_c_compare_irregwave}
 }
   \subfigure[Time-averaged powers in the system]{
  \includegraphics[scale = 0.3]{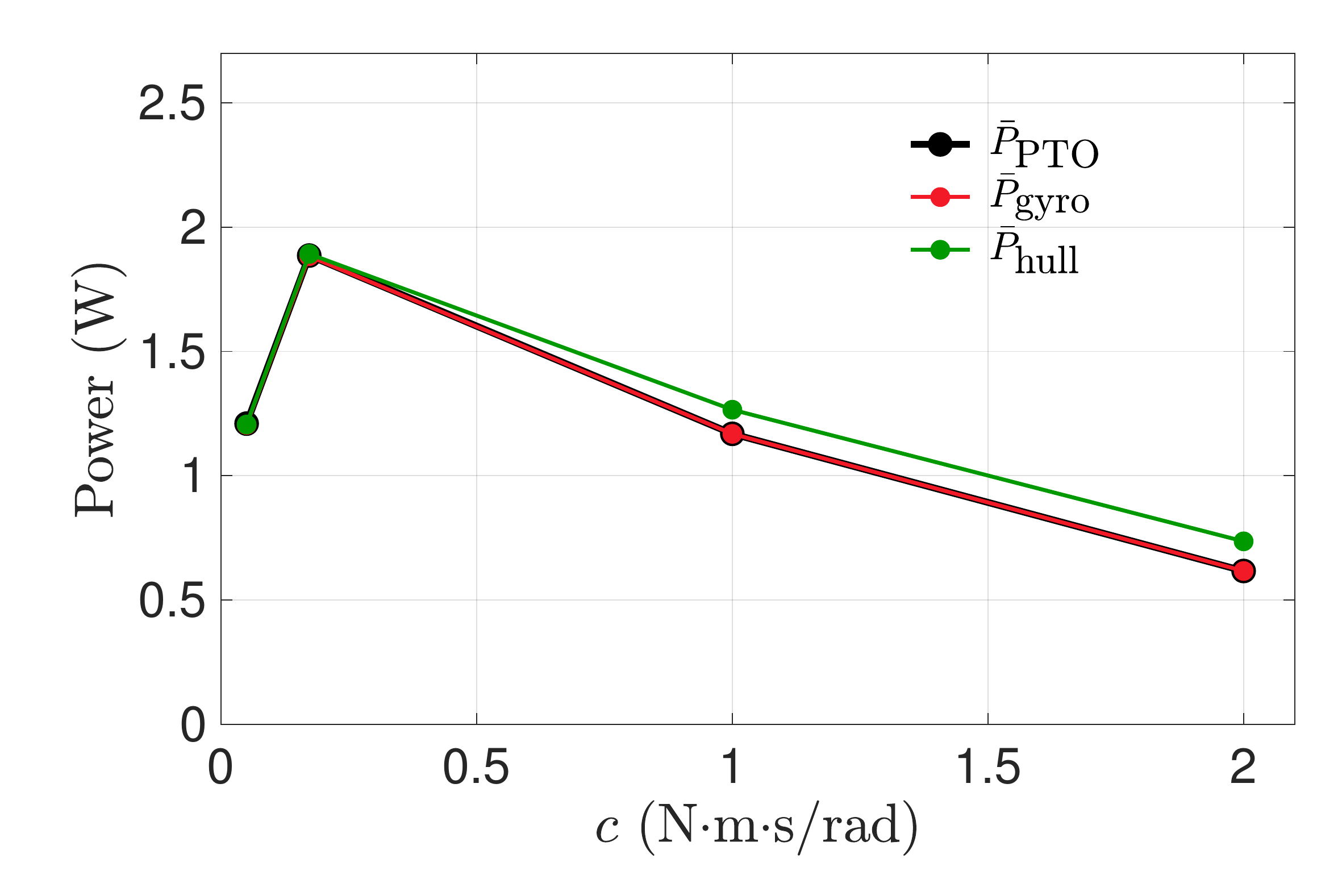}
  \label{fig_Powers_c_compare_irregwave}
  }
  \subfigure[Relative capture width (RCW)]{
  \includegraphics[scale = 0.3]{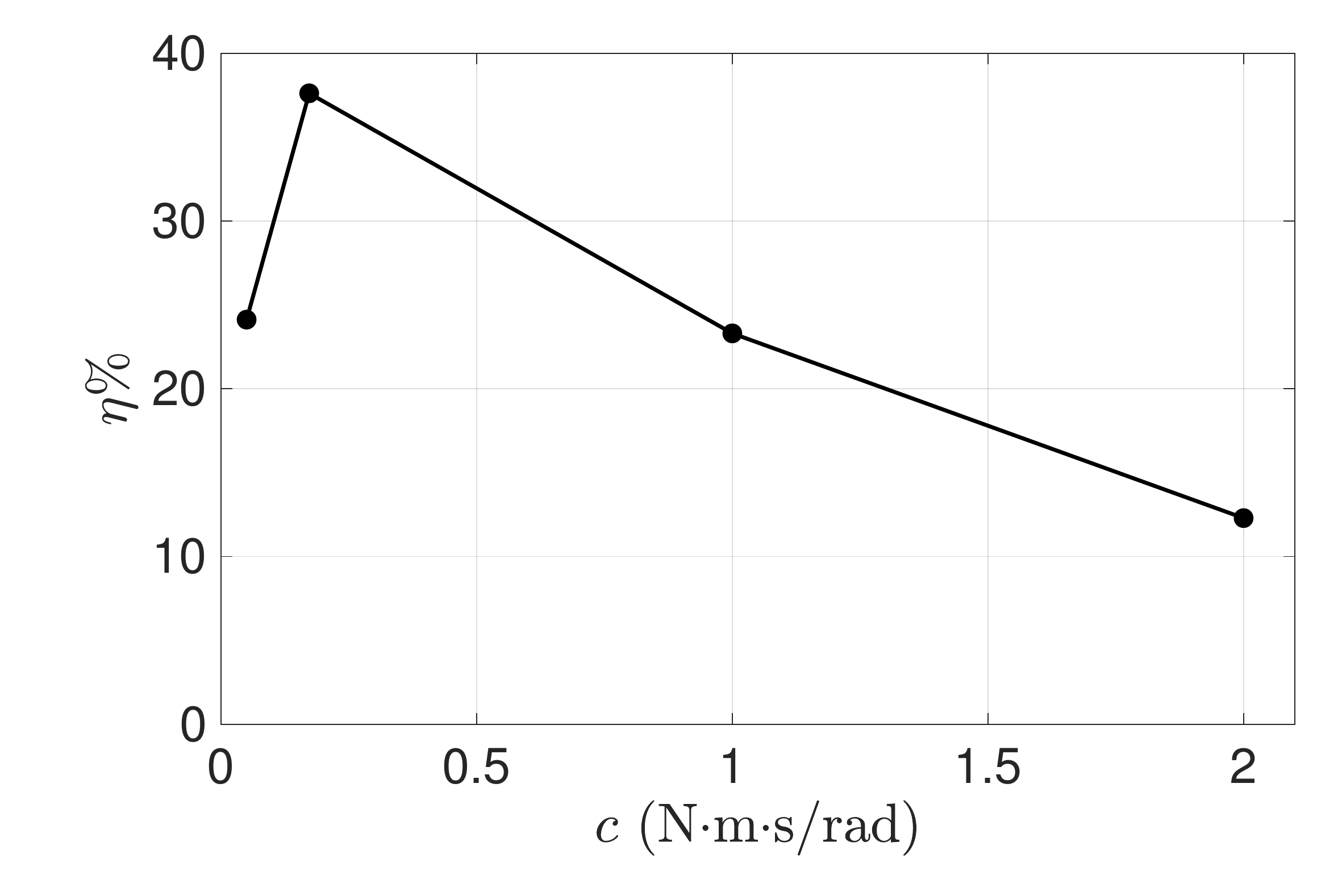}
  \label{fig_RCW_c_compare_irregwave}
 }
 \caption{Dynamics of the 2D ISWEC model for four different values of PTO damping coefficient $c$, with irregular wave properties $\cH_\text{s}$ = 0.1 m and $\cT_\text{p}$ = 1 s and $50$ wave components with frequencies $\omega_i$ in the range $3.8$ rad/s to $20$ rad/s. Temporal evolution of \subref{fig_delta_c_compare_regwave} hull pitch angle $\delta$, \subref{fig_e_c_compare_regwave} gyroscope precession angle $\varepsilon$, \subref{fig_Md_c_compare_regwave} pitch torque $\cM_\delta$, and \subref{fig_Me_c_compare_regwave} precession torque $\cM_\varepsilon$ for $c$ = 0.05 N$\cdot$m$\cdot$s/rad (\textcolor{black}{\textbf{-----}}, black), $c$ = 0.1724 N$\cdot$m$\cdot$s/rad (\textcolor{red}{\textbf{-----}}, red), $c$ = 1.0 N$\cdot$m$\cdot$s/rad (\textcolor{ForestGreen}{\textbf{-----}}, green), and $c$ = 2.0 N$\cdot$m$\cdot$s/rad (\textcolor{blue}{\textbf{-----}}, blue);
 \subref{fig_Powers_c_compare_regwave} comparison of time-averaged powers from the interval $t = 10$ s to $t = 20$ s for each value of $c$; \subref{fig_RCW_c_compare_irregwave} relative capture width $\eta$ for each value of $c$.
}
 \label{fig_c_compare_irregwave}
\end{figure}

Similar dynamics are observed when the ISWEC model is simulated in irregular wave conditions for four different values, $c$ = 0.05, 0.1724, 1.0 and 2.0 N$\cdot$m$\cdot$s/rad. The optimal damping coefficient value of $c = 0.1724$ is obtained from the theory. The results are compared in Fig.~\ref{fig_c_compare_irregwave} and the theoretically predicted optimum $c$ is verified. The response of the hull and gyroscope to irregular waves can be seen in Figs.~\ref{fig_delta_c_compare_irregwave} and \ref{fig_e_c_compare_irregwave}, respectively. The pitch torque and the precession torque are shown in Figs.~\ref{fig_Md_c_compare_irregwave} and \ref{fig_Me_c_compare_irregwave}, respectively. From Fig.~\ref{fig_Powers_c_compare_irregwave}, it is verified that the energy transfer pathway given by Eq.~\ref{eq_pathway} is satisfied. We note that the device efficiency is higher in irregular wave conditions as compared to regular wave conditions. This can be seen by comparing the maximum value of relative capture width for $\cH$ = 0.1 m in Figs.~\ref{fig_RCW_H_variation} and~\ref{fig_RCW_c_compare_irregwave}: $\eta_\text{max} = 24.36 \%$ vs. $\eta_\text{max} = 37.61 \%$, respectively. The power carried by irregular waves is approximately half that of regular waves when they have the same significant height and time period. Therefore for the prescribed device dimensions, the converter is more efficient in less energetic wave conditions.  

%%%%%%%%%%%%%%%%%%%%%%%%%%%%%
\subsubsection{Flywheel speed $\dot{\phi}$} \label{subsec_phi_variation}

Next, we conduct a parameter sweep of the flywheel speed $\dot{\phi}$ and investigate its effects on ISWEC dynamics.
The speed of the flywheel affects not only the amount of angular momentum $J\dot{\phi}$ generated in the gyroscope, but also the 
magnitude of the gyroscopic torques produced as seen in Eqs.~\eqref{eq_gyro_PTO} and~\eqref{eq_M_delta_simplified}. 
We consider four different flywheel speeds: $\dot{\phi} =  100$ RPM, $1000$ RPM, $4000$ RPM, and $8000$ RPM, with $\delta_0$ = 10$^\circ$ and the remaining gyroscope parameter are prescribed based on Table~\ref{tab_gyro_parameters_regularwave}. Recall that 
these values were obtained for $\dot{\phi} =  4000$ RPM in Table~\ref{tab_gyro_parameters_regularwave}.

\begin{figure}[]
 \centering
  \subfigure[Hull pitch angle]{
  \includegraphics[scale = 0.3]{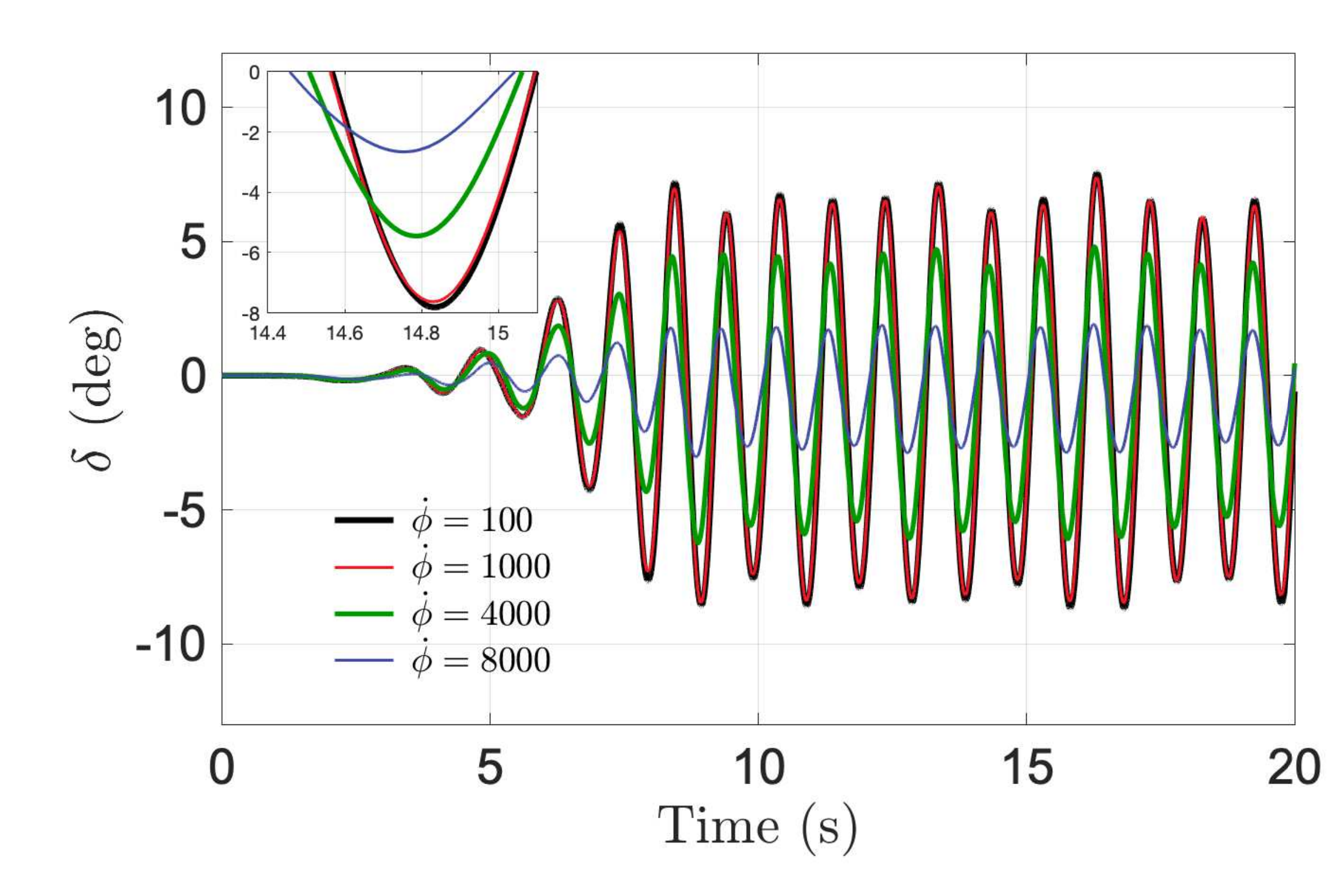}
  \label{fig_delta_rpm_comparison_regwave}
 }
   \subfigure[Gyroscope precession angle]{
  \includegraphics[scale = 0.3]{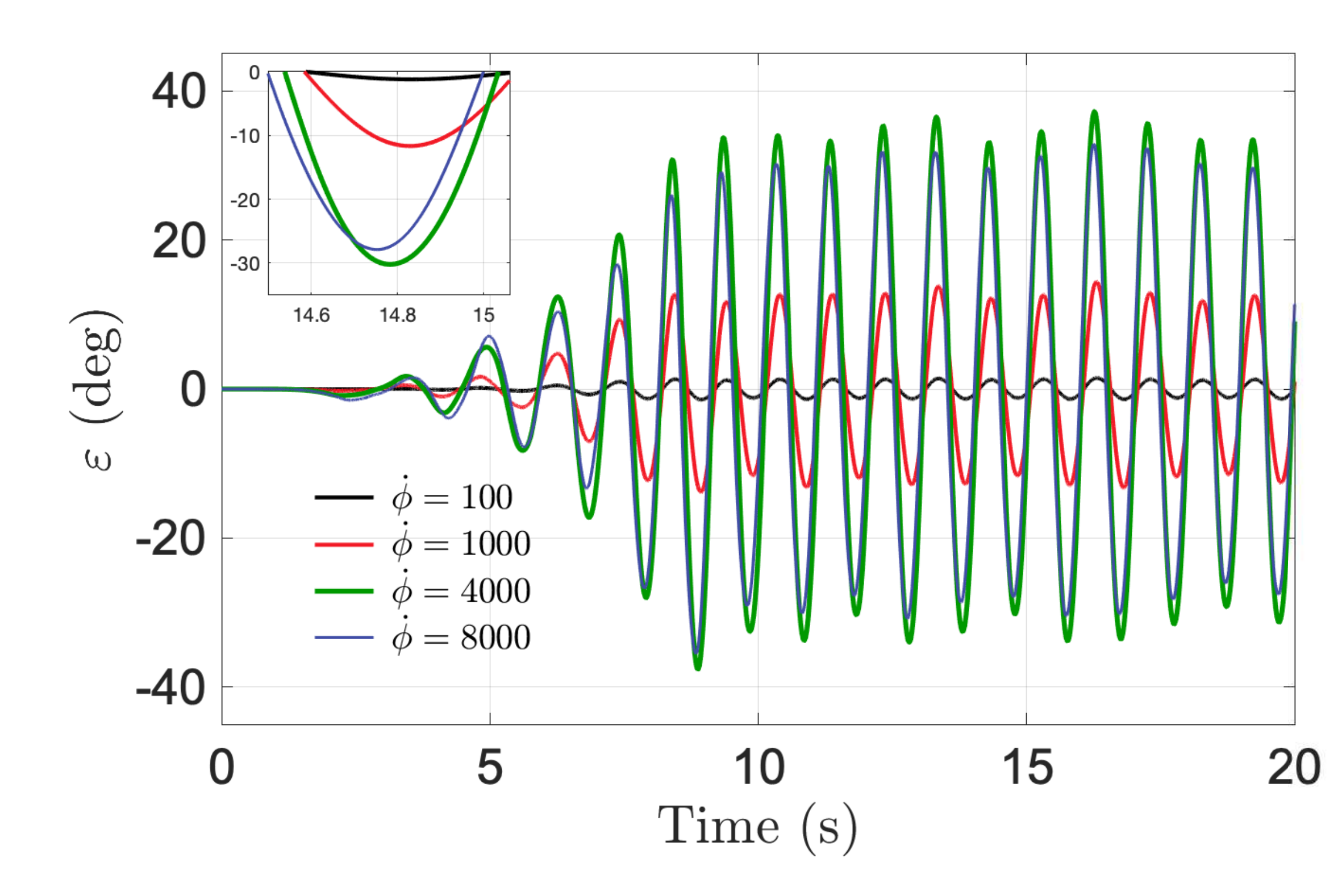}
  \label{fig_epsilon_rpm_comparison_regwave}
 }
  \subfigure[Powers]{
  \includegraphics[scale = 0.3]{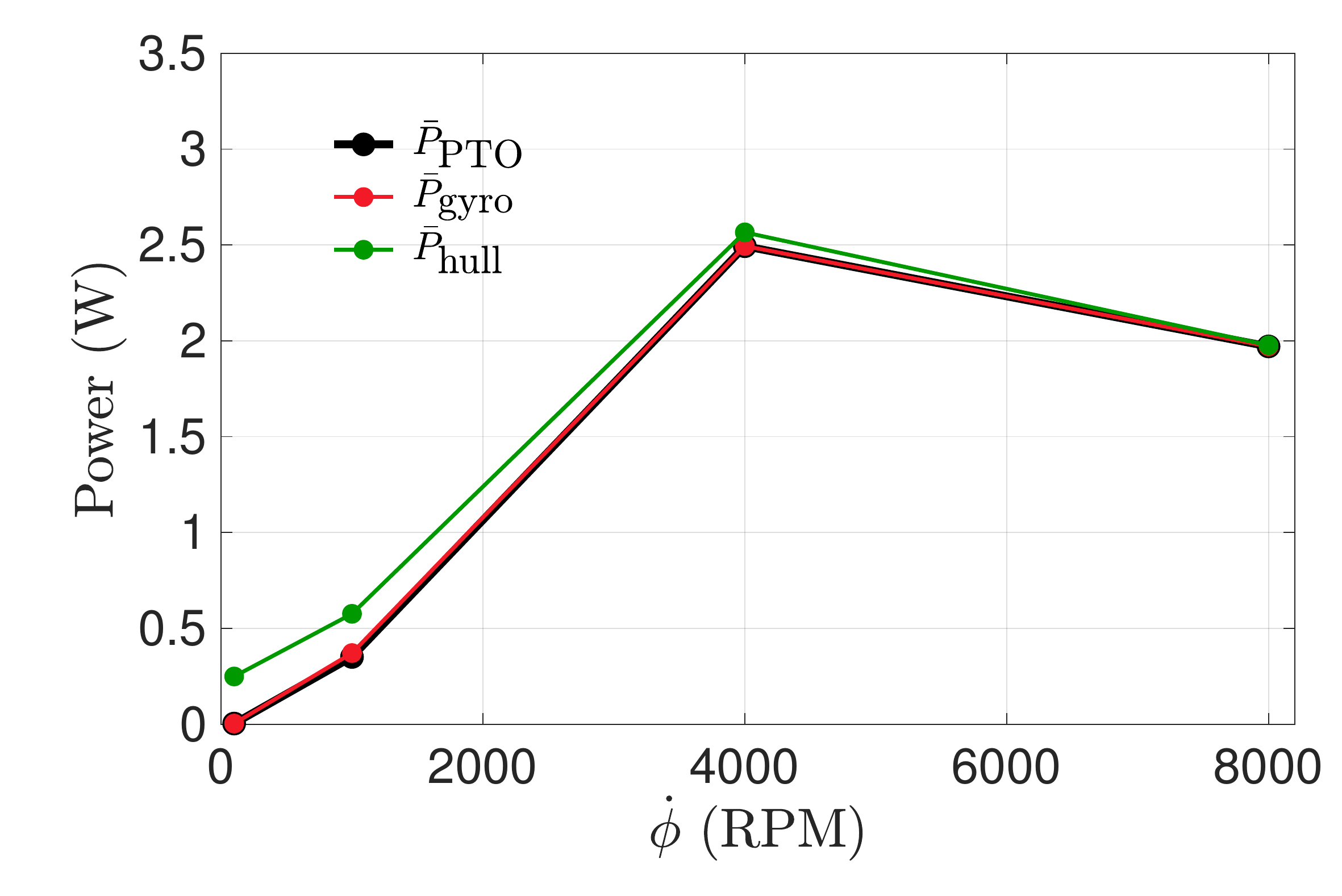}
  \label{fig_Powers_vs_rpm_regwave}
 }
 \caption{Dynamics of the 2D ISWEC model for four different values of flywheel speed $\dot{\phi}$. The regular wave properties are $\cH$ = 0.1 m and $\cT$ = 1 s. Temporal evolution of  \subref{fig_delta_rpm_comparison_regwave} hull pitch angle $\delta$, and \subref{fig_epsilon_rpm_comparison_regwave} gyroscope precession angle $\varepsilon$ for $\dot{\phi}$ = 100 RPM (\textcolor{black}{\textbf{-----}}, black), $\dot{\phi}$ = 1000 RPM (\textcolor{red}{\textbf{-----}}, red), $\dot{\phi}$ = 4000 RPM (\textcolor{ForestGreen}{\textbf{-----}}, green), and $\dot{\phi}$ = 8000 RPM (\textcolor{blue}{\textbf{-----}} , blue); \subref{fig_Powers_vs_rpm_regwave} comparison of time-averaged powers from the interval $t = 10$ s to $t = 20$ s for each value of $\dot{\phi}$.
}
 \label{fig_rpm_comparison}
\end{figure}

The results for a hull interacting with regular waves are shown in Fig.~\ref{fig_rpm_comparison}. It is seen that the maximum 
pitch angle decreases with increasing $\dot{\phi}$ (Fig.~\ref{fig_delta_rpm_comparison_regwave}), while a non-monotonic
relationship is seen between the maximum precession angle and $\dot{\phi}$ (Fig.~\ref{fig_epsilon_rpm_comparison_regwave}).
Time-averaged powers are shown in Fig.~\ref{fig_Powers_vs_rpm_regwave}, which again shows that Eq.~\eqref{eq_pathway}
is approximately satisfied. Power absorption is maximized at a flywheel speed of $\dot{\phi} = 4000$ RPM, which can be physically explained as follows. As $J\dot{\phi}$ increases, the gyroscopic system is able to generate significant precession torque which,
increases the absorption capacity of the PTO unit. However, this increased angular momentum also increases the pitch torque
opposing the hull, thereby limiting its pitching motion and reducing the power absorbed from the waves. 
These two competing factors leads to an optimum value of $\dot{\phi}$.

\begin{figure}[]
 \centering
  \subfigure[Hull pitch angle]{
  \includegraphics[scale = 0.3]{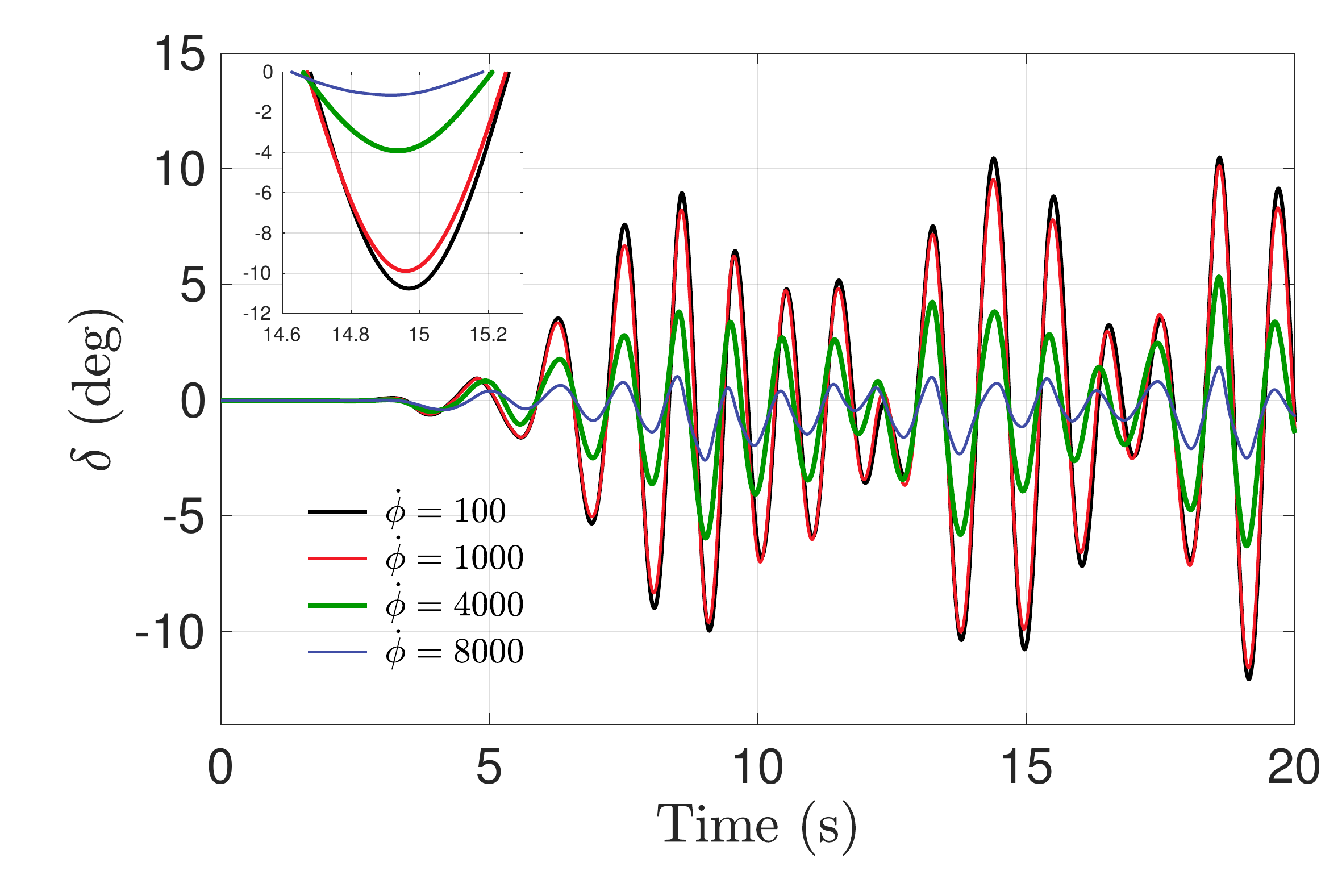}
  \label{fig_delta_rpm_comparison_irregwave}
 }
   \subfigure[Gyroscope precession angle]{
  \includegraphics[scale = 0.3]{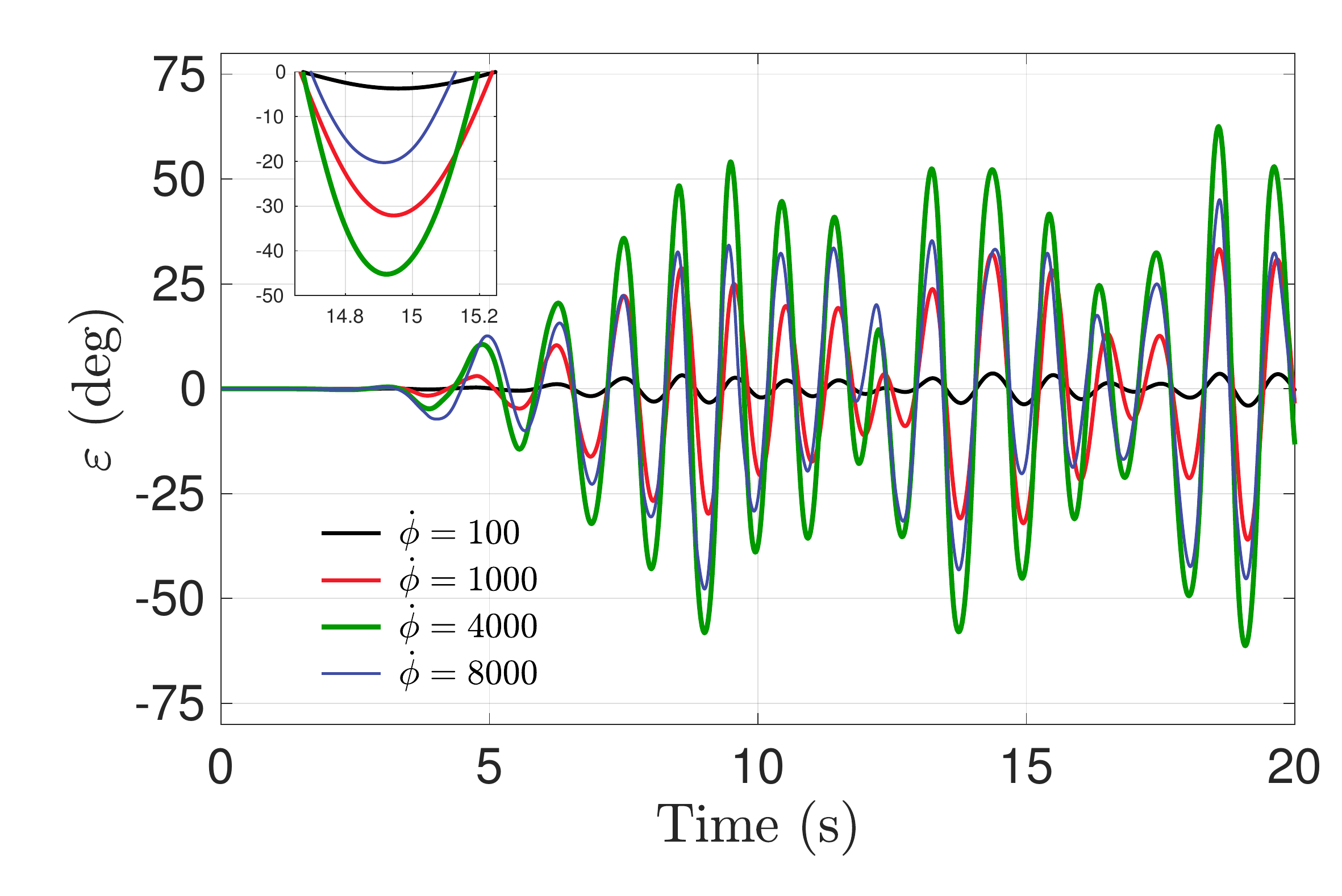}
  \label{fig_epsilon_rpm_comparison_irregwave}
 }
  \subfigure[Powers]{
  \includegraphics[scale = 0.3]{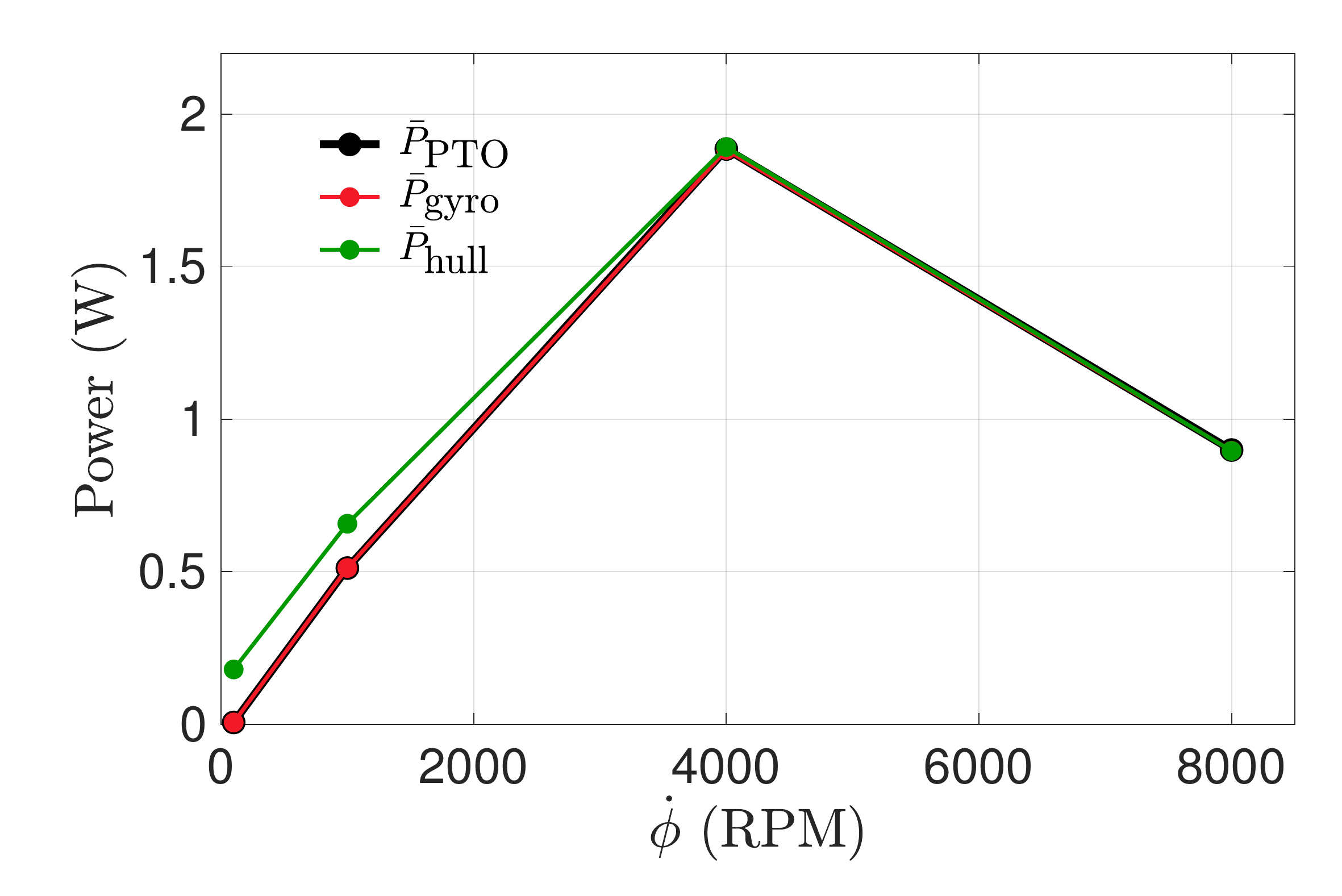}
  \label{fig_Powers_vs_rpm_irregwave}
 }
 \caption{Dynamics of the 2D ISWEC model for four different values of flywheel speed $\dot{\phi}$. The irregular wave properties are $\cH_\text{s}$ = 0.1 m and $\cT_\text{p}$ = 1 s and $50$ wave components with frequencies $\omega_i$ in the range $3.8$ rad/s to $20$ rad/s. Temporal evolution of  \subref{fig_delta_rpm_comparison_irregwave} hull pitch angle $\delta$, and \subref{fig_epsilon_rpm_comparison_irregwave} gyroscope precession angle $\varepsilon$ for $\dot{\phi}$ = 100 RPM (\textcolor{black}{\textbf{-----}}, black), $\dot{\phi}$ = 1000 RPM (\textcolor{red}{\textbf{-----}}, red), $\dot{\phi}$ = 4000 RPM (\textcolor{ForestGreen}{\textbf{-----}}, green), and $\dot{\phi}$ = 8000 RPM (\textcolor{blue}{\textbf{-----}} , blue); \subref{fig_Powers_vs_rpm_irregwave} comparison of time-averaged powers from the interval $t = 10$ s to $t = 20$ s for each value of $\dot{\phi}$.
}
 \label{fig_rpm_variation_irregwave}
\end{figure}

Similar dynamics are obtained when the ISWEC interacts with irregular waves for varying values of $\dot{\phi}$. The results are shown in Fig.~\ref{fig_rpm_variation_irregwave}. The comparison of pitch angle for various $\dot{\phi}$ values is shown in Fig.~\ref{fig_delta_rpm_comparison_irregwave} and of precession angle is shown in Fig.~\ref{fig_epsilon_rpm_comparison_irregwave}.  Eq.~\ref{eq_pathway} is again satisfied as seen from the time-averaged powers in Fig.~\ref{fig_Powers_vs_rpm_irregwave}. 

%%%%%%%%%%%%%%%%%%%%%%%%%%%%%
\subsubsection{Flywheel moment of inertia $J$ and $I$}\label{subsec_J_variation}

The angular momentum $J\dot{\phi}$ generated in the gyroscope can also be modified by varying the flywheel size via its moment of inertia components $J$ and 
$I$. First, we consider three different values $J = 0.0005$ kg$\cdot$m$^2$, $0.0058$ kg$\cdot$m$^2$ and $0.5$ kg$\cdot$m$^2$, which 
correspond to light, medium, and heavy weight gyroscopes, respectively. The $J = 0.0058$ value is obtained from theoretical estimates based on the prescribed $\delta_0$ and $\varepsilon_0$ values. A value of $I = 0.94 \times J$ is set for each case, and
the remaining gyroscope parameters are prescribed based on Table~\ref{tab_gyro_parameters_regularwave}.

\begin{figure}[]
 \centering
  \subfigure[Hull pitch angle]{
  \includegraphics[scale = 0.3]{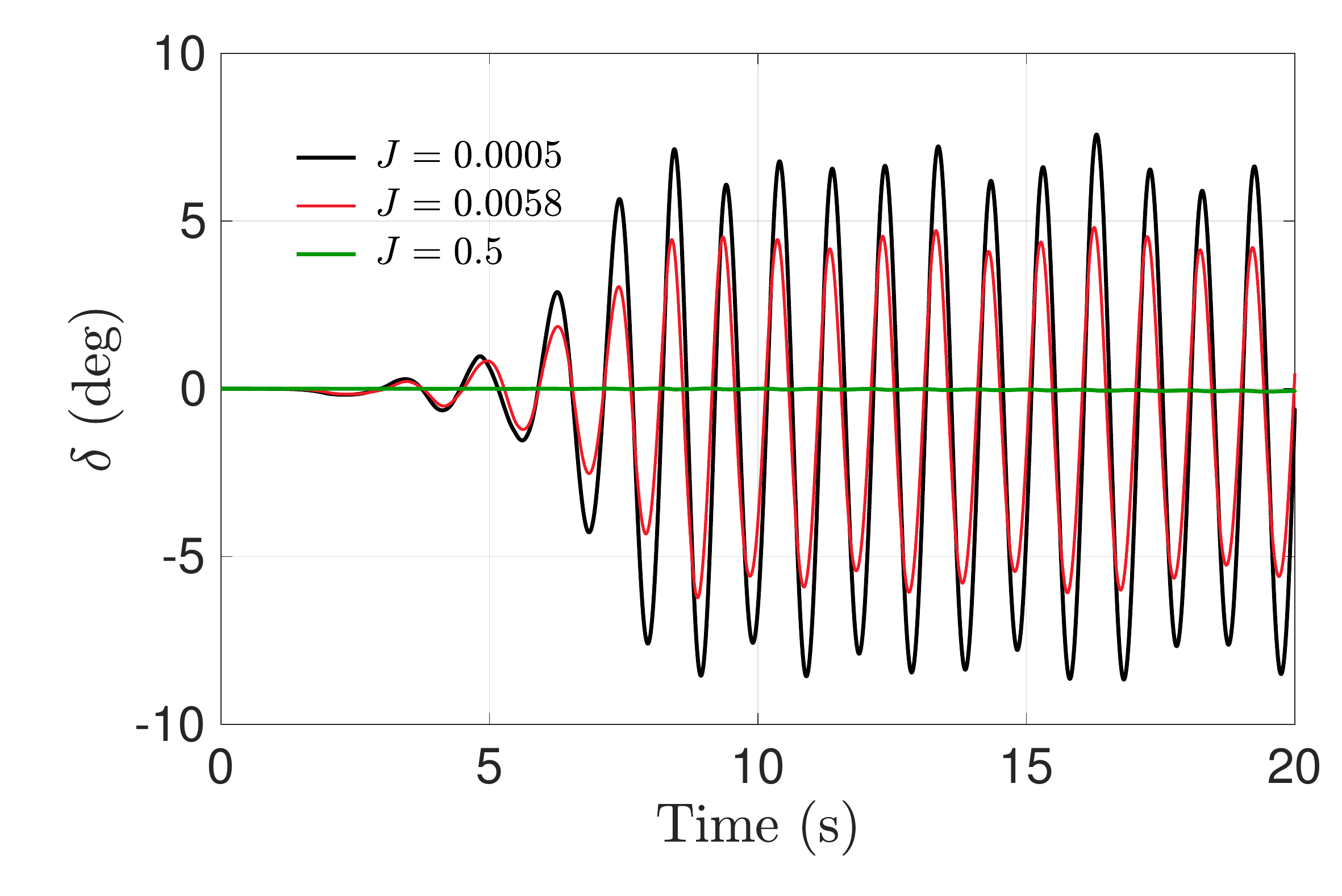}
  \label{fig_delta_J_comparison_regwave}
 }
   \subfigure[Gyroscope precession angle]{
  \includegraphics[scale = 0.3]{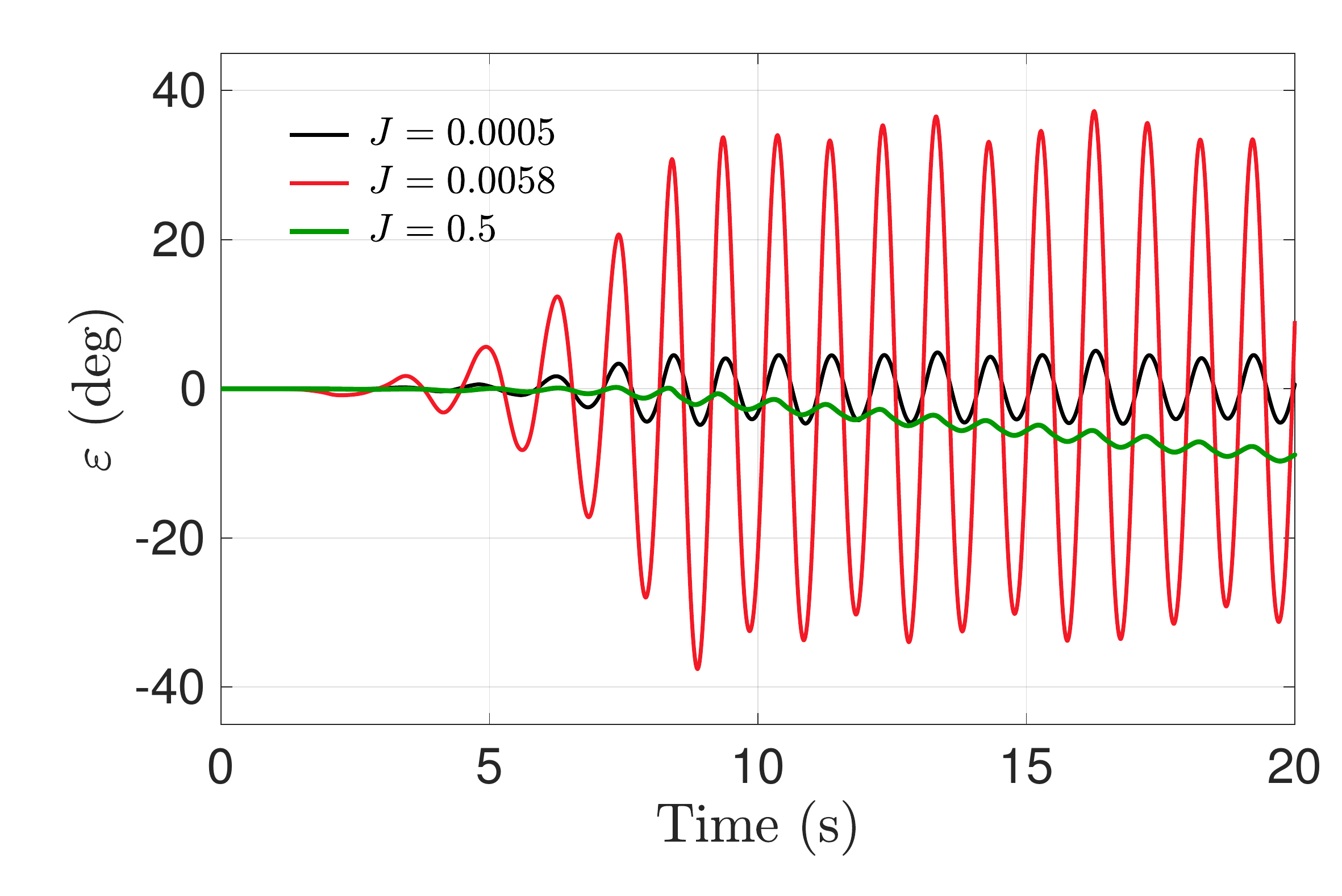}
  \label{fig_epsilon_J_comparison_regwave}
 }
  \subfigure[Pitch torque unloaded on the hull]{
  \includegraphics[scale = 0.3]{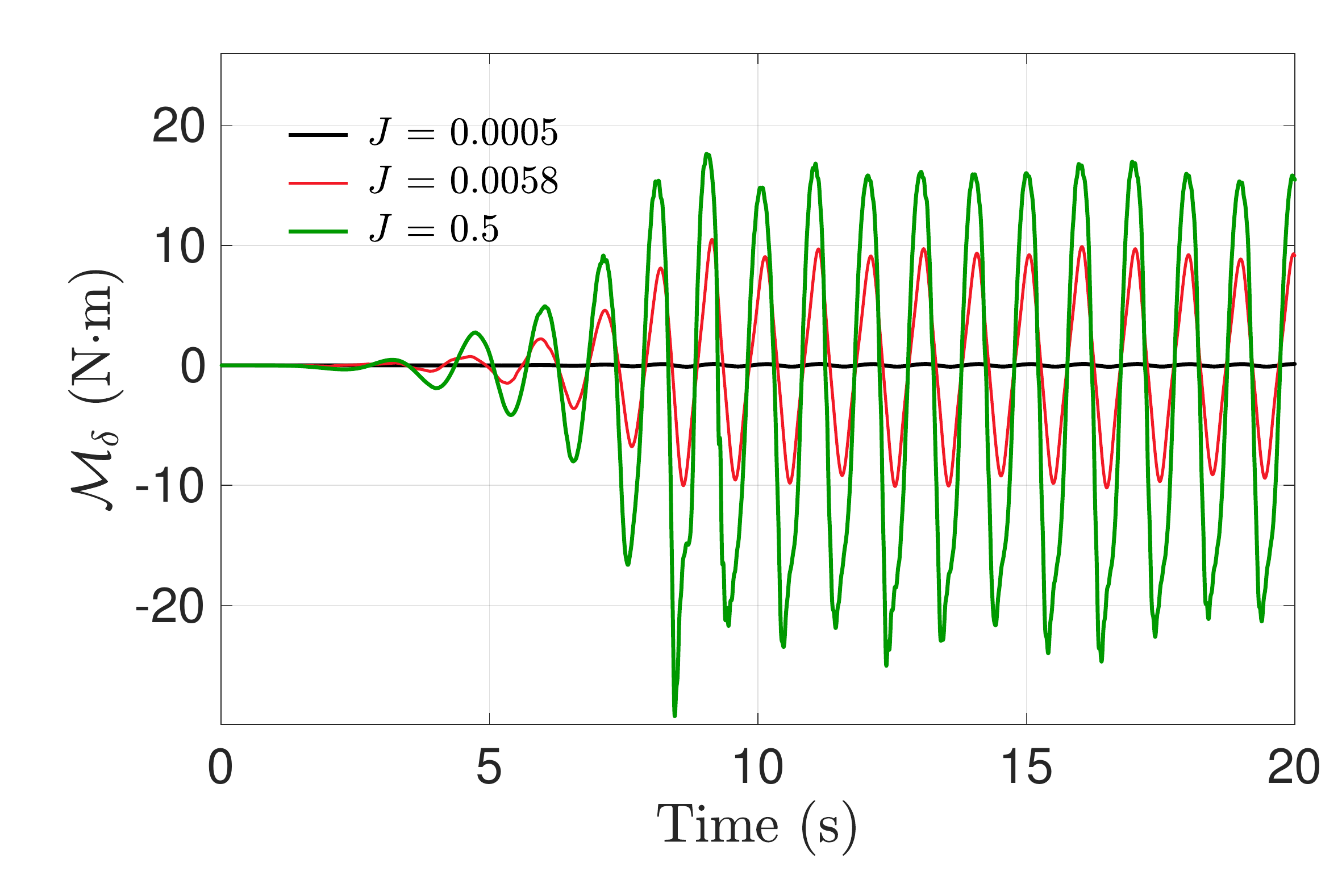}
  \label{fig_Md_J_comparison}
 }
  \subfigure[Precession torque on the PTO axis]{
  \includegraphics[scale = 0.3]{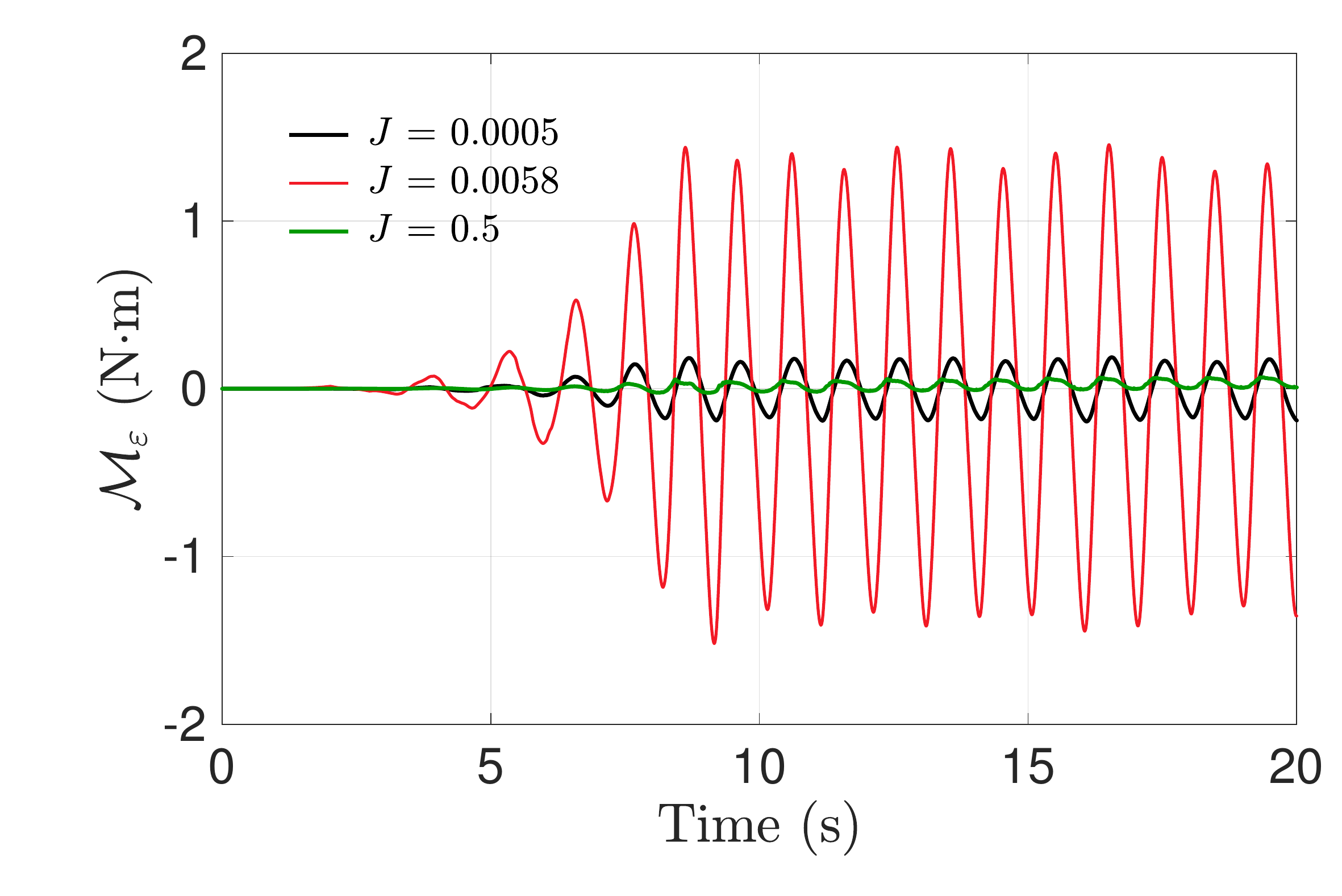}
  \label{fig_Me_J_comparison}
 }
 \caption{Dynamics of 2D ISWEC model for three different values of flywheel moment of inertia $J$. The regular wave properties are $\cH$ = 0.1 m and $\cT$ = 1 s. Temporal evolution of 
 \subref{fig_delta_J_comparison_regwave} hull pitch angle $\delta$,
 \subref{fig_epsilon_J_comparison_regwave} gyroscope precession angle $\varepsilon$,
 \subref{fig_Md_J_comparison} pitch torque $\cM_\delta$, and
 \subref{fig_Me_J_comparison} precession torque $\cM_\varepsilon$
 for
 $J$ = 0.0005 kg$\cdot$m$^2$ (\textcolor{black}{\textbf{-----}}, black),
 $J$ = 0.0058 kg$\cdot$m$^2$ (\textcolor{red}{\textbf{-----}}, red),
 and $J$ = 0.5 kg$\cdot$m$^2$ (\textcolor{ForestGreen}{\textbf{-----}}, green).
 For all cases, $I = 0.94 \times J$.
 }
 \label{fig_J_comparison}
\end{figure}

The results for a hull interacting with regular waves are shown in Fig.~\ref{fig_J_comparison}. It is seen that the light gyroscope produces 
insignificant precession angles and torques due to the lack of angular momentum generated by the flywheel. Moreover, the heavy
gyroscope produces even smaller $\cM_{\varepsilon}$ torque as it slowly drifts around the PTO axis; the proportional component of the control torque ($k \varepsilon$) is not strong enough to return the gyroscope to its mean position of $\varepsilon = 0^\circ$. 
Additionally, the light (heavy) weight gyroscope produces 
small (large) pitch torques $\cM_\delta$ opposing the hull, which explains the large (small) pitch amplitudes exhibited by the device.
Finally, it is seen that the medium weight gyroscope, with $J = 0.0058$ kg$\cdot$m$^2$ calculated from the procedure described in
Sec.~\ref{sec_PTO_params}, produces the largest precession amplitudes $\varepsilon$ and velocities $\dot{\varepsilon}$, leading
to high power absorption by the PTO unit.

\begin{figure}[]
 \centering
  \subfigure[Hull pitch angle]{
  \includegraphics[scale = 0.3]{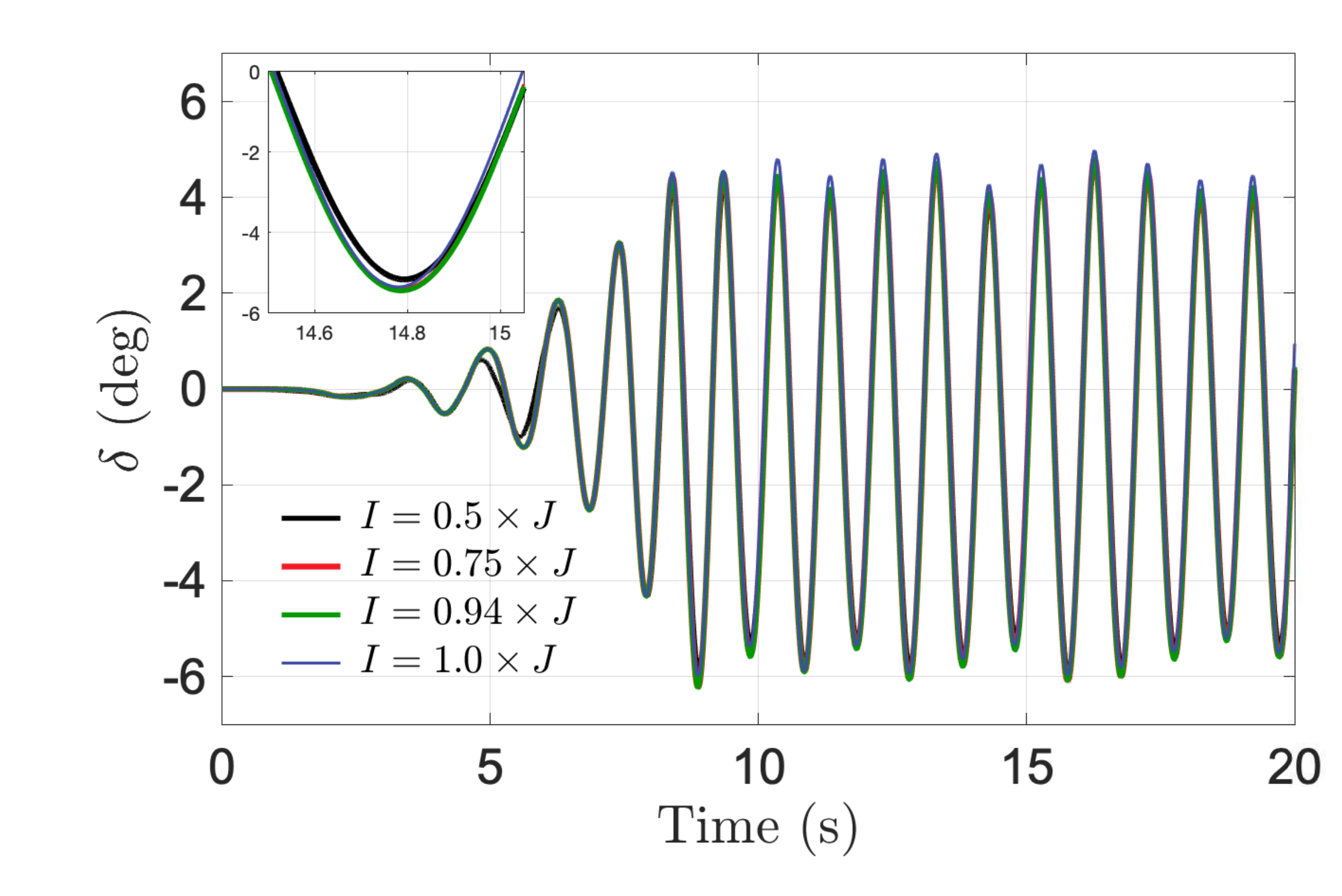}
  \label{fig_delta_I_variation}
 }
   \subfigure[Gyroscope precession angle]{
  \includegraphics[scale = 0.3]{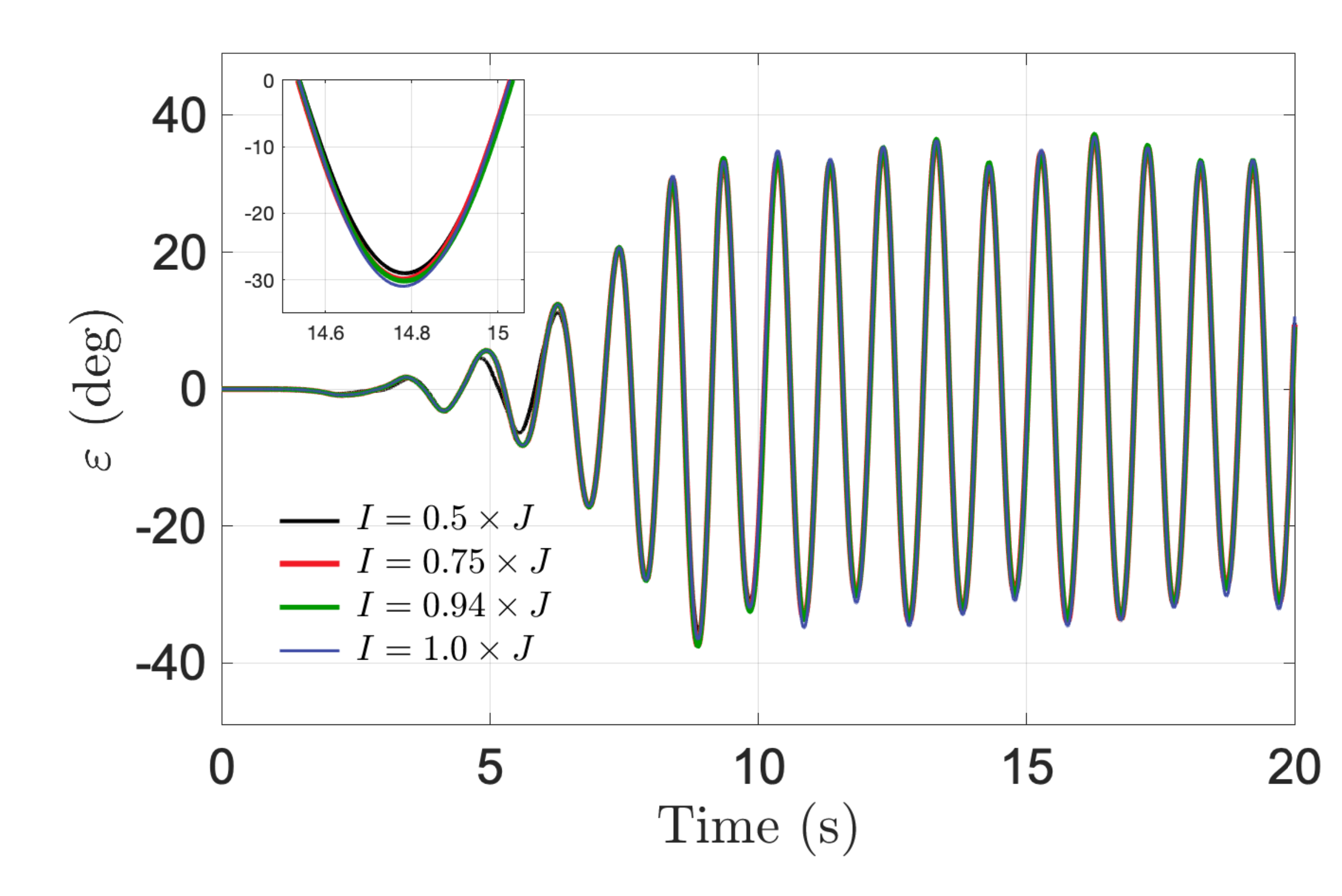}
  \label{fig_e_I_variation}
 }
  \subfigure[Powers]{
  \includegraphics[scale = 0.3]{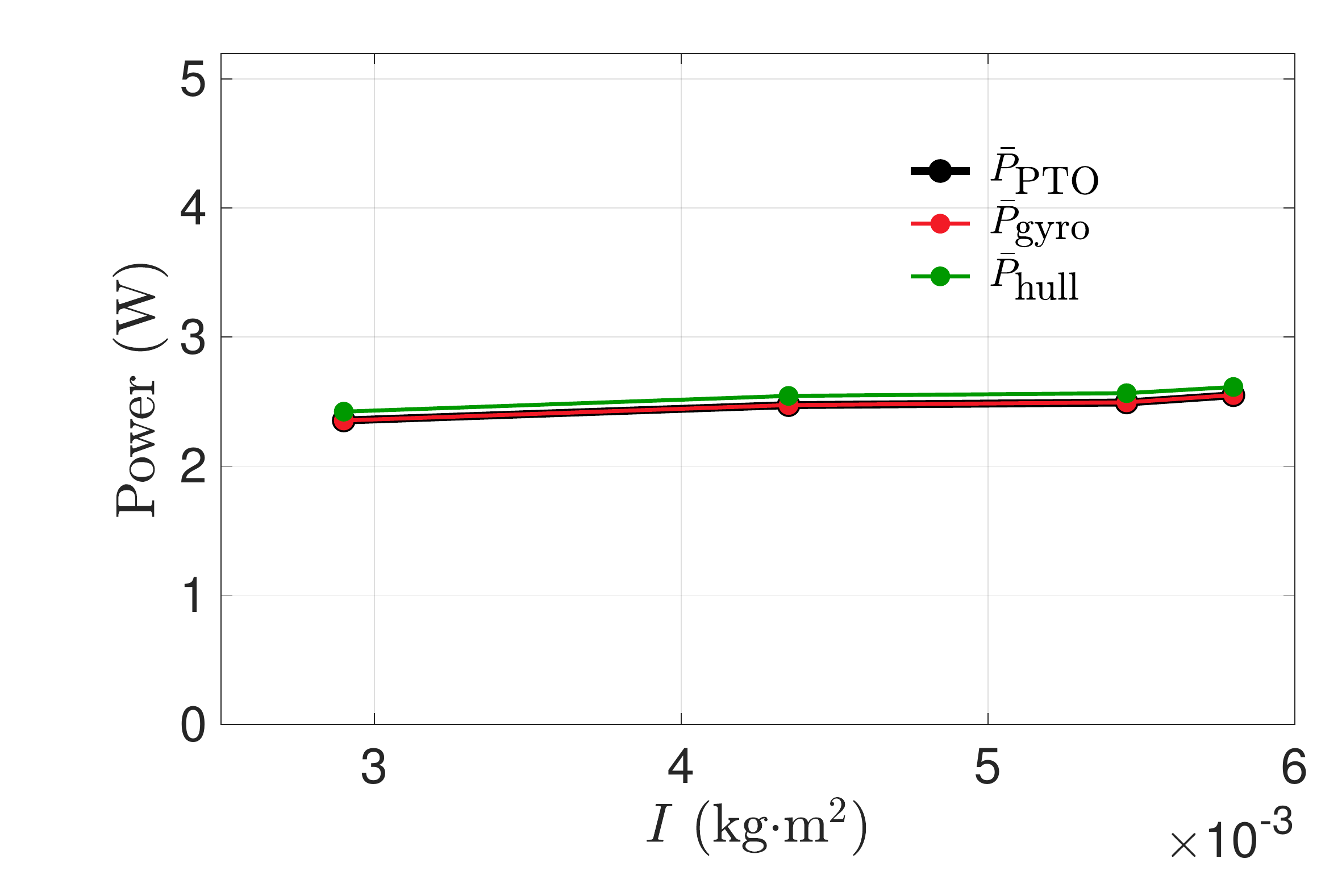}
  \label{fig_Powers_I_variation}
 }
 \caption{Dynamics of 2D ISWEC model for four different values of $I$. The regular wave properties are $\cH$ = 0.1 m and $\cT$ = 1 s.
 Temporal evolution of
 \subref{fig_delta_I_variation} hull pitch angle $\delta$, and
 \subref{fig_e_I_variation} gyroscope precession angle $\varepsilon$ for
 $I = 0.5 \times J$ (\textcolor{black}{\textbf{-----}}, black),
 $I = 0.75 \times J$ (\textcolor{red}{\textbf{-----}}, red), 
 $I = 0.94 \times J$ (\textcolor{ForestGreen}{\textbf{-----}}, green), and
 $I = 1.0 \times J$ (\textcolor{blue}{\textbf{-----}} , blue).
 \subref{fig_Powers_vs_rpm_regwave} Comparison of time-averaged powers from the interval $t = 10$ s to $t = 20$ s for each value of $I$.
 For all cases, $J$ = 0.0058 kg$\cdot$m$^2$.
}
 \label{fig_I_comparison}
\end{figure}

We also study the effect of varying $I$ while keeping $J = 0.0058$ kg$\cdot$m$^2$ fixed. We consider four different values 
$I = 0.5 \times J$, $I = 0.75 \times J$, $I = 0.94 \times J$ and $I = 1.0 \times J$, and the results for a device interacting with regular waves 
are shown in Fig.~\ref{fig_I_comparison}. It is seen that the dynamics of the hull and gyroscope and the system powers are not 
significantly affected by the choice of $I$.

\begin{figure}[]
 \centering
  \subfigure[Hull pitch angle]{
  \includegraphics[scale = 0.3]{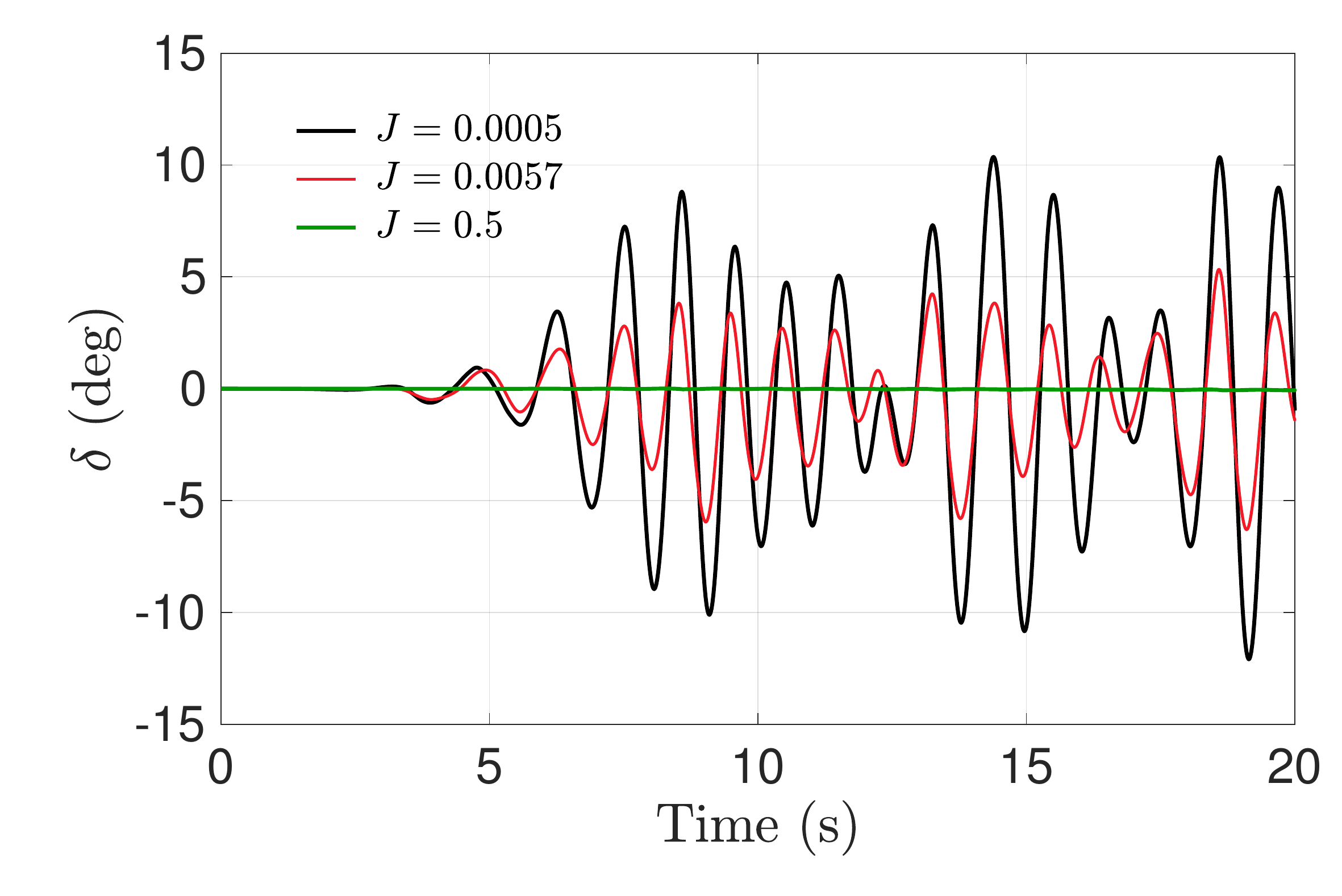}
  \label{fig_delta_J_comparison_irregwave}
 }
   \subfigure[Gyroscope precession angle]{
  \includegraphics[scale = 0.3]{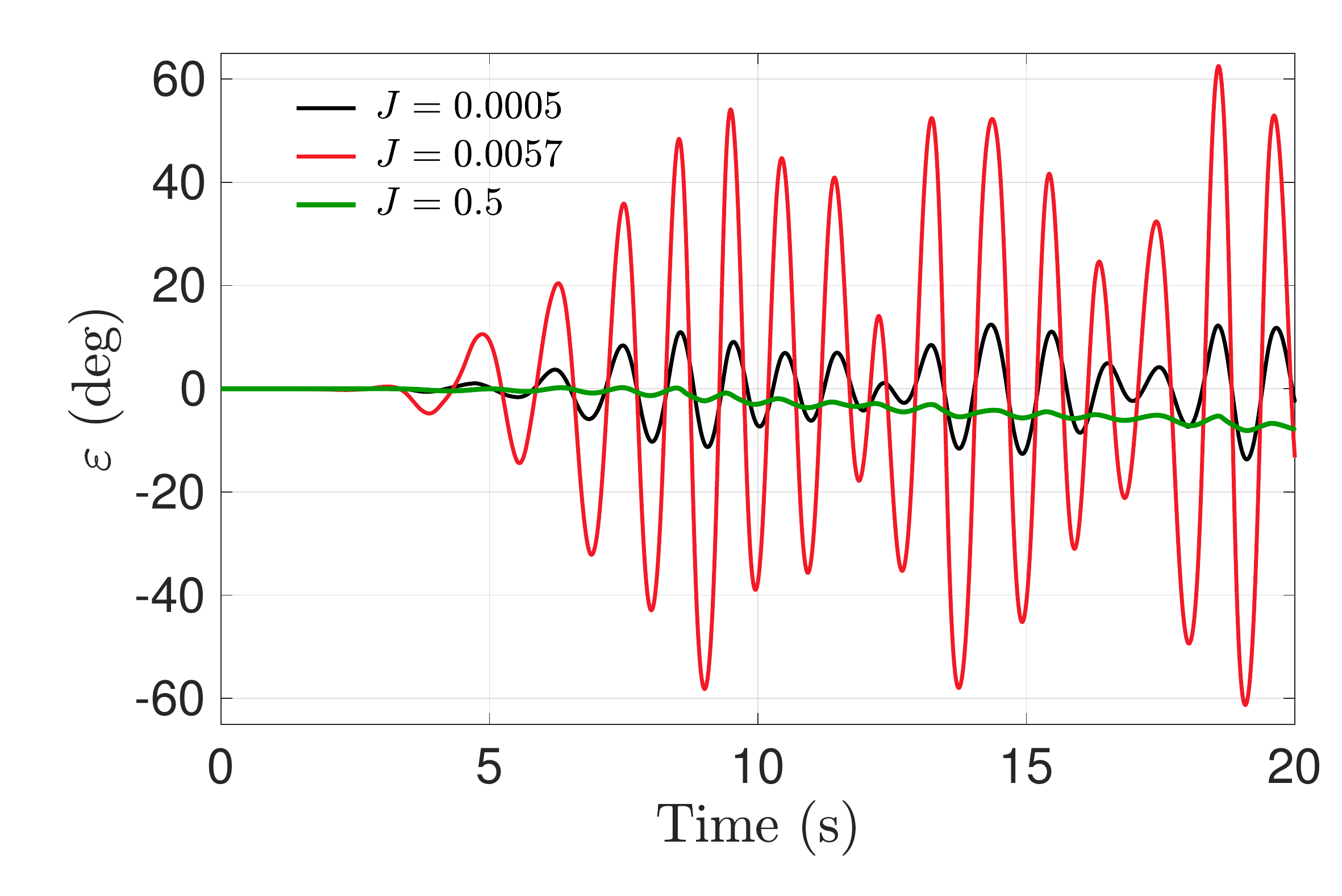}
  \label{fig_epsilon_J_comparison_irregwave}
 }
  \subfigure[Pitch torque unloaded on the hull]{
  \includegraphics[scale = 0.3]{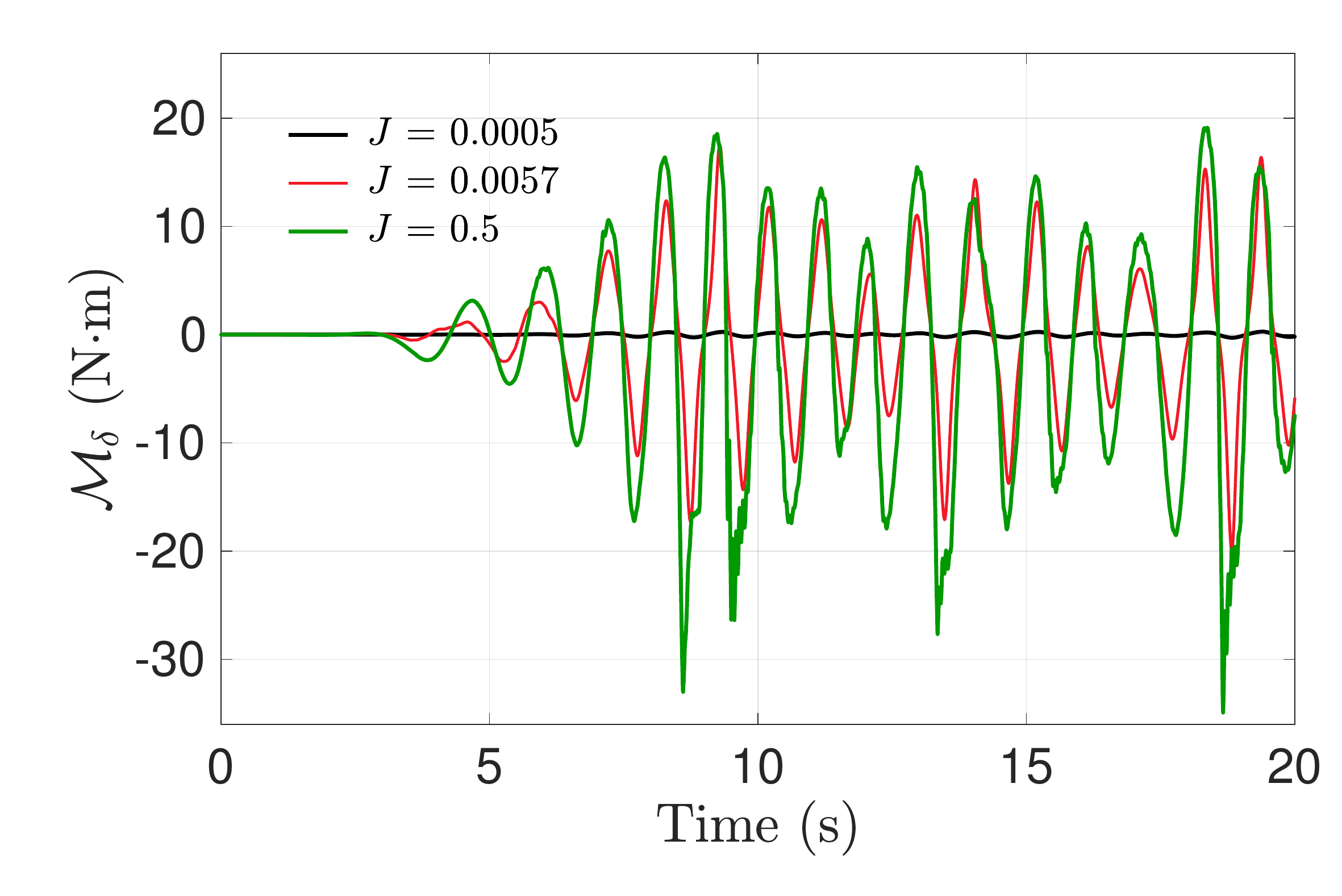}
  \label{fig_Md_J_comparison_irregwave}
 }
  \subfigure[Precession torque produced by the gyroscope]{
  \includegraphics[scale = 0.3]{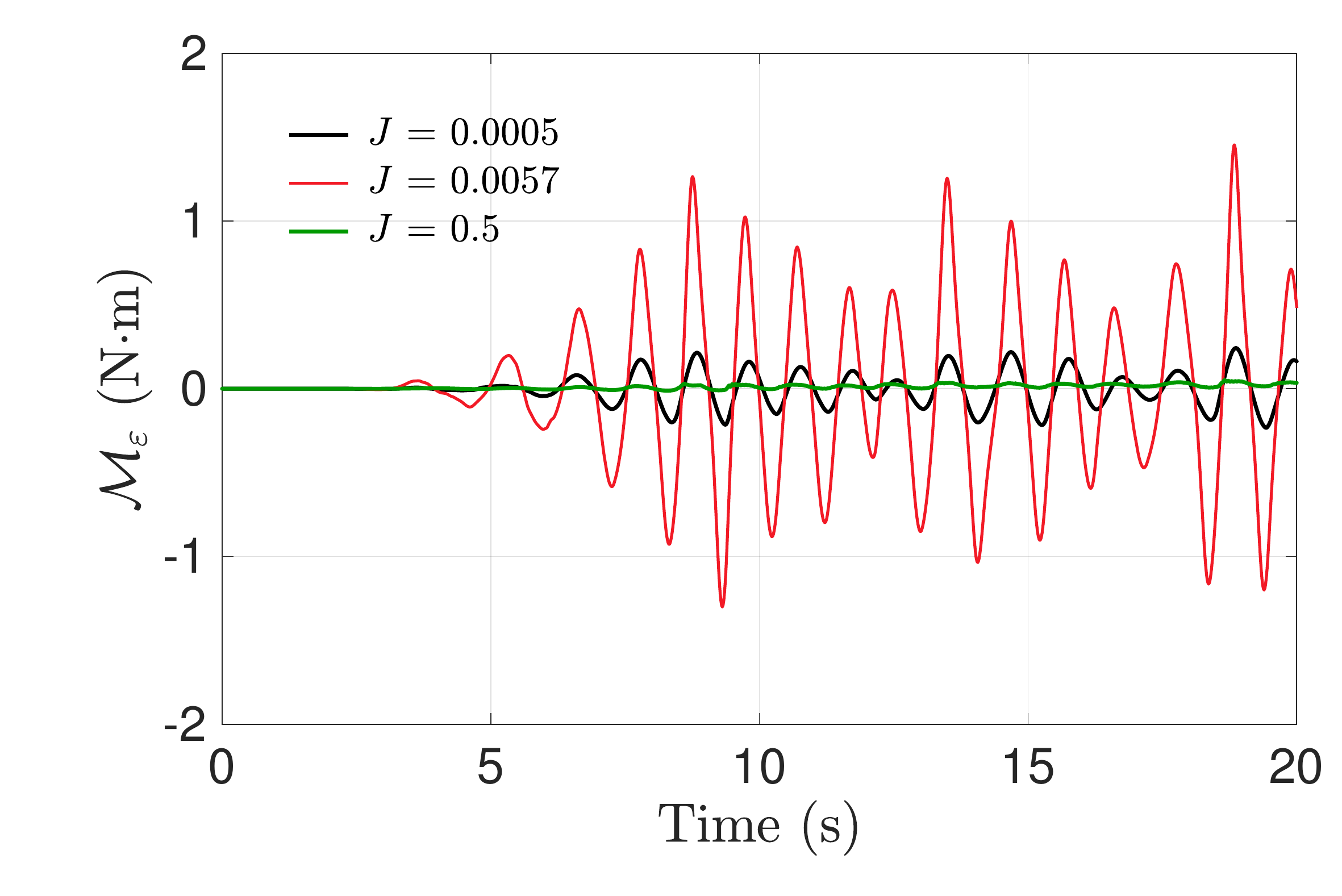}
  \label{fig_Me_J_comparison_irregwave}
 }
 \caption{Dynamics of 2D ISWEC model for three different values of flywheel moment of inertia $J$. The irregular wave properties are $\cH_\text{s}$ = 0.1 m and $\cT_\text{p}$ = 1 s and $50$ wave components with frequencies $\omega_i$ in the range $3.8$ rad/s to $20$ rad/s. Temporal evolution of 
 \subref{fig_delta_J_comparison_irregwave} hull pitch angle $\delta$,
 \subref{fig_epsilon_J_comparison_irregwave} gyroscope precession angle $\varepsilon$,
 \subref{fig_Md_J_comparison_irregwave} pitch torque $\cM_\delta$, and
 \subref{fig_Me_J_comparison_irregwave} precession torque $\cM_\varepsilon$
 for
 $J$ = 0.0005 kg$\cdot$m$^2$ (\textcolor{black}{\textbf{-----}}, black),
 $J$ = 0.0058 kg$\cdot$m$^2$ (\textcolor{red}{\textbf{-----}}, red),
 and $J$ = 0.5 kg$\cdot$m$^2$ (\textcolor{ForestGreen}{\textbf{-----}}, green).
 For all cases, $I = 0.94 \times J$.
 }
 \label{fig_J_comparison_irregwave}
\end{figure}

\begin{figure}[]
 \centering
  \subfigure[Hull pitch angle]{
  \includegraphics[scale = 0.3]{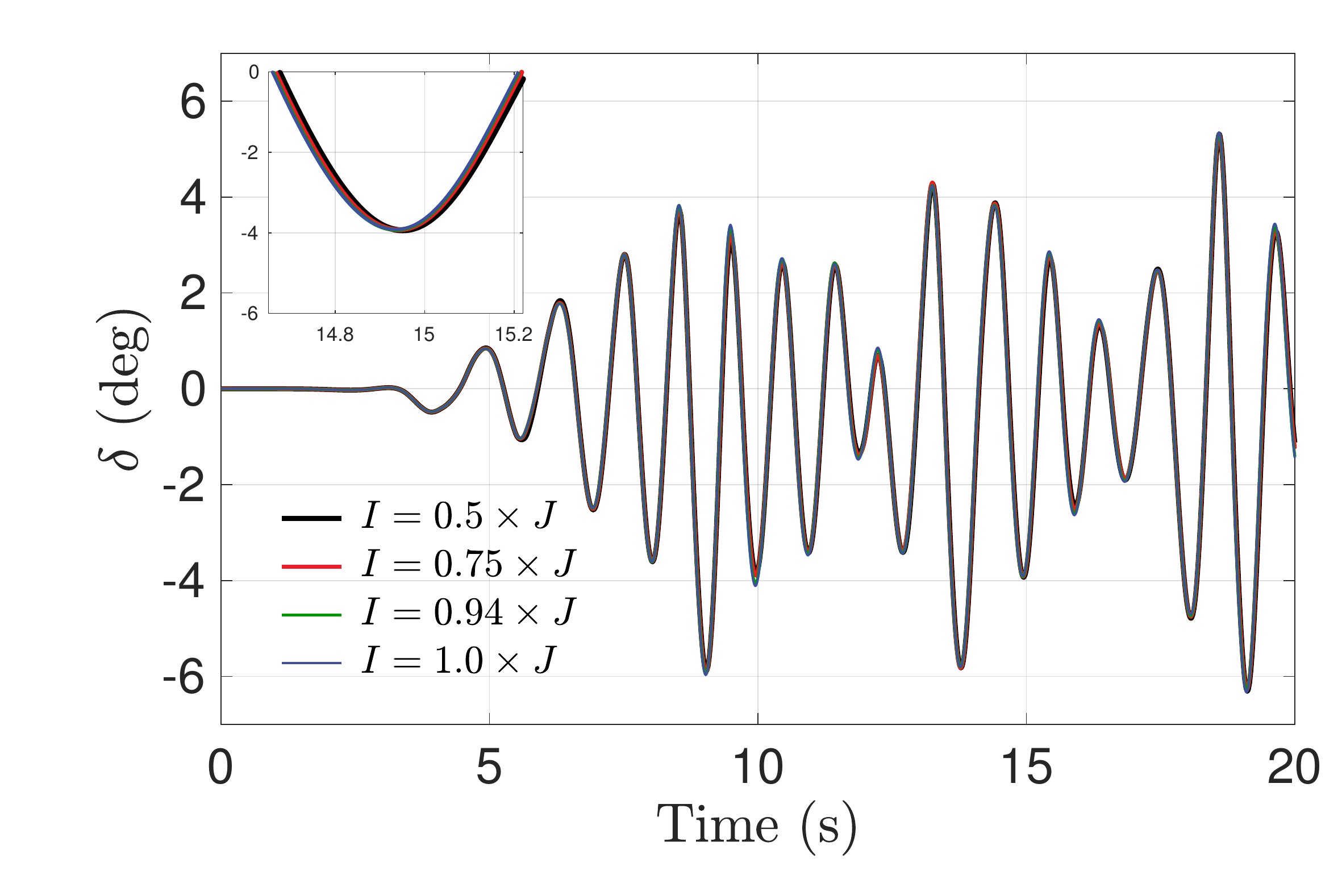}
  \label{fig_delta_I_variation_irregwave}
 }
   \subfigure[Gyroscope precession angle]{
  \includegraphics[scale = 0.3]{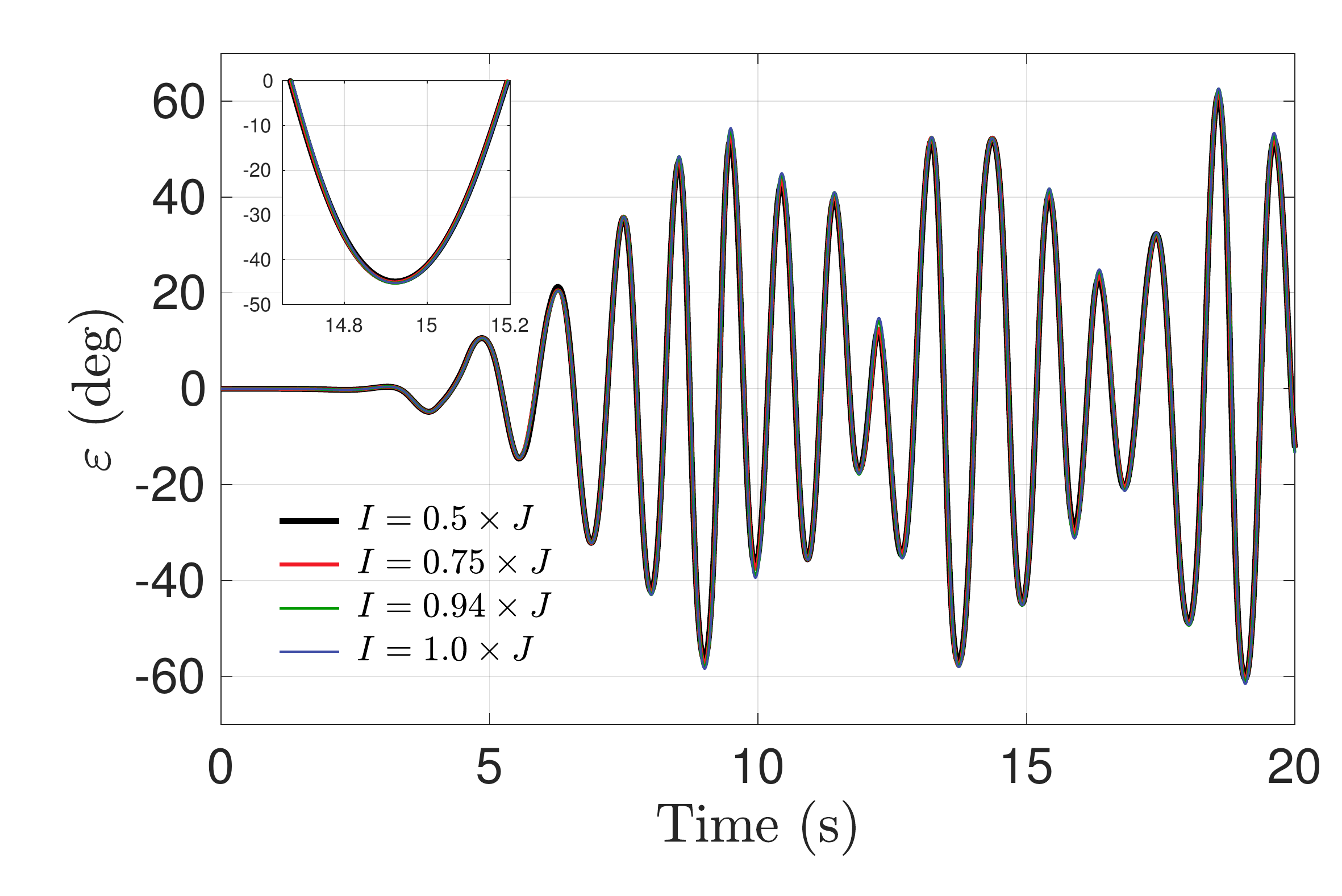}
  \label{fig_e_I_variation_irregwave}
 }
  \subfigure[Powers]{
  \includegraphics[scale = 0.3]{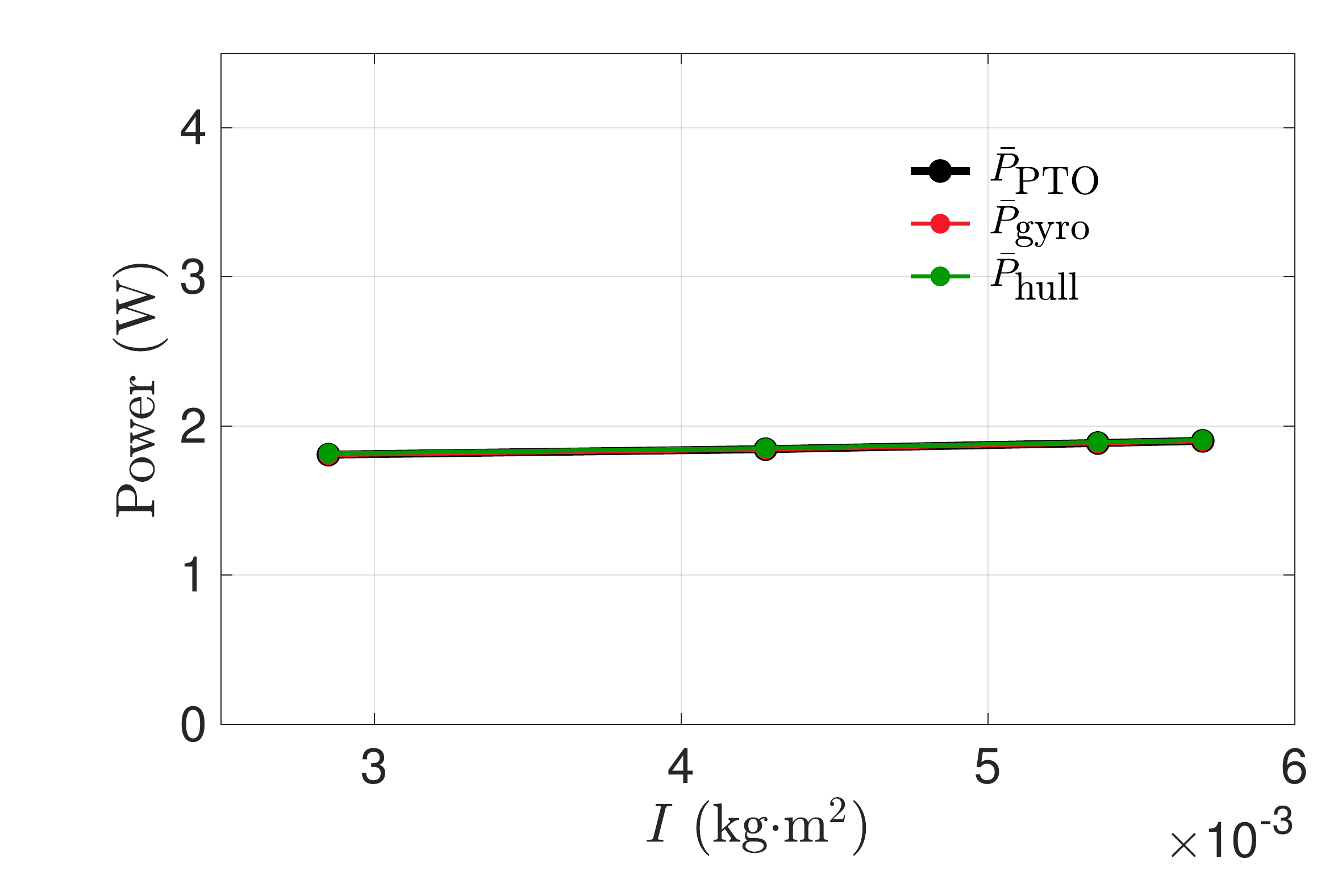}
  \label{fig_Powers_I_variation_irregwave}
 }
 \caption{Dynamics of 2D ISWEC model for four different values of $I$. The irregular wave properties are $\cH_\text{s}$ = 0.1 m and $\cT_\text{p}$ = 1 s and $50$ wave components with frequencies $\omega_i$ in the range $3.8$ rad/s to $20$ rad/s.
 Temporal evolution of
 \subref{fig_delta_I_variation_irregwave} hull pitch angle $\delta$, and
 \subref{fig_e_I_variation_irregwave} gyroscope precession angle $\varepsilon$ for
 $I = 0.5 \times J$ (\textcolor{black}{\textbf{-----}}, black),
 $I = 0.75 \times J$ (\textcolor{red}{\textbf{-----}}, red), 
 $I = 0.94 \times J$ (\textcolor{ForestGreen}{\textbf{-----}}, green), and
 $I = 1.0 \times J$ (\textcolor{blue}{\textbf{-----}} , blue).
 \subref{fig_Powers_I_variation_irregwave} Comparison of time-averaged powers from the interval $t = 10$ s to $t = 20$ s for each value of $I$. For all cases, $J$ = 0.0058 kg$\cdot$m$^2$.
}
 \label{fig_I_comparison_irregwave}
\end{figure}

Similarly, ISWEC dynamics with irregular waves are studied for three different values of $J$. Results for varying $J$ values are compared in Fig.~\ref{fig_J_comparison_irregwave}, which are qualitatively similar to the results obtained with regular waves. The effect of varying $I$ with respect to $J$ is also simulated, and the results are shown in Fig.~\ref{fig_I_comparison_irregwave}. It is seen that the hull pitch and the gyroscope precession angles are relatively insensitive to variations in $I$.  It is seen that the powers are relatively constant across different $I$ values under irregular wave conditions as well.

%%%%%%%%%%%%%%%%%%%%%%%%%%%%%%%%%
\subsubsection{PTO stiffness coefficient $k$}\label{subsec_k_variation}

Finally, we study the effect of varying the PTO stiffness coefficient $k$ on the dynamics of the ISWEC device. This term appears as a restoring 
torque $k \varepsilon$ in the precession angle Eq.~\eqref{eq_gyro_PTO_simple} and acts to drive the gyroscope's oscillation about
its mean position $\varepsilon$ = 0$^\circ$. The oscillation frequency is directly influenced by $k$ and can be chosen to ensure a 
resonant condition is attained between the gyroscope and the incoming waves, thus maximizing the power absorbed by the system.

\begin{figure}[]
 \centering
  \subfigure[Hull pitch angle]{
  \includegraphics[scale = 0.3]{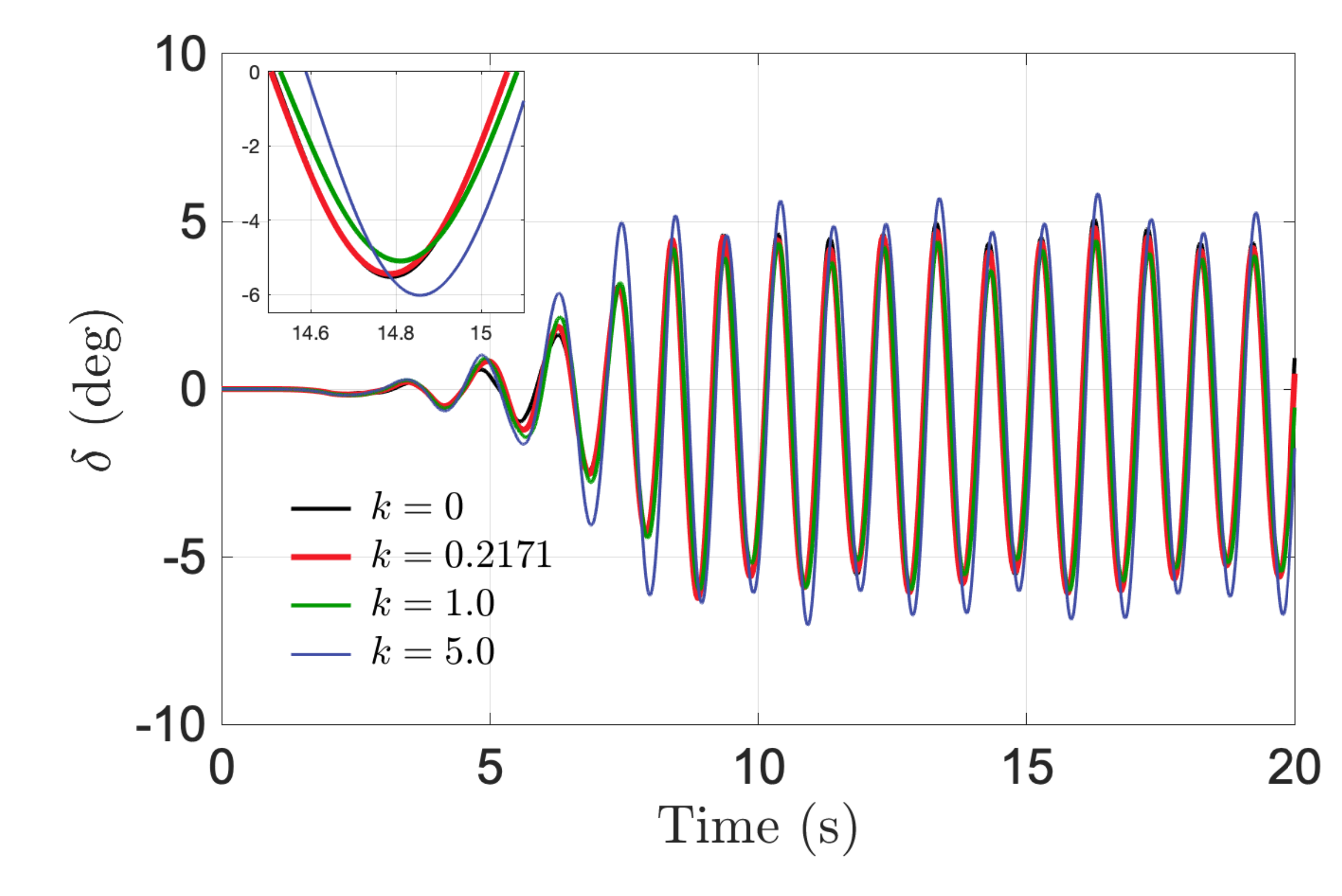}
  \label{fig_delta_k_variation_regwave}
 }
   \subfigure[Gyroscope precession angle]{
  \includegraphics[scale = 0.3]{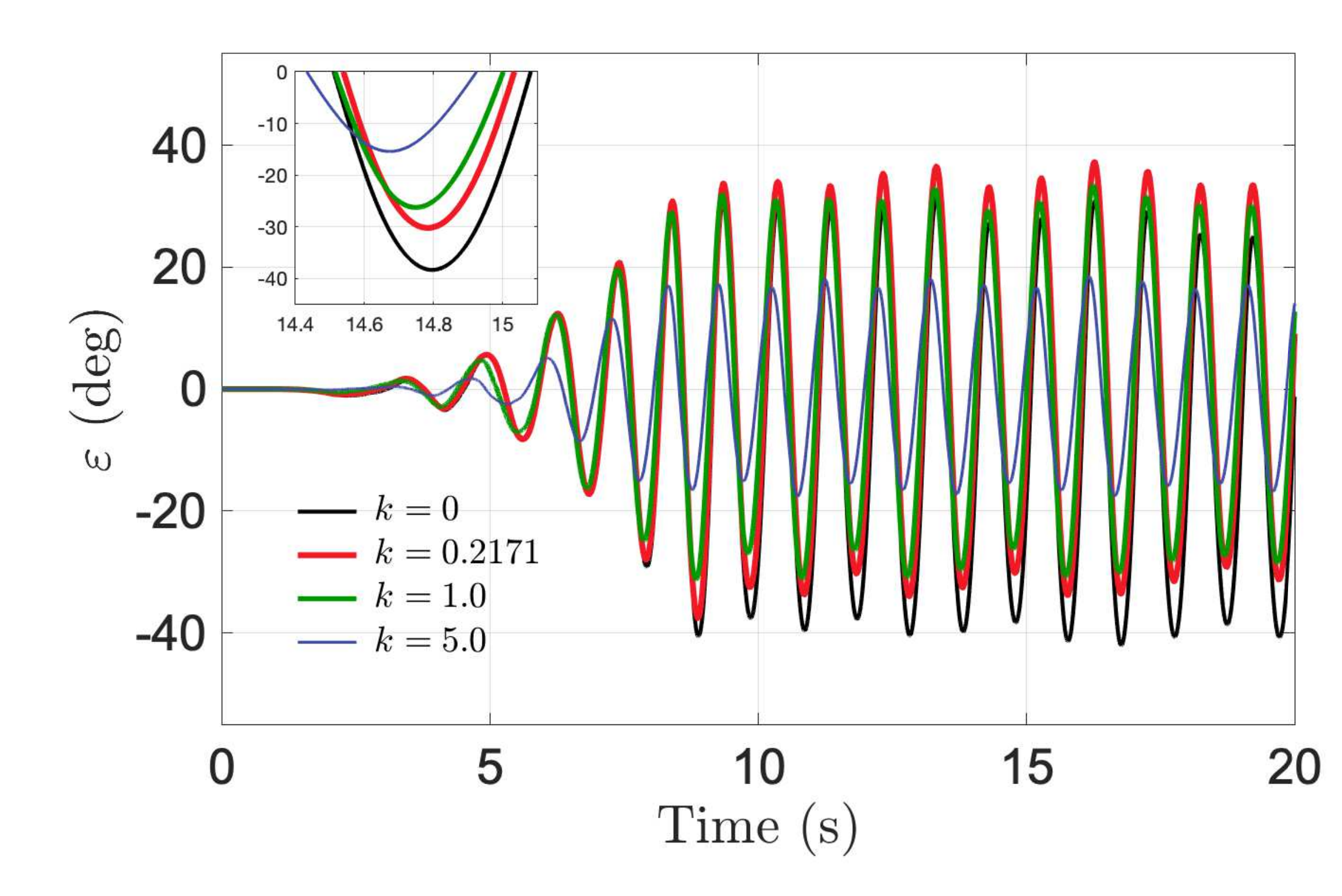}
  \label{fig_e_k_variation_regwave}
 }
  \subfigure[Pitch torque unloaded on the hull]{
  \includegraphics[scale = 0.3]{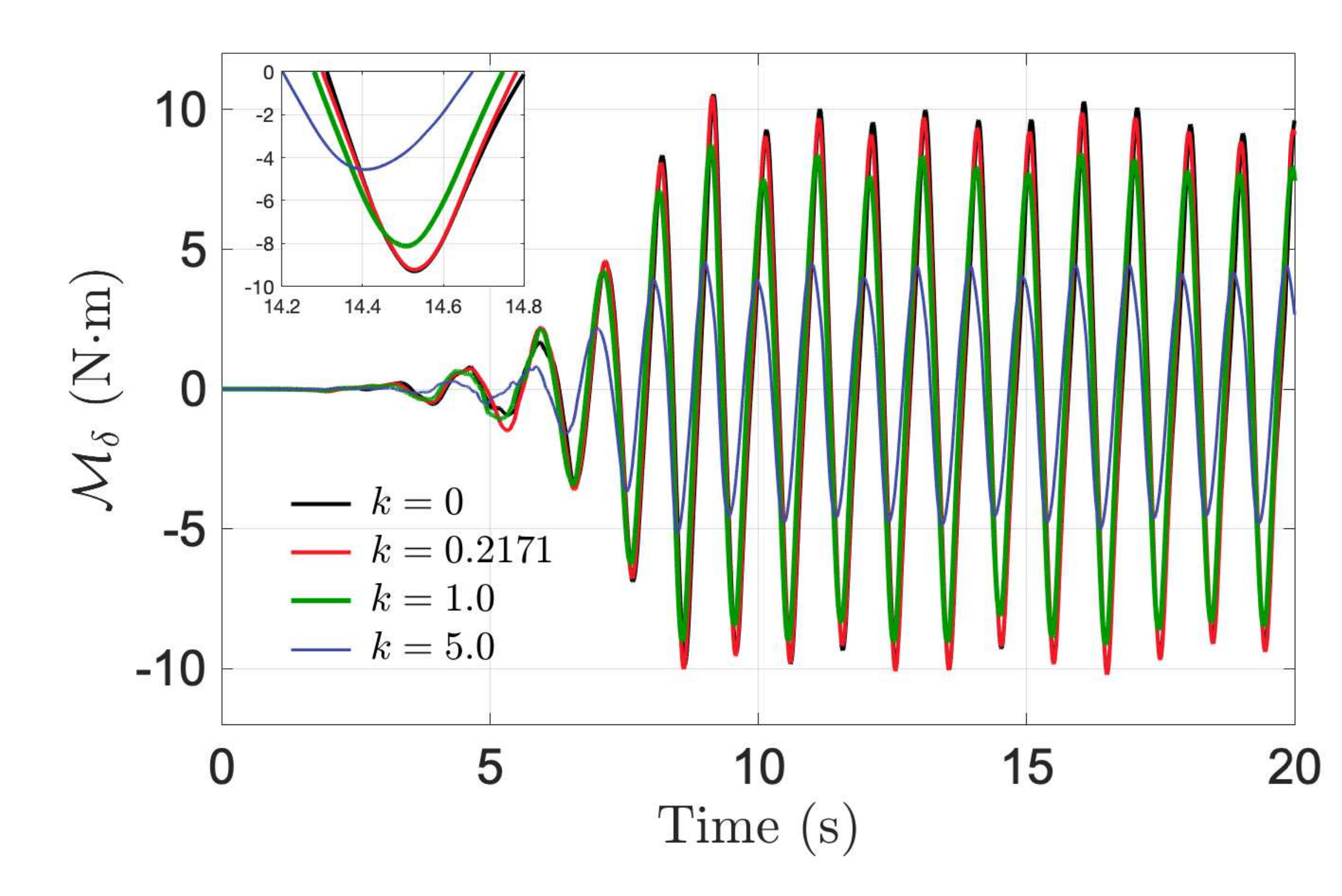}
  \label{fig_Md_k_variation_regwave}
 }
  \subfigure[Precession torque produced by the gyroscope]{
  \includegraphics[scale = 0.3]{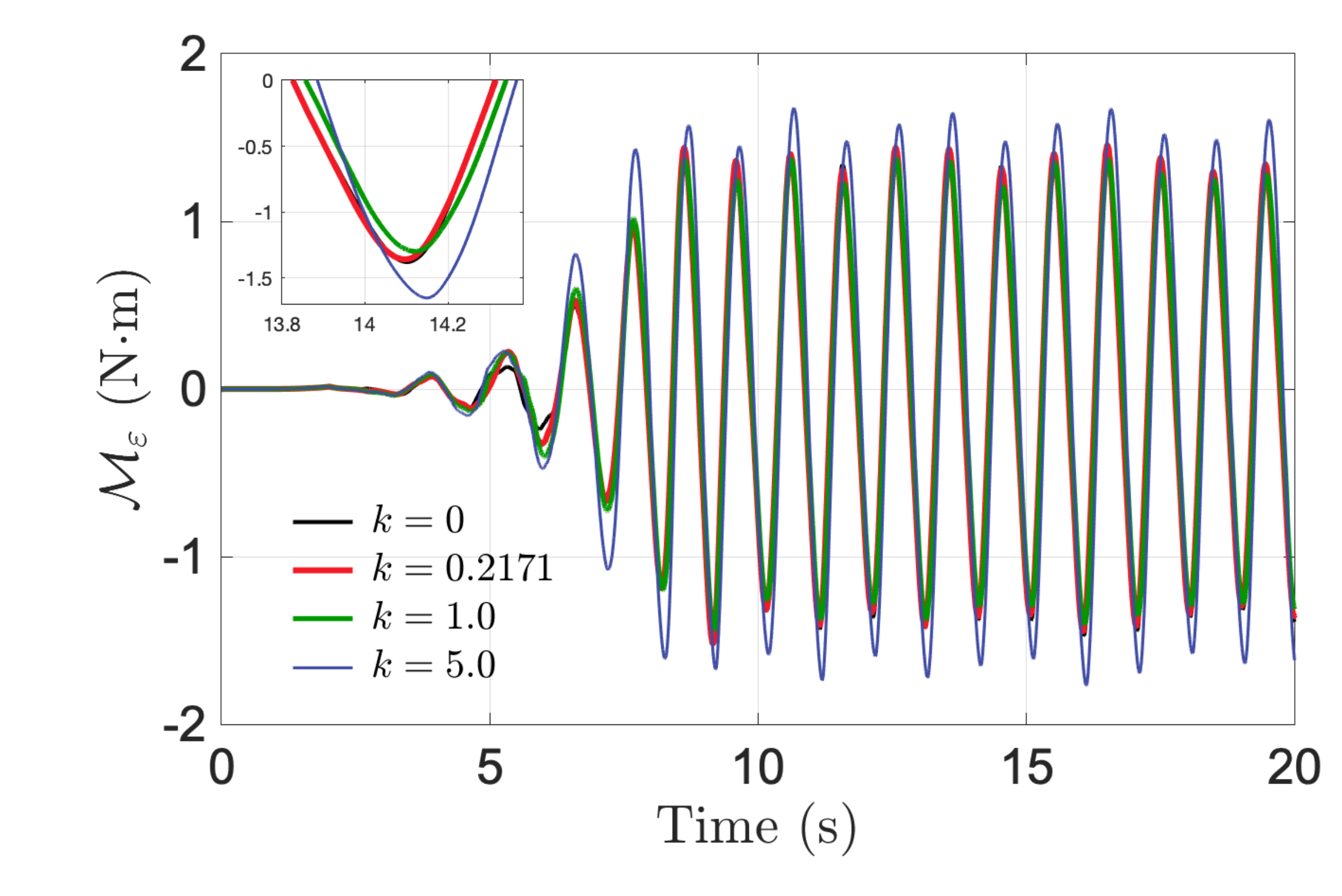}
  \label{fig_Me_k_variation_regwave}
 }
  \subfigure[Powers]{
  \includegraphics[scale = 0.3]{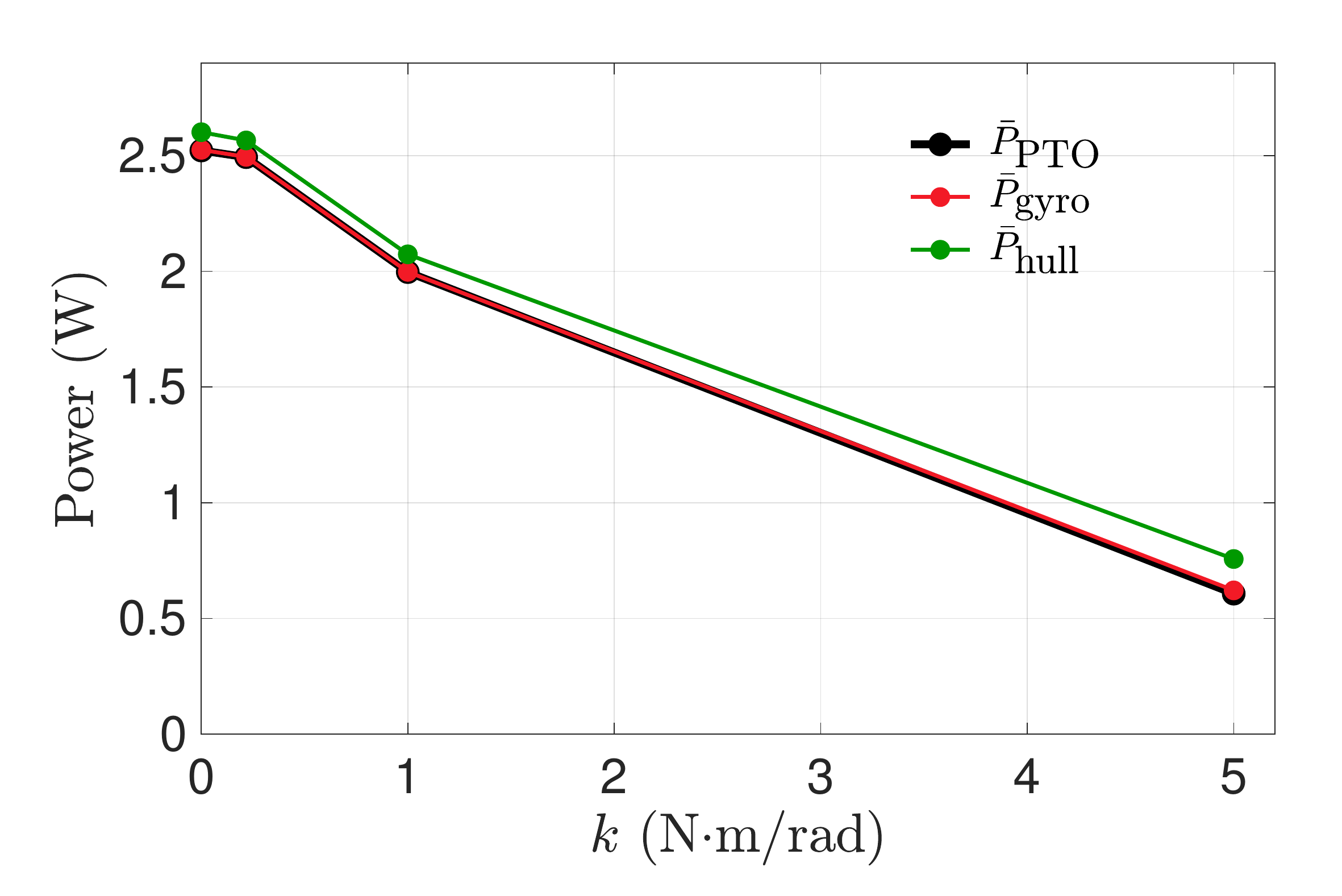}
  \label{fig_Powers_k_variation}
 }
 \caption{Dynamics of 2D ISWEC model for four different values of PTO stiffness $k$. The regular wave properties are $\cH$ = 0.1 m and $\cT$ = 1 s.
Temporal evolution of
\subref{fig_delta_k_variation_regwave} hull pitch angle $\delta$,
 \subref{fig_e_k_variation_regwave} gyroscope precession angle $\varepsilon$,
 \subref{fig_Md_k_variation_regwave} pitch torque $\cM_\delta$, and
 \subref{fig_Me_k_variation_regwave} precession torque $\cM_\varepsilon$ for
 $k$ = 0 N$\cdot$m/rad (\textcolor{black}{\textbf{-----}}, black),
 $k$ = 0.2171 N$\cdot$m/rad (\textcolor{red}{\textbf{-----}}, red), 
 $k$ = 1.0 N$\cdot$m/rad (\textcolor{ForestGreen}{\textbf{-----}}, green), 
 and $k$ = 5.0 N$\cdot$m/rad (\textcolor{blue}{\textbf{-----}}, blue);
 \subref{fig_Powers_k_variation} comparison of time-averaged powers from the interval $t = 10$ s to $t = 20$ s for each value of $k$.
}
 \label{fig_k_variation}
\end{figure}

We consider four different values of $k$ = 0.0 N$\cdot$m/rad, 0.2171 N$\cdot$m/rad, 1.0 N$\cdot$m/rad, and 5.0 N$\cdot$m/rad,
with the remaining gyroscope parameter chosen according to Table~\ref{tab_gyro_parameters_regularwave}. The $k = 0.2171$ value is obtained from theoretical considerations provided in Sec.~\ref{sec_PTO_params}. The results for a hull
interacting with regular waves are shown in Fig.~\ref{fig_k_variation}.
As $k$ increases, the maximum precession angle $\varepsilon$ and velocity $\dot{\varepsilon}$ decreases leading to decreased
power absorption by the device. The increased PTO stiffness value tends to keep the gyroscope close to its zero-mean position, which reduces the hull-gyroscope coupling.  This can be observed from the lowered values of $\cM_\delta$ torques in Fig.~\ref{fig_Md_k_variation_regwave}. As a consequence, the hull pitching motion increases, as seen in Fig.~\ref{fig_delta_k_variation_regwave}.

\begin{figure}[]
 \centering
  \subfigure[Hull pitch angle]{
  \includegraphics[scale = 0.3]{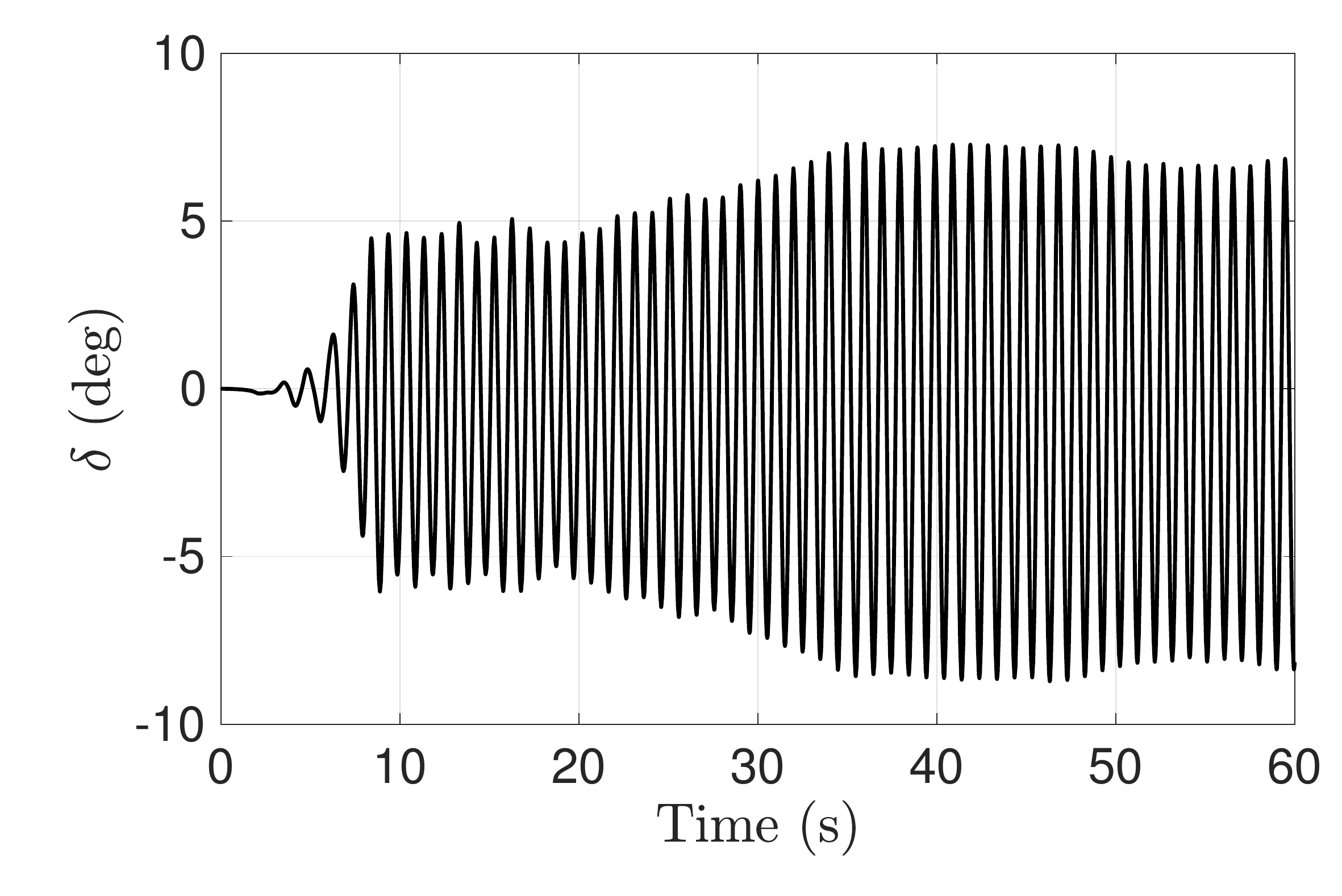}
  \label{fig_delta_k0}
 }
   \subfigure[Gyroscope precession angle]{
  \includegraphics[scale = 0.3]{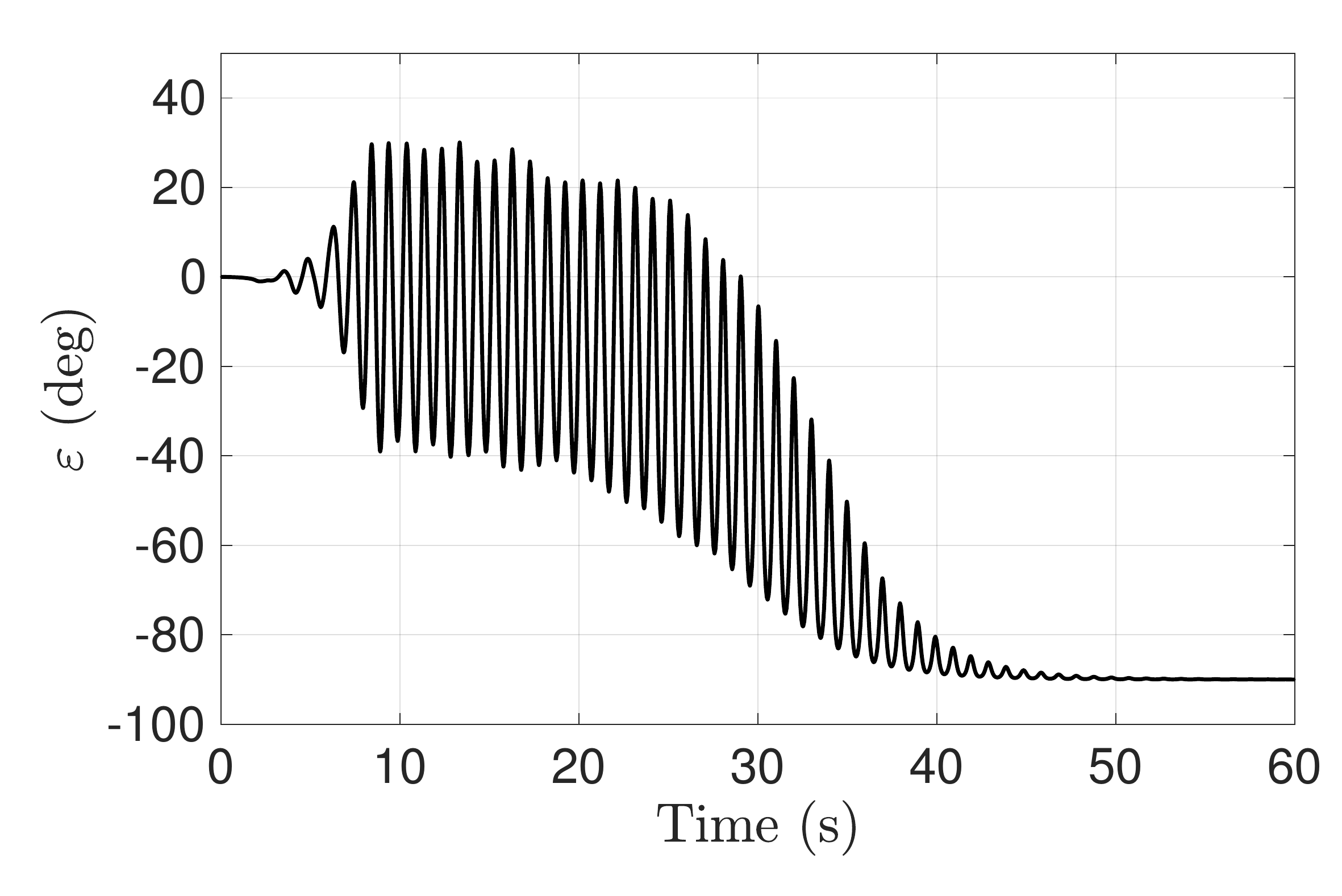}
  \label{fig_e_k0}
 }
  \subfigure[Pitch torque unloaded on the hull]{
  \includegraphics[scale = 0.3]{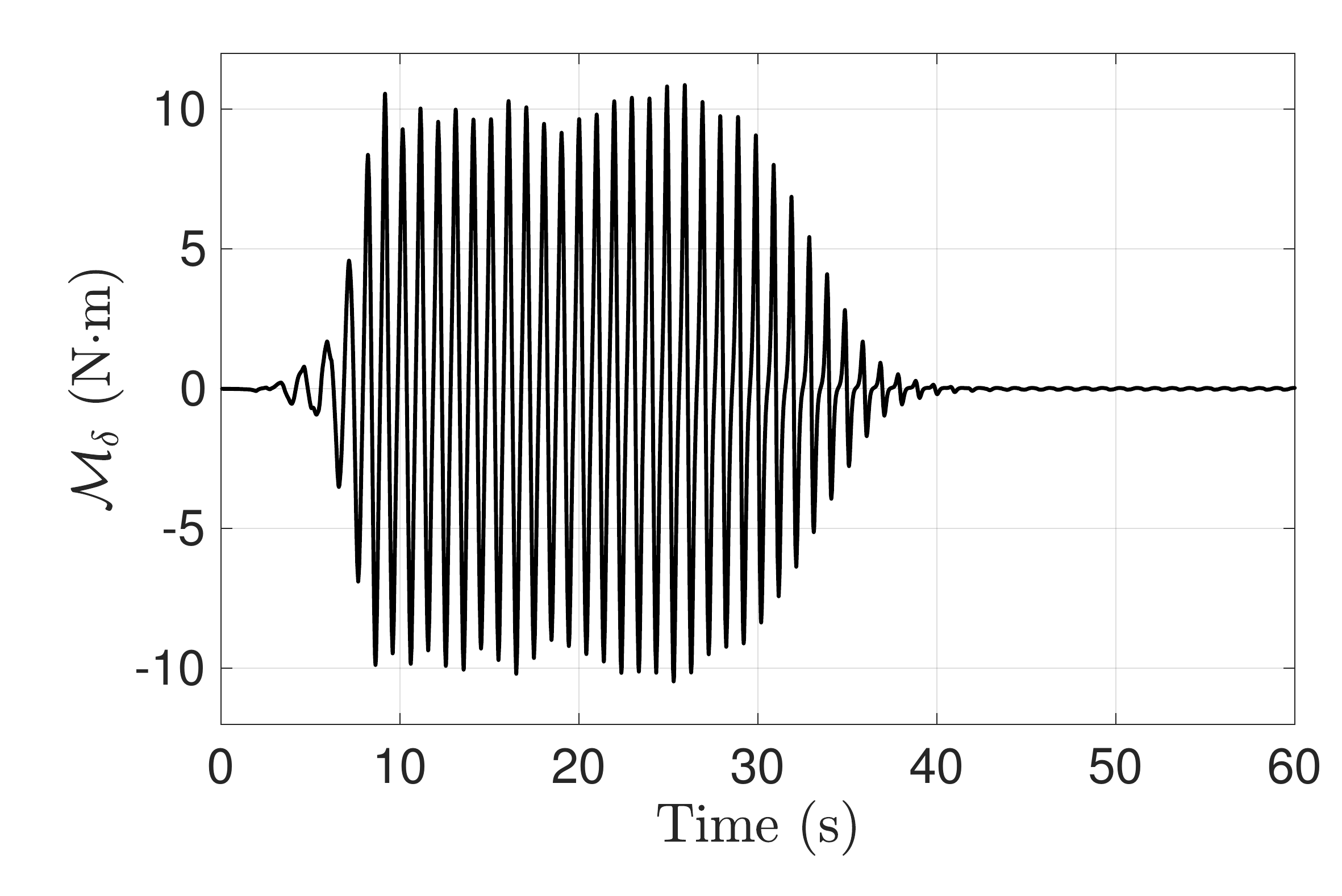}
  \label{fig_Md_k0}
 }
  \subfigure[Precession torque produced by the gyroscope]{
  \includegraphics[scale = 0.3]{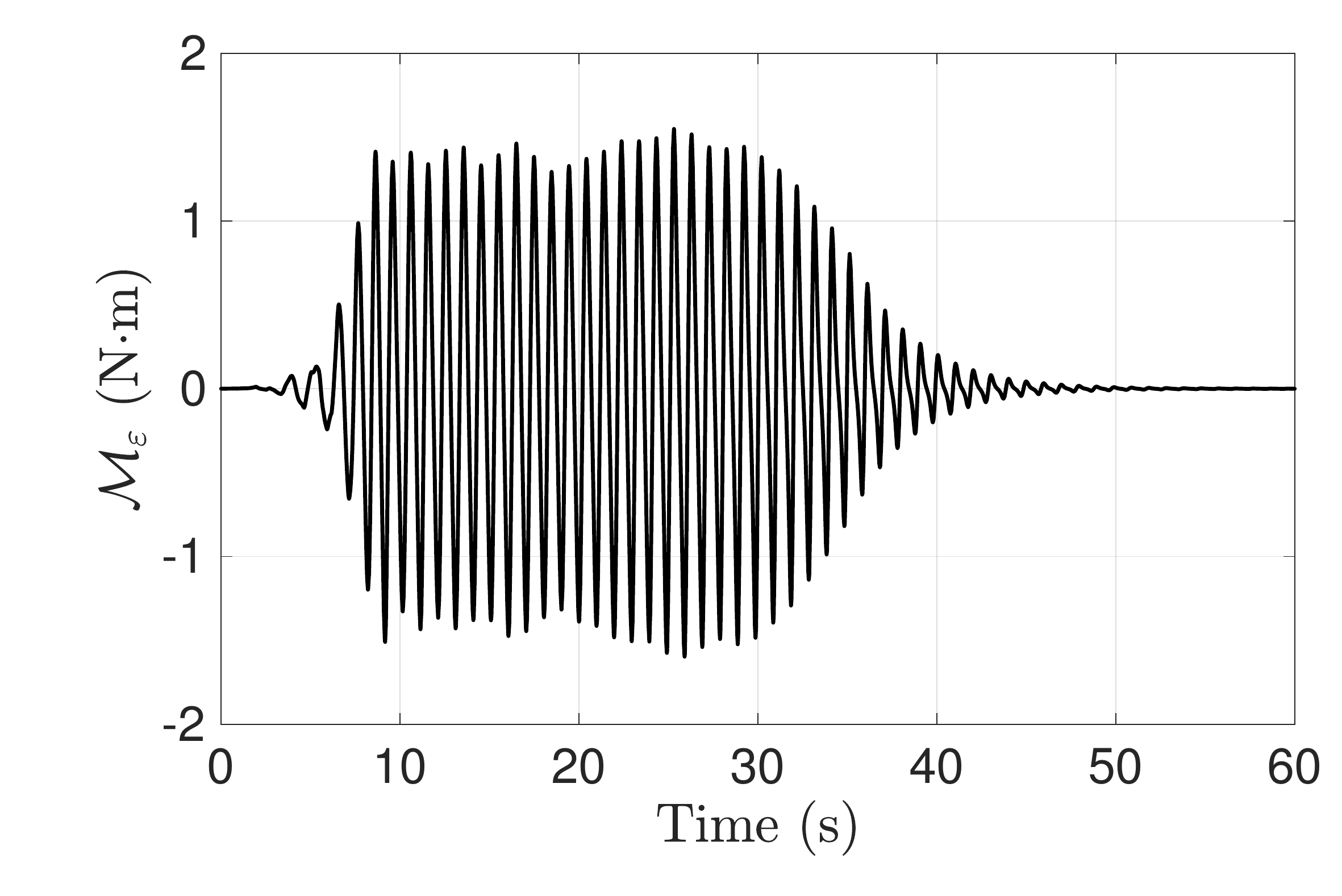}
  \label{fig_Me_k0}
 }
 \caption{Long-term dynamics of the 2D ISWEC model for $k = 0$ PTO stiffness: \subref{fig_delta_k0} hull pitch angle $\delta$, \subref{fig_e_k0} gyroscope precession angle $\varepsilon$, \subref{fig_Md_k0} pitch torque $\cM_\delta$, and \subref{fig_Me_k0} precession torque $\cM_\varepsilon$.
 The regular wave properties are $\cH$ = 0.1 m and $\cT$ = 1 s.
}
 \label{fig_k0_dynamics}
\end{figure}

The $k = 0$ case warrants additional discussion.
When the PTO stiffness is zero, the gyroscope attains a larger maximum precession amplitude and generates more power than the $k > 0$ cases over the time period $t = 10$ s and $t = 20$ s. However, Fig.~\ref{fig_k0_dynamics} shows the long-term dynamics for $k = 0$;
it is seen that the gyroscope is unable to sustain its precession oscillation as it eventually falls to one side ($\varepsilon = -90^\circ$) and remains there. At this configuration, the gyroscope yaw axis and the hull pitch axis are aligned, and the precession effect is lost. 
As these gyroscopic oscillations vanish, the torques tend towards zero, the hull exhibits unrestrained pitch oscillation, and no power is generated.

\begin{figure}[]
 \centering
  \subfigure[Hull pitch angle]{
  \includegraphics[scale = 0.3]{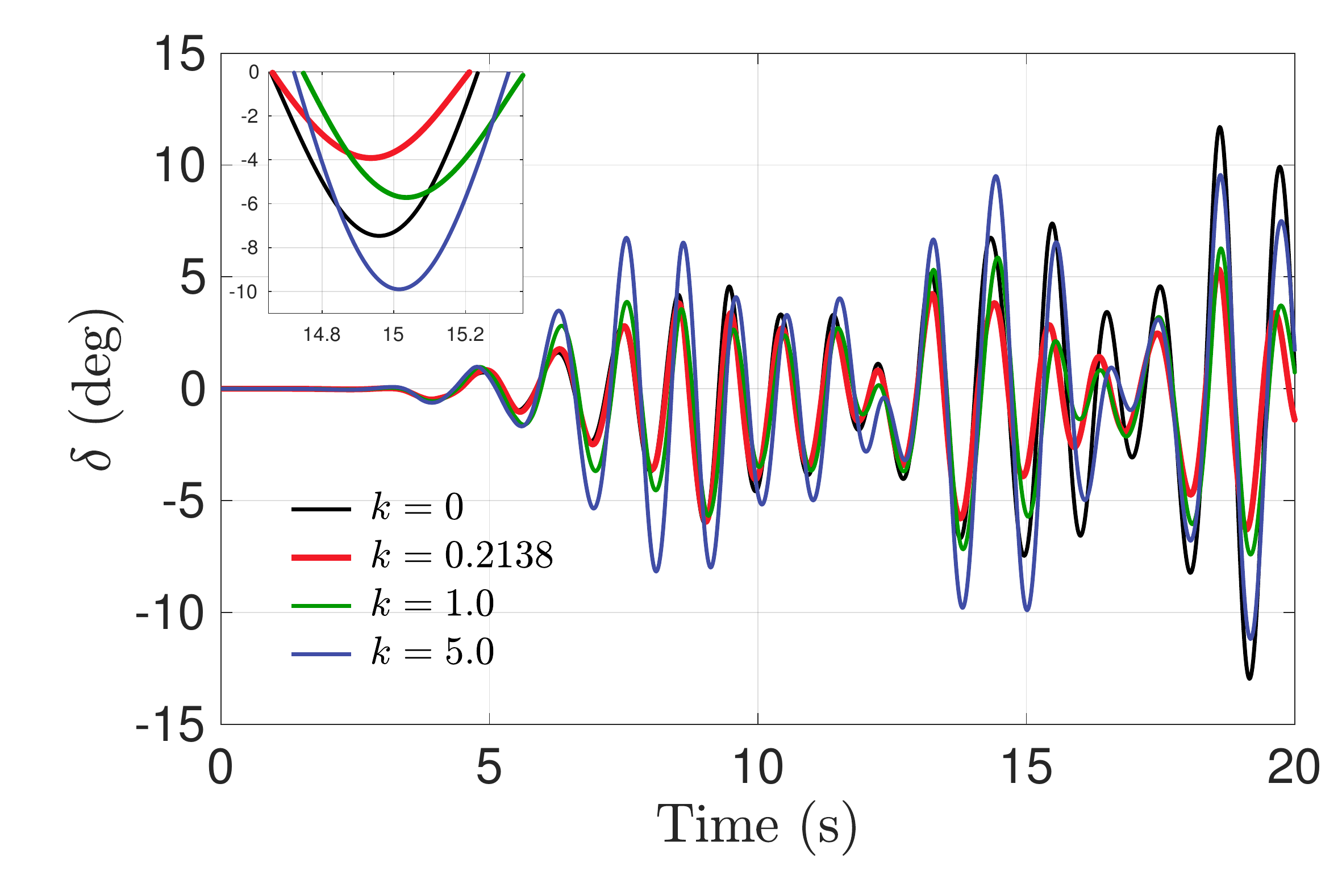}
  \label{fig_delta_k_variation_irregwave}
 }
   \subfigure[Gyroscope precession angle]{
  \includegraphics[scale = 0.3]{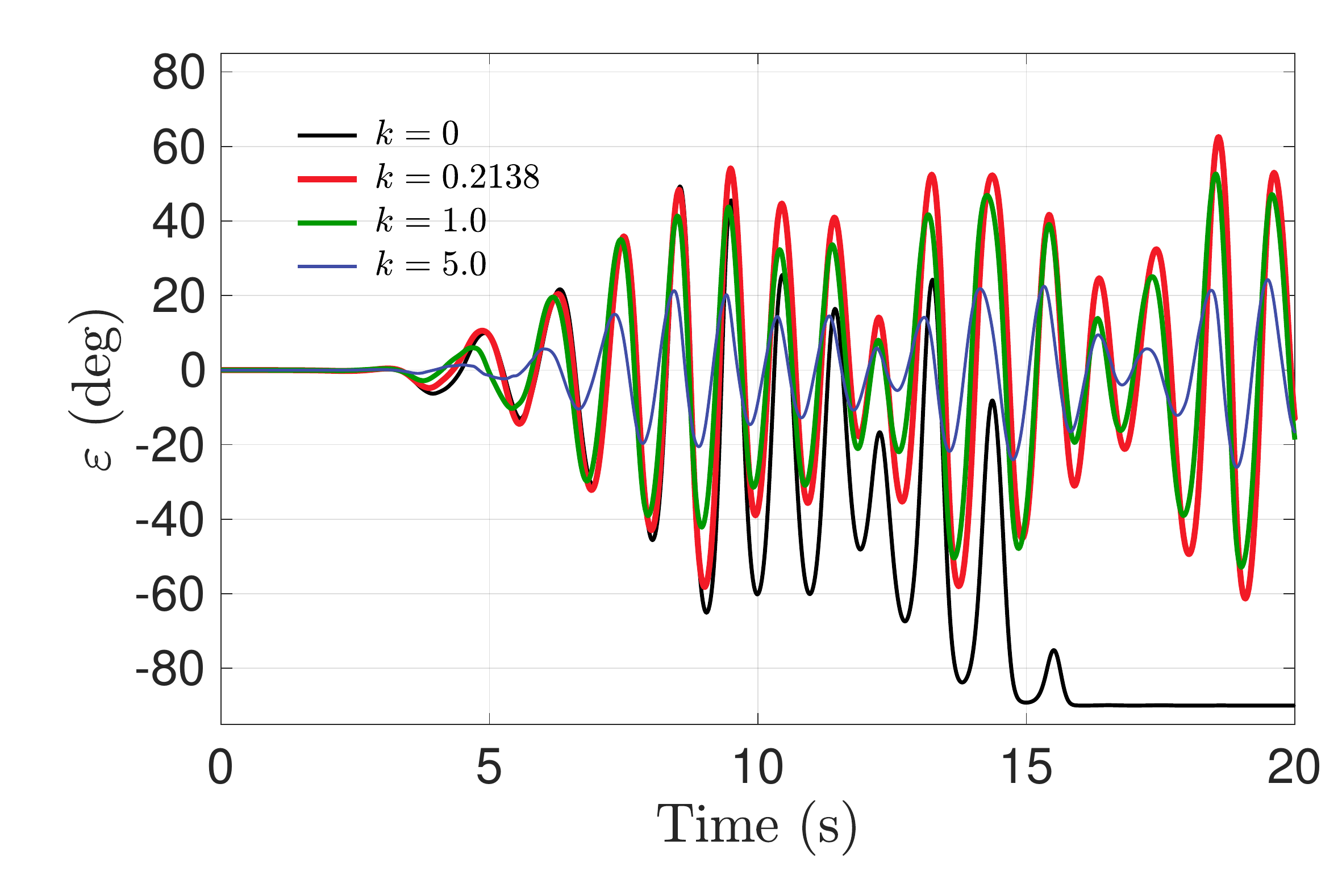}
  \label{fig_e_k_variation_irregwave}
 }
  \subfigure[Pitch torque unloaded on the hull]{
  \includegraphics[scale = 0.3]{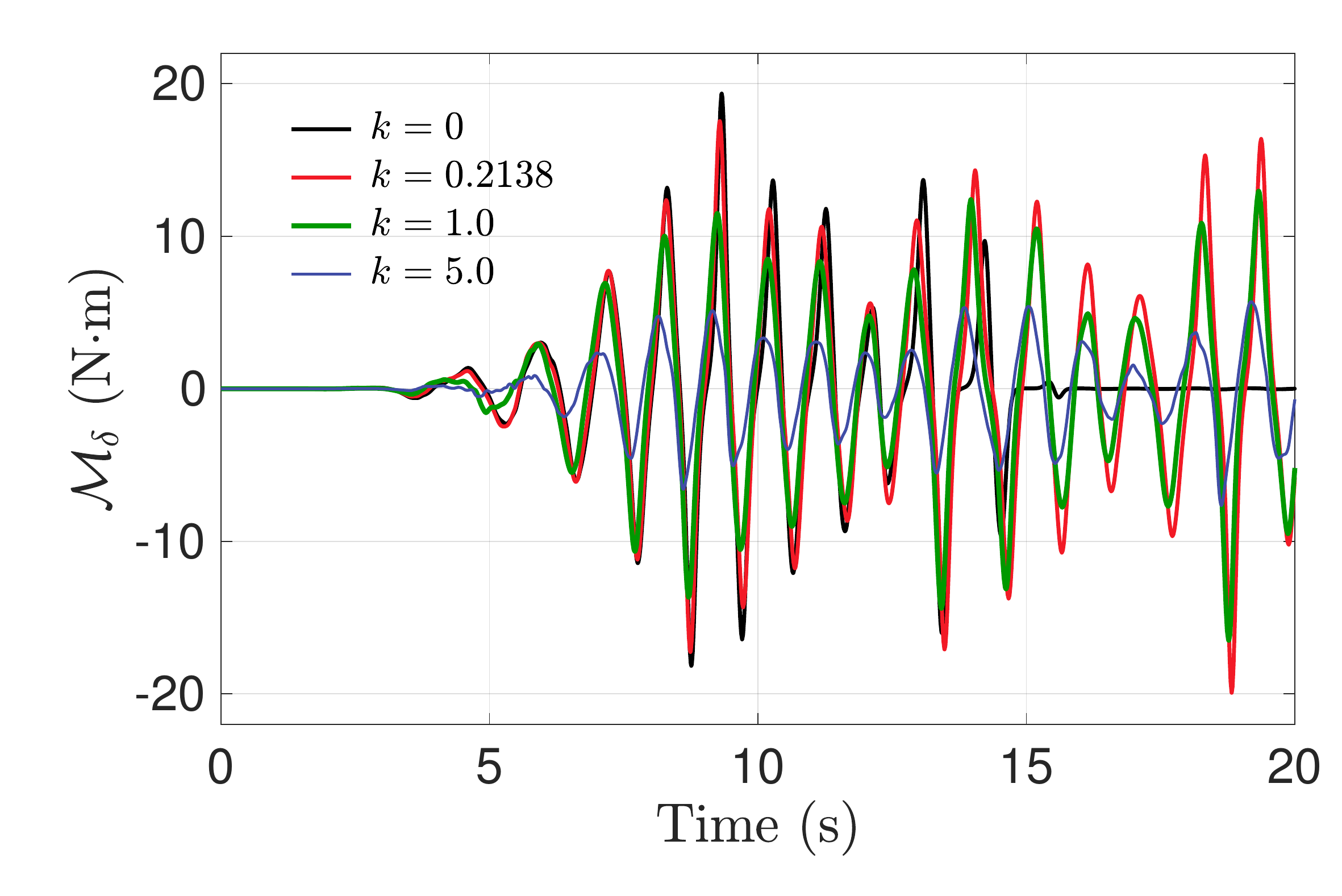}
  \label{fig_Md_k_variation_irregwave}
 }
  \subfigure[Precession torque produced by the gyroscope]{
  \includegraphics[scale = 0.3]{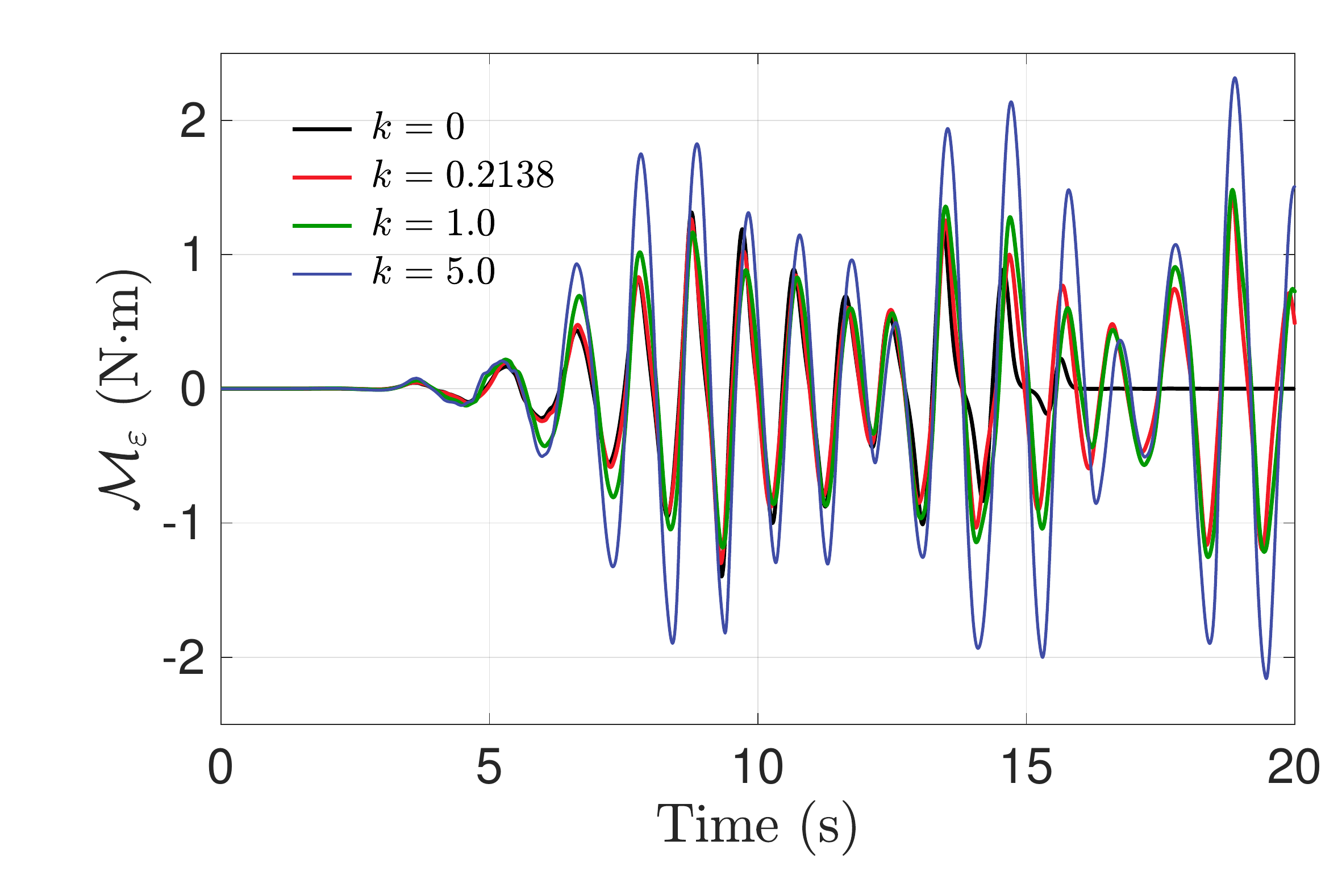}
  \label{fig_Me_k_variation_irregwave}
 }
  \subfigure[Powers]{
  \includegraphics[scale = 0.3]{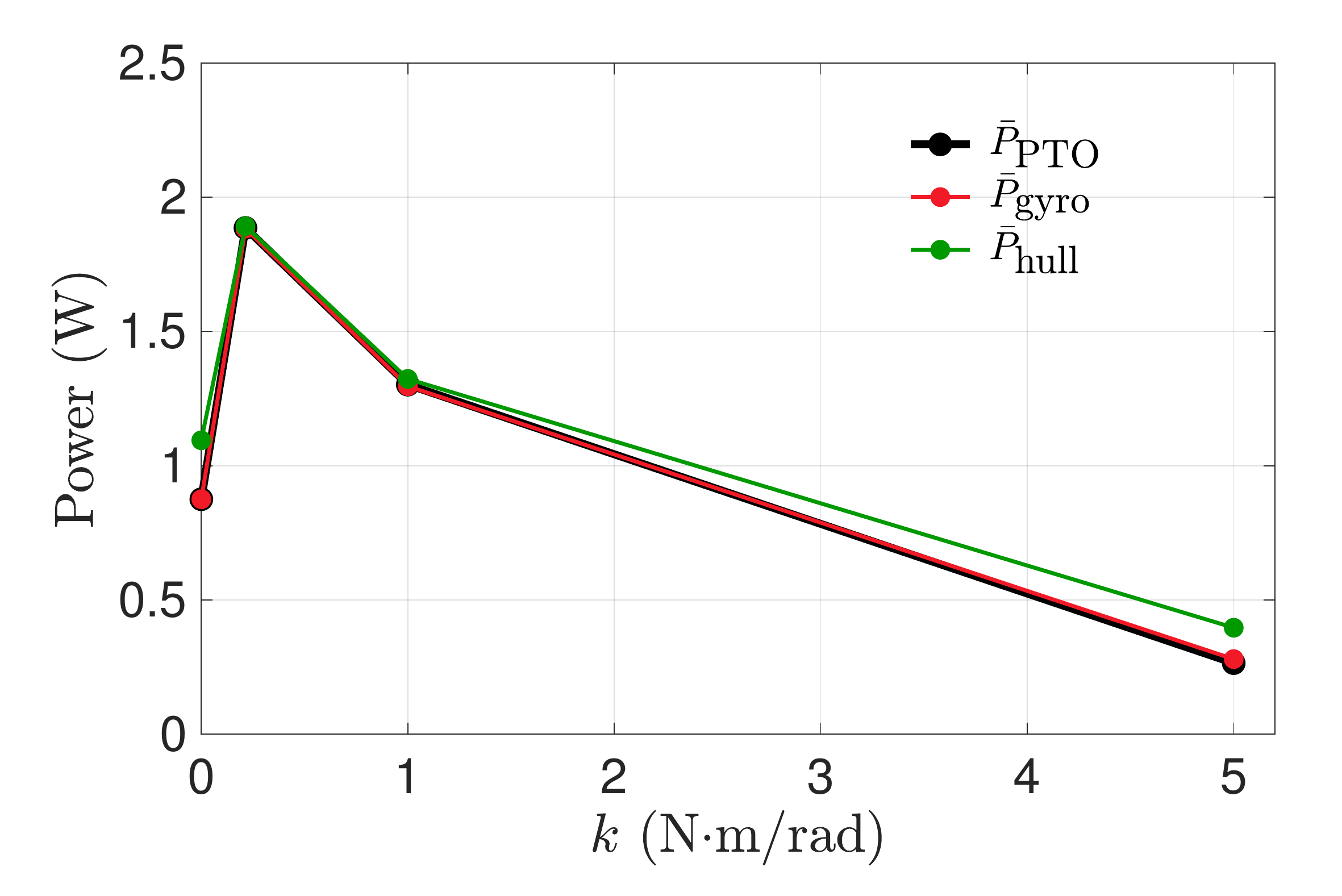}
  \label{fig_Powers_k_variation_irregwave}
 }
 \caption{Dynamics of 2D ISWEC model for four different values of PTO stiffness $k$. The irregular wave properties are $\cH_\text{s}$ = 0.1 m and $\cT_\text{p}$ = 1 s and $50$ wave components with frequencies $\omega_i$ in the range $3.8$ rad/s to $20$ rad/s.
Temporal evolution of
\subref{fig_delta_k_variation_irregwave} hull pitch angle $\delta$,
 \subref{fig_e_k_variation_irregwave} gyroscope precession angle $\varepsilon$,
 \subref{fig_Md_k_variation_irregwave} pitch torque $\cM_\delta$, and
 \subref{fig_Me_k_variation_irregwave} precession torque $\cM_\varepsilon$ for
 $k$ = 0 N$\cdot$m/rad (\textcolor{black}{\textbf{-----}}, black),
 $k$ = 0.2138 N$\cdot$m/rad (\textcolor{red}{\textbf{-----}}, red), 
 $k$ = 1.0 N$\cdot$m/rad (\textcolor{ForestGreen}{\textbf{-----}}, green), 
 and $k$ = 5.0 N$\cdot$m/rad (\textcolor{blue}{\textbf{-----}}, blue);
 \subref{fig_Powers_k_variation_irregwave} comparison of time-averaged powers from the interval $t = 10$ s to $t = 20$ s for each value of $k$.
}
 \label{fig_k_variation_irregwave}
\end{figure}

Next, we simulate ISWEC dynamics with irregular waves using four different values of $k$ = 0 N$\cdot$m/rad, 0.2138 N$\cdot$m/rad, 1.0 N$\cdot$m/rad, and 5.0 N$\cdot$m/rad. The results are compared in Fig.~\ref{fig_k_variation_irregwave}, and are qualitatively similar to the those obtained with regular waves.  Similar behavior of the ISWEC with $k$ = 0 is observed --- the gyroscope is unable to oscillate and falls to one side ($\varepsilon = -90^\circ$) and produces vanishing precession effects. However, with irregular waves the precession effects are lost much sooner compared to the regular waves case.

%%%%%%%%%%%%%%%%%%%%%%%%%%%%%%%%%
\subsection{Hull length to wavelength ($L/\lambda$) variation}\label{subsec_lambda_variation}

\begin{figure}[]
 \centering
  \subfigure[Hull pitch angle]{
  \includegraphics[scale = 0.3]{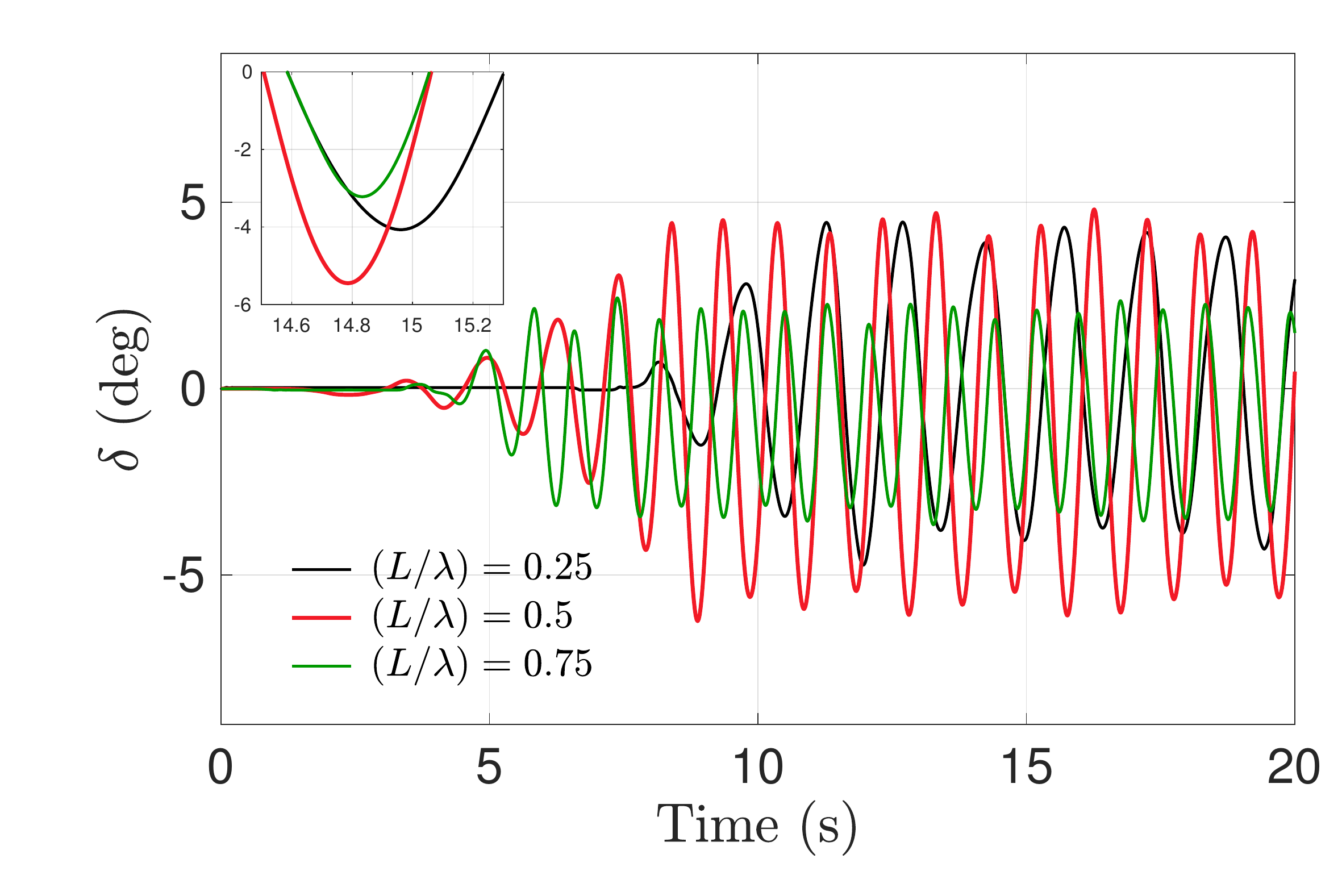}
  \label{fig_delta_L_by_lambda_var}
 }
   \subfigure[Gyroscope precession angle]{
  \includegraphics[scale = 0.3]{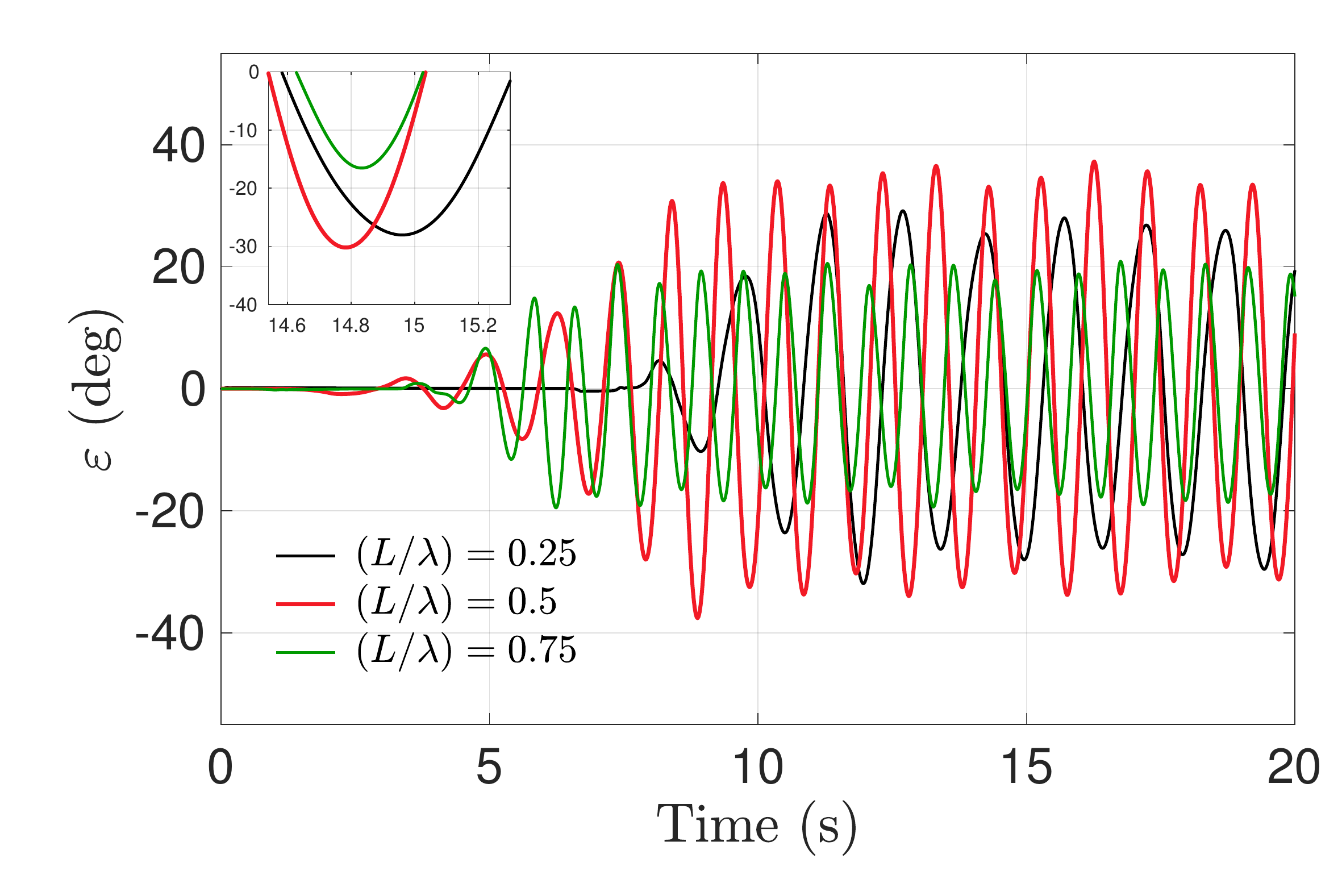}
  \label{fig_e_L_by_lambda_var}
 }
  \subfigure[RCW]{
  \includegraphics[scale = 0.3]{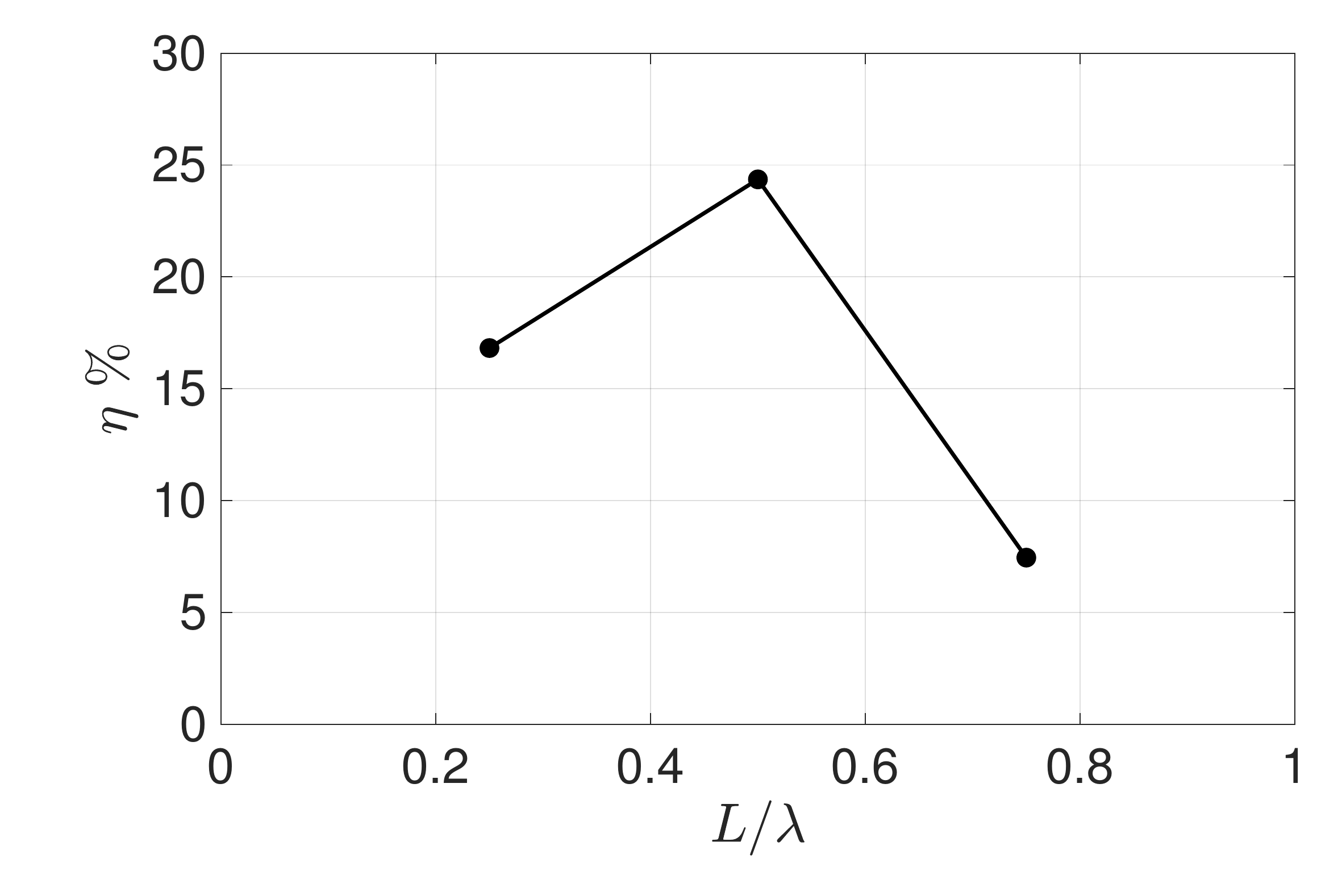}
  \label{fig_RCW_L_by_lambda_var}
 }
 \caption{Dynamics of 2D ISWEC model for three different hull length to wavelength ratios $L/\lambda$.
 The regular wave height is $\cH = 0.1$ m, while its period $\cT$ is calculated based on the dispersion relation given
 by Eq.~\eqref{eq_explicit_dispersion_relation} as wavelength $\lambda$ is varied.
 Temporal evolution of
 \subref{fig_delta_L_by_lambda_var} hull pitch angle $\delta$ and 
 \subref{fig_e_L_by_lambda_var} gyroscope precession angle $\varepsilon$ for 
  $L/\lambda$ = 0.25 (\textcolor{black}{\textbf{-----}}, black), $L/\lambda$ = 0.5 (\textcolor{red}{\textbf{-----}}, red), and $L/\lambda$ = 0.75 (\textcolor{ForestGreen}{\textbf{-----}}, green);
  \subref{fig_RCW_L_by_lambda_var} RCW $\eta$ computed using time-averaged powers from the interval $t = 10$ s to $t = 20$ s for each value of $L/\lambda$.
}
 \label{fig_L_by_lambda_var}
\end{figure}

In this section, we study the effect of hull length to wavelength ratio ($L/\lambda$) on ISWEC dynamics. We select three ratios $L/\lambda$ = 0.25, 0.5 and 0.75 for this analysis. The length of the hull is kept constant at $L$ = 0.7665 m, and the wavelength of the 
regular water waves is varied. The PTO and gyroscope parameters used in the three simulations are presented in Table~\ref{tab_L_by_lambda_variation}. Results consist of temporal evolution of the hull pitch and gyroscope precession angles in Figs.~\ref{fig_delta_L_by_lambda_var} and~\ref{fig_e_L_by_lambda_var}, respectively. It is observed that the hull pitch is maximum when $ \lambda/ 3 \leq L \leq \lambda/2$, as discussed in Sec.~\ref{sec_hull_shape}.
%This configuration is optimum to obtain higher pitch amplitude of the hull. 
As a consequence, the gyroscope precesses more and the conversion efficiency of the device increases (see Fig.~\ref{fig_RCW_L_by_lambda_var}). 

\begin{table}[]
 \centering
 \caption{Calculated values of PTO and gyroscope parameters for different $L/\lambda$ ratios using $L$ = 0.7665 m, $\delta_0$ = 10$^\circ$, $\varepsilon_0$ = 70$^\circ$, and $\dot{\phi}$ = 4000 RPM. $I = 0.94 \times J$ for all cases. Regular water waves with $\cH$ = 0.1 m are simulated. The rated power of the device $\bar{P}_\text{R}$ is taken to be the available wave power $\bar{P}_\text{wave}$ for these calculations. Units: $\lambda$ is in m, $c$ is in N$\cdot$m$\cdot$s/rad, $J$ and $I$ are in kg$\cdot$m$^2$ and $k$ is in N$\cdot$m/rad.}
  \rowcolors{2}{}{gray!10}
\begin{tabular}{c c c c c}
\toprule
\multirow{2}{*}{$L/\lambda$} & \multirow{2}{*}{$\lambda$} & \multicolumn{3}{c}{PTO and gyroscope parameters}  \\
\cline{3-5}
		   &                             & $c$      	 & $J$      	& $k$      	\\
\midrule
0.25             & 3.0659                 & 1.3491  & 0.0225  & 0.3705   \\
0.5               & 1.5456                 & 0.3473  & 0.0058  & 0.2171   \\
0.75             & 1.0219                 & 0.1773  & 0.0029  & 0.1679  \\
\bottomrule
 \end{tabular}
 \label{tab_L_by_lambda_variation}
\end{table}

%%%%%%%%%%%%%%%%%%%%%%%%%%%%%%%%%
\subsection{Device protection during inclement weather conditions}\label{subsec_device_protection}

\begin{figure}[h!]
 \centering
  \subfigure[Hull pitch angle]{
  \includegraphics[scale = 0.3]{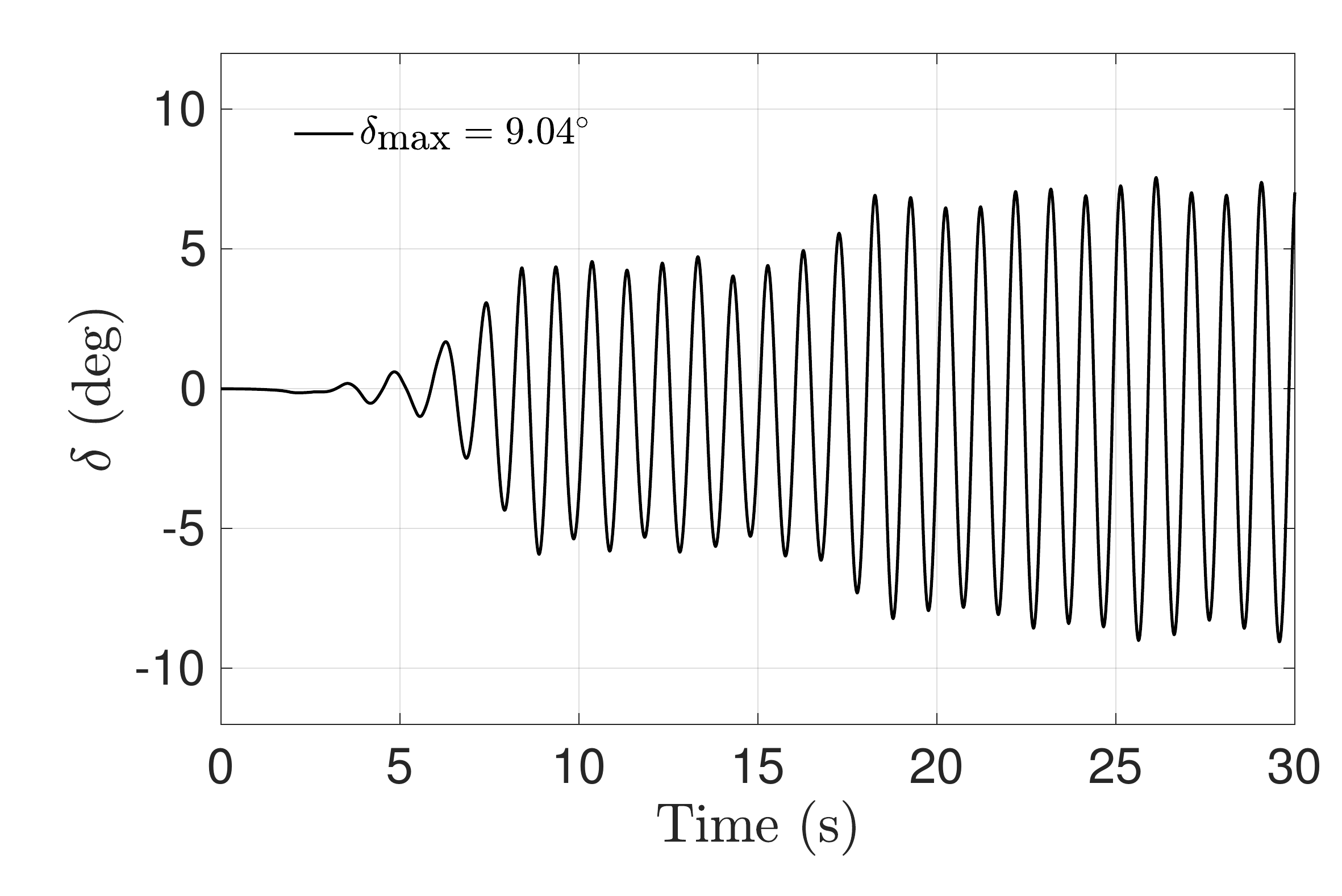}
  \label{fig_delta_rpm_reduction}
 }
   \subfigure[Gyroscope precession angle]{
  \includegraphics[scale = 0.3]{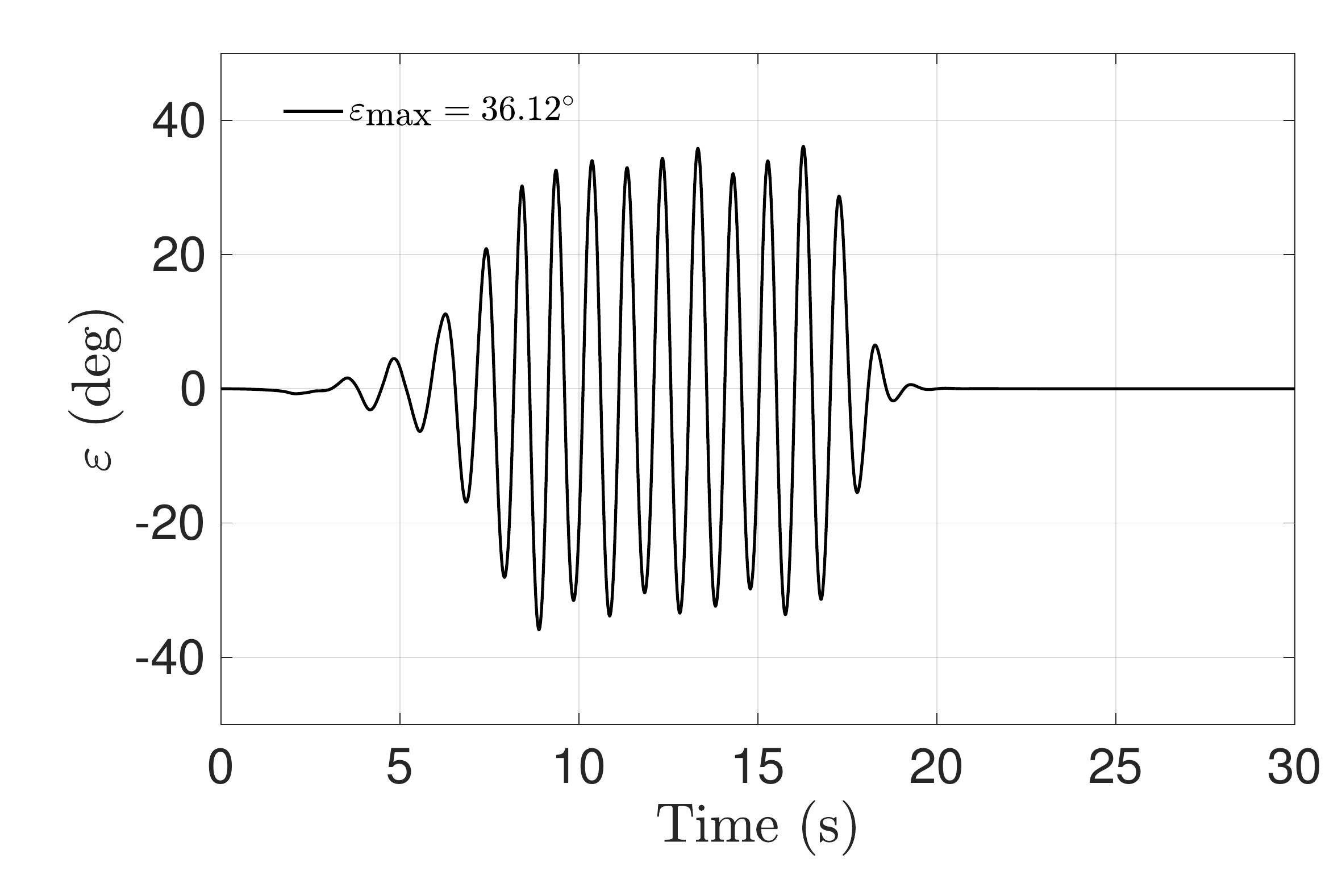}
  \label{fig_e_rpm_reduction}
 }
  \subfigure[Pitch torque unloaded on the hull]{
  \includegraphics[scale = 0.3]{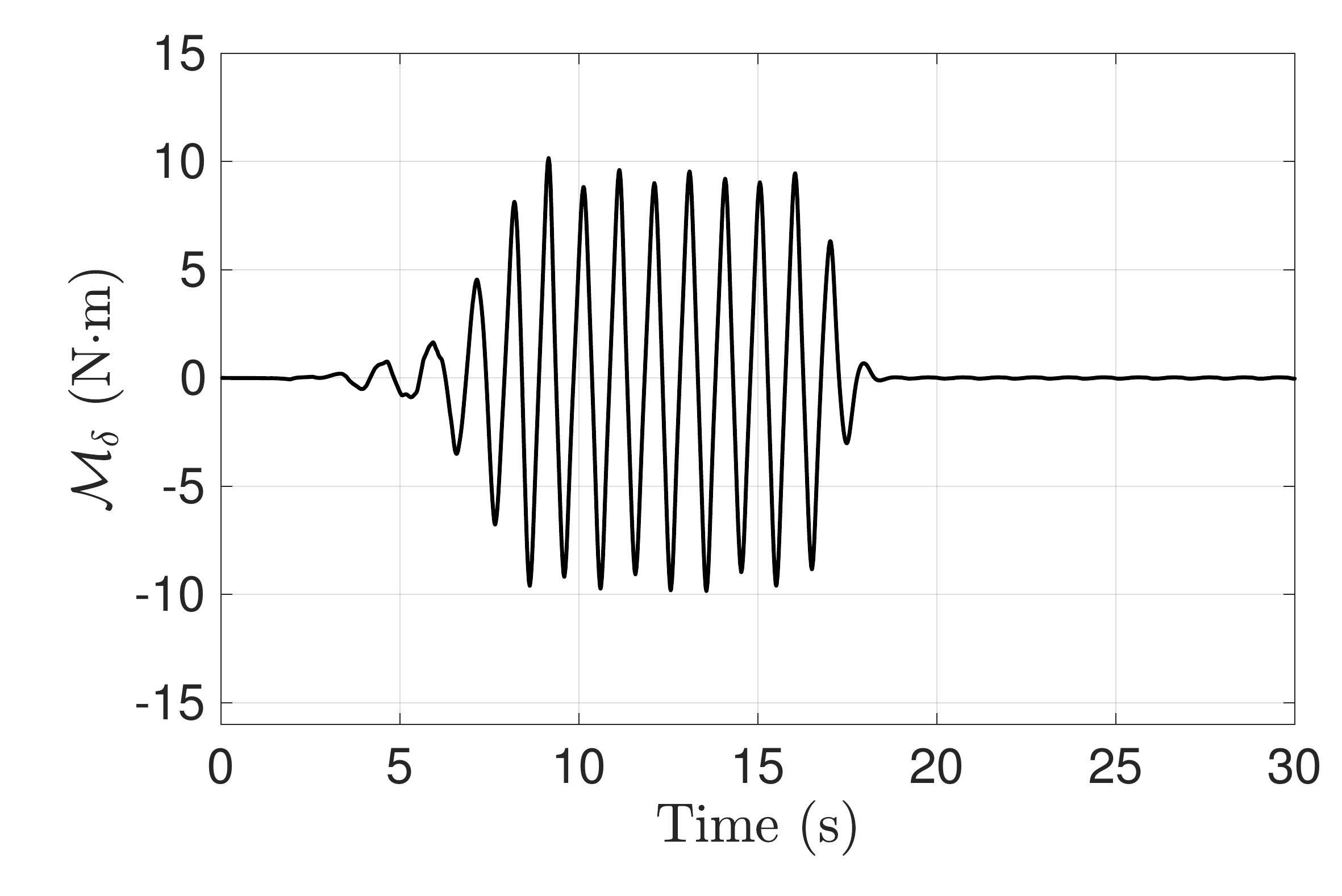}
  \label{fig_Md_rpm_reduction}
 }
   \subfigure[Precession torque on PTO axis]{
  \includegraphics[scale = 0.3]{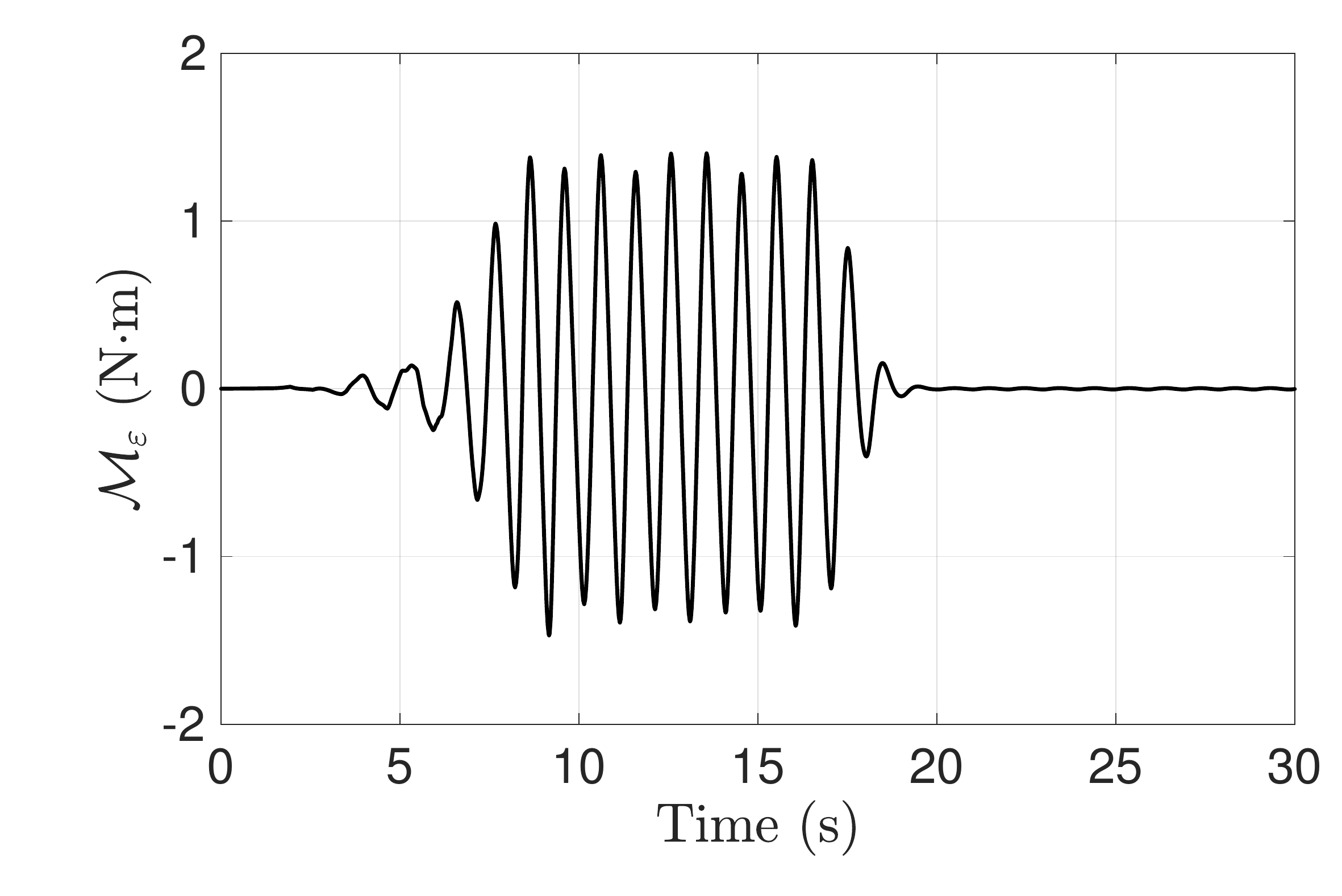}
  \label{fig_Me_rpm_reduction}
 }
   \subfigure[Flywheel speed]{
  \includegraphics[scale = 0.3]{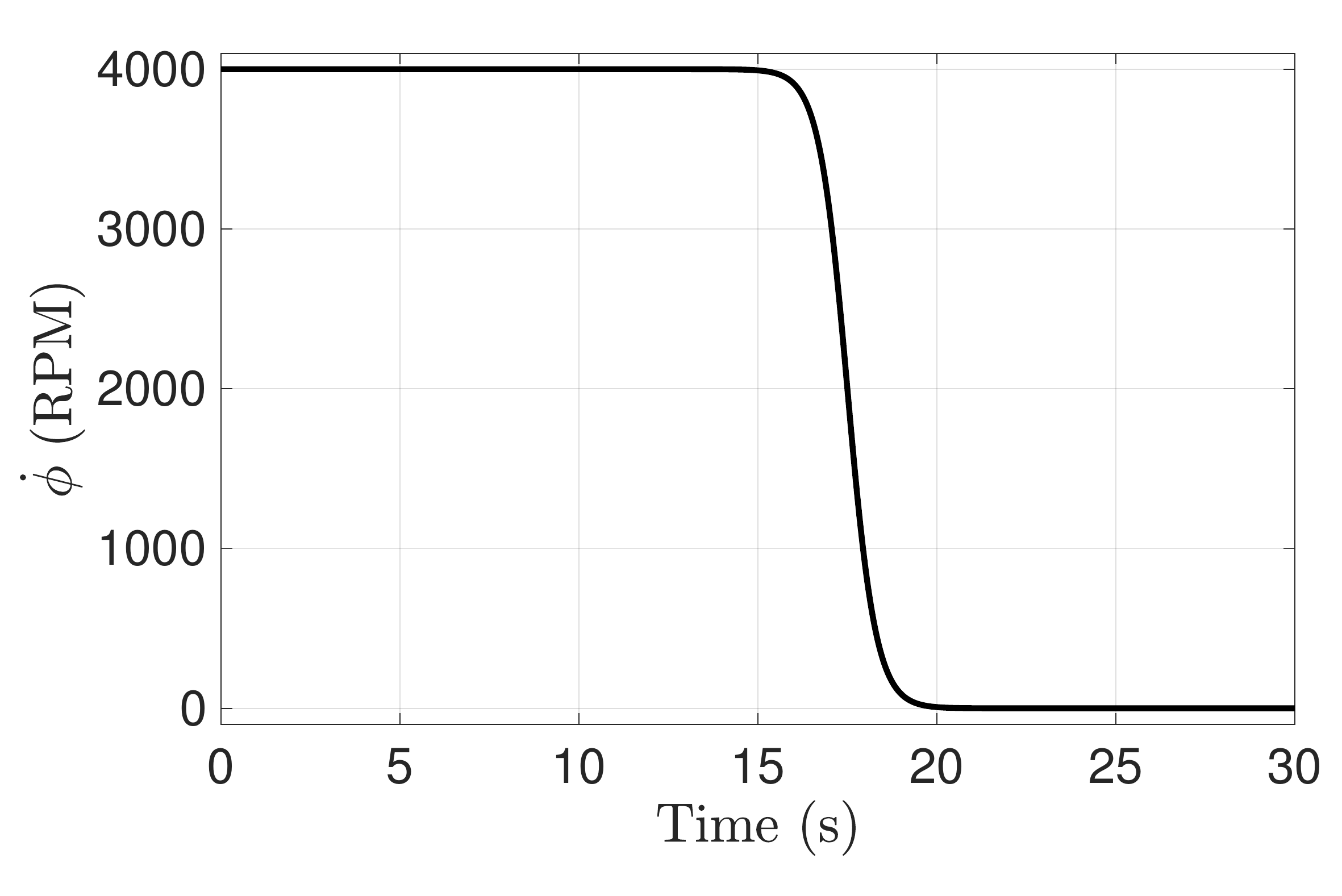}
  \label{fig_rpm_reduction_func}
 }
 \caption{Dynamics of 2D ISWEC model as the flywheel speed $\dot{\phi}$ is reduced from 4000 RPM to 0 RPM amidst steady operation.
The regular wave properties are $\cH$ = 0.1 m and $\cT$ = 1 s.
Temporal evolution of~\subref{fig_delta_rpm_reduction} hull pitch angle $\delta$;~\subref{fig_e_rpm_reduction} gyroscope precession angle $\varepsilon$;~\subref{fig_Md_rpm_reduction} pitch torque $\cM_\delta$; ~\subref{fig_Me_rpm_reduction} precession 
torque $\cM_\varepsilon$; and~\subref{fig_rpm_reduction_func} temporal variation of flywheel speed.
}
 \label{fig_flywheel_speed_reduction}
\end{figure}

The ISWEC hull houses costly electro-mechanical components that need to be protected during harsh, stormy weather conditions. During inclement weather, the hull and gyroscope dynamics can be chaotic, which may damage the system components. To protect the housed components, the gyroscope needs to be turned off. This can be done by reducing the flywheel speed to zero using remote human-machine interfaces. The combined hull-gyroscope system then behaves like a single floating entity. In this section, we simulate the dynamics of the ISWEC device as the flywheel speed is reduced to zero amidst steady operation. We simulate this scenario with regular water waves of  $\cH$ = 0.1 m and $\cT$ = 1 s. To reduce the flywheel speed from 4000 RPM to 0 RPM, we use the following relation

\begin{equation}
	\label{eq_reduce_rpm}
	\dot{\phi}(t) = 4000 \cdot (1 - f(t))/2,
\end{equation}
in which $f(t)$ is a function that smoothly transitions from -1 to 1 in the transition time interval $\Delta T$. The function $f$ is given by
\begin{equation}
	\label{eq_smooth_tanh_func}
	f = \tanh\left( \frac{2 \pi(t - T_\text{half})}{\Delta T} \right),
\end{equation}
in which $T_\text{half}$ = $T_\text{start}$ + $\Delta T$/2.  In our simulation, we set $T_\text{start}$ = 15 s and $\Delta T$ = 5 s. Fig.~\ref{fig_rpm_reduction_func} shows the smooth transition of the flywheel speed towards zero in 5 s. When the gyroscope is turned off, the precession effects cease and the system attains a mean zero position, thus protecting the device. This is seen in Figs.~\ref{fig_e_rpm_reduction},~\ref{fig_Md_rpm_reduction}, and~\ref{fig_Me_rpm_reduction}, which show that $\varepsilon$, $\cM_\delta$, and $\cM_\varepsilon$ are reduced to zero, respectively. As the gyroscopic effects vanish, the hull is observed to be oscillating with greater pitch amplitude (Fig.~\ref{fig_delta_rpm_reduction}).

%%%%%%%%%%%%%%%%%%%%%%%%%%%%%%%%
\section{Conclusions}
\label{sec_conclusions}
In this study, we systematically investigated the wave-structure interaction dynamics of the inertial
sea wave energy converter (ISWEC) technology. Our computational model is based on the incompressible
Navier-Stokes equations and employs a fictitious domain Brinkman penalization (FD/BP) approach to handle the fluid-structure
coupling. The dynamics of the ISWEC hull and gyroscope system were coupled to this CFD solver to enable
fully-resolved 1-DOF simulations of the device. To emulate realistic operating conditions of the device, a numerical
wave tank was used to generate both regular waves based on fifth-order Stokes theory and irregular waves based on
the JONSWAP spectrum. We performed Froude scaling analysis of the full-scale ISWEC model to determine the required
parameters for our 1:20 scaled-down two- and three-dimensional simulations.

Our numerical investigation demonstrated that the 2D model was sufficient to accurately simulate the hull's pitching motion,
and to predict the power generation/absorption capability of the converter. We showed that setting the prescribed hull
pitch angle parameter $\delta_0$ close to the maximum wave steepness will maximize the device's relative capture width
(i.e. power generation efficiency). A comprehensive parameter sweep demonstrated that
the device achieves peak performance when the gyroscope specifications are chosen based on the reactive control theory described in Sec.~\ref{sec_PTO_params}. It was also shown that a proportional control of the PTO control torque is required 
to generate continuous precession effects of the gyroscope, without which the gyroscope tends to align with the hull pitch axis. 
Under this scenario, the device does not generate any power. We also showed that although the yaw torque in the gyroscope reference 
frame is small, it is of the same order of magnitude as the pitch torque induced on the hull in an inertial reference frame. 
Therefore, the yaw torque on the hull should be considered in the design phase of these devices to avoid any misalignment of the converter from the main wave direction. Our simulations also verify that the hull length to wavelength ratio should be between one-half and one-third to achieve high conversion efficiency. Throughout our parameter study, we numerically verify the theoretical power transfer pathway between the 
water waves and the hull, the hull and the gyroscope, and the gyroscope to the PTO unit for both regular and irregular wave environments. Although the power transfer is derived for ISWEC devices in this work, an analogous relationship could be derived for heaving or surging point absorbers. Finally, we investigated the dynamics of the ISWEC system as the flywheel speed is reduced to zero to emulate device protection during inclement weather conditions.

By making use of high performance computing, our work demonstrates that it is feasible to use fully-resolved simulations to interrogate the device physics and dynamics of wave energy converters. They can also be used as a design tool to explore the parametric space for further optimization of such devices.

%%%%%%%%%%%%%%%%%%%%%%%%%%%%%%%%

\section*{Acknowledgements}
A.P.S.B~acknowledges helpful discussions with Giovanni Bracco and Giuliana Mattiazzo over the course of this 
work.  A.P.S.B.~acknowledges support from NSF award OAC 1931368. NSF XSEDE and SDSU Fermi compute resources 
are particularly acknowledged.  
%%%%%%%%%%%%%%%%%%%%%%%%%%%%%%%%

\appendix
\renewcommand\thesection{\Alph{section}}
\section{Two degrees of freedom ISWEC model} \label{sec_appendix}

\begin{figure}[]
  \centering
  \subfigure[Hull pitch]{
    \includegraphics[scale = 0.28]{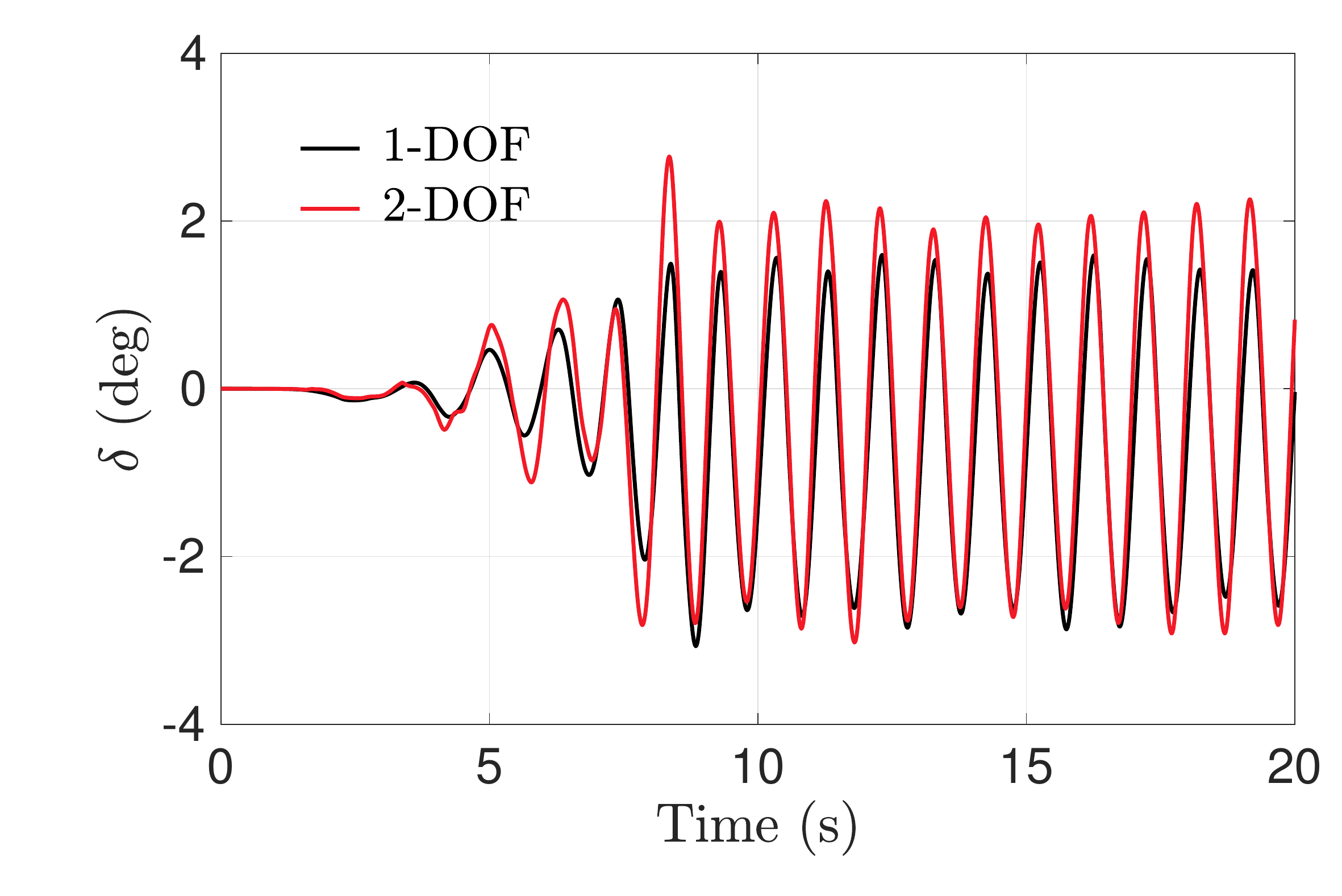}
    \label{fig_delta_1DOF_2DOF}
  } 
  \subfigure[Gyroscope precession]{
    \includegraphics[scale = 0.28]{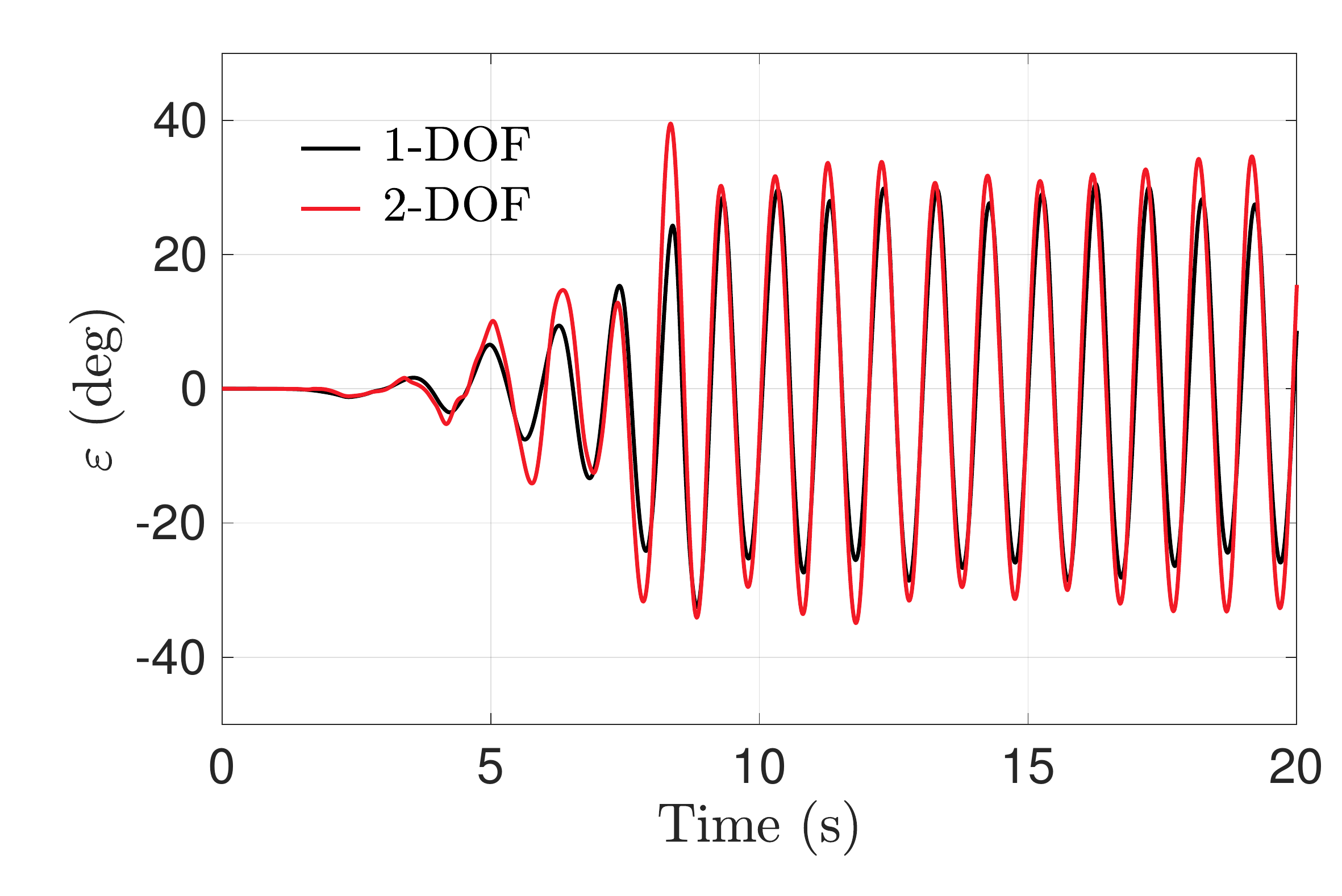}
    \label{fig_e_1DOF_2DOF}
  }
    \subfigure[Hull heave]{
    \includegraphics[scale = 0.28]{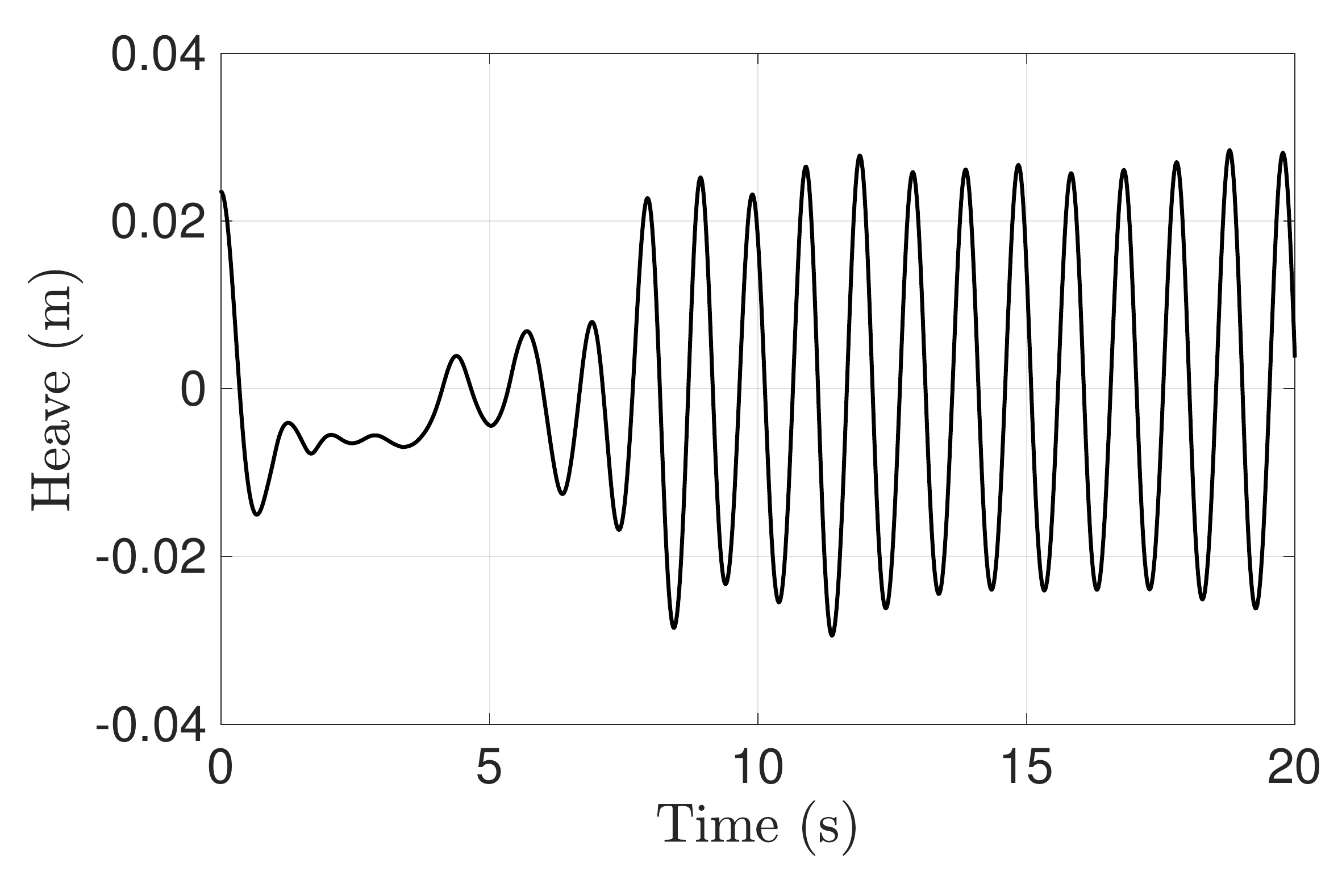}
    \label{fig_y_heave}
  }
     \subfigure[Powers in the system (2-DOF model)]{
    \includegraphics[scale = 0.28]{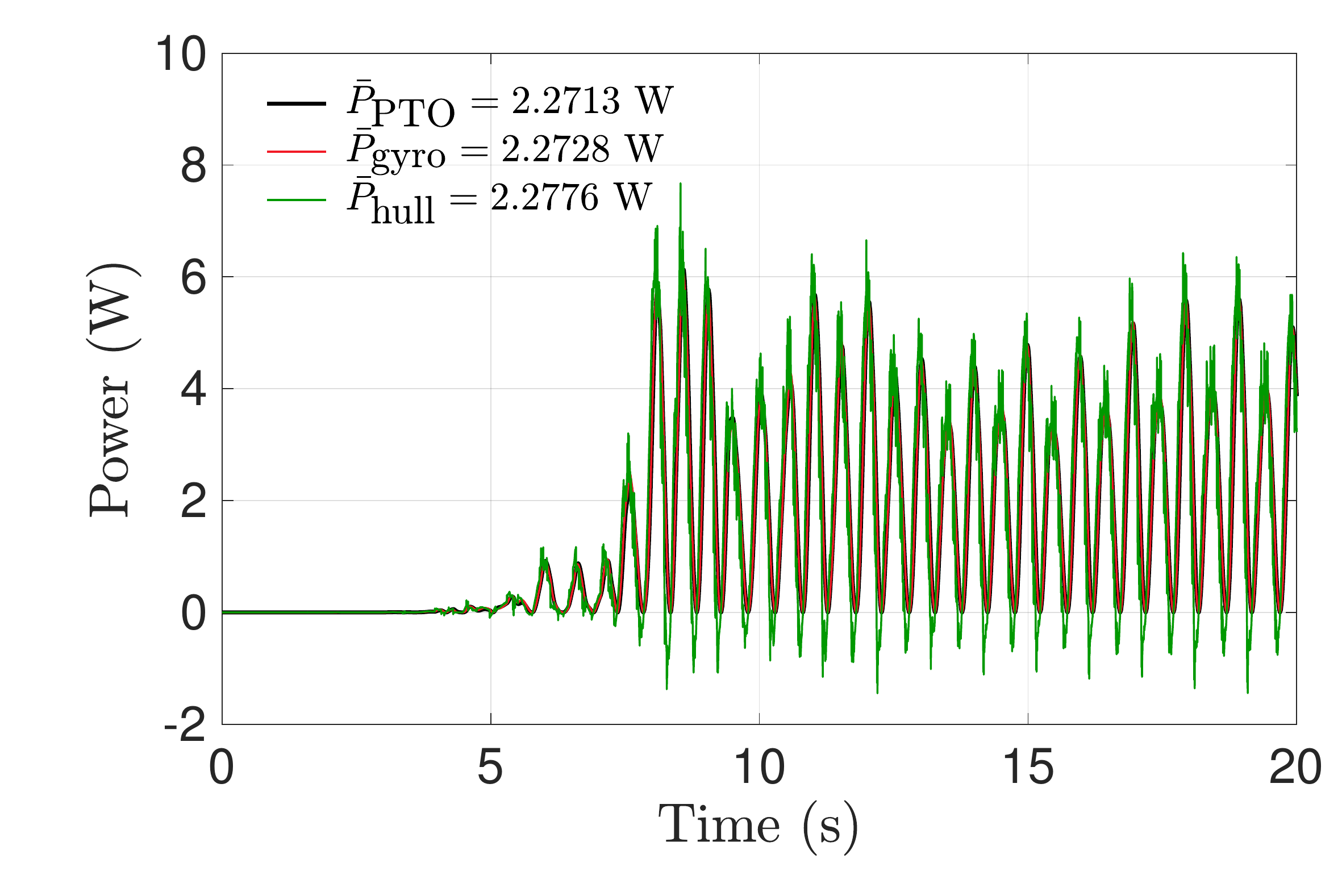}
    \label{fig_Power_vs_t_2DOF}
  }   
\caption{Comparison of 2-DOF (pitch and heave) and 1-DOF (pitch only) ISWEC models for ~\subref{fig_delta_1DOF_2DOF} hull pitch angle $\delta$, and~\subref{fig_e_1DOF_2DOF} gyroscope precession angle $\varepsilon$.~\subref{fig_y_heave} Hull heave displacement, and~\subref{fig_Power_vs_t_2DOF} power at various levels for the 2-DOF ISWEC model. Fifth-order regular water waves are generated with $\cH$ = 0.1 m, $\cT$ = 1 s and $\lambda$ = 1.5456 m,
satisfying the dispersion relation given by Eq.~\ref{eq_dispersion_relation}.
A maximum ISWEC pitch angle $\delta_0$ = 5$^\circ$ and a maximum gyroscope precession angle of $\varepsilon_0$ = 70$^\circ$ are used. 
The gyroscope parameters are: damping coefficient $c$ = 0.3473 N$\cdot$m$\cdot$s/rad, moment of inertia 
 $J = 0.0116$ kg$\cdot$m$^{2}$, and PTO stiffness $k$ = 0.4303 N$\cdot$m/rad.
The speed of the flywheel is $\dot{\phi} = 4000$ RPM, and $I = 0.94 \times J = 0.0109$ kg
 $\cdot$m$^{2}$.}
  \label{fig_2dof}
\end{figure}

We compare the hull and gyroscope dynamics obtained using two degrees of freedom (pitch and heave) and one degree of freedom (pitch only) ISWEC models. The same case from Sec.~\ref{subsec_grid_convergence} is simulated using the two models on a medium grid resolution. Figs.~\ref{fig_delta_1DOF_2DOF} and~\ref{fig_e_1DOF_2DOF} show the comparison of hull pitch angle $\delta$ and gyroscope precession angle $\varepsilon$, respectively. As can be seen in Fig.~\ref{fig_2dof}, including an additional heave degree of freedom only marginally affects the rotational motion of the hull and gyroscope, and consequently the power output of the device. Fig.~\ref{fig_y_heave} shows the heave dynamics of the hull about its mean $z$-location. The heave amplitude is approximately one-tenth of the hull height for the prescribed wave characteristics. Although the heave motion is not negligible in this case, it nonetheless does not significantly affect the rotational dynamics. Moreover, in a real device, the heave (along with the surge and sway) motion is constrained to a certain extent by the mooring system.  Finally, Fig.~\ref{fig_Power_vs_t_2DOF} shows that the power transfer equation is satisfied even for the 2-DOF ISWEC model.  

%%%%%%%%%%%%%%%%%%%%%%%%%%%%%%%%%
%%%%%%%%%%%%%%%%%%%%%%%%%%%%%%%%%
\section*{Bibliography}
\begin{flushleft}
 \bibliography{ISWEC_paper_biblography}
\end{flushleft}

\end{document}